%% file: mythesis.tex
\documentclass[12pt,a4paper,twoside]{report}
\usepackage[top=3.0cm, bottom=3.0cm, left=3.8cm, right=2.8cm]{geometry}
\usepackage{fancyhdr}
\usepackage{bbold}
\usepackage{dcolumn}
\usepackage{epsfig}
\usepackage{bbm}
\usepackage{amsfonts}
\usepackage{amssymb}
\usepackage{amsmath}
\usepackage{amsthm}
\usepackage{amsbsy} 
\usepackage{amscd} 
\usepackage{notoccite}
\usepackage{latexsym}
\usepackage{bm}
\usepackage{verbatim}
\usepackage{graphicx}
\usepackage{slashed}
\usepackage{hyperref}
\usepackage[square,comma,sort&compress,numbers]{natbib}
\usepackage{mciteplus}
 \usepackage{titlesec}
\usepackage{float}
\restylefloat{table}
\usepackage{simplewick}
\usepackage{natbib}
\usepackage{tabularx}
\usepackage{xcolor}
\hypersetup{
 colorlinks=true,
 breaklinks=true,
 citecolor=green,
 linkcolor=blue,
 urlcolor=blue,
linkbordercolor={1 1 1}, 
citebordercolor={1 1 1} 
}
\usepackage{breakurl}

\newcolumntype{d}{D{.}{.}{-1}}

\titleformat{\chapter}[display]
  {\normalfont\huge\bfseries}{\chaptertitlename\ \thechapter}{20pt}{\Huge}

\expandafter\ifx\csname package@font\endcsname\relax\else
 \expandafter\expandafter
 \expandafter\usepackage
 \expandafter\expandafter
 \expandafter{\csname package@font\endcsname}%
\fi
\newcolumntype{C}[1]{>{\centering\arraybackslash}p{#1}}

\def\be{\begin{equation}}
\def\ee{\end{equation}}

\def\bea{\begin{eqnarray}}
\def\eea{\end{eqnarray}}
\def\gsim{\ \rlap{\raise 2pt\hbox{$>$}}{\lower 2pt \hbox{$\sim$}}\ }
\def\lsim{\ \rlap{\raise 2pt\hbox{$<$}}{\lower 2pt \hbox{$\sim$}}\ }

\newcommand{\dcp}{\delta_{CP}}
\newcommand{\nova}{NO$\nu$A}
\newcommand{\dmmm}{\Delta_{\mu\mu} }
\newcommand{\tmm}{\theta_{\mu\mu}}
\newcommand{\mlj}{\Delta_{31}}
\newcommand{\mkj}{\Delta_{21}}
\newcommand{\nubar}{\overline{\nu}}
\newcommand{\piS}{$\pi$S}
\newcommand{\muDS}{$\mu$DS}
\newcommand{\nS}{$n$S}
\newcommand{\cnv}{\v{C}erenkov}

   \makeatletter
   \def\cleardoublepage{\clearpage\if@twoside \ifodd\c@page\else%
        \hbox{}%
        \thispagestyle{empty}
        \newpage%
        \if@twocolumn\hbox{}\newpage\fi\fi\fi}
    \makeatother

\begin{document}



\thispagestyle{empty}
\input{./Frontpage}
\cleardoublepage


\newpage
\thispagestyle{empty}
\input{./Declaration}

\cleardoublepage

\newpage
\thispagestyle{empty}
\input{./Certificate}
\cleardoublepage

\pagenumbering{roman}
\setcounter{page}{1}
\addcontentsline{toc}{chapter}{Acknowledgements}
\input{./acknowledge}
\cleardoublepage

\newpage
\phantomsection
\addcontentsline{toc}{chapter}{Abstract}
\input{./abstract}

\cleardoublepage

 
 
 
 \newpage
 \phantomsection
 \addcontentsline{toc}{chapter}{Contents}
 \tableofcontents
 \cleardoublepage
 
 \newpage
 \phantomsection
 \addcontentsline{toc}{chapter}{List of Tables}
 \listoftables
 \cleardoublepage
 
 \newpage
 \phantomsection
 \addcontentsline{toc}{chapter}{List of Figures}
 \listoffigures
 \cleardoublepage

 \clearpage
 \pagenumbering{arabic}
 \setcounter{page}{1}
 
  \newpage
  \chapter{Introduction}
  \label{chap:intro}
  \input{intro}

  \cleardoublepage
  
 \newpage
 
  \chapter{Neutrino Oscillation}
  \label{chap:oscillation}
  \input{chap2_oscillation}

  \cleardoublepage
 
  \newpage
 
 \chapter{Probing Neutrino Oscillation Parameters in Future Experiments}
 \label{chap:sens}
 \input{cp}

 \input{LBNO}
 \input{LBNE}
  \input{chap3_sens}

 \cleardoublepage
 
 \newpage
 
 \chapter{Neutrino Mass Matrices}
 \label{chap:matrix}
 \input{chap4_intro}

 \input{two_zero}

\input{chap4_matrix}
 \cleardoublepage
 
 \newpage
 
 \chapter{Conclusion: Present Aspects and Future Prospects}
 \label{chap:concl}
 \input{Conclusion}

 \cleardoublepage
 
 \newpage

 
 \appendix 
 \chapter{Expressions for Neutrino Oscillation Probabilities}
 \label{app:decay}
  \input{appendix1}
  \cleardoublepage
 
  \newpage
 \chapter{Calculation of Events and $\chi^2$ Analysis}
 \label{app:ws}
 \input{appendix2}

 \cleardoublepage

 \newpage
 \chapter{Extraction of the Sterile Mixing Parameters}
 \label{app:loop}
 \input{appendix3}


 \cleardoublepage
 \phantomsection
 \addcontentsline{toc}{chapter}{Bibliography}
 \bibliography{neutosc}{}
 \bibliographystyle{apsrev4-1}
 
 

\end{document}

%% file: Frontpage.tex
\begin{center}
{\LARGE \emph{\textbf{Present Aspects and Future Prospects of Neutrino Mass and Oscillation 
}}}\\
\vspace*{1.0cm}
{\large \textbf{A THESIS}}\\
\vspace*{0.2cm}
{\large \it\textbf{submitted for the Award of Ph.D. degree of}}\\
\vspace*{0.2cm}
{\Large \textbf{MOHANLAL SUKHADIA UNIVERSITY}} \\
\vspace*{0.5cm}
{\large \it \textbf{in the}} \\
\vspace*{0.2 cm}
{\large \it \textbf{Faculty of Science}} \\
\vspace*{0.2 cm}
{\large \it \textbf{by} } \\
\vspace*{0.2 cm}
{\Large \textbf{Monojit Ghosh}} \\
\vspace*{0.4cm}
\begin{figure}[htb]
\centering
 \includegraphics[width=3.5cm,bb=154 247 459 553]{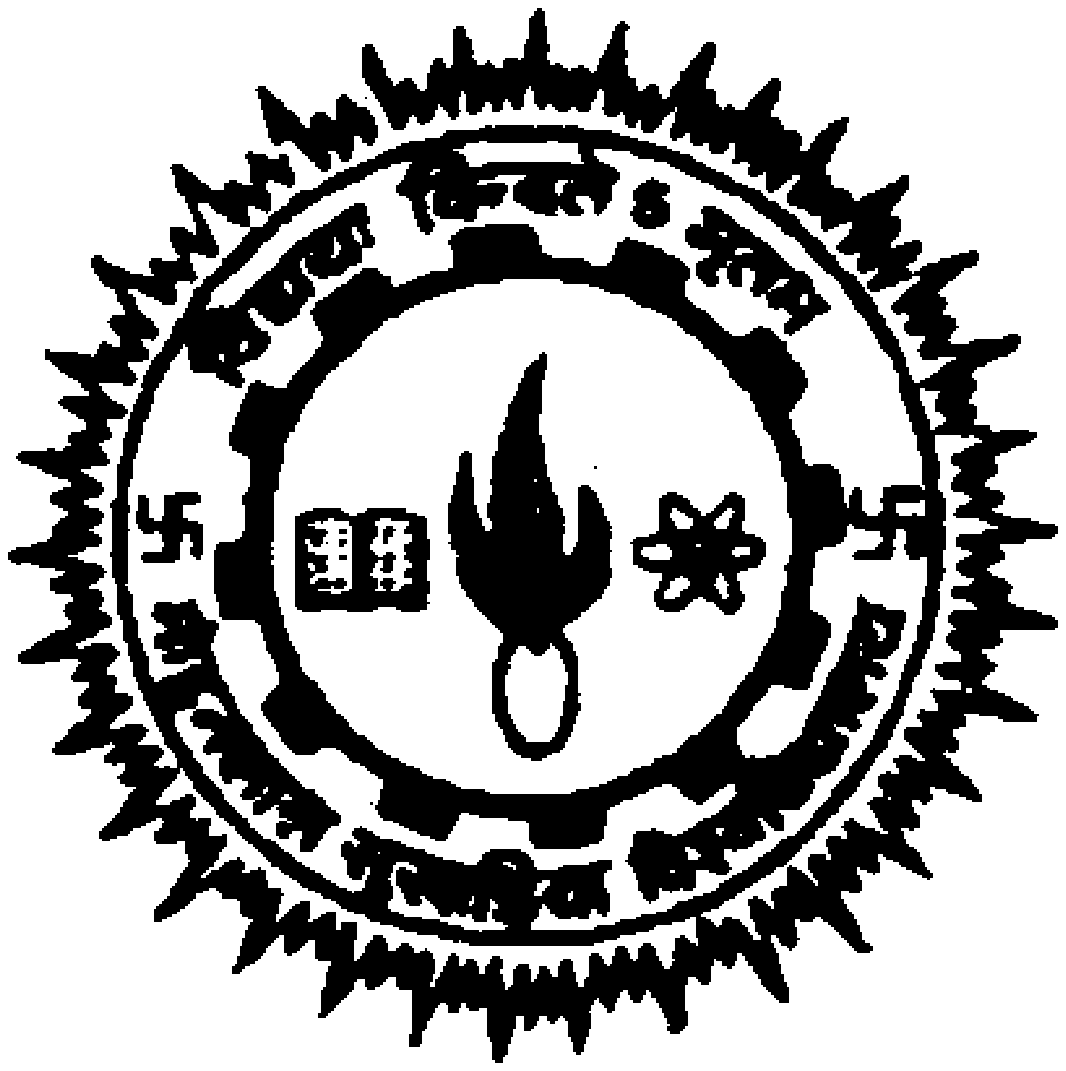}
\end{figure}
\vspace*{ 0.4 cm}    
{\large \emph{\textbf{Under the Supervision of}}}  \\
\vspace*{ 0.2 cm}
{\Large \textbf{Prof. Srubabati Goswami}} \\
{\textbf{Professor}} \\
{\textbf{Theoretical Physics Division} } \\
{\textbf{Physical Research Laboratory}} \\
{\textbf{Ahmedabad, India. }} \\
\vspace*{0.5cm}
{\Large{\textbf{DEPARTMENT OF PHYSICS}}}\\
\vspace*{ 0.1 cm} 
{\Large \textbf{MOHANLAL SUKHADIA UNIVERSITY}}\\
\vspace*{ 0.1 cm} 
{\Large \textbf{UDAIPUR}}\\
\vspace*{ 0.1 cm}
{\Large \textbf{Year of submission: 2015 }}

\end{center}

%% file: Declaration.tex
\vspace*{2.0 cm}
\begin{center}
{\LARGE \it \textbf{DECLARATION}} \\
\end{center}
\vspace*{1.0 cm}

\hspace*{0.6cm}{\it \large \quad\ I, {\bf Mr. Monojit Ghosh}, S/o Mr. Monoranjan Ghosh, resident of
RN-011, PRL student hostel campus, Thaltej, Ahmedabad 380058, hereby declare that the research work 
incorporated in the present thesis entitled, {\bf ``Present Aspects and Future Prospects of Neutrino Mass and Oscillation''} is 
my own work and is original. This work (in part or in full) has not been submitted to any University 
for the award of a Degree or a Diploma. I have properly acknowledged the material collected 
from secondary sources wherever required. I solely own the responsibility for the originality of the entire content.}

\vspace*{3.0cm}
\begin{flushleft}
{\large{\bf Date: }}\\
\end{flushleft}

\begin{flushright}
{\Large (Monojit Ghosh)}\\ 
\end{flushright}

%% file: Certificate.tex
\vspace*{0.5cm}
\begin{center}
{\LARGE \it \textbf{CERTIFICATE}} \\
\end{center}
\vspace*{0.1cm}

\hspace*{0.6cm} {\large I feel great pleasure in certifying that the thesis entitled,
\textbf{``Present Aspects and Future Prospects of Neutrino Mass and Oscillation''} embodies a record of the results of investigations
carried out by Mr. Monojit Ghosh under my guidance. 
He has completed the following requirements as per Ph.D regulations of the University.\\
\hspace*{0.6cm}(a) Course work as per the university rules.\\
\hspace*{0.6cm}(b) Residential requirements of the university.\\
\hspace*{0.6cm}(c) Regularly submitted six monthly progress reports.\\
\hspace*{0.6cm}(d) Presented his work in the departmental committee.\\
\hspace*{0.6cm}(e) Published minimum of one research papers in a refereed research journal.\\
\hspace*{0.6cm}I am satisfied with the analysis, interpretation of results and
conclusions drawn. I recommend the submission of thesis.}

\vspace*{0.2cm}
\begin{flushleft}
{\large{\bf Date: }}
\end{flushleft}

\vspace*{0.2cm}
\begin{flushright}
{\large Prof. Srubabati Goswami}\\
{\large (Thesis Supervisor)}\\
{\large Professor, THEPH,}\\
{\large Physical Research Laboratory,}\\
{\large Ahmedabad - 380 009}\\
\end{flushright}


\begin{flushleft}
\vspace*{0.2cm} \large{Countersigned by} \\
\large{Head of the Department}
\end{flushleft}

%% file: acknowledge.tex
\begin{center}
{\LARGE \textbf{Acknowledgements}}
\end{center}
\vspace{0.5cm}
{\em 
I want to take this opportunity to express my sincere and deep gratitude to my supervisor Prof. Srubabati Goswami, for her
invaluable guidance, support and involvement throughout my work. 
Her immense knowledge, perspective and expertise in the subject have helped me during my research and writing of the thesis.
Discussing physics with her was fun. We had a great compatibility in understanding the views of each other. 
She always encouraged me to have a deep and critical understanding of the subject.
Due to her friendly nature I could discuss any of my problems with her and she was always there to help me.
Her continuous encouragement and motivation gave me confidence to continue research in this field. 
I could not have imagined having a better advisor for my Ph.D study.

I would specially like to thank Dr. Bijaya Sahoo, Prof. Hiranmaya Mishra, Prof. Jitesh Bhatt, Dr. Namit Mahajan, Dr. Partha Konar,
Prof. Raghavan Rangarajan and Prof. Subhendra Mohanty 
for taking courses on quantum mechanics, particle physics, quantum field theory and mathematical methods.
My special thanks goes to my thesis expert Prof. S. Mohanty and the academic committee
for reviewing my work thoroughly.
I also thank Prof. Amol Dighe, Prof. Raj Gandhi, Prof. Sandhya Choubey, Dr. Sanjib Agarwala and Prof. Uma Shankar 
for various physics discussions. I would also like to thank Prof J.W.F. Valle for
giving me the opportunity to visit IFIC, Valencia.
I thank him and Prof. Mariam Tortolla for very useful discussions at IFIC.

I specially thank Dr. Shivani Gupta with whom I have carried out my first research project.
I learned a lot of things from her during my initial days of research.
I offer my heartiest thanks to Dr. Pomita Ghoshal and Dr. Sushant Raut whose contributions have shaped my knowledge and understanding
of neutrino oscillation physics. I thank them for answering my infinite number of questions with patience.
I can never forget those long stimulating discussion sessions in my supervisor's office. I also want to thank Animesh Chatterjee,
Moon Moon Devi and Reetanjali Moharana with whom I have spent sleepless nights to finish our project at WHEPP 2013.
It has been a nice experience to work with all of you.

I thank my senior Subrata da, my batch mate Gulab and my juniors Chandan Gupta, Chandan Hati, Newton and Vishnu for various useful discussions. I also
want to express my gratitude to Abhay, Kuldeep Suthar, Tanmoy Mondal and Ujjal with whom I shared my office space at PRL.
They helped me greatly in my research work. A special thanks to Ujjal for a thorough proofreading of my thesis. 

My stay at PRL hostel will be memorable due to the presence of all the students including all my seniors, juniors, project associates, project students
and my batchmates. The festival celebrations, the sports events, our hundred kilometer long bike trips, hangouts,
mind boggling discussions on cinema, politics, literature
and many other things will remain unforgettable and will stay close to my heart throughout my life.
I want thank you all:
Aadhi, Anjali, Arko, Avdhesh, Bhavya, Damu, Dillip, Dinesh, Gaurav Tomar, Girish Chakravarty, Lata, Lekshmy, Midhun, Naveen, Nigam, Priyanka, Reddy, 
Tanmoy Chattopadhay, Upendra, Wageesh, Yashpal,
Abhishek da, Akhilesh, Amrendra, Amzad da, Aravind Saxena, Aravind Thakur, Arun Awasthi, Asish, Bhaswar da, Chinmay da,
Fazlul da, Ketan, Koushik da, Moumita di, Pankaj, Prashant, Rabiul da, Shashi, Siddhartha da, Soumya, Srinu, Suman da, Sunil, Sushanta, Tanushree di, Tapas da, 
Vimal, Vineet, Zeen,  
Ananta, Gaveshna, Ila, Mansi, Nabyendu da, Ranjita, Shweta, Sudip da,  
Alok, Anirban, Arun Pandey, Chitrabhanu, Gaurav Jaiswal, Girish Kumar, Guru, Ikshu, Manu, Sanjay, Shradha, 
Apurv, Ashim, Bivin, Deepak, Dipti, Jiniya, Lalit, Pankaj, Rahul, Sukanya, Venky,  
 Ali, Jabir, Kuldeep Pandey, Kumar, Navpreet, Prahlad, Rukmani, Rupa, Satish, 
Wriju,
Balaji, Bose, Dharak, Ejaz, Jayesh Khunt, Nitesh, Ranjit, Rishi, Santosh, Sharad, Sneha, Subha Anand, Vijayan,   
Abhishek, Ankur, Dipanweeta, Jayesh Agarwal, Rashika, Ria, Sameer, Sudha and Vasu 
for spending such beautiful times with me.  

I am grateful to all the PRL library, computer center, dispensary and
administration staff and also to the staff members of Theoretical Physics Division of
PRL for their sincere support.

I also acknowledge Biswajit, Dipankar, Kamakhya, Sabyasachi, Suprabh and Tuhin  
whom I met in different conferences and became very good friends.
We discussed physics whenever we meet and spent precious time together.
I could not forget the warm hospitality of Biswajit when we went for the DAE 2014 meeting.

I would also like to thank few of my university friends Anupam, Arghya, Barun, Kalpana, Koushik, Soubhik, Sudip and Suman
who frequently keep on asking that how my work is going on and 
supported me in difficult times. I also want to thank few of my school friends  Sandipan, Subho and Tamoghna who were
always there to greet me whenever I went home.

I express my deepest gratitude to my family: my parents and my brother for supporting me throughout and for their unconditional love.
They always stood beside and encouraged me to pursue a career that I like. This thesis would never be possible without their constant support.

And last but not the least I would also like to thank Anushmita,
for proof reading of my thesis partially and motivating me by the following Hemingway's advice on writing at the time when I was struggling 
to write my thesis by thinking if a 10 page paper takes several months to write then how it will be possible to manage a thesis consisting of 200 pages:

\begin{center}
``Advice on how to write? Sit on a typewriter and start typing. That's the easiest and the hardest part."
\end{center}

 \vspace{0.4cm}
 \begin{flushright}
 {\bf  Monojit}
 \end{flushright}
 
}

%% file: abstract.tex
\begin{center}
{\Large \textbf{ABSTRACT}}
\end{center}
\vspace{0.5cm}
Neutrinos are neutral, spin-$\frac{1}{2}$ particles which undergo only weak interactions.  
The experimentally observed phenomenon of neutrino oscillation establishes the fact that neutrinos are massive 
and there is mixing between different neutrino flavours.
This constitutes the first unambiguous
hint towards the physics Beyond Standard Model (BSM). In the BSM theories, 
the neutrino mass terms in the Lagrangian lead to the non-diagonal neutrino mass matrix in the flavour basis 
which depends on neutrino mass and mixing parameters.
Thus knowledge of the neutrino oscillation parameters and understanding 
the underlying symmetries of the neutrino mass matrix are very important as they
can give an insight to the new physics beyond Standard Model. Therefore the measurement of different oscillation parameters and studying
the structure of the neutrino mass matrix are some of the main goals in neutrino physics at present.

Currently the paradigm of neutrino oscillation between three flavours is well established from different experiments and 
the oscillation parameters are getting measured with continued precision.
The current unknowns in the neutrino oscillation physics in the standard three generation framework are: the neutrino mass hierarchy, octant of $\theta_{23}$ and 
the leptonic phase $\dcp$. There are many ongoing/future experiments where these unknown oscillation parameters can be probed.
These experiments utilize different sources and detectors as well as different baselines along which oscillations can develop.
As the oscillation probabilities depend differently on the parameters
in different oscillation experiments, combination of different experiments can often be useful.
Apart from three flavours oscillations there are also evidences of oscillation involving sterile neutrinos.
In the presence of a sterile neutrino
there
will be new mixing angles and phases contributing to the oscillation of the neutrinos. This gives rise to the possibility
that the well understood phenomenological behaviour of the neutrino mass matrix may change in the presence of sterile neutrinos. 

In this thesis we have studied the potential of present/future neutrino oscillation experiments and synergy between them 
to determine the unknown parameters in the neutrino sector 
in the light of current experimental results. 
We consider the beam based experiments T2K, \nova, LBNO, LBNE, the atmospheric experiment INO@ICAL and the ultra high energy neutrino experiment IceCube for our analysis.
We find that the data from atmospheric neutrino experiment ICAL can significantly improve the CP sensitivity of the long-baseline experiments T2K/\nova\ 
in their unfavourable parameter space.
To improve the sensitivity beyond what can be achieved in T2K/\nova,
it is important to study the physics potential of the
proposed long-baseline experiments LBNO and LBNE in view of the current experiments.
We have shown that the required exposure of LBNO and LBNE
in determining neutrino mass hierarchy, octant of
$\theta_{23}$ and $\dcp$ can be reduced significantly when data from T2K, \nova\ and ICAL are added to them. We have also explored the possibility to constrain
the CP phase $\dcp$ by analyzing the IceCube data in terms of
the flavour compositions of the ultra
high energy neutrinos. 

We have also studied 
the phenomenological consequences of texture zeros in the neutrino mass matrices in the presence of a sterile neutrino. 
We have carried out a detailed analysis of the two-zero and one-zero textures in the 3+1 scenario which
involves three active neutrinos and one sterile neutrino.
We find that in the 3+1 picture,
conclusions differ significantly as compared to the standard 3 generation case.  The allowed two-zero textures in the $3 \times 3$ structure
are more phenomenologically constrained as compared to the $4 \times 4$ structure. The correlations between the different oscillation parameters are also very different
when the texture zero conditions between the 3 generation and 3+1 generations are compared.

\vspace{0.5cm}

{\bf Keywords} : Neutrino Physics, Neutrino Oscillation, Long-Baseline Neutrino Experiments, Atmospheric Neutrino Experiments,
Ultra High Energy Neutrinos, Leptonic CP Phase, Neutrino Mass Matrix, Sterile Neutrino, Texture Zero

%% file: intro.tex

\section{The Origin of Neutrinos}

The physics of neutrinos started with Pauli's ``Neutrino Hypothesis", but the origin of neutrinos can be traced back to
the late 19th century (1896) when Becquerel discovered radioactivity.
In radioactivity, nucleus of an unstable atom loses energy by emitting alpha ($\alpha$) particles, beta ($\beta$) particles and gamma ($\gamma$) rays.
As in the
mechanism of $\alpha$-particle emission, it was believed that $\beta$-decay is also governed by the two-body process
\begin{equation} \nonumber
N_0(A,Z) \rightarrow N(A,Z+1) + e^-, 
\end{equation}
and energy of the electron is given by the small differences in  masses of the nuclei. However, measurements of the electron energy spectra did not match this expectation.
By late 1920 it was confirmed that this emission gives a continuous spectra for the electron. This posed a puzzle since a two-body decay would 
imply a fixed energy line for the electrons. To overcome this Niels Bohr suggested that the energy in the microworld was conserved only on an average, not on an 
event-by-event basis. In 1930 Pauli postulated the ``Neutrino Hypothesis" to save the principle of conservation of energy in $\beta$-decay. 
He suggested that the continuum spectra might be due to one more “invisible” light neutral particle involved in the $\beta$-decay.  
With three particles involved, the electron would be able to take any momentum from 
zero to the maximum allowed value, the balance being taken care of by the other light “invisible” particle. 
In 1933 Fermi formulated his theory of $\beta$-decay based on Pauli's hypothesis. At that time the existence of Pauli's 
“invisible” particle was accepted and the name neutrino was coined. It was postulated that all $\beta$-decays were due to the same basic underlying process, 
\begin{equation} \nonumber
n \rightarrow p + e^- +\bar{\nu}. 
\end{equation}
To satisfy the conservation of angular momentum, neutrinos must be spin-1/2 particles obeying the Fermi-Dirac Statistics.
But this theory was firmly established only in 1953 when
Clyde Cowan and Frederick Reines detected this weakly interacting particle experimentally \cite{Cowan:1992xc,Reines:1956rs}. 
In their experiment, electron type antineutrinos ($\bar{\nu}_e$) coming out of nuclear reactors were detected.
Now it is well established that there are three types of neutrinos.
In 1962, Leon M. Lederman, Melvin Schwartz and Jack Steinberger showed the existence of 
the muon neutrino ($\nu_\mu$) \cite{Danby:1962nd} and the first detection of
tau neutrino ($\nu_\tau$) interactions was announced in the summer of 2000 by 
the DONUT collaboration at Fermilab \cite{Kodama:2000mp}. 

\section{Neutrinos in Standard Model}

The discoveries of different fundamental particles (including neutrinos) in the middle of the 20th century necessitated the 
formulation of a basic theory to understand the properties of these
particles and how they interact. In the effort to unify the electromagnetic, weak, and strong forces, 
a theory known as the Standard Model (SM) \cite{Weinberg:1967tq} of particle physics
was developed throughout the latter half of the 20th century\footnote{
The chronological development of the SM can be found in this link \cite{sm_timeline}.}.
The current formulation was finalised in the mid-1970s upon experimental confirmation of the existence of quarks.
Mathematically, SM is a non-abelian gauge theory based on the symmetry group $ U(1)_Y \times SU(2)_L \times SU(3)_c $. In this model the left handed fermion fields
are SU(2) doublets and right-handed fermions are SU(2) singlets: 
\begin{center}
            $ Q=\begin{pmatrix}
                u\\
                d \\
               \end{pmatrix}_L $,
             $ L=\begin{pmatrix}
                   \nu_e\\
                    e \\
                  \end{pmatrix}_L $ and $u_R$, $d_R$, $e_R$.
\end{center}
Here $Q$ and $L$ denote the quark and lepton fields respectively belonging to the first generation. 
In totality SM has three generations of quarks (first generation : $u$, $d$; second generation: $c$, $s$; third generation: $t$, $b$) 
and three generations of leptons (first generation : $e$, $\nu_e$; second generation: $\mu$, $\nu_\mu$; third generation: $\tau$, $\nu_\tau$).
Each of the six quarks have three $SU(3)$ colour charges: red, green and blue.
The $W^{\pm}$ and $Z$ bosons are the mediators of the weak force, photon is the carrier of the electromagnetic force and the strong
force is mediated by the gluons.
In this model neutrinos interact with the other leptonic fields weakly via the exchange of $W^{\pm}$ and $Z$ bosons.
The interactions mediated by the $W^{\pm}$ boson are called charge current (CC) interactions and the interactions mediated by the $Z$ boson
are called neutral current (NC) interactions.
In SM one can count
the number of light neutrino species that have
the usual electroweak interactions in the following manner: 
SM allows $Z$ boson to decay to the invisible $\nu \bar{\nu}$ pairs. 
This invisible decay width is the difference between the total decay width of $Z$ and the visible decay width of $Z$. 
Visible decay width of $Z$ boson is referred as the sum
of its partial widths of decay into quarks and charged leptons.
From the LEP (Large Electron-Positron collider) data, the ratio of the invisible decay width of $Z$ and 
the decay width of $Z$ to the charged leptons ($\Gamma_{{\rm inv}}/\Gamma_{ll}$) is 
measured as $5.943 \pm 0.016$. The SM value for the ratio of the partial widths to neutrinos and to charged leptons ($\Gamma_{\nu\nu}/\Gamma_{ll}$) is $1.99125 \pm 0.00083$. 
From this ($\Gamma_{{\rm inv}}/\Gamma_{ll} = N_{\nu} \frac{\Gamma_{\nu\nu}}{\Gamma_{ll}}$) 
the number of the light active neutrino species $N_{\nu}$ can be calculated to be $2.9840 \pm 0.0082$ \cite{ALEPH:2005ab}. 
This is consistent with the fact that
experiments have also discovered only three light active neutrinos.
Though SM is a mathematically self-consistent model and has demonstrated huge and continued success in providing
predictions which could be confirmed experimentally,
but there are certain drawbacks. One of them is the mass of the neutrinos. In SM, the masses of the fermions and gauge bosons are zero before the symmetry breaking of the group.
After the spontaneous symmetry breaking, the gauge bosons acquire mass via Higgs mechanism. This same Higgs mechanism is also responsible for the masses of the fermions.
The mass term
of the fermions arise from the Yukawa term which is written as:
$ -y \bar{\psi}_L \psi_R \langle \phi \rangle $,
where $y$ is the Yukawa coupling,  $\psi_L$, $\psi_R$ are the left-handed and right-handed fermionic fields respectively and $\langle \phi \rangle $ is the
vacuum expectation value (VEV) of the Higgs field. 
In SM there are no right-handed neutrinos.
With no suitable 
right-handed partner, it is impossible to write a gauge invariant mass term for them in SM and thus neutrinos remain massless. 
The absence of right-handed neutrinos in SM is
motivated by observation of parity violation in weak interactions. 
As a solution of the $\tau-\theta$ puzzle, in 1956 Lee and Yang conjectured that parity is violated in 
weak interactions \cite{Lee:1956qn}.  
The violation of parity in weak interactions has been first observed in Wu's experiment.  
When the nuclear spins of $^{60}$Co were aligned by an external magnetic field
, an asymmetry in the direction of the emitted electrons were observed \cite{Wu:1957my}. 
The decay process under consideration was
\begin{equation} \nonumber
 ^{60}\text{Co} \rightarrow~ ^{60}\text{Ni} + e^- + \bar{\nu}_e.
\end{equation}
It was found that nuclear spin of the electron was always opposite to its momentum. In other words the observed correlation between the nuclear spin and 
the electron momentum is only explained by the presence of $e_L$ and $\bar{\nu}_R$. The absence of ``mirror image states" $\bar{\nu}_L$ and $\nu_R$
indicated
a clear violation of parity. 
In 1958, Goldhaber, Grodzins and Sunyar experimentally measured that neutrinos are left-handed and antineutrinos are right-handed \cite{Goldhaber:1958nb}.

Although the neutrinos are massless in the SM, the experimentally observed phenomenon of ``neutrino oscillation" dictates that neutrinos have non-zero mass.


\section{Neutrino Oscillation}

Neutrino oscillation originally conceived by Bruno Pontecorvo in the 1950's is a quantum mechanical interference phenomenon in which 
a neutrino created with a specific lepton flavour ($\nu_e$, $\nu_\mu$ or $\nu_\tau$) can later be measured to have a different flavour \cite{Pontecorvo:1957cp}. 
This occurs if  neutrinos have masses and mixing. In that case the flavour eigenstates
and the mass eigenstates are not the same.
Neutrinos are produced according to the gauge Lagrangian in their flavour or gauge eigenstates ($\nu_\alpha$). The
mass eigenstates or the propagation eigenstates ($\nu_i$) are related to these as 
\begin{equation}
 |\nu_\alpha \rangle = U_{\alpha i} |\nu_i \rangle,
\end{equation}
with $\alpha = e, \mu, \tau$ and $i$ = 1, 2, 3. Here $U$ is the unitary mixing matrix known as the Pontecorvo-Maki-Nakagawa-Sakata (PMNS) matrix. 
The probability that a neutrino of flavour $\nu_\alpha$ gets transformed into a flavour $\nu_\beta$ ($\nu_\alpha \rightarrow \nu_\beta$) after a time interval $t$ 
is given by the amplitude squared $|\langle \nu_\beta(t)|\nu_\alpha \rangle|^2$. For oscillation of the three flavours of neutrinos in vacuum,
the probability of flavour transition $\nu_\alpha \rightarrow \nu_\beta$ can be expressed as
\footnote{We will give the derivation of this expression in Chapter ~\ref{chap:oscillation}.}
\begin{eqnarray} \label{N_gen_2}
 P_{\alpha \beta}  = \delta_{\alpha \beta} &-& 4  \sum_{i < j} \text{Re} \big( U_{\alpha i} U_{\beta j} U_{\alpha j}^* U_{\beta i}^* \big) \sin^2 \{\Delta_{ij}L/4E\} \\ \nonumber
                  &+& 2 \sum_{i > j} \text{Im} \big( U_{\alpha i} U_{\beta j} U_{\alpha j}^* U_{\beta i}^* \big) \sin\{2 \Delta_{ij}L/4E\},
 \end{eqnarray}
where $\Delta_{ij} = m_i^2 - m_j^2$ and $i$, $j$ runs from 1 to 3.
In this expression we clearly see that the oscillatory terms depend on the mass squared differences of the neutrinos.
Neutrino oscillation is also characterised by the energy of the neutrinos $E$ and the baseline $L$ associated with it and the dependence goes as $L/E$.
The oscillation probability is maximum when $L/E$ is of the order of $\Delta_{ij}$.
The above expression corresponds to oscillation in vacuum. For neutrinos
traveling in matter, the interaction potential due to matter modifies the neutrino masses and mixing.
We will discuss this in detail in the next chapter.

Now let us discuss briefly about the parametrisation
of the unitary PMNS matrix $U$. We know that any general $N \times N$ unitary matrix consists of $N^2$ number of independent parameters having
$N(N-1)/2$ number of angles and $N(N+1)/2$ number of phases. But among $N(N+1)/2$ phases not all are physical. It can be shown that a total $(2N-1)$ number
of phases can be absorbed in the $2N$ number of fields in the Lagrangian (For $N$ generation of fermions, there will be $N$ generation of charged leptons and $N$ generation of
neutrinos) and this gives total number of physical phases as \footnote{If neutrinos are Majorana particle then there will be $(N-1)$ more phases. But
this phases can not be probed in neutrino oscillation.} $(N-1)(N-2)/2$.
So for three generations of neutrinos, the mixing matrix is parametrised by three mixing angles : $\theta_{12}$, $\theta_{23}$ and $\theta_{13}$, 
one phase: the Dirac type phase $\dcp$ in the following way,
 \begin{equation}
  U = R_{23} \tilde{R}_{13} R_{12}, 
 \end{equation}
 where $ R_{ij} $ are the orthogonal rotation matrices corresponding to rotations in the $i-j$ plane. For instance
  \begin{eqnarray}
  R_{23} = 
 \begin{pmatrix}
  1 & 0 & 0 \\
 0 & c_{23} & s_{23} \\
 0 & -s_{23} & c_{23}
 \end{pmatrix}, \hspace{3 mm}
 \tilde{R}_{13} = 
 \begin{pmatrix}
 c_{13} & 0 & s_{13}e^{-i \delta_{CP}} \\
 0 & 1 & 0 \\
 -s_{13} e^{-i \delta} & 0 & c_{13}
 \end{pmatrix}. 
 \end{eqnarray}
 From which it follows that
 \begin{equation}
 U=
  \begin{pmatrix}
   c_{12} c_{13} & s_{12} c_{13} & s_{13} e^{-i\delta_{CP}} \\
 -s_{12} c_{23} - c_{12} s_{23} s_{13} e^{i\delta_{CP}} & c_{12} c_{23} - s_{12} s_{23} s_{13} e^{i\delta_{CP}} & s_{23} c_{13} \\
 s_{12} s_{23} - c_{12} c_{23} s_{13} e^{i\delta_{CP}} & -c_{12} s_{23} - s_{12} c_{23} s_{13}  e^{i\delta_{CP}} & c_{23} c_{13}
  \end{pmatrix}, \qquad
 \end{equation}
 where $ c_{ij} = \cos\theta_{ij} $ and $ s_{ij} = \sin\theta_{ij} $. 
The three flavour neutrino oscillation also involves two mass squared differences: the solar mass squared difference: $\Delta_{21} =m_2^2 - m_1^2$ and 
the atmospheric mass squared difference: $\Delta_{31} = \pm (m_3^2 - m_1^2)$. 

Here it is important to note that the phenomenon of neutrino oscillation can only probe the mass squared differences of the neutrinos but not their absolute masses. 
There are  tritium beta decay experiments which measure the absolute mass of neutrinos. The combined data of Troitsk \cite{Aseev:2011dq} and Mainz \cite{Kraus:2004zw} 
experiments give the upper bound of electron neutrino mass as $< 1.8$ eV. The KATRIN experiment \cite{Priester:2015bfa} which will be operational in 2016 
is expected to improve on this bound. 
There are also weak bounds on muon neutrino mass and tau neutrino mass coming from pion and tau decay 
as $< 0.17$ MeV \cite{Assamagan:1995wb} and $< 18.2$ MeV \cite{Barate:1997zg} respectively. 
The neutrinoless double beta decay ($0 \nu \beta \beta$) experiments \cite{Rodejohann:2012xd} 
which can probe Majorana nature of the neutrinos can also put constraint on the effective Majorana neutrino mass \footnote{The averaged electron, muon and tau neutrino masses
are given by $m_{\nu_\alpha} = \sqrt{\sum_{i=1}^3 | U_{\alpha i}^2 | m_i^2}$ where $\alpha=e, \mu, \tau$ 
and the effective Majorana neutrino mass is given by $m_{{\rm eff}}=\sum_{i=1}^3 U_{e i}^2 m_i$.}.
An upper bound on the sum of active neutrino masses as 0.23 eV \cite{Ade:2013zuv} comes from cosmology. 
From the neutrino
oscillation experiments we know that the two mass squared differences which govern the oscillation of the three generations of neutrinos are of the order of $10^{-5}$ eV$^2$ 
and $10^{-3}$ eV$^2$ \cite{global_fogli}. 
Thus the oscillation data together with the cosmological bound signify that the neutrino masses are much smaller than the masses of the charged leptons.

\section{Evidences of Neutrino Oscillation}

\begin{figure}[ht!]
\vspace{0.3 in}
\begin{center}
\includegraphics[scale=0.2]{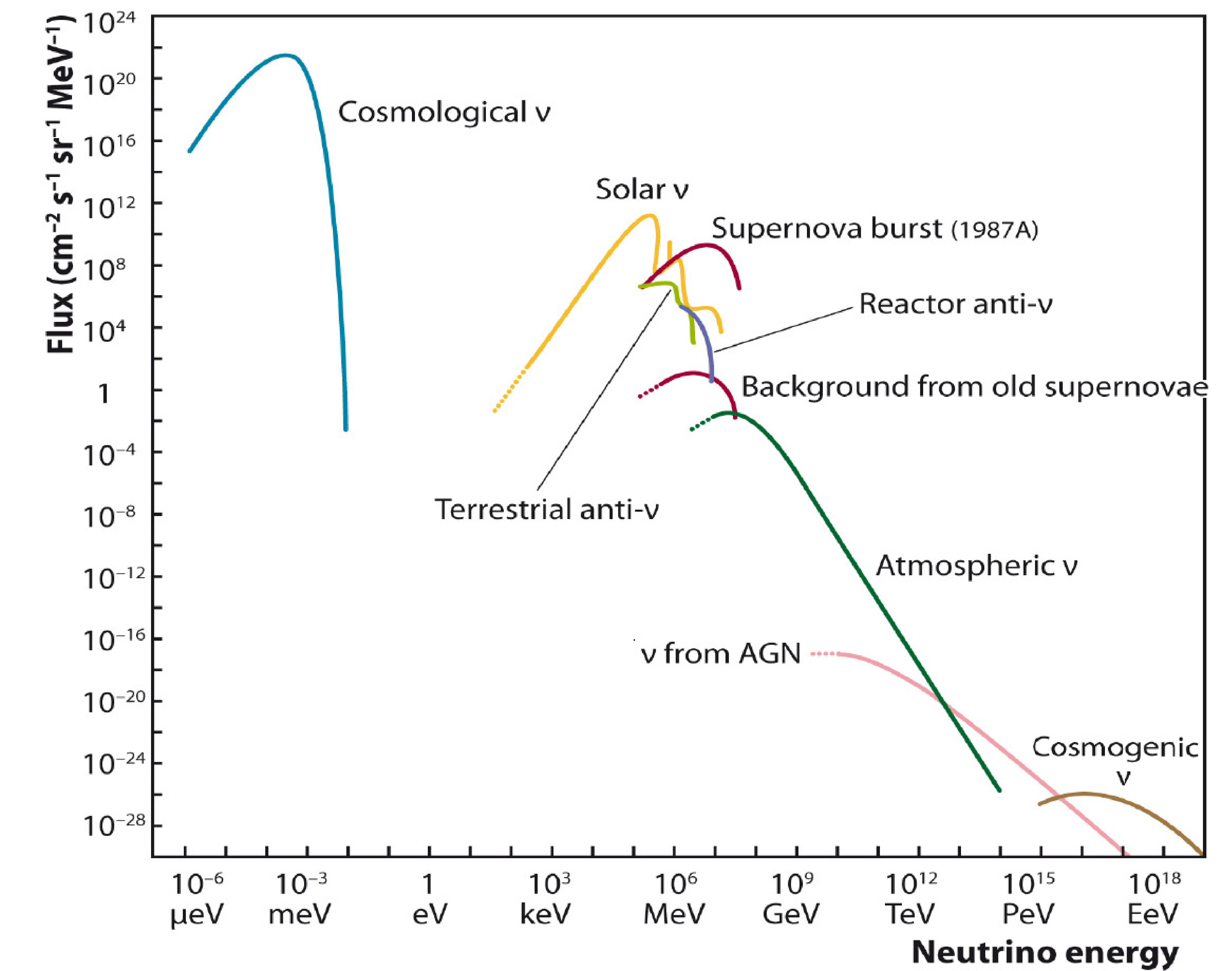}
\caption{Fluxes of neutrinos at different energies. The figure is taken from Ref. \cite{Spiering:2012xe}.}
\label{fig:figure0}
\end{center}
\end{figure}

Neutrinos can originate from different sources having energy ranging from few eV to PeV. From Fig. \ref{fig:figure0},
we can see that among all the sources, the relic neutrinos, which were decoupled from the other particles at the very early stage of the universe,
have the smallest energy but maximum flux. They are the most abundant particles in the universe after the photons. Most of the solar neutrinos
are generated from the $p$-$p$ fusion inside the sun whereas reactor and geo-neutrinos originate from the 
beta decay process and all of these neutrinos have energy in the MeV range. The neutrinos coming from the supernova explosions 
are generated through the electron capture of nuclei and free protons as well as through pair production. They
also have energy in the MeV range. The interactions of the cosmic rays with the atmospheric nuclei produce neutrinos in the GeV range and 
the neutrinos coming from the extragalactic sources fall in the energy range of TeV. The neutrinos produced in the man-made accelerators
can have energy in MeV or GeV.
The highest energy cosmogenic neutrinos are produced due to 
interaction of the ultrahigh energy cosmic rays with cosmological photon backgrounds. They could also be produced in the
interactions of accelerated protons with surrounding medium.

Among these different sources, evidences of neutrino oscillation have come from solar, atmospheric, accelerator and reactor neutrino experiments.
Below we discuss about the production mechanism of neutrinos in these type of experiments and describe how different experiments contributed to establish the phenomenon
of neutrino oscillation on a firm footing. In this context we will also discuss the production mechanism of the ultra high energy neutrinos. 
Though the main aim of the ultra high energy neutrino experiments is to study the interaction and production mechanism of the neutrinos from various astrophysical sources,
data from these experiments can also be used to constrain the oscillation parameters.

\subsection{Solar Neutrinos}

Solar neutrinos are produced by thermo-nuclear fusion reactions occurring at the core of the Sun. The underlying process is, 
\begin{equation} \nonumber
 4 p \rightarrow \alpha + 2 e^+ + 2 \nu_e .
\end{equation}
This occurs through proton-proton ($pp$) chain and CNO cycle producing a large number of neutrinos in MeV energy range.
Detecting these neutrinos 
at Earth was important to study the theory of stellar structure and evolution, which is the basis of the
standard solar model (SSM).
This was the aim of the pioneering 
experiment by Davis and collaborators using radiochemical Chlorine ($^{37} {\rm Cl}$) detector \cite{Davis:1968cp}.
This was sensitive to only electron neutrinos.
However, it was found that the observed neutrino flux
is only about one third of the solar-model predictions \cite{Bahcall:2004pz,Bahcall:2004mq}.
This deficit constitutes ``the solar-neutrino problem".  It was difficult to explain this deficit within the 
SSM and there were attempts to explain the discrepancy by proposing that the models of the Sun were wrong \cite{Bahcall:1998wm,Dar:1996nf,Cumming:1996gv}. 
Many model independent solutions were proposed \cite{Bludman:1993tk,Hata:1994qs,Hata:1997di,Parke:1994xt,Heeger:1996se}.
The phenomenon of neutrino oscillations was also considered as one of the possible solutions.
In 1981-82, the real time neutrino-electron scattering experiment, Kamiokande \cite{Fukuda:1996sz}, became operational
which confirmed the deficit observed in the chlorine experiment and also proved that the detected neutrinos actually came from the Sun. 
Later in 1990 Gallium based experiments, 
with a lower energy threshold (and thus sensitive to the $pp$ neutrinos) like 
GALLEX \cite{Anselmann:1992um}
and SAGE \cite{Abdurashitov:1994bc} (later GNO \cite{Altmann:2005ix}) corroborated the fact that the measured neutrino signal was indeed smaller than the SSM prediction.
The importance of the Ga experiments lies in the detection of the primary $pp$ neutrinos thereby confirming the basic hypothesis of stellar energy generation. 
Super-Kamiokande (SK), an upgraded version of the Kamiokande experiment \cite{Hosaka:2005um} further confirmed the solar neutrino deficit with enhanced statistics.  
But the real breakthrough in solar neutrino physics was due to the advent of the SNO \cite{Ahmad:2002jz,Aharmim:2005gt} experiment.
Because of its sensitivity to both charge current (CC) and neutral current (NC) interactions, it measured simultaneously the contributions from only electron neutrinos and
from all three active flavours respectively.  
By measuring the CC/NC ratio as less than one,
SNO established the presence of $\nu_\mu$ and $\nu_\tau$ flavours in the solar neutrino flux.
The NC measurement also confirmed that the measured total neutrino flux was in very good agreement with the SSM predictions. 
These results clearly showed that neutrinos change their flavour during their way
from the production point in the Sun to the detector and the phenomenon of neutrino oscillation 
emerged as the clear solution to the solar neutrino problem.

\subsection{Atmospheric Neutrinos}

Apart from the solar neutrinos, 
neutrino oscillation has also been observed in the atmospheric neutrino experiments.
Atmospheric neutrinos were first detected in the mines of Kolar Gold Fields of India \cite{Achar:1965ova}
and at the same time in a gold mine of South Africa \cite{Reines:1965qk}.
Atmospheric neutrinos result from the interaction of cosmic rays with atomic nuclei in the Earth's atmosphere, 
creating pions and kaons, which are unstable and produce neutrinos when they decay in the following manner:
\begin{eqnarray} \nonumber
 \pi^{\pm}(K^{\pm}) &\rightarrow& \mu^{\pm} + \nu_\mu (\bar{\nu}_\mu ), \\ \nonumber
 \mu^{\pm} &\rightarrow&  e^{\pm} + \nu_e(\bar{\nu}_e) + \bar{\nu}_\mu (\nu_\mu ). \\ \nonumber
\end{eqnarray}
From this decay chain, the expected number of muon neutrinos are about twice
that of electron neutrinos. However water \cnv\ detectors like Kamiokande \cite{atmosanomaly}, IMB \cite{Casper:1990ac,BeckerSzendy:1992hq} 
and iron calorimeter detector Sudan2 \cite{Goodman:1994uc} reported results contrary to this expectation. To reduce the
uncertainties in the absolute flux values, these experiments presented results in terms of the double ratio $R$ 
\begin{eqnarray}
 R = \frac{(N_\mu/N_e)_{{\rm obs}}}{(N_\mu/N_e)_{{\rm MC}}},
\end{eqnarray}
where MC denotes Monte-Carlo simulated ratio. 
The above experiments found the value of $R$ to be significantly less than one.
This became known as
the ``Atmospheric Neutrino Anomaly". 
However two other iron calorimeter detectors Fr\'{e}jus \cite{Berger:1989wy,Berger:1990rd} and Nusex \cite{Aglietta:1988be}
found results consistent with the theoretical expectations. The reduction in $R$ can be explained by either $\nu_\mu \rightarrow \nu_e$
or $\nu_\mu \rightarrow \nu_\tau$ oscillations or both. Apart from altering the flavour content of the atmospheric neutrino flux, oscillations
can induce the following effect. If the oscillation length is much larger than the height of the atmosphere but smaller than the diameter of the
Earth then neutrinos coming from the opposite side of the Earth (upward going neutrinos) will have significant oscillations. This will create a 
non uniform zenith angle dependence in the observed data. The high statistics SK experiment indeed found this zenith angle dependence in their multi-GeV data
establishing neutrino oscillation on a firm footing.


\subsection{Accelerator Neutrinos}

Neutrino oscillations have also been observed from the neutrinos produced in particle accelerators. 
Neutrino beams produced at a particle accelerator offer the greatest control over the neutrinos being studied.
In accelerators, neutrino beam can be produced in two methods: through the decay of pions at rest (DAR) and the decay of pions in flight (DIF). In both the methods, 
high intensity protons are collided with a fixed target to produce charged pions.
In DAR mechanism, the resulting $\pi^-$ are being absorbed and $\pi^+$ are brought to rest and then they decay in the following manner
\begin{eqnarray} \nonumber
 \pi^+ &\rightarrow& \mu^+ + \nu_\mu, \\ \nonumber
 \mu^+ &\rightarrow&  e^+ + \nu_e + \bar{\nu}_\mu, \\ \nonumber 
\end{eqnarray}
to produce $\bar{\nu}_\mu$ having maximum allowed energy of 52.8 MeV. The main aim of these type of experiments is to observe clean oscillations 
in the $\bar{\nu}_\mu \rightarrow \bar{\nu}_e$ channel as there is no intrinsic $\bar{\nu}_e$ background from the source.
In the DIF mechanism, the pions decay while traveling in the decay pipe to produce neutrinos and muons. The muons are absorbed 
and thus one gets pure neutrino or antineutrino beams depending on the polarity of the charged pions. The neutrinos produced in this fashion
are essentially beams of $\nu_\mu$ and $\bar{\nu}_\mu$ with energy ranging from few tens of MeV to GeV.
There were several experiments having baselines of the order of few tens of meters\footnote{For a comprehensive list see for instance Ref. \cite{Choubey:2001zk}.} looking
for neutrino oscillations using neutrinos produced in the accelerators. All of them gave null result except the LSND experiment \cite{Aguilar:2001ty}.
LSND reported an excess of events in both $\nu_\mu \rightarrow \nu_e$ and $\bar{\nu}_\mu \rightarrow \bar{\nu}_e$ oscillations.
It has observed the neutrino events in DIF mode and antineutrino events in DAR mode.
The MiniBooNE \cite{Tayloe:2010qga} experiment at Fermilab was proposed to test the LSND results using a different $L$ and $E$ but the same $L/E$ ratio as LSND.
It is found that the antineutrino data of MiniBooNE is consistent with the LSND observations\footnote{Note that
the oscillation results of LSND and MiniBooNE can not be explained in the three generation neutrino framework and
require the existence of sterile neutrinos.}.
There are also accelerator experiments like K2K \cite{Ahn:2006zza} and MINOS \cite{Adamson:2011qu} 
which have studied the oscillations of the neutrinos in the GeV energy range.
These are the long-baseline experiments having baselines around several hundreds of kilometers. K2K has
observed neutrino oscillations via muon neutrino disappearance channel ($\nu_\mu \rightarrow \nu_\mu$) and MINOS
has observed events in both appearance ($\nu_\mu \rightarrow \nu_e$) and disappearance measurements. 
For these experiments, the neutrino beam power of the accelerators were around few hundreds of KW.
The ongoing long-baseline experiment T2K \cite{Abe:2015awa} has observed oscillated muon and electron neutrino events at the far detector located 295 km
away from the neutrino source. Recently the \nova\ experiment at Fermilab also given its first results which also show a clear evidence of neutrino oscillation \cite{nova_recent}. 
To have enough sensitivity of the sub-dominant electron neutrino appearance channel, 
T2K/\nova\ as well as the other future generation long-baseline experiments are designed to have beam power of the order of MW. 
Because of this very high beam power, this type of
experiments are often termed as ``superbeam" experiments. The high beam power of these experiments also allow to obtain enough statistically significant number of
signal events over the expected backgrounds.
These accelerator based long-baseline experiments have confirmed the oscillations of the atmospheric neutrinos as the associated $L/E$ in these cases are such that 
the oscillations are governed by the atmospheric mass square difference $\Delta_{31}$.

\subsection{Reactor Neutrinos}

Another major source of the man-made neutrinos from where oscillations have been observed are the nuclear reactors.
In reactors, antineutrinos of the energy around few MeV are produced by the nuclear fission processes.
Because of this low energy such experiments are sensitive to only $\bar{\nu}_e \rightarrow \bar{\nu}_e$ oscillations i.e.,
they look for a diminution in the $\bar{\nu}_e$ flux.
Many experiments have searched for oscillation of the reactor neutrinos by detecting the oscillated electron antineutrino events via inverse beta decay (IBD). 
The measurement of oscillation parameters in the nuclear reactors mainly suffers due to the uncertainties in
the strength of the sources, the detector efficiency and the cross sections for neutrino interactions.
Thus one needs a good knowledge of the flux. The uncertainties can be minimized by the inclusion of a near detector.
The earlier experiments like  ILL-Grenoble \cite{Kwon:1981ua}, Rovno \cite{Kuvshinnikov:1990ry}, Savannah River \cite{Greenwood:1996pb}, 
Gosgen \cite{Zacek:1985kr}, Krasnoyarsk \cite{Vidyakin:1994ut}, BUGEY \cite{Declais:1994su} 
searched for oscillations of the reactors antineutrinos at distances $< 100$ m from the reactor core. 
But all these experiments got null results \footnote{
Recent study of reactor antineutrino spectra show a 3\% enhancement in the fluxes as compared to the previous calculation. 
With this new re-evaluated fluxes, the ratio of observed event rate to predicted rate for the $< 100$ m reactor experiments shifts from 0.976 to 0.943, giving rise to reactor
neutrino anomaly \cite{Mention:2011rk}. This deficit could not be explained in three flavour framework and the presence of sterile neutrinos were evoked as a possible explanation.}. 
The next generation longer baseline reactor experiments
CHOOZ \cite{Apollonio:1999ae,Apollonio:1997xe} and Palo Verde \cite{Boehm:2000vp,Boehm:1999gk} looked for oscillations at a distance of 1 km
but they also did not report any evidence of neutrino
oscillations.
The KamLAND \cite{Eguchi:2002dm} experiment, started in 2002, was the first to observe oscillations of the antineutrinos coming out of nuclear reactors.
As the baseline of KamLAND was 180 km, it was sensitive to oscillations governed by the mass squared difference $10^{-5}$ eV$^2$ which is relevant for the flavour conversion
of the solar neutrinos. 
Thus
KamLAND confirmed the oscillations of the solar neutrinos using a man-made neutrino source.
Recently the observations of oscillation in the reactor experiments DOUBLE-CHOOZ \cite{dchooz2},  
RENO \cite{reno} and Daya Bay \cite{dayabay} have established the non zero value of $\theta_{13}$ with significant confidence level.
These experiments have baselines of few kilometers and thus sensitive to oscillations governed by atmospheric mass squared difference of the order of $ 10^{-3} $ eV$^2$.

\subsection{Ultra High Energy Neutrinos}

Ultra high energy (UHE) neutrino telescopes were planned to study the neutrinos
from distant astrophysical sources \cite{hulth2006ultra}.  
Currently envisaged astrophysical sources of high energy 
cosmic neutrinos include for instance, active galactic nuclei (AGN) \cite{Nellen:1992dw} and gamma ray 
burst (GRB) fire balls \cite{Waxman:1997ti}. The production of
high energy cosmic neutrinos
from sources other than the AGN’s and GRB’s are also possible \cite{Gaisser:1994yf}.
In those sources protons are accelerated to very high energies 
by the Fermi acceleration mechanism \cite{Baring:1997ka}. The interactions of these
protons with soft photons or matter from the source can
give UHE neutrinos. These neutrinos travel a long distance from their source to reach the Earth.
The oscillation of this very high energy neutrinos are averaged out due to the long distance and their final flavour composition
depends on the initial sources of the neutrinos as different sources can have different initial flavour composition.
Recently the IceCube \cite{Aartsen:2013jdh} collaboration has reported the results of an all-sky search for UHE neutrino events which was
conducted during May 2010 to May 2013. 
They have detected a total of 37 neutrino events of extraterrestrial origin at $5.7 \sigma$ confidence level.
These events fall in the energy range between 30 to 2000 TeV.
For these observed 37 events, the expected cosmic ray muon background was $8.4 \pm 4.2$ and the backgrounds from atmospheric neutrinos were $6.6^{+5.9}_{-1.6}$ events.
These results are consistent with the framework of neutrino oscillations over the astronomical distances.
The recent IceCube observation of a 2.3 PeV event correspond to the highest-energy neutrino interaction ever observed \cite{icecubewebsite}.

\section{Neutrino Mass Matrix}

The observation of non-zero neutrino mass via neutrino oscillations necessitates an extension of the Standard Model. 
A successful model for neutrino mass needs to explain how neutrinos get their mass as well as why the mass
is so tiny. It also requires to explain the observed mixing pattern among the neutrinos.
One can simply extend the SM by adding right-handed neutrinos and generate the Dirac neutrino masses in a similar fashion as that of the charged leptons and 
quarks. But to obtain neutrino mass in the sub eV range, one requires a very small value of the Yukawa coupling i.e., of the order of $10^{-12}$.
Introduction of such small coupling constants is generally considered unnatural
and one must find a symmetry reason for such smallness. The most elegant way to generate small
neutrino mass naturally is the See-Saw mechanism which relates the smallness of neutrino masses to new physics at high scale. 
In See-Saw mechanism neutrino mass originates from the dimension five operator \cite{Weinberg:1979sa,Weinberg:1980bf} $\frac{k}{M_R}LL\phi\phi$, 
where $L$ is the lepton doublet, $\phi$ is the Higgs doublet, $k$ is the dimensionful coupling constant
and $M_R$ is the scale of the beyond Standard Model (BSM) physics. 
This operator can be realized at the tree level by three ultraviolet completions which are known as 
Type I \cite{Minkowski:1977sc,GellMann:1980vs,Glashow:1979nm,Mohapatra:1979ia}, 
Type II \cite{Magg:1980ut,Lazarides:1980nt,Mohapatra:1980yp,Schechter:1981cv} and Type III \cite{Foot:1988aq} See-Saw.
In Type I See-Saw, SM is extended by heavy right-handed singlet neutrinos. In Type II and Type III See-Saw, scalar triplets and
fermion triplets are added to the SM respectively.
The most economical case among these three is the Type I See-Saw where 
after the spontaneous symmetry breaking, the neutral component of the Higgs doublet acquires a vacuum expectation value (VEV)
$v$ and the light neutrino mass is obtained as
\begin{eqnarray}
 m_\nu \sim v^2/M_R.
\end{eqnarray}
To have neutrino mass around 0.1 eV, one needs $M_R$ around $10^{14}$ GeV which is close to the scale of the Grand Unified Theories (GUT).
See-Saw mechanism predicts the Majorana nature of the neutrinos which implies that the neutrinos are their own antiparticles.
This mechanism also predicts violation of lepton number by two units.
In Type I See-Saw the light neutrino mass matrix which is a $3 \times 3$ complex matrix in 3 generation, is given by
\begin{eqnarray}
 M_\nu = M_D M_R^{-1} M_D^T,
\end{eqnarray}
where $M_D$ is the Dirac mass arising from the Yukawa term  $ y_{\nu} \bar{L}_L \nu_R \phi$ and $M_R$ is the Majorana mass coming from the
Majorana mass term $\bar{\nu}_R^C \nu_R$. 
In general the neutrino mass matrix $M_\nu$ in flavour basis is not diagonal and
the complex symmetric $ 3\times 3 $ low energy mass matrix is given by
\begin{eqnarray}
M_{\nu}&=&V^*M_{\nu}^{{\rm diag}}V^{\dagger}, \\
        &=&
 \begin{pmatrix}
  m_{ee} & m_{e \mu} & m_{e \tau} \\
  m_{\mu e} & m_{\mu \mu} & m_{\mu \tau} \\
  m_{\tau e} & m_{\tau \mu} & m_{\tau \tau}
 \end{pmatrix}, \qquad
 \label{mass_matrix}
\end{eqnarray}
where, $M_{\nu}^{{\rm diag}} = {\rm diag}({m_1,m_2,m_3})$
and $V=U.P$ denotes the leptonic mixing matrix in a basis 
where the charged lepton mass matrix is diagonal.
$U$ is the PMNS matrix described earlier and $P$ is the diagonal phase matrix of Majorana phases written as  
\begin{equation}
P=\text{diag}(1,e^{i \alpha}, e^{i {(\beta+\delta)}}).
\end{equation}
As the elements of this matrix are functions of oscillation parameters (including the Majorana phases), 
the structure of the low energy mass matrix can be constrained using present experimental data.

\section{Unknown Oscillation Parameters and Future Prospects: Three Generations}

In the last two decades there has been a tremendous progress in the determination of the parameters that describe neutrino oscillation of the three
active neutrinos. The solar neutrino experiments and KamLAND have measured the parameters $\theta_{12}$ and $\Delta_{21}$ with considerable precision.
The measurements of $\theta_{23}$ and $|\Delta_{31}|$ come from atmospheric neutrino experiments, MINOS and T2K. The reactor experiments have measured
the value of $\theta_{13}$ with appreciable precision.
We will discuss the present constraints on oscillation parameters from the global analysis of the world neutrino data in the next chapter.
At present the unknown oscillation parameters are: 
(i) the sign of $\Delta_{31}$ or the neutrino mass hierarchy,  
(ii) the octant of $\theta_{23}$ (i.e., whether $\theta_{23} < 45^\circ$ or $ > 45^\circ$) and 
(iii) CP violation in leptonic sector and the precision of $\dcp$. 
Apart from these, the following unresolved issues are also of interest:
(i) the absolute mass of the neutrinos,  
(ii) the exact nature of the neutrinos i.e., Dirac or Majorana,  
(iii) the mechanism of generation of neutrino masses and explanation of their smallness,
(iv) non standard interaction (NSI) of the neutrinos, 
(v)  non-unitary neutrino mixing
and
(vi) CPT violation in neutrino oscillation etc.

The measurement of various oscillation parameters are important not only to understand
the exact nature of neutrino oscillation but also for building models in BSM scenario. Many BSM models can be accepted or rejected
depending upon their prediction of different oscillation parameters. So a precise measurement of the oscillation parameters can guide towards a successful BSM theory.
Determination of $\dcp$ can also give clue in understanding the present matter-antimatter asymmetry of the universe.
The matter-antimatter asymmetry of the universe can be explained by the process of baryogenesis. 
But the baryogenesis in SM is not sufficient to explain the observed baryon asymmetry of the universe.
One option to create additional baryon asymmetry is via leptogenesis in which the decay of heavy right handed neutrinos (for instance those belonging to the See-Saw models)
can create lepton asymmetry which can be converted to baryon asymmetry.
Different studies show that under certain conditions,
it may be possible to connect the leptonic CP phase $\delta_{CP}$ to leptogenesis \cite{Joshipura:2001ui}.

There are various current ongoing/future upcoming neutrino oscillation experiments dedicated for determining the remaining unknown oscillation parameters.
Below we mention some of the major projects.
The beam based long-baseline experiments T2K and NO$\nu$A \cite{nova}, which are both taking data at present,
adopted the off-axis technique which gives a narrow flux at the oscillation maxima to
reduce backgrounds at the high energy tail.
The experiment T2K itself does not have hierarchy sensitivity and \nova\ has hierarchy sensitivity in a limited range of $\dcp$ space.
But their main aim is to measure the leptonic phase $\dcp$. 
The experiments 
LBNE\footnote{Recently there are discussions on converging the LBNO and LBNE projects into a 
combined initiative called DUNE \cite{lbnf}.} \cite{lbne_interim2010} and 
LBNO \cite{lbno_eoi} will make use of the on-axis broad band flux to probe oscillation over a wide energy range.
Due to the comparatively longer baseline and higher statistics, LBNO and LBNE experiments can measure all the three above mentioned unknowns 
with significant confidence level.
The DAE$\delta$LUS experiment \cite{Alonso:2010fs} proposes to replace the antineutrinos of the superbeam experiments 
by the low energy antineutrinos from muon decay at rest 
and using Gd-doped water \cnv\ detector.
This approach will give larger
antineutrino event sample as compared to the conventional superbeam technique.
The superbeam experiment at the ESS facility \cite{Baussan:2013zcy} proposed to study the physics at the second oscillation maximum
for obtaining significant sensitivity towards establishing CP violation.

The atmospheric neutrino experiments ICAL@INO \cite{Ahmed:2015jtv} will consist of a magnetized iron calorimeter detector
for studying neutrino and antineutrino events separately.
These type of detectors are sensitive to
muons and they have good energy and direction measurement capability.
The Hyper-Kamiokande \cite{hk} and PINGU \cite{pingu} experiments will have large volume water \cnv\ detectors.
These detectors can measure energy and direction of both electrons and muons but do not have the charge identification capability.
The aim of these experiments are mainly to determine the neutrino mass hierarchy.

In reactor experiments one can have hierarchy sensitivity by using the oscillation interference effect between $\Delta_{31}$ and $\Delta_{32}$.
The primary goal of the medium baseline reactor neutrino experiments JUNO \cite{Li:2014qca} and RENO-50 \cite{Kim:2014rfa}
is to determine the mass hierarchy using liquid scintillator detector. These experiments require the precise measurement of
the oscillation spectrum with an excellent energy
resolution.

Apart from these above mentioned neutrino oscillation experiments, there are also experiments
whose primary aim is not to determine the oscillation parameters but still it is possible to probe different oscillation parameters in these experiments.
The $0\nu\beta\beta$ and ultra high energy neutrino experiments are example of such experiments.
The main aim of the ongoing ultra high energy neutrino detector IceCube at south pole is to 
understand the origins and acceleration mechanisms of high-energy cosmic rays. But it is also possible to probe different oscillation parameters at IceCube.    
So a comprehensive phenomenological study regarding the potential of the various neutrino oscillation experiments 
towards completing the gaps in oscillation physics
is extremely relevant at this point.

\section{Sterile Neutrinos: Beyond Three Generations}

Another intriguing aspect of current oscillation picture is the existence of light sterile neutrino.
Neutrino oscillation in the standard three flavour picture is now well established from
different oscillation experiments. However,   
the reported observations of $\bar{\nu}_\mu$ - $\bar{\nu}_e$
oscillations in the LSND experiment \cite{Athanassopoulos:1996jb,Athanassopoulos:1997pv,Aguilar:2001ty} and recent confirmation of this by the MiniBooNE
experiment \cite{Aguilar-Arevalo:2013pmq,AguilarArevalo:2009yj} with oscillation frequency governed by a mass-squared difference around 1 eV$^2$
cannot be accounted for in the above framework. These results motivate the introduction
of at least one extra neutrino of mass  of the order of eV to account for the three independent mass
scales governing solar, atmospheric and LSND oscillations. As we already know that the LEP data on measurement of
$Z$-line shape dictates that there can be only three light neutrinos
with standard weak interactions,
the fourth light neutrino, if it exists must be a Standard Model singlet or sterile.
Recently this hypothesis garnered additional support from
(i) disappearance of electron
antineutrinos in reactor experiments with recalculated fluxes \cite{Mention:2011rk} and (ii) deficit of electron
neutrinos measured in the solar neutrino detectors GALLEX and SAGE using radioactive
sources \cite{Giunti:2010zu}. The recent ICARUS results \cite{Antonello:2012pq} however, did not find any evidence for the LSND
oscillations. But this does not completely rule out the LSND
parameter space and small
active-sterile mixing still remains allowed.
There are also constraint about existence of an extra relativistic species from the CMB anisotropy measurements
\cite{Komatsu:2010fb,Dunkley:2010ge,Keisler:2011aw,Archidiacono:2011gq,Hinshaw:2012aka,Hou:2012xq} 
which prefer the the effective neutrino number
to be greater than three. Recently the combined data of Planck, WMAP polarization and the high multipole results
gives $N_{{\rm eff}} = 3.36^{+0.68}_{-0.64}$ at 95$\%$ C.L \cite{Ade:2013zuv}.
Clearly this data do not completely rule out the existence of a fourth neutrino species.
Thus, the situation with sterile neutrinos
remains quite interesting and many future experiments are proposed/planned to test these
results and reach a definitive conclusion \cite{Abazajian:2012ys}.
In view of these,
the study of different phenomenological implications of sterile neutrinos assume an
important role. Note that the results in the 3 generation scenario can differ significantly in the presence of sterile neutrino.

\section{Thesis Overview}

In this thesis first we have studied the potential of long-baseline experiments T2K, NO$\nu$A and atmospheric neutrino experiment ICAL@INO to discover CP violation
in the leptonic sector.
We have also studied the role of the three above mentioned experiments to economise the configuration of future proposed long-baseline experiments
LBNO and LBNE in determining the remaining unknowns in neutrino oscillation.
We have used the recent IceCube data to put constrain over $\delta_{CP}$ as well as various
astrophysical sources. Finally we have studied the structure of the low energy neutrino mass matrix in flavour basis
in terms of texture zeros in the presence of one extra light sterile neutrino. The plan of the thesis goes as follows. 

In Chapter ~\ref{chap:oscillation}, we give an overview of neutrino oscillation in vacuum and matter elaborating on
how matter effect modifies the mass and mixing parameters.
We will give derivations of the relevant expressions of the oscillation probabilities.
Then we will describe
the present status of the oscillation parameters. We will also review the parameter degeneracy in neutrino oscillation and describe how the
physics capability of different long-baseline experiments are constrained due to the parameter degeneracy. We will end this chapter by giving a short description
about the current/future oscillation experiments which we have studied in this thesis. 

In Chapter \ref{chap:sens} we will discuss how the various neutrino oscillation parameters can be
probed in future oscillation experiments. This chapter will contain the main results of our neutrino oscillation analysis and
will be organised as follows: 
First we will discuss the CP sensitivity of the T2K and \nova\ by taking their projected exposures.
Next we discuss how atmospheric neutrino experiment ICAL can improve the CP sensitivity of T2K and \nova.
We further extend this study taking different exposures and gauge the capability of these setups to discover CP violation and also in measuring the precision of $\dcp$.

Next we study how the different setups of the LBNO project can be economised by using current/upcoming facilities T2K, \nova\ and ICAL. For our analysis
we consider three prospective LBNO setups – CERN-Pyh\"{a}salmi (2290 km), CERN-Slanic
(1500 km) and CERN-Fr\'{e}jus (130 km) and emphasize on the advantage of exploiting the synergies offered by T2K, \nova\ and ICAL in evaluating
the adequate exposure which is the minimum exposure required in
each case for determining the remaining unknowns of neutrino oscillation i.e hierarchy, octant and $\dcp$ at a given confidence level.

Then we will carry out a similar analysis as described above, for the LBNE project at Fermilab. Apart from finding
the adequate exposure of LBNE in conjunction with T2K, \nova\ and ICAL, 
we will also quantify the effect of the 
proposed near detector on systematic errors, examine the role 
played by the second oscillation cycle in furthering the physics reach of 
LBNE and present an optimisation study of 
the neutrino-antineutrino running.

Finally we will study how 
the recent data of IceCube can constrain the leptonic CP violating phase $\delta_{CP}$.
We also use this data to impose constraints on the sources of the neutrinos. 


In Chapter \ref{chap:matrix}, we will discuss the structure and properties of 
neutrino mass matrices and its phenomenological consequences in terms of texture zeros with sterile neutrinos and
compare our results with the three generation case.
First we will consider
the two-zero textures of the
low energy neutrino mass matrix in presence of one additional sterile neutrino. 
We discuss the mass spectrum and the parameter 
correlations that
we find in the various textures. We also present the effective mass governing neutrinoless
double beta decay as a function of the lowest mass. Next
we will study the phenomenological implications of the one-zero textures of the same
neutrino mass matrices in the presence of a sterile neutrino.
We study the
possible correlations between the sterile mixing angles and the Majorana phases to give a zero element in the mass matrix.

We will summarize and present the impact of our work in the last chapter.

%% file: chap2_oscillation.tex

\section{Overview}

In this chapter we discuss the salient features of neutrino oscillation phenomena.
As mentioned in the introduction, neutrino oscillation is described by the transition probability from one flavour to another. 
This is a function of neutrino mass squared differences, mixing angles and the Dirac type CP phase. 
To understand the dependence of the oscillation probability on different oscillation parameters,
one needs to derive the analytic expressions for the same.
In the first section of this chapter we will give the derivations of the
expressions for the oscillation probability in different scenarios.
For the determination of the remaining unknowns in the oscillation sector, one needs to use the information from the past/present experiments
as inputs. Thus in the 
next section, we discuss the current status of the oscillation parameters by comparing global analysis of the world neutrino data as obtained by
different groups.
Next we discuss what are the difficulties in 
measuring the unknowns and what are the future facilities that are aimed towards determination of these.
This leads us to the discussion about the parameter degeneracies in view of the current oscillation data.
We present the hierarchy-$\dcp$ degeneracy and the octant-$\dcp$ degeneracy in detail.
In the next section we will give
the salient features of the present/future oscillation experiments whose physics potential
we have studied in this thesis.


\section{Derivation of Oscillation Probability}

In this section we will give the derivations of the expressions for neutrino oscillation probabilities in vacuum and matter 
and show the dependence of the oscillation probabilities on
different oscillation parameters. 
For neutrinos propagating in vacuum, it is possible to derive exact analytic expressions.
For matter one needs to solve the propagation equation using the relevant density profile.
For matter of constant density, exact expressions can be derived for the two flavour case. 
For three flavours, 
the probability expressions even in constant matter density can be derived only under certain approximations. 
We will start this section by deriving the vacuum oscillation probability for two flavours
which is important to understand the basic mechanism of neutrino oscillation. We also give expressions for
generalised $N$ flavour oscillation from which we can easily calculate the three flavour expression.
For the matter case, first we will derive the exact two flavour expression in constant density matter and show how the matter effect can modify the vacuum
mass and mixing. We will end this section by describing how under the ``one mass scale dominance" (OMSD) 
and the $\alpha-s_{13}$ (double expansion in $\alpha(=\Delta_{21}/\Delta_{31})$ and $\sin\theta_{13}$) approximations one can derive the 
expressions for probability
of the three generation neutrinos in matter of constant density. 
We will also discuss the validity condition of these approximations.

\subsection{Two Flavour Oscillation in Vacuum}

First let us consider only the first two generations of neutrinos $\nu_e$ and $\nu_\mu$. 
In this case the mixing matrix $U$ will be $2 \times 2$ unitary matrix parametrised by one mixing angle $\theta$. The relation between
the flavour eigenstates and the mass eigenstates can be written as
\begin{eqnarray}
\begin{pmatrix}
  \nu_e  \\
  \nu_\mu \\
 \end{pmatrix} 
        &=&
 \begin{pmatrix} 
  \cos\theta & \sin\theta \\
   -\sin\theta & \cos\theta \\
 \end{pmatrix}
 \begin{pmatrix}
  \nu_1  \\
  \nu_2 \\
 \end{pmatrix}. 
 \label{mixing_matrix}
\end{eqnarray}
The time evolution of the state $\nu_e$ after time $t$ is given by
\begin{eqnarray}
| \nu_e(t) \rangle = e^{- i E_1 t} \cos\theta |\nu_1\rangle + e^{- i E_2 t} \sin\theta |\nu_2\rangle,
\end{eqnarray}
where $E_1$ and $E_2$ are the energies of the mass eigenstates $\nu_1$ and $\nu_2$ having mass $m_1$ and $m_2$. The energy $E_i$ can be written as 
(in the units of $c=1$)
\begin{eqnarray}
 E_i^2 &=& p^2 + m_i^2 \\
   E_i   &\approx& p + \frac{m_i^2}{2p}. 
 \label{energy}
\end{eqnarray}
Note that in this plane wave treatment of neutrino oscillation, we have assumed that all the massive neutrinos are of equal momentum
\footnote{A more realistic approach is to consider the wave packet treatment as
real localised particles are described by superpositions of plane waves \cite{bsc}.
However for the oscillation scenarios which are considered in this thesis, wave packet effects have no practical consequences when neutrinos are
relativistic \cite{NP_kim}.
} $p$.
The survival probability of the electron neutrino
$\nu_e$ is given by
 \begin{eqnarray}
   P_{ee} &=& |\langle \nu_e|\nu_e(t)\rangle|^2 \\
           &=& (\cos^2\theta e^{- i E_1 t} + \sin^2\theta e^{- i E_2 t})(\cos^2\theta e^{ i E_1 t} + \sin^2\theta e^{ i E_2 t}) \\ \label{pee_1} 
           &=& 1 - \sin^22\theta \sin^2\{(E_2 - E_1)t/2\}. 
\end{eqnarray}
Using Eq. \ref{energy} in Eq. \ref{pee_1} and remembering the fact that in the relativistic limit $p \approx E$ and $t \approx L$ (for $c=1$ and $\hbar=1$) we obtain
\begin{eqnarray}
  P_{ee} &=& 1 - \sin^22\theta \sin^2\{\Delta_{21}L/4E\} \\ \label{pee_2}
         &=& 1 -  \sin^22\theta \sin^2\{1.27 \Delta_{21}L/E\},
\end{eqnarray}
where in Eq. \ref{pee_2},  $\Delta_{21} = m_2^2 - m_1^2$ is in eV$^2$, $L$ is in km and $E$ is in GeV. The conversion probability i.e., the transition probability
from $\nu_e \rightarrow \nu_\mu$ can be obtained from Eq. \ref{pee_2} as
\begin{eqnarray}
 P_{e \mu} &=& 1 - P_{ee} \\ \label{pemu}
           &=& \sin^22\theta \sin^2\{1.27 \Delta_{21}L/E\}.
\end{eqnarray}
From Eq. \ref{pemu} it is clear that the oscillatory behaviour
of the neutrinos is embedded in the term containing $\Delta_{21}$ and this term will go to zero when either the masses $m_1$ and $m_2$ are equal or when both of them are zero. Thus
neutrino oscillation requires non-degenerate and non-zero masses of neutrinos. Another important feature of Eq. \ref{pemu} is that, this expression is not sensitive 
to the octant of $\theta$ (i.e., $\theta < 45^\circ$ or $> 45^\circ$) and the sign of $\Delta_{21}$
as the transformation defined by $\theta \rightarrow \pi/2 - \theta$ and $\Delta_{21} \rightarrow - \Delta_{21}$,
leaves this equation unaltered. 

Here it is important to note that neutrino oscillation is also characterised by the value of $L$ and $E$ under consideration. 
For a combination of $L$ and $E$ such that the oscillatory term 
$\Delta_{21}L/4E$ goes to zero, there will be no oscillation. 
On the other hand if we consider a very high value of $L$, then there will be very large number of oscillation cycles
at smaller values of $E$. Hence
the oscillation
will be averaged out and probabilities will not depend explicitly on the masses of the neutrinos any more. 
This is the case for ultra high energy neutrinos which travel a large distance in vacuum
to reach the Earth. The maximum oscillation of the neutrinos can be obtained under the condition
\begin{eqnarray}
 1.27 \Delta_{21}L/E = n \pi/2,
\end{eqnarray}
where $n=1$ correspond to first oscillation maxima. This is the case for accelerator based long-baseline neutrino experiments. Here the accelerators are designed such that
the neutrino flux peaks at the energies where the oscillation is maximum. For example in the T2K experiment, the distance from source to detector is 295 km and
the neutrino flux peaks at 0.6 GeV. Putting this number in the above equation, for $n=1$ we get the value of the mass squared difference as $2.5 \times 10^{-3}$ eV$^2$ which
is close to the current best-fit of the atmospheric mass squared difference.

Before generalising the above formula for $N$ flavours, we would like to give another alternative method to derive the same oscillation formula. 
This formalism will be important 
at the time of deriving the oscillation formula in matter. The time dependent Schr\"{o}dinger equation in the mass basis can be written as
\begin{eqnarray}
 i \frac{\partial \nu_i}{\partial t} &=& H_M \nu_i,
  \label{mass_basis}
  \end{eqnarray}
where $H_M$ is the effective Hamiltonian in the mass basis and $\nu_i$ is the mass eigenstate.
For the case of two generations of neutrinos this can be written as
 \begin{eqnarray}
 H_M =  
  \begin{pmatrix} 
   E_1 & 0 \\
    0 & E_2 \\
  \end{pmatrix}.
 \end{eqnarray}
Using Eq. \ref{energy} one gets
 \begin{eqnarray}
  H_M = E I + \frac{1}{2E}
  \begin{pmatrix}
   m_1^2 & 0 \\
   0 & m_2^2 \\
  \end{pmatrix},
 \end{eqnarray}
where $I$ is the $2 \times 2$ identity matrix.
Here we would like to mention that as the common diagonal terms affect both the neutrino flavours in the same way, they
do not contribute in the final expressions of probability. Thus we can always add or subtract any diagonal term from the effective Hamiltonian $H$.
Using Eq. \ref{mixing_matrix}, we convert the Eq. \ref{mass_basis} into flavour basis and obtain the following equation for the two flavour scenario
\begin{eqnarray}
 i \frac{\partial}{\partial t}
\begin{pmatrix}
  \nu_e  \\
  \nu_\mu \\
 \end{pmatrix} 
        &=&  H_F 
 \begin{pmatrix}
  \nu_e  \\
  \nu_\mu \\
 \end{pmatrix},
\label{evolution}
\end{eqnarray}
where $H_F$ is the effective Hamiltonian in the flavour basis and is given by
\begin{eqnarray}
  H_F &=& U^\dagger  \label{evolution_1}
  \begin{pmatrix} 
   m_1^2/2E & 0 \\
    0 & m_2^2/2E \\
  \end{pmatrix}
   U. 
  \end{eqnarray}
Subtracting $\frac{m_1^2 +m_2^2}{4E}$ from the diagonal elements, the above equation can be simplified to  
  \begin{eqnarray}
    H_F &=& \frac{1}{4E}
  \begin{pmatrix} 
   - \Delta_{21} \cos2\theta & \Delta_{21} \sin2\theta \\
    \Delta_{21} \sin2\theta & \Delta_{21} \cos2\theta  \label{evolution_2}
\end{pmatrix},    
 \end{eqnarray}
and Eq. \ref{evolution} can be explicitly written as
\begin{eqnarray} \label{coupled_1}
  i \frac{\partial \nu_e}{\partial t} &=& -a \nu_e + b \nu_\mu, \\ 
  i \frac{\partial \nu_\mu}{\partial t} &=& b \nu_e + a \nu_\mu, \label{coupled_2}
\end{eqnarray}
where
\begin{eqnarray}
 a = \frac{\Delta_{21} \cos2\theta}{4E}, \hspace{3 mm} b = \frac{\Delta_{21} \sin2\theta}{4E}.
\end{eqnarray}
Solving the two coupled differential Eqs. \ref{coupled_1} and \ref{coupled_2} we obtain
\begin{eqnarray}
 \nu_e(t) &=& A_1 e^{-i \omega t} + A_2 e^{i \omega t}, \\
 \nu_\mu(t) &=& B_1 e^{-i \omega t} + B_2 e^{i \omega t},
\end{eqnarray}
with the condition $|\nu_e(t)|^2 + |\nu_\mu(t)|^2 =1 $, where $\omega^2 = a^2 + b^2 = \left(\frac{\Delta_{21}}{4E}\right)^2$.
Using initial conditions $\nu_e(0) = 1$ and $\nu_\mu(0) = 0$, we obtain
\begin{eqnarray}
 A_1 = \sin^2\theta, ~ A_2 = \cos^2\theta, ~ B_1 = \sin\theta \cos\theta, ~ B_2 = - \sin\theta \cos\theta.
\end{eqnarray}
This gives the transition probability $\nu_e \rightarrow \nu_\mu$ as
\begin{eqnarray}
 P_{e \mu} =  |\nu_\mu(t)|^2 = \sin^22\theta \sin^2\{1.27 \Delta_{21}L/E\}.
\end{eqnarray}

\subsection{$N$ Flavour Oscillation in Vacuum}

In this section we derive the vacuum oscillation probability for a generalised $N$ flavour oscillation scenario. For $N$ flavours, the mixing matrix $U$ is a $N \times N$
unitary matrix parametrised by $N(N-1)/2$ number of mixing angles and $(N-1)(N-2)/2$ number of phases. At time $t=0$, the flavour eigenstates are written as
\footnote{In some references, for instance \cite{bilenky,giunti}, the convention
$ |\nu_\alpha \rangle  = \sum_{i=1}^N U_{\alpha i}^* | \nu_{i} \rangle$ is used. 
}
\begin{eqnarray}
  |\nu_\alpha(0) \rangle  = \sum_{i=1}^N U_{\alpha i} | \nu_{i} \rangle.
  \label{relation_1}
\end{eqnarray}
Here the index $\alpha$ corresponds to all the $N$ flavours of neutrinos. After time $t$, the flavour states will evolve to
\begin{eqnarray}
 |\nu_\alpha(t) \rangle  = \sum_i^N U_{\alpha i} e^{- i E_i t} | \nu_i \rangle.
\end{eqnarray}
The oscillation probability $\nu_\alpha \rightarrow \nu_\beta$ ($P_{\alpha \beta}$) is given by
\begin{eqnarray}
 P_{\alpha \beta} &=& \left\vert \langle \nu_\beta|\nu_\alpha(t)\rangle \right\vert^2 \\
                  &=& \left\vert\sum_{i=1}^N U_{\alpha i} U_{\beta i}^* e^{-i E_i t} \right\vert^2 \\
                  &=& \sum_{i=1}^N \sum_{j=1}^N (U_{\alpha i} U_{\beta i}^* e^{-i E_i t})(U_{\alpha j}^* U_{\beta j} e^{ i E_j t}) \\
                  &=& \sum_{i=j} |U_{\alpha i}|^2 |U_{\beta i}|^2 + \sum_{i \neq j} U_{\alpha i} U_{\beta j} U_{\alpha j}^* U_{\beta i}^* e^{-i(E_i - E_j)t}.
\label{N_gen}                 
\end{eqnarray}
Using the relation
\begin{eqnarray}
 \left\vert \sum_i U_{\alpha i} U_{\beta i}^* \right\vert^2 = \sum_i |U_{\alpha i}|^2 |U_{\beta i}|^2 + \sum_{i \neq j} U_{\alpha i} U_{\beta j} U_{\alpha j}^* U_{\beta i}^*,
\end{eqnarray}
in Eq. \ref{N_gen} we obtain
\begin{align}
 P_{\alpha \beta} =  &\left\vert \sum_i U_{\alpha i} U_{\beta i}^* \right\vert^2 - \sum_{i \neq j} U_{\alpha i} U_{\beta j} U_{\alpha j}^* U_{\beta i}^*  \\ \nonumber
                     &+  \sum_{i \neq j} U_{\alpha i} U_{\beta j} U_{\alpha j}^* U_{\beta i}^* e^{-i(E_i - E_j)t}.
\end{align}
Using the unitarity relation
\begin{eqnarray}
  \sum_i U_{\alpha i} U_{\beta i}^* = \delta_{\alpha \beta},
  \label{unitary}
 \end{eqnarray}
we obtain
\begin{align}
 P_{\alpha \beta} = ~ &\delta_{\alpha \beta}
   - \bigg[\sum_{i<j} U_{\alpha i} U_{\beta j} U_{\alpha j}^* U_{\beta i}^* + \sum_{i<j} U_{\alpha i}^* U_{\beta j}^* U_{\alpha j} U_{\beta i} \bigg] \\ \nonumber
&+ \bigg[\sum_{i<j} U_{\alpha i} U_{\beta j} U_{\alpha j}^* U_{\beta i}^* e^{-i(E_i - E_j)t} + \sum_{i<j} U_{\alpha i}^* U_{\beta j}^* U_{\alpha j} U_{\beta i} e^{i(E_i - E_j)t} \bigg] \\ \label{N_gen_1} 
                 = ~ &\delta_{\alpha \beta} - 2  \sum_{i<j} \text{Re} \big(U_{\alpha i} U_{\beta j} U_{\alpha j}^* U_{\beta i}^* \big) \\ \nonumber
                  &+ 2  \sum_{i<j} \text{Re} \big( U_{\alpha i} U_{\beta j} U_{\alpha j}^* U_{\beta i}^* \big) \cos(E_i - E_j)t \\ \nonumber
                  &+ 2  \sum_{i<j} \text{Im} \big(U_{\alpha i} U_{\beta j} U_{\alpha j}^* U_{\beta i}^* \big) \sin(E_i - E_j)t. 
\end{align}
In  deriving Eq. \ref{N_gen_1} we used the following fact that if $z$ is a complex number then
\begin{eqnarray}
 z + z^* &=& 2 ~ \text{Re}(z), \\
 z -z^* &=& 2 ~ i ~ \text{Im}(z).
\end{eqnarray}
Thus the final form of the oscillation probability for $N$ generations is
\begin{align} \label{N_gen_2}
 P_{\alpha \beta} = ~ & \delta_{\alpha \beta} - 4 \sum_{i<j} \text{Re} \big( U_{\alpha i} U_{\beta j} U_{\alpha j}^* U_{\beta i}^* \big) \sin^2 \{\Delta_{ij}L/4E\} \\ \nonumber
                  & + 2 \sum_{i<j}  \text{Im} \big( U_{\alpha i} U_{\beta j} U_{\alpha j}^* U_{\beta i}^* \big) \sin\{2 \Delta_{ij}L/4E\},
 \end{align}
 where $\Delta_{ij}=m_i^2 - m_j^2$. Using the Eq. \ref{N_gen_2}, it is now straightforward to derive the corresponding expressions for three flavours. 

\subsection{Two Flavour Oscillation in Matter}

\begin{figure}[ht!]
\begin{center}
\vspace{0.3 in}
\includegraphics[scale=0.5]{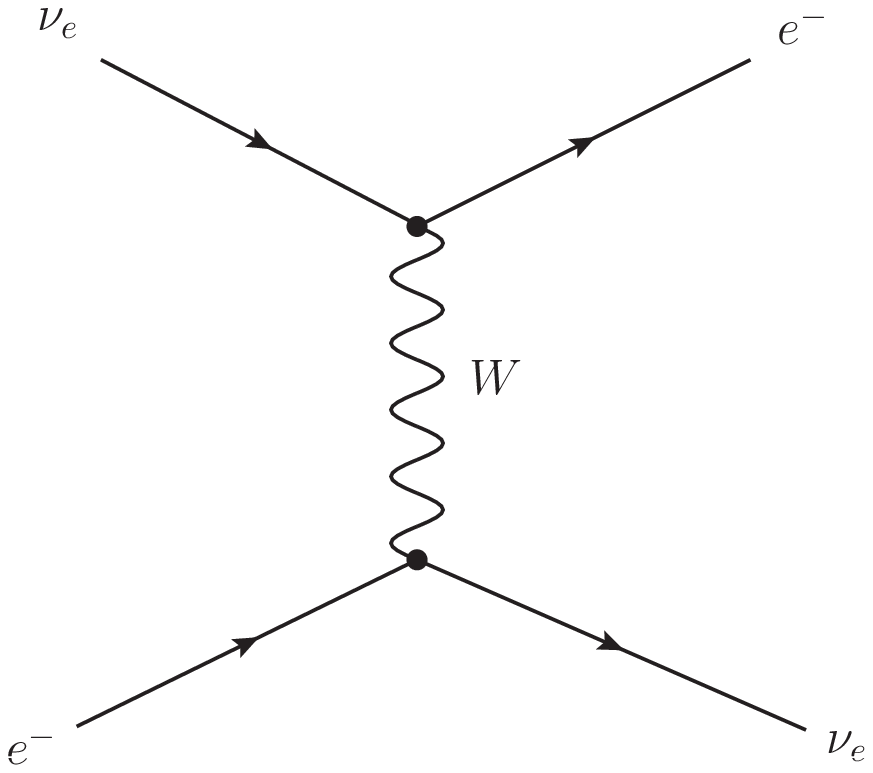}
\includegraphics[scale=0.5]{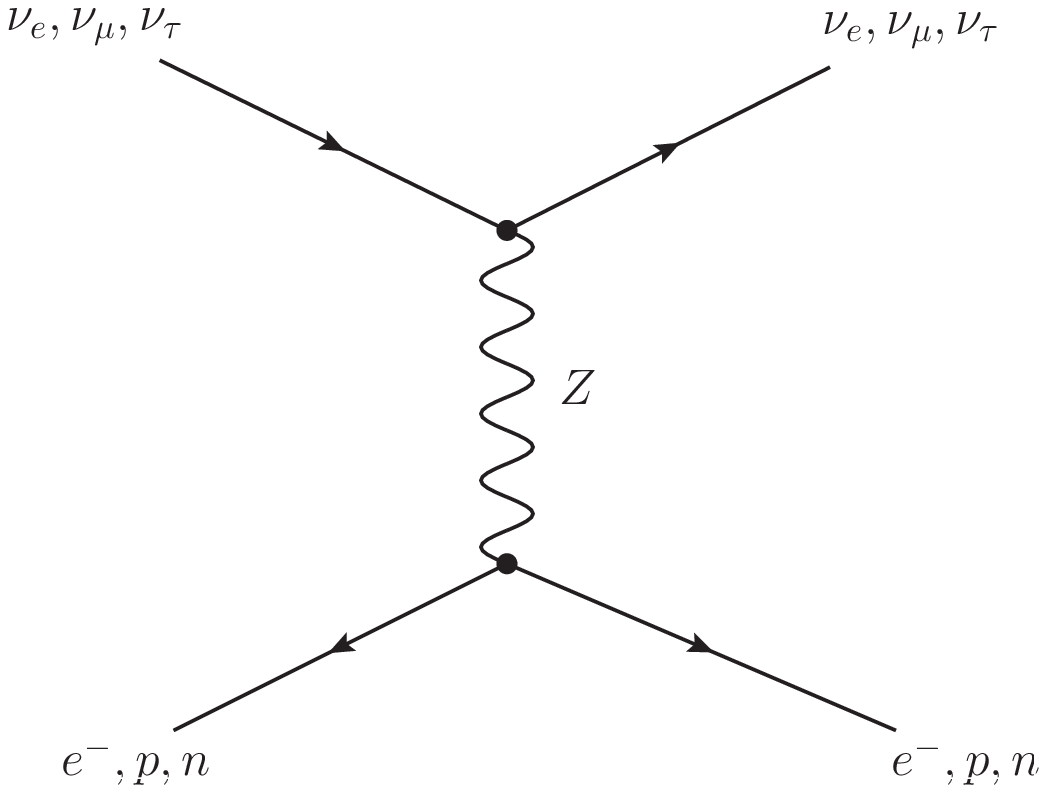}
\caption{CC and NC interactions of neutrinos.}
\label{fig:figure1}
\end{center}
\end{figure}

Neutrino propagation in matter modifies the neutrino oscillation probabilities. When active neutrino flavours traverse through matter, their
evolution equation is affected by the potentials due to the interactions with the medium through coherent
forward elastic weak charge current (CC) and neutral current (NC) scatterings. The charge current interactions affect only $\nu_e$
since normal matter consists of electron, proton and neutron
but the neutral current
interactions affect all the three active neutrinos. These interactions can be represented by the Feynman diagrams given in Fig. \ref{fig:figure1}.
\vspace{5 mm}

As the NC scattering potential modifies the propagation equation for all the neutrinos in the same way, it does not have any effect on the final expressions of 
neutrino oscillation probabilities. The CC interaction affects only the electron neutrinos and it modifies the probability expression significantly. 
The effective Hamiltonian for
the CC interaction can be written as 
\begin{eqnarray}
 H_{{\rm eff}} =  \frac{G_F}{\sqrt{2}} \big[\bar{e} \gamma_\mu (1 - \gamma_5) \nu_e \big] \big[\bar{\nu}_e \gamma^\mu (1-\gamma_5) e \big],
\end{eqnarray}
where $G_F$ is the Fermi constant. Using the Fierz transformation we obtain
\begin{eqnarray}
 H_{{\rm eff}} = \frac{G_F}{\sqrt{2}} \big[\bar{e} \gamma_\mu (1 - \gamma_5) e \big] \big[\bar{\nu}_e \gamma^\mu (1-\gamma_5) \nu_e \big].
\end{eqnarray}
The interaction potential is given by the average of the effective Hamiltonian over the electron background i.e.,
\begin{eqnarray}
 \bar{H}_{{\rm eff}} = \frac{G_F}{\sqrt{2}} \langle \bar{e} \gamma_\mu (1 - \gamma_5) e \rangle \big[\bar{\nu}_e \gamma^\mu (1-\gamma_5) \nu_e \big].
\end{eqnarray}
In the non-relativistic limit using the explicit forms of Dirac spinors one can show that \cite{Pal:1991pm,Kuo:1989qe}
\begin{eqnarray}
 \langle \bar{e} \gamma_\mu \gamma_5 e \rangle &\sim& \text{spin}, \\
 \langle \bar{e} \gamma_i e \rangle  &\sim& \text{velocity}, \\
 \langle \bar{e} \gamma_0 e \rangle &=& N_e,
\end{eqnarray}
where $N_e$ is the electron number density of the medium.
In the rest frame of unpolarised electrons only $N_e$ term is non-zero and thus we obtain
\begin{eqnarray}
 \bar{H}_{{\rm eff}} &=& \sqrt{2} G_F N_e \bar{\nu}_{eL} \gamma^0 \nu_{eL} \\
               &=& v_{CC} \bar{\nu}_{eL} \gamma^0 \nu_{eL} \\
               &=& v_{CC} j_\nu,
\end{eqnarray}
where $\nu_{eL} = \frac{1 - \gamma_5}{2} \nu_{e}$, $j_\nu = \bar{\nu}_{eL} \gamma^0 \nu_{eL}$ and $v_{CC}$ is the interaction potential given by
\begin{eqnarray}
 v_{CC} = \sqrt{2} G_F N_e.
\end{eqnarray}

For antineutrinos, we have to consider the charge conjugate field $\nu_{eL}^C$ i.e.,
\begin{eqnarray}
 j_\nu^C &=& \bar{\nu}_{eL}^C \gamma^0 \nu_{eL}^C \\
         &=& - \nu_{eL}^T C^{-1} \gamma^0 C \bar{\nu}_{eL}^T, 
\end{eqnarray}
where $C$ is the charge conjugation operator and we have used the fact that
\begin{eqnarray}
 \nu_{eL}^C &=& C \bar{\nu}_{eL}^T,  \\
 \bar{\nu}_{eL}^C &=& - \nu_{eL}^T C^{-1}.
\end{eqnarray}
 Using the property
 \begin{eqnarray}
  C^{-1} \gamma^0 C = (-\gamma^0)^T,
 \end{eqnarray}
we obtain
\begin{eqnarray}
 j_\nu^C  &=&  \nu_{eL}^T (\gamma^0)^T  \bar{\nu}_{eL}^T  \\
          &=& - \bar{\nu}_{eL} \gamma^0  \nu_{eL},
\end{eqnarray}
and thus for antineutrinos the effective Hamiltonian becomes
\begin{eqnarray}
 \bar{H}_{{\rm eff}} &=& -\sqrt{2} G_F N_e \bar{\nu}_{eL} \gamma^0 \nu_{eL},
\end{eqnarray}
which gives
\begin{eqnarray}
\bar{v}_{CC}  = -\sqrt{2} G_F N_e,
\end{eqnarray}
for antineutrinos.

With the inclusion of the potential $v_{CC}$,
the evolution Eq. \ref{evolution} becomes
\begin{eqnarray}
i \frac{\partial}{\partial t}
\begin{pmatrix}
  \nu_e  \\
  \nu_\mu \\
 \end{pmatrix} 
     =  H_F^{{\rm matt}} 
   \begin{pmatrix}
  \nu_e  \\
  \nu_\mu \\
 \end{pmatrix},
\end{eqnarray}
with
 \begin{eqnarray}
  H_F^{{\rm matt}} =
  \begin{pmatrix} 
    - \frac{\Delta_{21}}{4E} \cos2\theta + v_{CC} & \frac{\Delta_{21}}{4E} \sin2\theta \\
     \frac{\Delta_{21}}{4E} \sin2\theta & \frac{\Delta_{21}}{4E} \cos2\theta \\
 \end{pmatrix}. 
 \label{matter}
 \end{eqnarray}
By defining
\begin{eqnarray}
 A = 2\sqrt{2} G_F N_e E,
\end{eqnarray}
and subtracting $\frac{A}{4E}$ from the diagonal elements,
Eq. \ref{matter} simplifies to
 \begin{eqnarray}
   H_F^{{\rm matt}} = \frac{1}{4E}
  \begin{pmatrix} 
      A- \Delta_{21} \cos2\theta  & \Delta_{21} \sin2\theta \\
      \Delta_{21} \sin2\theta &  -A +\Delta_{21} \cos2\theta \\
  \end{pmatrix}. 
  \end{eqnarray}
 The energy eigenvalues of $H_F^{{\rm matt}}$ are obtained by diagonalising the above:
\begin{eqnarray}
E_{1,2} = \frac{1}{4E}\left[A \pm \sqrt{(-A + \Delta_{21} \cos2\theta)^2 +(\Delta_{21} \sin2\theta)^2}\right].
\end{eqnarray}
Now remembering the fact that $E_2 - E_1 = (m_2^2 - m_1^2)/2E$, we obtain the modified mass squared difference in the presence of matter as
\begin{eqnarray}
\Delta_{21}^M  = \sqrt{(-A + \Delta_{21} \cos2\theta)^2 +(\Delta_{21} \sin2\theta)^2}.
\label{new_mass}
\end{eqnarray}

The above equation shows how the masses are modified in the presence of the matter term $A$. Now we will see how the mixing is being modified.
Let us assume that the modified mixing angle in the presence of matter is $\theta_M$ and we call the modified mixing matrix as $U_M$. The
matrix $H_F^{{\rm matt}}$ which is now in flavour basis can be converted into mass basis by the 
transformation $H_M^{{\rm matt}} = U_M^\dagger H_F^{{\rm matt}} U_M$. Setting the off-diagonal terms as zero we obtain
\begin{eqnarray}
 \tan2\theta_M = \frac{\Delta_{21} \sin2\theta}{-A + \Delta_{21} \cos2\theta},
\label{new_mixing}
\end{eqnarray}
and the expression for the probability for $P_{e \mu}$ becomes \footnote{In this derivation, we have used the constant matter density approximation.}
\begin{eqnarray}
 P_{e \mu} = \sin^22\theta_M \sin^2(1.27 \Delta_{21}^M L/E).
\end{eqnarray}

Note that, the expression for the vacuum oscillation probability
was not sensitive to the sign of $\Delta_{21}$ and octant of $\theta$ but due the modification in mass and mixing,
the expression is sensitive to both of them. Another interesting phenomenon in this case is the MSW (Mikheyev-Smirnov-Wolfenstein) resonance. This happens when
\begin{eqnarray}
 \Delta_{21} \cos2\theta  &=& A  \\
  &=& 0.76 \times 10^{-4} \bigg[\frac{\rho}{{\rm gm/cc}}\bigg] \bigg[\frac{E}{{\rm GeV}} \bigg] {\rm eV}^2.
\end{eqnarray}
If this condition is satisfied then we see that the mixing angle becomes 
maximal\footnote{For matter density of 4.15 gm/cc, which is relevant for baseline of 7000 km, resonance occurs at 7.5 GeV for a mass squared difference of $10^{-3}$ eV$^2$
and $\sin^22\theta = 0.1$.} i.e., $\pi/4$. 
This leads to the possibility of total transitions between the two flavours. 
Since for neutrinos $A$ is positive, resonance can only occur for $\Delta_{21} > 0$ and $\theta < \pi/4$ or $\Delta_{21} < 0$ and $\theta > \pi/4$. 
For antineutrinos the resonance condition 
is given by $\Delta_{21} > 0$ and $\theta > \pi/4$ or $\Delta_{21} < 0$ and $\theta < \pi/4$. 
From this it is clear that the enhancement of the neutrino and antineutrino probabilities depend on the sign of $\Delta_{21}$
and octant of $\theta$. Thus the experimental observation of this resonance effect can lead to the determination of the same.

\subsection{Three Flavour Oscillation in Matter: The OMSD Approximation}

In this section we discuss how the probability expressions can be derived, for three generations in matter of constant density. 
As we have mentioned earlier, in this case it is difficult to find exact
analytic expressions for the probabilities. 
In this section we will use the one mass scale dominance (OMSD) approximation \cite{Choubey:2003yp} in deriving the same. For the three generation
scenario the effective Hamiltonian in the flavour basis takes the following form
\begin{eqnarray}
 H_F^{{\rm matt}} = U 
 \begin{pmatrix}
  0 & 0 & 0 \\
  0 & \Delta_{21}/2E & 0 \\
  0 & 0 & \Delta_{31}/2E \\
 \end{pmatrix} 
 U^\dagger
 +
\begin{pmatrix}
  \sqrt{2} G_F N_e & 0 & 0 \\
  0 & 0 & 0 \\
  0 & 0 & 0 \\
 \end{pmatrix}.  
\end{eqnarray}
The OMSD approximation implies that the measured small mass squared difference $\Delta_{21}$ can be neglected as compared to $\Delta_{31}$. 
Under this approximation, the effects of the solar mixing angle $\theta_{12}$ and of the CP violating phase in
$U$ become inconsequential and $U$ simply becomes
\begin{eqnarray}
 U = R_{23} R_{13} = 
 \begin{pmatrix}
1 & 0 & 0 \\
0 & c_{23} & s_{23} \\
0 & -s_{23} & c_{23} \\
 \end{pmatrix}
\begin{pmatrix}
c_{13} & 0 & s_{13} \\
0 & 1 & 0 \\
-s_{13} & 0 & c_{13} \\
 \end{pmatrix}. 
\end{eqnarray}
Using this, the energy eigenvalues of $H_F^{{\rm matt}}$ can be obtained as
\begin{eqnarray}
 E_{1,3} &=& \frac{1}{4E}\left[\Delta_{31} + A \pm \sqrt{(\Delta_{31} \cos2\theta_{13} - A)^2 + (\Delta_{31} \sin2\theta_{13})^2} \right], \\
 E_2 &=& 0.
\end{eqnarray}
In this approximation the modified mixing matrix $U_M$ can be written as
\begin{eqnarray}
 U_M = R_{23} R_{13}^M.
\end{eqnarray}
Thus matter effect do not modify the mixing angle $\theta_{23}$. This can be qualitatively understood from the fact that 
matter effect only modifies the evolution equation for $\nu_e$ and mixing of $\nu_e$ with the mass eigenstates states does not involve the mixing angle 
$\theta_{23}$. Again in a similar manner as described in the two flavour case,
the relation between the modified mixing angle $\theta_{13}^M$ and the vacuum angle $\theta_{13}$ can be derived as
\begin{eqnarray}
 \tan2\theta_{13}^M = \frac{\Delta_{31} \sin2\theta_{13}}{\Delta_{31} \cos2\theta_{13} - A}.
\end{eqnarray}
In this case the expression for transition probability  $P_{\alpha \beta}$ can be derived from Eq. \ref{N_gen_2} by replacing $U$ by $U_M$.
Below we write the probability formula derived under the OMSD approximation for the transition $\nu_e \rightarrow \nu_\mu$:
\begin{eqnarray} \label{OMSD_ap}
 P_{e \mu} &=& \sin^2\theta_{23} \sin^22\theta_{13}^M \sin^2(\Delta_{31}^M L/4E), \\ \label{OMSD}
 P_{\mu \mu} &=& 1 - \cos^2\theta_{13}^M \sin^22\theta_{23} \sin^2\left(\frac{\Delta_{31} + A + \Delta_{31}^M}{4E}\right)L \\ \nonumber
             && ~ - \sin^2\theta_{13}^M \sin^2\theta_{23} \sin^2\left(\frac{\Delta_{31} + A - \Delta_{31}^M}{4E}\right)L \\ \nonumber
             && ~ - \sin^4\theta_{23} \sin^22\theta_{13}^M \sin^2\left(\frac{\Delta_{31}^M L}{4E}\right),
\end{eqnarray}
with
\begin{eqnarray}
\Delta_{31}^M = \sqrt{(\Delta_{31} \cos2\theta_{13} - A)^2 + (\Delta_{31} \sin2\theta_{13})^2}.
\end{eqnarray}
We will give the expression for the $P_{ee}$ channel in the appendix.

Let us now briefly discuss the validity condition of the OMSD approximation. The condition on the neutrino energy and baseline for the OMSD 
approximation to be valid is 
\begin{eqnarray}
 \frac{\Delta_{21} L}{E} \ll 1.
\end{eqnarray}
This corresponds to $L/E \ll 10^4$ km/GeV which is mainly the case for the atmospheric neutrinos. OMSD approximation also needs large values of $\theta_{13}$ because
the terms appearing with $\Delta_{21}$ can only be dropped if they are small compared to the leading order term containing $\theta_{13}$. We will discuss this
point again after discussing the $\alpha-s_{13}$ approximation which we will use in the next subsection to derive the most general three flavour oscillation expression
in matter.

\subsection{Three Flavour Oscillation in Matter: The $\alpha-s_{13}$ Approximation}

In this subsection we will give the derivation of the approximate three flavour 
probability expressions using the series expansion method \cite{akhmedov} in a constant matter density. 
We will study expansions in terms of
the mass hierarchy parameter $\alpha = \Delta_{21}/\Delta_{31}$ and mixing parameter $s_{13} = \sin\theta_{13}$ keeping terms up to second order.
The effective Hamiltonian in flavour basis can be written as
 \begin{eqnarray}
  H_F^{{\rm matt}} = \frac{\Delta_{31}}{2E} \big[ U \text{diag}(0, \alpha, 1) U^\dagger + \text{diag}(\hat{A}, 0, 0) \big],  
 \end{eqnarray}
 where $\hat{A} = A/\Delta_{31}$. In order to derive the double expansion, we write the above Hamiltonian as
 \begin{eqnarray}
  H_F^{{\rm matt}} = \frac{\Delta_{31}}{2E} R_{23} U_{\delta} M U_{\delta}^\dagger R_{23}^T,
 \end{eqnarray}
 where $U_{\delta} = \text{diag}(1, 1, e^{i \delta_{CP}})$. We define, 
\begin{eqnarray}
H_F^{\prime {{\rm matt}}} &=& \frac{\Delta_{31}}{2E} M \\
  &=& \frac{\Delta_{31}}{2E} \left[R_{13} R_{12} \text{diag}(0, \alpha, 1) R_{12}^T R_{13}^T + \text{diag} (\hat{A}, 0, 0)\right] \\
    &=&
    \begin{pmatrix}
     s_{12}^2 c_{13}^2 \alpha + s_{13}^2 + \hat{A} & \alpha c_{12} c_{13} s_{12} & s_{13} c_{13} (1 - \alpha s_{12}^2) \\
    s_{12} c_{12} c_{13} \alpha & \alpha c_{12}^2 & -\alpha c_{12} s_{12} s_{13} \\
    s_{13} c_{13} (1 - \alpha s_{12}^2) & - s_{12} c_{12} s_{13} \alpha & \alpha s_{12}^2 s_{13}^2 + c_{13}^2 \\
    \end{pmatrix}.
\end{eqnarray}
Diagonalisation is performed using perturbation theory up to second order in the small parameters $\alpha$ and $s_{13}$ i.e.,
\begin{eqnarray}
 M = M^0 + M^1 + M^2, 
\end{eqnarray}
where
\begin{eqnarray}
  M^0 = \text{diag}(\hat{A}, 0, 1) = \text{diag}(\lambda_1^0, \lambda_2^0, \lambda_3^0), 
\end{eqnarray}
\begin{eqnarray}
  M^1 = 
  \begin{pmatrix}
   \alpha s_{12}^2 & \alpha s_{12} c_{12} & s_{13} \\
   \alpha s_{12} c_{12} & \alpha c_{12}^2 & 0 \\
   s_{13} & 0 & 0 \\
  \end{pmatrix},
  \end{eqnarray}
  \begin{eqnarray}
  M^2 = 
  \begin{pmatrix}
   s_{13}^2 & 0 & - \alpha s_{13} s_{12}^2 \\
   0 & 0 & - \alpha s_{13} s_{12} c_{12} \\
   - \alpha s_{13} s_{12}^2 & - \alpha s_{13} s_{12} c_{12} & -s_{13}^2 \\
  \end{pmatrix}.
\end{eqnarray}
For eigenvalues we write
\begin{eqnarray}
 \lambda_i = \lambda_i^0 + \lambda_i^1 + \lambda_i^2,
\end{eqnarray}
and for the eigenvectors we write
\begin{eqnarray}
 v_i = v_i^0 + v_i^1 + v_i^2.
\end{eqnarray}
Since $M^0$ is already diagonal we have
\begin{eqnarray}
 v_i^0 = e_i,
\end{eqnarray}
i.e.,
\begin{eqnarray}
 v_1^0 = 
 \begin{pmatrix}
  1 \\
  0 \\
  0 \\
 \end{pmatrix}, \hspace{3 mm} 
 v_2^0 = 
 \begin{pmatrix}
  0 \\
  1 \\
  0 \\
 \end{pmatrix}, \hspace{3 mm}
 v_3^0 = 
 \begin{pmatrix}
  0 \\
  0 \\
  1 \\
 \end{pmatrix}.
\end{eqnarray}
Now the first and second order corrections to the eigenvalues are given by
\begin{eqnarray}  \label{eigenvalue_1}
 \lambda_i^1 &=& M_{ii}^1 = \langle v_i^0|M^1|v_i^0\rangle, \\
 \lambda_i^2 &=& M_{ii}^2 + \sum_{j \neq i} \frac{(M_{ii}^1)^2}{\lambda_i^0 - \lambda_j^0}, \label{eigenvalue_2}
\end{eqnarray}
and the corrections to the eigenvectors are given by \footnote{These expressions include the normalization factors \cite{QM_MV}.}
\begin{eqnarray} \label{eigenvector_1}
 v_i^1 &=& \sum_{j \neq i} \frac{M_{ij}^1}{\lambda_i^0 - \lambda_j^0} e_j, \\ \label{eigenvector_2}
 v_i^2 &=& \sum_{j \neq i} \frac{1}{\lambda_i^0 - \lambda_j^0} \big[M_{ij}^2 + (M^1 v_i^1)_j - \lambda_i^1 (v_i^1)_j \big] e_j.
\end{eqnarray}
Using Eqs. \ref{eigenvalue_1} and \ref{eigenvalue_2} and keeping in mind the fact that $E_i = \frac{\Delta_{31}}{2E} \lambda_i$, 
we obtain the following expressions for energy eigenvalues 
\begin{eqnarray} \label{energy_1}
 E_1  &=& \frac{\Delta_{31}}{2E} \big(\hat{A} + \alpha s_{12}^2 + s_{13}^2 \frac{\hat{A}}{\hat{A}-1} + \alpha^2 \frac{\sin^22\theta_{12}}{4\hat{A}} \big), \\ \label{energy_2}
 E_2 &=& \frac{\Delta_{31}}{2E} \big( \alpha c_{12}^2 - \alpha^2 \frac{\sin^22\theta_{12}}{4\hat{A}} \big), \\ \label{energy_3}
 E_3 &=& \frac{\Delta_{31}}{2E} \big( 1 - s_{13}^2 \frac{\hat{A}}{\hat{A} - 1} \big),
\end{eqnarray}
 and using Eqs. \ref{eigenvector_1} and \ref{eigenvector_2} we get the three eigenvectors as
 \begin{eqnarray}
 v_1 &=& 
 \begin{pmatrix}
  1 \\
  \frac{\alpha \sin2\theta_{12}}{2\hat{A}} +\frac{\alpha^2 \sin4\theta_{12}}{4\hat{A}^2} \\
  \frac{s_{13}}{\hat{A} - 1} - \frac{\hat{A} \alpha s_{13} s_{12}^2}{(\hat{A} -1)^2} \\
 \end{pmatrix}, \hspace{3 mm} 
 v_2 = 
 \begin{pmatrix}
  -\frac{\alpha \sin2\theta_{12}}{2\hat{A}} - \frac{\alpha^2 \sin4\theta_{12}}{4 \hat{A}^2} \\
  1 \\
  \frac{\alpha s_{13} \sin2\theta_{12} (\hat{A}+1)}{2\hat{A}} \\
 \end{pmatrix},   \\ \nonumber
 \text{and} \hspace{3 mm}
 v_3 &=& 
 \begin{pmatrix}
  -\frac{s_{13}}{\hat{A} - 1} + \frac{\hat{A} \alpha s_{13} s_{12}^2}{(\hat{A} - 1)^2} \\
  \frac{\hat{A} \alpha s_{13} \sin2\theta_{12}}{2(\hat{A} - 1)} \\
  1 \\
 \end{pmatrix}.
\end{eqnarray}
 With these, the modified mixing matrix in matter is given  by
\begin{eqnarray}
 U_M = R_{23} U_{\delta} W,
\end{eqnarray}
with $W = (v_1, v_2, v_3)$. 

Now it is straightforward to obtain the expressions for oscillation probabilities from Eq. \ref{N_gen_2} using the elements
of $U_M$ and expressions derived in the Eqs. \ref{energy_1}, \ref{energy_2} and \ref{energy_3}. Below we write down the expressions corresponding to the
transition probability $\nu_\mu \rightarrow \nu_e$\footnote{Note that $P_{\mu e}$ is the time reversal state of $P_{e \mu}$. 
Thus the expression of $P_{\mu e}$ can be obtained by replacing $\dcp$ of $P_{e \mu}$ by $-\dcp$.}  and the leading order term for $\nu_\mu \rightarrow \nu_\mu$: 
\begin{align} \label{alpha_s13}
 P_{\mu e} = ~ & 4 s_{13}^2 s_{23}^2 \frac{\sin^2(\hat{A} - 1) \Delta}{(\hat{A} - 1)^2} \\ \nonumber
           &+ 2 \alpha s_{13} \sin2\theta_{12} \sin2\theta_{23} \cos(\Delta + \delta_{CP}) \frac{\sin\hat{A}\Delta}{\hat{A}} \frac{\sin(\hat{A} - 1)\Delta}{\hat{A} - 1} \\ \nonumber
           &+ \alpha^2 \sin^22\theta_{12} c_{23}^2 \frac{\sin^2\hat{A}\Delta}{\hat{A}^2}, \label{alpha_s13_disap} \\
 P_{\mu \mu} = ~ & 1 - \sin^22\theta_{23} \sin^2\Delta + \text{higher order terms},            
\end{align}
where $\Delta = \Delta_{31}/4E$. For antineutrinos, the relevant formula can be obtained by $\hat{A} \rightarrow -\hat{A}$ and $\delta_{CP} \rightarrow -\delta_{CP}$.
The expressions for IH can be obtained by replacing $\Delta$ by $-\Delta$ and $\hat{A}$ by $-\hat{A}$. 
We will give the full expressions for $P_{\mu \mu}$ and $P_{ee}$ in the appendix.

Formally this calculation is based upon the approximation $\alpha, s_{13} \ll 1$ and no explicit assumptions about the values of $L/E$ are made. However, we
remark that the series expansion formula is no longer valid as soon as $\alpha \Delta = \Delta_{21} L/4E$ becomes order of unity, i.e., when the oscillatory
behaviour is governed by the mass squared difference $\Delta_{21}$. As this is not the case for the current generation long-baseline experiments, we will
use this formula to understand the oscillation physics for the same. But this condition can occur for very long baselines and/or very low energies where 
these equations will no longer be applicable.

After discussing the validity of the $\alpha-s_{13}$ approximation, let us go back to the validity of the OMSD approximation. We have already discussed 
the fact that the OMSD approximation will fail if $\theta_{13}$ is too small. If we compare the first and second terms of Eq. \ref{alpha_s13}, it is
obvious that the first term can be safely neglected in comparison to second if
\begin{eqnarray}
\sin\theta_{13} \gg \alpha.
\end{eqnarray}
Now using current best-fit values of $\Delta_{21}$ and $\Delta_{31}$, the above condition translates to 
\begin{eqnarray}
 \sin\theta_{13} \gg 0.03,
\end{eqnarray}
which is consistent with the current value of this parameter.

Note that, as the OMSD approximation is exact in $\theta_{13}$, the physics near the resonance region can be explained better using this
approximation as compared to the $\alpha-s_{13}$ approximation. 
For this reason one can use the OMSD approximation to understand the oscillation results of
the atmospheric neutrino experiments. For the baselines involved in these experiments the MSW resonance effect is relevant.
On the other hand the $\alpha-s_{13}$ approximation is appropriate for
explaining the physics of the current generation long-baseline experiments for probing the sub-leading effect of $\dcp$. 
One can check this by solving
the full three flavour neutrino propagation equation numerically assuming the Preliminary
Reference  Earth  Model  (PREM)  density  profile  for
the  Earth \cite{Gandhi:2004md} and comparing this with the various analytic expressions. 

\begin{figure}[ht!]
\begin{center}
\vspace{0.3 in}
\includegraphics[scale=0.9]{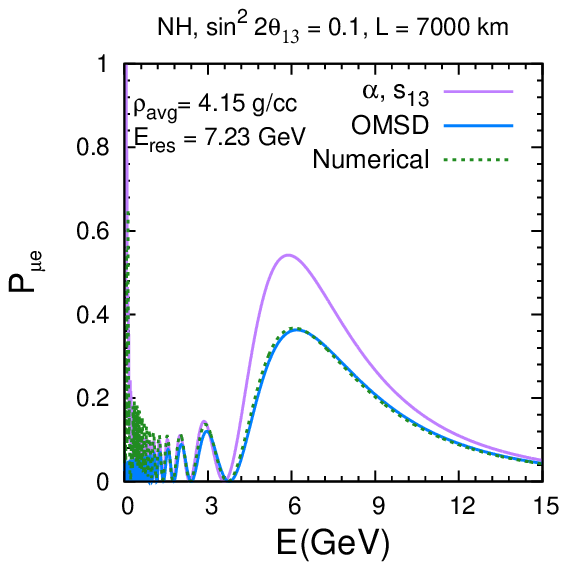}
\hspace{-0.8 in}
\includegraphics[scale=0.9]{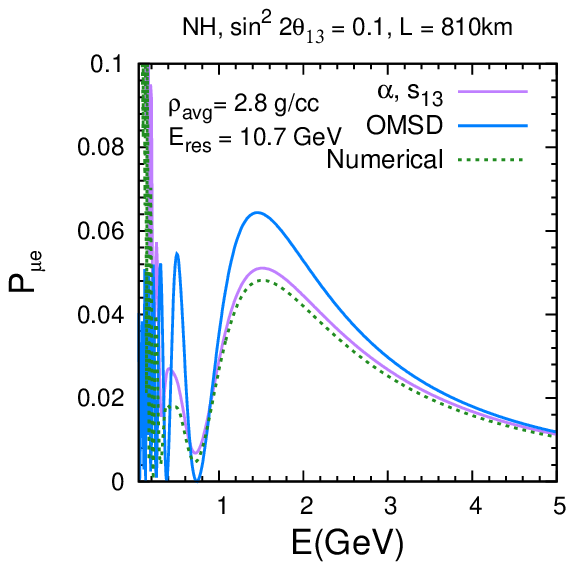}
\caption{Comparison between OMSD and $\alpha-s_{13}$ approximation.}
\label{compare}
\end{center}
\end{figure}

 As for example, in Fig. \ref{compare} we have plotted $P_{\mu e}$ as a function of energy for the baselines 7000 km and 810 km.
From the figure we can notice that for $L=7000$ km and $E=7.23$ GeV which is the MSW region relevant for the case of the atmospheric neutrinos, 
the OMSD approximation matches perfectly with the numerical calculation \footnote{
We have done the numerical estimation using the GLoBES \cite{globes1}(General Long Baseline Experiment Simulator) software taking PREM density profile.}. 
On the other hand for the baseline of 810 km,
where the first oscillation maximum lies very far from the resonance energy,
the $\alpha-s_{13}$
approximation gives better estimation than the OMSD approximation.


\section{Current Status of the Oscillation Parameters}

In this section we discuss the current status of neutrino oscillation parameters.
The mass and mixing parameters that describe the oscillation of the three generation neutrinos,
are divided into three categories (except the leptonic phase $\dcp$): the solar neutrino parameters
i.e., $\theta_{12}$, $\Delta_{21}$, the atmospheric neutrino parameters i.e., $\theta_{23}$, $\Delta_{31}$ and reactor neutrino parameter i.e., $\theta_{13}$.
The parameters are termed like this because the oscillation in the respective sectors are governed mainly by these parameters.
Specifically the parameters $\theta_{12}$ and $\Delta_{21}$ are mainly constrained from the solar neutrino experiments and 
KamLAND reactor data.
The accelerator based long-baseline experiments (MINOS, T2K) 
constrain the  parameters $|\Delta_{31}|$, $\theta_{23}$, and $\delta_{CP}$.
The parameters $|\Delta_{31}|$ and $\theta_{23}$ are also constrained from Super-Kamiokande.
The reactor data (Daya-Bay, RENO and Double-Chooz) constrain $\theta_{13}$ 
and $\Delta_{ee}\big(=s_{12}^2 \sin^2(\frac{\Delta_{32} L}{4E}) + c_{12}^2 \sin^2(\frac{\Delta_{31} L}{4E})\big)$.
Note that, as non-zero value of $\theta_{13}$ affects both solar 
and atmospheric oscillation results, it plays an important role in the global fit of world neutrino data.

\begin{figure}[ht!]
\begin{center}
\vspace{0.3 in}
\includegraphics[scale=0.5]{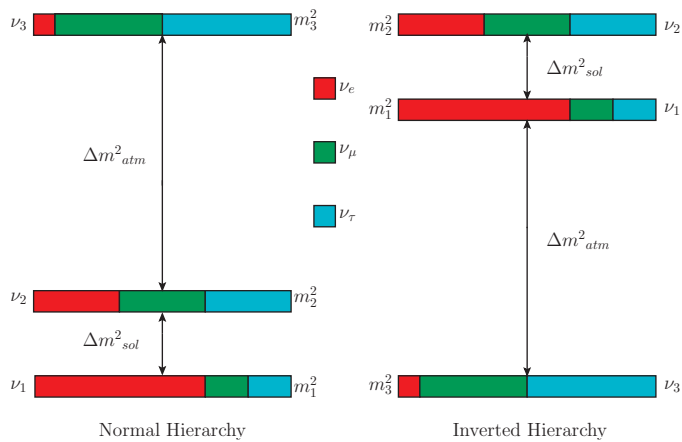}
\caption{Two possible mass orderings of neutrinos.}
\label{fig:figure2}
\end{center}
\end{figure}

At present, one of the major unknowns in the three flavour oscillation picture is the
sign of the atmospheric mass squared difference $\Delta_{31}$ or
the neutrino mass hierarchy.
The $+$ve sign of $\Delta_{31}$ corresponds to $m_3 > m_1$ which is known as normal hierarchy (NH) and 
$-$ve sign of $\Delta_{31}$ implies $m_3 < m_1$ which is known as inverted hierarchy\footnote{Note that
apart from these two possible mass orderings, neutrino mass spectrum can also be quasi-degenerate (QD) 
i.e., $m_1 \approx m_2 \approx m_3$.}(IH) (shown in Fig. \ref{fig:figure2}).
The second unknown in this sector is the octant of $\theta_{23}$. 
If $\theta_{23}$ is less than $45^\circ$,
then the octant of $\theta_{23}$ is lower (LO) and if $\theta_{23}$ is greater than $45^\circ$ then the octant of $\theta_{23}$ is higher (HO). 
The last remaining unknown in the three flavour framework is the value
of the leptonic phase $\delta_{CP}$. 

\begin{figure}[ht!]
\begin{center}
\vspace{0.3 in}
\includegraphics[scale=1.0]{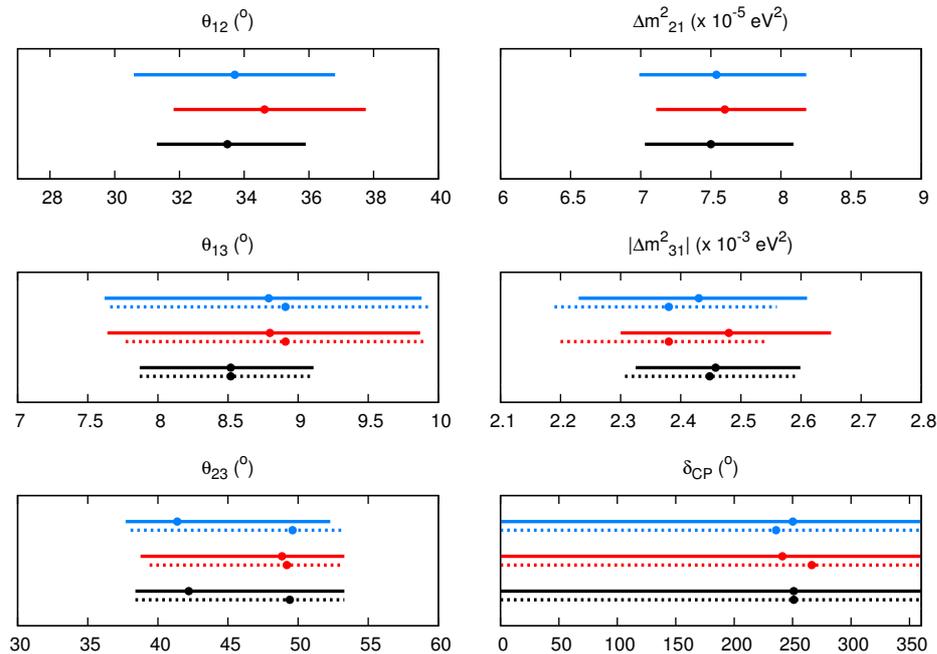}
\caption{Comparison of the best-fit values and $3\sigma$ ranges of 
the oscillation parameters. The figure is taken from Ref. \cite{sruba_ichep14}.}
\label{fig:figure3}
\end{center}
\end{figure}

Next we discuss 
the current status of these
parameters in detail.
Currently there are three groups doing the global analysis of the world neutrino data. We have given the results of the latest global analysis by the Nu-fit
group \cite{global_nufit} in Table \ref{global}  and compared the results of different groups in Fig. \ref{fig:figure3} in terms of best-fit values
and $3\sigma$ ranges.
The blue, red and black lines correspond to the analysis by the Bari group \cite{global_fogli}, IFIC group \cite{global_valle} and Nu-fit group respectively.
The solid (dashed) line corresponds to NH (IH).
\begin{table}[t]
\begin{center}
  \begin{tabular}{|ccc|}
    \hline
    parameter & present value & precision  
    \\
    \hline
    $\frac{\Delta m^2_{21}}{10^{-5}~\mathrm{eV}^2}$
    &  7.50$^{+0.19}_{-0.17}$ & 2.3\%  \\[1.5mm] 
    $\sin^2\theta_{12}$
    & $0.304^{+0.012}_{-0.012}$ & 4\%\\[2.5mm]  
    $\frac{|\Delta m^2_{31}|}{[10^{-3}~\mathrm{eV}^2}$
    & +2.458$^{+0.002}_{-0.002}$ 
    & 2\% 
    \\[4.5mm]
    $\frac{|\Delta m^2_{32}|}{[10^{-3}~\mathrm{eV}^2}$
    & -2.458$^{+0.002}_{-0.002}$ 
    & 2\% 
    \\[4.5mm]
    $\sin^2\theta_{23}$
    &
0.451$^{+0.001}_{-0.001} \oplus 0.577^{+0.027}_{-0.035}$
& 7.5\%
    \\[4mm]
    $\sin^2\theta_{13}$
    &
$0.0219^{+0.0010}_{-0.0011}$
   &
    5\% \\[4mm]
    $\delta_{CP}$
   &
   \begin{tabular}{c}
 $0.80\pi$ (NH) \\
     $-0.03\pi$ (IH)
   \end{tabular}
   &
   $0-2\pi$  \\
       \hline
\end{tabular}
\caption{The best-fit values and $3\sigma$ ranges of
 neutrino oscillation parameters from global analysis by the Nu-fit group \cite{global_nufit}. }
 \label{global}
\end{center}
\end{table}
From Fig. \ref{fig:figure3} we can see that except $\theta_{23}$, the global analysis results for the other parameters, are consistent among the three groups.
For the mixing angles, best-fit values of $\theta_{12}$ and $\theta_{13}$ are around $34^\circ$ and just below
$9^\circ$ respectively. For mass squared differences, the best-fit values of $\Delta_{21}$ and $|\Delta_{31}|$ come around $7.5 \times 10^{-5}$ eV$^2$ and $2.4 \times 10^{-3}$ eV$^2$
respectively. The analysis also shows that, at this moment both the hierarchies give equally good fit to the data. Regarding the phase $\delta_{CP}$, we can see that the current
data signal a best-fit value around $250^\circ$. This hint is mainly driven by the T2K appearance channel measurement and data from the reactors measuring $\theta_{13}$. 
But this signal is not statistically significant as at $3\sigma$, the full $\delta_{CP}$ range becomes allowed. 

Now let us discuss the 
case of $\theta_{23}$ in detail. In their analysis, the Bari group has fitted $\theta_{23}$ separately for NH and IH. When the data from long-baseline,
solar and KamLAND
are combined, they get the best-fit of $\theta_{23}$ in the higher octant for both NH and IH. But after addition of the reactor data, whose main effect is to reduce the
$\theta_{13}$ uncertainty, the best-fit for NH shifts to the lower octant.
The conclusions after adding the Super-Kamiokande atmospheric data are different before and after the Neutrino 2014 conference.
Before Neutrino 2014, due to the addition of the Super-Kamiokande data, the best-fit for IH shifted to the
lower octant but after Neutrino 2014 addition of Super-Kamiokande data only increases the significance of the fit for the lower octant in NH
and for IH the best-fit remains in the higher octant \cite{sruba_ichep14}. 
Thus according to results from the global analysis by the Bari group, $\theta_{23}$ seems to prefer the LO for NH and and HO for IH.
This result is different with their previous analysis result \cite{global_fogli_old}, where for both the hierarchy best-fit was obtained in the lower octant. 
But the conclusions are somewhat different according to the analysis of IFIC group, which also do separate fits for NH and IH.  From the analysis
of the solar, KamLAND and accelerator data, they found the best-fit to come in the lower octant for both NH and IH.
By the addition of reactor data, the best-fit values for both
the hierarchies move to the
higher octant. In this case the addition of Super-Kamiokande data do not have any effect. Thus the analysis of IFIC group shows that $\theta_{23}$ prefers HO for both
the hierarchies. Now let us come to the results of Nu-fit group. The Nu-fit group does not analyse NH and IH separately. 
In their analysis they get the global best-fit of $\theta_{23}$ in the higher octant and for IH. 
For NH, a local minima is obtained. So in conclusion we can say that, from the present available information of the world neutrino data, there is yet
no clear hint about the octant of $\theta_{23}$ and more data from present/future experiments are expected to address this issue. 

Now let us discuss a bit more about of the present unknowns in terms of oscillation probabilities. 

\begin{itemize}
 \item As discussed earlier, matter effect plays a key role in the determination of neutrino mass hierarchy.
 Note that for the solar neutrinos, the MSW resonance effect in the core of the Sun implies that 
 the solar mixing angle $\theta_{12}$ is in the first octant and this gives $m_2 > m_1$ \cite{Goswami:2003ek}. 
 For the determination of the sign of $\Delta_{31}$, one needs to study the Earth matter effect of neutrino oscillation.
 In Table \ref{resonance}, we list the resonance energies for different baselines in the Earth as given in \cite{ushier}. From these we understand that 
 the energies of the current generation long-baseline experiments
 are very far from the resonance energies to observe the MSW resonance effect. But still
 these experiments can have hierarchy sensitivity coming from the electron neutrino appearance channel depending on the length of the baseline.
 At present T2K and \nova\ are the examples of such long-baseline experiments which have given data
 corresponding to appearance channel. But
 the comparatively smaller baseline of T2K does
 not allow it to measure neutrino mass hierarchy and
 the first results from \nova\ \cite{nova_recent} are also not yet statistically significant for getting a clear hint. 
 On the other hand the atmospheric neutrino experiments study the oscillation of the neutrinos having energy
 ranging from 1-10 GeV with the maximum available baseline equals to the diameter of the Earth.
 So it is possible to determine neutrino mass hierarchy
 in the atmospheric neutrino experiments via MSW resonance effect in both appearance and disappearance channel 
 \footnote{In atmospheric neutrinos the hierarchy sensitivity can also come from the
  disappearance channel. This is a sharp contrast in comparison with the long-baseline experiments because in this case, the larger matter effects
 also give hierarchy sensitivity for the disappearance channel.}. 
 But for the current Super-Kamiokande experiment, at this moment there is not enough statistics
 for obtaining any clear hint about the true mass hierarchy.

 \begin{table}
\begin{center}
  \begin{tabular}{|c|c|}
    \hline
    Baseline(Km) &  Resonance Energy (GeV)  
    \\
    \hline
    3000  & 9.4 \\ [1.5mm] 
    \hline
    5000  & 8.7 \\[2.5mm]  
    \hline
    7000  & 7.5  \\[4.5mm]
    \hline
    10000 & 6.6  \\[4.5mm]
    \hline
\end{tabular}
\caption{Resonance energies at different baselines assuming average PREM density profile as calculated in \cite{ushier}.}
\label{resonance}
\end{center}
\end{table}

\item Now let us discuss why the octant of $\theta_{23}$ is still unknown. 
Similar to that of hierarchy, the octant sensitivity of the long-baseline experiments come from only appearance channel (cf. Eqs. \ref{alpha_s13} and \ref{alpha_s13_disap}) 
and in the atmospheric experiments octant sensitivity arise from both appearance and disappearance channel (cf. Eqs. \ref{OMSD_ap} and \ref{OMSD}).
From the probability expressions in matter, we see that
the leading order term of the appearance channel depends on $\sin^2\theta_{23}$ and
thus the appearance channel probability is an increasing function of $\theta_{23}$. 
But in the disappearance channel probability, the leading order term $\sin^22\theta_{23}$ gives
equal probability\footnote{It is important to note that in the appearance channel, the octant sensitive term $\sin^2\theta_{23}$ appears with $\theta_{13}$.
Thus the precise measurement of $\theta_{13}$ also improves the octant sensitivity \cite{Minakata:2002jv}.
} for $\theta_{23}$ and $\pi/2 - \theta_{23}$.
But note that, though the disappearance channel do not give any octant sensitivity for the long-baseline experiments but 
this channel is useful for the precision measurement of $\theta_{23}$. 
Thus in principle it is possible to determine the octant of $\theta_{23}$ from the combination of appearance and disappearance channels in the
long-baseline experiments if $\theta_{23}$ is not very close to maximal. 
On the other hand, for atmospheric neutrinos, due to matter effect, the disappearance channel also contribute in the octant measurement. 
But at present the available statistics
is not enough to predict the correct octant of $\theta_{23}$.

\item In the previous section we have seen that under OMSD approximation the conversion probability for $\nu_e$ do not involve
$\delta_{CP}$ and in the $\alpha-s_{13}$ approximation, $\delta_{CP}$ appears in the sub leading term $\alpha$. So we understand that
$\delta_{CP}$ is a sub-leading effect in neutrino oscillation and thus it is the most difficult parameter to probe in the experiments. 
Though the atmospheric
neutrino experiments can have CP sensitivity in principle but they are not expected to measure $\delta_{CP}$ 
as we will show in the next chapter that their CP sensitivity is compromised due 
to their dependence on the direction of the
incoming neutrinos. On the other hand as reactor experiments are sensitive only to electron appearance channel which do not depend on $\delta_{CP}$, they are also
not capable of measuring this parameter.
So it is expected that the first hint of
$\delta_{CP}$ will come from the appearance channel measurement of the long-baseline experiments.
As we have already discussed, the current T2K data gives a hint of $\dcp$ about ($-90^\circ$), but this needs to be confirmed from the further
runs of T2K as well as data from future experiments.

\end{itemize}

Apart from the reasons mentioned above, the measurement of these above unknowns are also restricted due to the presence of parameter degeneracy, which we will discuss now.


\section{Degeneracies in Oscillation Parameters}

In this section we will discuss the degeneracies between the different oscillation parameters. The 
parameter degeneracy in the context of neutrino oscillations implies getting the same value of probability for different sets of
oscillation parameters. 
The sensitivity of an experiment depends on the number of events which are the functions of neutrino oscillation probabilities. 
This implies that in the presence of degeneracies, different sets of parameters can give equally good fit to the data making it difficult to determine the 
actual values of the parameters unambiguously. So a clear understanding of various degeneracies, their dependence on different oscillation parameters 
and their resolution
are very important for the determination of the unknown parameters.
Previously when $\theta_{13}$ was unknown, three types of degeneracies have been discussed widely in the literature: 
(i) the intrinsic ($\theta_{13}$, $\delta_{CP}$) degeneracy \cite{intrinsic}, (ii) hierarchy-$\dcp$ degeneracy \cite{Minakata:2001qm}, and
(iii) the degeneracy of octant of $\theta_{23}$ \cite{lisi}. 
The intrinsic
degeneracy of the $P_{\mu e}$ channel refers to the same value of probability coming from a different $\theta_{13}$ and $\dcp$ value and can be expressed as
\begin{eqnarray}
 P_{\mu e}(\theta_{13}, \dcp) = P_{\mu e}(\theta_{13}^{\prime}, \dcp^{\prime}).
 \label{intrinsic}
\end{eqnarray}
The hierarchy-$\dcp$
degeneracy of the $P_{\mu e}$ channel leads to wrong hierarchy solutions arising due to a
different value of $\dcp$
other than the true value. This degeneracy can be expressed mathematically as
\begin{eqnarray}
 P_{\mu e}(\text{NH}, \delta_{CP}) = P_{\mu e}(\text{IH}, \delta_{CP}^\prime).
 \label{hier_dcp}
\end{eqnarray}
The intrinsic octant degeneracy of the $P_{\mu \mu}$ channel refers to the
clone solutions occurring for $\theta_{23}$ and $\pi/2 - \theta_{23}$ and expressed as
\begin{eqnarray}
 P_{\mu \mu}(\theta_{23}) = P_{\mu \mu}(\pi/2 - \theta_{23}).
 \label{octant}
\end{eqnarray}
These above mentioned degeneracies together (Eqs. \ref{intrinsic}, \ref{hier_dcp} and \ref{octant}) gave rise to a total eight fold degeneracy \cite{barger}.
It is important to note that there is no intrinsic octant degeneracy in the $P_{\mu e}$ channel as the dependence of $\theta_{23}$ in the leading order term of 
$P_{\mu e}$ channel goes as $\sin^2\theta_{13} \sin^2 \theta_{23}$. But due to the presence
of the $\sin^2\theta_{13}$ term, continuous allowed regions were obtained in the $\theta_{13} - \theta_{23}$ plane for a given value of $\dcp$ 
when $\theta_{13}$ was not known very precisely. 
There were several proposals on how to break these degeneracies to have a clean measurement of the neutrino oscillation parameters
\cite{Donini:2002rm,degeneracy1,twobase5,Narayan:1999ck,twobase1,twobase2,twobase3,twobase6,synergynt,
Huber:2003pm,menaparke,Mena:2005ek}.
It is shown that the intrinsic degeneracy can be removed to a large extent by using spectral information \cite{twobase2}.
At present, the measurement of the non-zero precise value of $\theta_{13}$ from the reactor
experiments resolves the degeneracies associated with $\theta_{13}$. The intrinsic degeneracy is largely resolved and
the octant sensitivity of the appearance channel has greatly improved. But due to the completely unknown value of $\dcp$, the hierarchy-$\dcp$
degeneracy still persists and there are also degenerate solutions arising due to different values of $\theta_{23}$ and $\dcp$. This degeneracy is
referred as octant-$\dcp$ degeneracy \cite{suprabhoctant}. 
We will now discuss the behaviour of these degeneracies in detail in the next subsections.

\subsection{The Hierarchy-$\delta_{CP}$ Degeneracy}
\label{bimagic}

\begin{figure}[ht!]
\begin{center}
\vspace{0.3 in}
\includegraphics[scale=0.9]{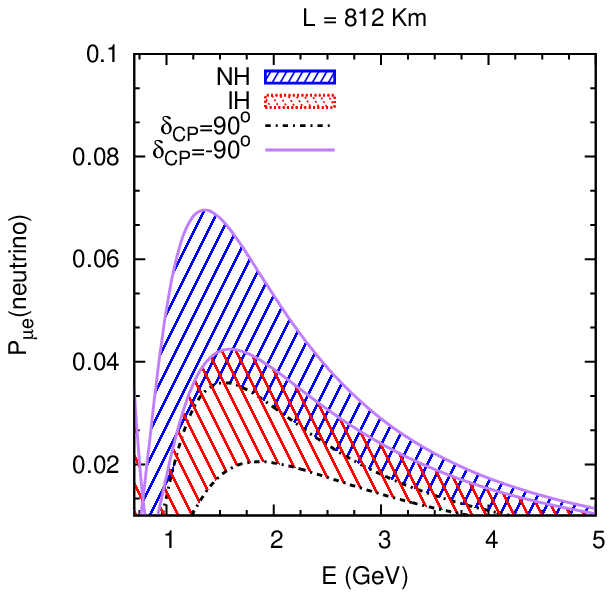}
\hspace{-0.8 in}
\includegraphics[scale=0.9]{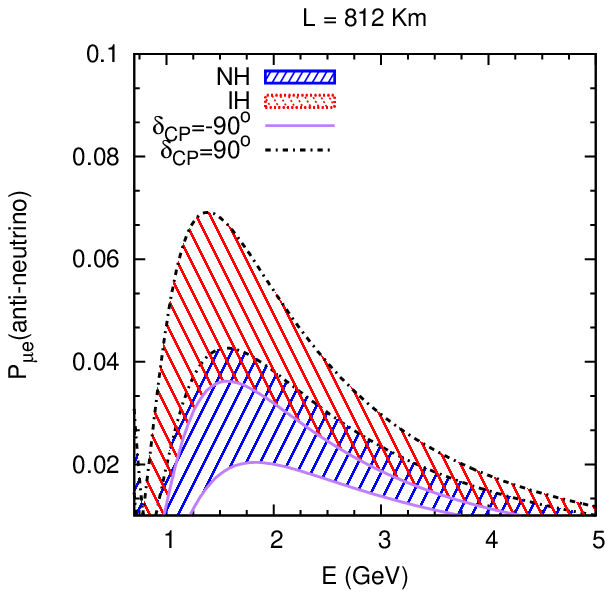}
\caption{$P_{\mu e}$ vs energy for L=812 km. The blue band corresponds to NH and the red band corresponds to IH.}
\label{fig:figure4}
\end{center}
\end{figure}

As discussed above the hierarchy-$\delta_{CP}$ degeneracy of the appearance channel is defined by
\begin{eqnarray}
 P_{\mu e}(\text{NH}, \delta_{CP}) = P_{\mu e}(\text{IH}, \delta_{CP}^\prime),
\end{eqnarray}
i.e., for a given octant, the probability for NH can be the same as that of probability for IH, for a different value of $\delta_{CP}$. To understand this degeneracy,
in Fig. \ref{fig:figure4} we have plotted the appearance channel probability vs energy for a baseline of 812 km.
At this baseline the constant matter density approximation holds good and we can use Eq. \ref{alpha_s13} to understand the behaviour
of Fig. \ref{fig:figure4}.
The left panel in Fig. \ref{fig:figure4} is for neutrinos and the right panel corresponds to antineutrinos. In this plot $\theta_{23}$ is fixed at $39^\circ$. 
The blue band corresponds to NH and the red band corresponds to IH. 
Here the values of $\Delta_{31}$ are taken as $2.4 (-2.4) \times 10^{-3}$ eV$^2$ corresponding to NH (IH). 
The width
of the bands is due to the variation of $\delta_{CP}$ from $-180^\circ$ to $+180^\circ$. 
As matter effect enhances the neutrino probability for NH and antineutrino probability for IH,
in these plots we see that for the neutrinos, NH probability is higher than IH and for the antineutrinos, IH probability is higher than NH. 
In the neutrino probability,
we observe that for each band (either NH or IH) $\delta_{CP}=-90^\circ$ corresponds to the highest point in the probability and $\delta_{CP}=90^\circ$
is the lowest point in the probability. This behaviour is opposite for the antineutrinos. This can be understood by having a close look at the Eq. \ref{alpha_s13}.
 At the oscillation maximum, $\Delta = 90^\circ$. Thus the $\delta_{CP}$ dependent term for neutrinos becomes maximum for $\delta_{CP} = -90^\circ$ and minimum for 
$\delta_{CP} = 90^\circ$. For antineutrinos as $\delta_{CP}$ changes its sign this behaviour gets reversed. In both the plots we
observe that there is an overlap between the NH band at $\delta_{CP}=+90^\circ$ and IH band at $\delta_{CP}=-90^\circ$. If the true values of the 
parameters fall in the overlapping region then we will have degenerate solutions. Based upon this observation if we divide the total range of $\delta_{CP}$
into two half-planes i.e., the lower half-plane (LHP: $-180^\circ < \delta_{CP} < 0^\circ$) and upper half-plane (UHP:  $0^\circ < \delta_{CP} < 180^\circ$),
then we see that the combination of hierarchy and $\delta_{CP}$ given by NH-LHP and IH-UHP lies far away from the overlapping regions. Hence these combinations do not
suffer from the hierarchy-$\dcp$ degeneracy and thus LHP (UHP) is the favourable half-plane for NH (IH). 
On the other hand the combination of NH-UHP and IH-LHP lies very close to the 
overlapping area and hence can give rise to hierarchy-$\delta_{CP}$ degeneracy. Thus UHP (LHP) is the unfavourable half-plane for NH (IH) in view of this degeneracy.
Here it is important to note that the favourable and unfavourable half-planes remain same in both neutrinos and antineutrinos.

\begin{figure}[ht!]
\begin{center}
\vspace{0.3 in}
\includegraphics[scale=0.9]{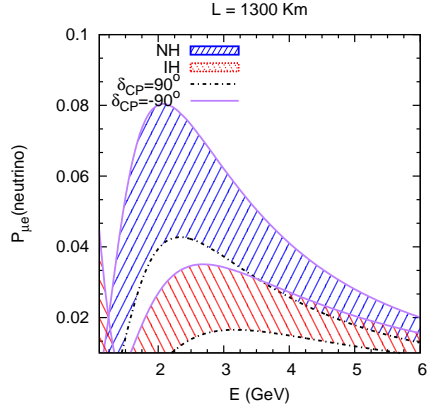}
\caption{Similar plot as that of Fig. \ref{fig:figure4} but for 1300 km baseline.}
\label{1300}
\end{center}
\end{figure}

One of the ways to overcome the hierarchy-$\dcp$ degeneracy is to look for neutrino oscillations at higher baselines. At longer baselines, there will be more
matter enhancement of the oscillation probabilities and this will cause more separation between the NH and IH bands. This is shown in Fig. \ref{1300}, where we can
see that near the oscillation maxima, there is no overlap between the NH and IH bands for the 1300 km baseline.

\begin{figure}[ht!]
\begin{center}
\vspace{0.3 in}
\includegraphics[scale=0.9]{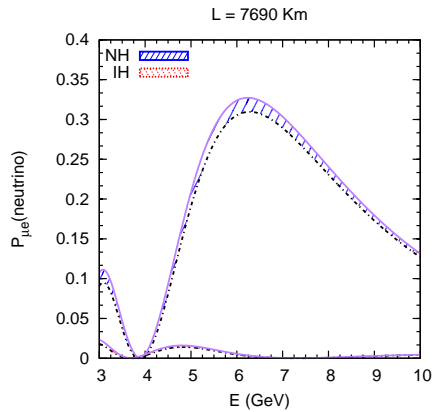}
\caption{Similar plot as that of Fig. \ref{fig:figure4} but for 7690 km baseline.}
\label{7690}
\end{center}
\end{figure}

\begin{figure}[ht!]
\begin{center}
\vspace{0.3 in}
\includegraphics[scale=0.9]{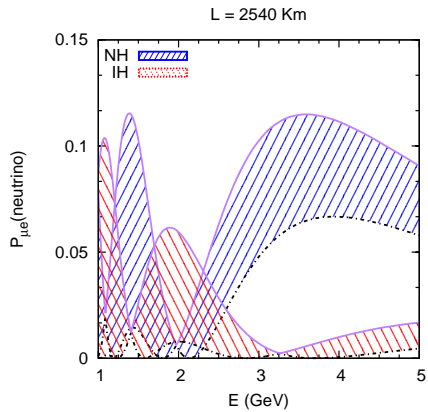}
\caption{Similar plot as that of Fig. \ref{fig:figure4} but for 2540 km baseline.}
\label{2540}
\end{center}
\end{figure}

Another elegant method to get rid of this degeneracy is the following: 
If one puts $\sin^2\hat{A}\Delta=0$ in Eq. \ref{alpha_s13}, then the probability expression becomes free from
the $\dcp$ term and thus there is no hierarchy-$\dcp$ degeneracy. This condition translates to $L = 7690$ km for both NH and IH
and this baseline is referred as the magic baseline \cite{Smirnov:2006sm}. This is reflected in Fig. \ref{7690}, where one can see that 
for $L = 7690$ km, the hierarchy-$\dcp$ degeneracy is completely absent. One also notices that, the width of the $\dcp$ band is very narrow. 
In spite of having this {\it magical} property, practically it is a very difficult task to design a neutrino oscillation experiment for a baseline of 7690 km.
As the flux of the neutrinos fall as $1/L^2$, one needs a very powerful collimated neutrino beam to have enough statistics at such a large distance
\footnote{There were several proposals to explore the physics at magic baseline using neutrino factories \cite{magic2} and beta beams \cite{Agarwalla:2006vf}.}. 
Another problem with the magic baseline is that,
it has no CP sensitivity. 
These problems lead to the question that can there be a shorter baseline where the hierarchy-$\dcp$ degeneracy can be removed and at the same time 
one can also have CP sensitivity. These questions were answered in \cite{bnlhs} by observing that 
if one puts $\sin^2(\hat{A} - 1) = 0$ in Eq. \ref{alpha_s13}, then also the CP sensitive term goes to zero. 
Unlike the magic baseline, this condition also depends on hierarchy.
If we now demand that there is no $\dcp$ dependence for IH and at the same time there is a probability maxima in NH, then one gets the following conditions
\begin{eqnarray}
 (1 + \hat{A} ) &>& n\pi ~~ {\rm for} ~~ n>0, \\ \nonumber
 (1 - \hat{A} ) &>& (m-1/2)\pi ~~ {\rm for} ~~ m>0.
\end{eqnarray}
Solving these two equations simultaneously one obtain $L \approx 2540$ km and $E \approx 3.3$ GeV for $n=m=1$. Now on the other hand if one demands that
there will be no $\dcp$ dependence for NH and a probability maxima in IH, then one obtains \cite{bimagic}
\begin{eqnarray}
 (1 - \hat{A} ) &>& n\pi ~~ {\rm for} ~~ n>0, \\ \nonumber
 (1 + \hat{A} ) &>& (m-1/2)\pi ~~ {\rm for} ~~ m>0,
\end{eqnarray}
and solutions of these two equations again yield $L \approx 2540$ km but $E \approx 1.9$ GeV for $n=1$, $m=2$. 
These features are shown in Fig. \ref{2540}, where one can see that for
$L=2540 $ km
there is no CP sensitivity in IH and probability maxima in NH at $E = 3.3$ GeV and at $E = 1.9$ GeV there is no CP sensitivity in NH and probability maxima in IH.
Thus for a baseline of $L=2540$ km,
one can have hierarchy sensitivity for both NH and IH with CP dependence in one hierarchy and with no CP dependence for the opposite hierarchy
though at different energies. This baseline was termed as the 
bi-magic baseline \cite{bnlhs,bnlhs_long,bimagic}. In the next chapter, while discussing the physics potential of the LBNO experiment
we will see that there is an exceptional hierarchy sensitivity at the CERN-Pyh\"{a}salmi baseline of $L = 2290$ km due to its proximity to the bi-magic baseline.

Note that in the atmospheric neutrinos experiments, oscillation takes place over the baselines ranging from 100 km to 12000 km experiencing huge Earth matter effects.
For this reason the effect of hierarchy-$\dcp$ degeneracy is less pronounced for the case of atmospheric neutrinos.

\subsection{The Octant-$\delta_{CP}$ Degeneracy}

\begin{figure}[ht!]
\begin{center}
\vspace{0.3 in}
\includegraphics[scale=0.9]{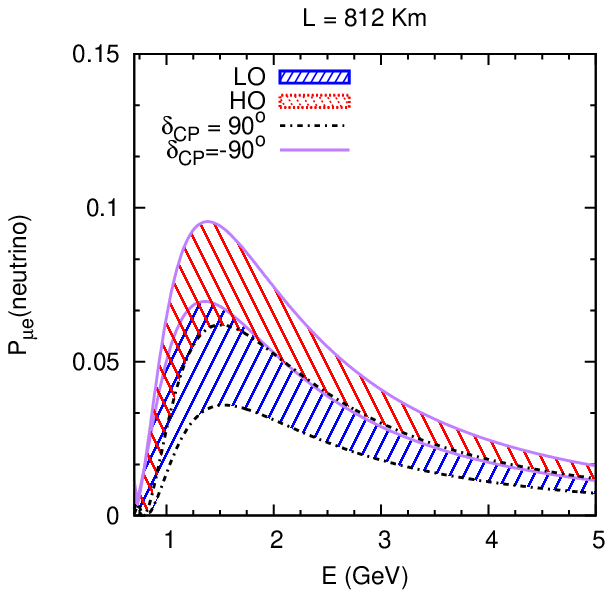}
\hspace{-0.8 in}
\includegraphics[scale=0.9]{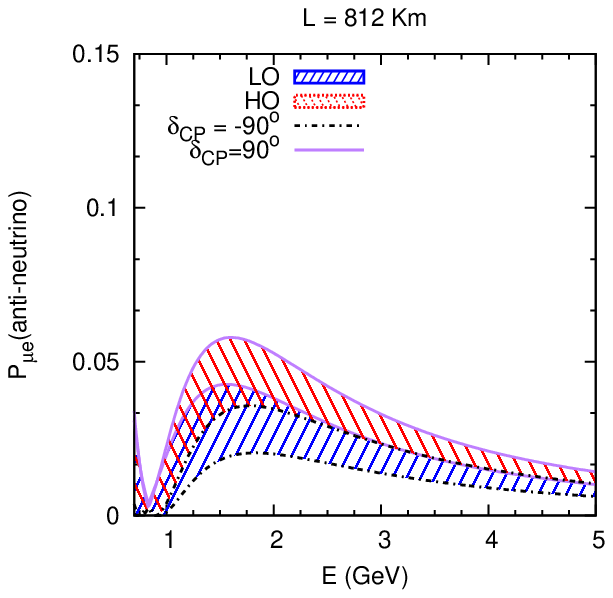}
\caption{$P_{\mu e}$ vs energy for L=812 km. The blue band corresponds to LO and the red band corresponds to HO.}
\label{fig:figure5}
\end{center}
\end{figure}

The octant-$\delta_{CP}$ degeneracy can be expressed mathematically in the following way
\begin{eqnarray}
P_{\mu e}(\text{LO}, \delta_{CP}) = P_{\mu e}(\text{HO}, \delta_{CP}^\prime), 
\end{eqnarray}
i.e., for a given hierarchy, the appearance channel probability for lower octant can be the same as the probability for the higher octant corresponding to a
different value of $\delta_{CP}$. To understand how this degeneracy behaves in neutrino and antineutrino probabilities, in Fig. \ref{fig:figure5}, we show the
appearance channel probability as a function of energy by fixing the hierarchy to be NH. This figure corresponds to a baseline of 812 km. 
The left panel is for neutrinos and
the right panel is for antineutrinos. The blue band corresponds to LO and red band corresponds to HO. 
In these plots LO (HO) corresponds to $\theta_{23} = 39^\circ (51^\circ)$.
Here also the width of the bands are due to the variation of $\delta_{CP}$ from $-180^\circ$ to $+180^\circ$. As the leading order term in the oscillation probability
depends on $\sin^2\theta_{23}$, we see that for both neutrinos and antineutrinos, the probability for HO is higher than LO. 
As discussed earlier, 
for neutrinos, in each band, 
$\delta_{CP}=-90^\circ$ corresponds to maximum point in the probability and $\delta_{CP}=90^\circ$ corresponds to the minimum point in the probability while opposite
is true for antineutrinos.
Now from the probability figure we observe that, for neutrinos the overlap is around LO-$\delta_{CP} = -90^\circ$ and HO-$\delta_{CP} = 90^\circ$
and for antineutrinos the overlap is around LO-$\delta_{CP} = 90^\circ$ and HO-$\delta_{CP} = -90^\circ$. 
So in this case for neutrinos, UHP (LHP) is the favourable half-planes for LO (HO), where it is less probable to have the octant-$\delta_{CP}$ degeneracy and
LHP (UHP) is the unfavourable half-planes for LO (HO) where one has the octant-$\delta_{CP}$ degeneracy. But this situation gets reversed in the antineutrinos i.e.,
the favourable regions for neutrinos become unfavourable in antineutrinos and the unfavourable regions for neutrino become favourable in antineutrinos.
This is the main difference between the hierarchy-$\delta_{CP}$ and octant-$\delta_{CP}$ degeneracy, where in the former case, the data from antineutrinos do not help
to resolve this degeneracy but for the later case, it is possible to resolve this degeneracy with balanced neutrino and antineutrino runs.

Here we would like to mention that as the atmospheric neutrinos consists of both neutrinos and antineutrinos, the octant-$\dcp$ degeneracy is
less for the case of atmospheric neutrino experiments. 
In this case the sub-leading terms of the disappearance channel\footnote{Which is proportional to $\sin^4\theta_{23}$. We have given the 
full matter expressions of $P_{\mu \mu}$ in the appendix.} also provide octant sensitivity
due to large matter effects. In this context it is important to note that in the atmospheric neutrinos
the uncertainty in $\theta_{13}$
do not affect the measurement of $\theta_{23}$.
This is because near the resonance region the $\sin^22\theta_{13}^M$ term in Eq. \ref{OMSD} becomes unity.
\paragraph*{}
From the above discussions we understand how the presence of hierarchy-$\delta_{CP}$ and octant-$\delta_{CP}$ degeneracy, can severely
affect the determination of the unknown oscillation
parameters. The hierarchy-$\delta_{CP}$ degeneracy gives rise to wrong hierarchy solutions, whereas octant-$\delta_{CP}$ gives rise to
wrong octant solutions. These two types of degeneracies together give rise to a generalised hierarchy-$\theta_{23}$-$\dcp$ degeneracy defined as
\begin{eqnarray}
 P_{\mu e}({\rm NH}, \theta_{23}, \dcp) = P_{\mu e}({\rm IH}, \theta_{23}^\prime, \dcp^\prime), 
\end{eqnarray}
which can give wrong hierarchy-wrong octant solution.
As hierarchy, octant and $\delta_{CP}$ are interlinked, the determination of one quantity depends on the information of the other 
quantities and the presence of degeneracies makes it difficult to extract the correct information as the true solution can often be faked by other degenerate solutions.
In \cite{suprabhoctant,t2knova} it was shown that the hierarchy and octant sensitivity of the long-baseline experiments suffers from these above mentioned degeneracies 
due to the unknown value of $\dcp$.
In the next chapter when we will study the physics potential of the various oscillation experiments, we will show how the lack of knowledge of hierarchy 
and octant, affect the CP measurement capabilities of the long-baseline experiments.


\section{Salient Features of the Present/Future Generation Oscillation Experiments}

In this section we discuss the main features of the currently running/upcoming oscillation experiments T2K, \nova, LBNO, LBNE and INO
which are expected to throw light on the three yet undetermined parameters of neutrino oscillation.
We will also briefly describe the IceCube experiment at south pole. Though the aim of the IceCube experiment
is to study the physical processes associated with the ultra high energy neutrinos of astrophysical origin, but one can also put constraint on the various
oscillation parameters by analysing its data. 
These are the experiments whose potentials have been studied in this thesis for determination of neutrino mass hierarchy, octant of $\theta_{23}$ and 
the leptonic CP phase $\dcp$.

\subsection{T2K}

T2K (Tokai to Kamioka) is a long-baseline experiment in Japan looking for neutrino oscillations in both appearance and disappearance channels \cite{t2k}.
T2K uses Super-Kamiokande (SK) as the far detector, which is a water \cnv\ detector located in the Kamioka 
Observatory, Gifu, Japan, at a distance of 295 km from
J-PARC high-intensity proton accelerator. The beam power is
0.75 MW with mean neutrino energy as 0.76 GeV. The neutrino beam is directed 2.5 degrees off-axis from the SK detector,
in order to produce a narrow-band flux at 0.6 GeV.
The experiment also includes two near detectors at a distance 280 m (INGRID and ND280).  
Construction of the neutrino beamline started in April 2004. After successful installation of the accelerator and neutrino beamline in 
2009, T2K began accumulating neutrino beam data for physics analysis in January 2010. But it was interrupted for one year due to the Great East Japan
Earthquake in 2011. Up to now T2K has collected data corresponding to the exposure of $6.6 \times 10^{20}$ POT (Protons on Target) in neutrino mode
and currently taking data in antineutrino mode. The recent measurements of the oscillation parameters in T2K can be found in \cite{Abe:2015awa}.
We summarize the details of data taking of T2K in Table \ref{t2k_data}.

\begin{table}
\begin{center}
  \begin{tabular}{|ccc|}
    \hline
    Run Period & Dates & POT  
    \\
    \hline
    Run 1 & Jan. 2010 - Jun. 2010 & $0.32 \times 10^{20}$  \\ [1.5mm] 
    Run 2 & Nov. 2010 - Mar. 2011 & $1.11 \times 10^{20}$  \\[2.5mm]  
    Run 3 & Mar. 2012 - Jun. 2012  & $1.58 \times 10^{20}$ \\[4.5mm]
    Run 4 & Oct. 2102 - May 2013   & $3.56 \times 10^{20}$  \\[4.5mm]
    \hline
    Total & Jan. 2010 - May 2013 & $6.57 \times 10^{20}$ \\
       \hline
\end{tabular}
\caption{The data taking period of T2K in neutrino mode as given in \cite{Abe:2015awa}.}
\label{t2k_data}
\end{center}
\end{table}

\subsection{NO$\nu$A}

The NuMI Off-axis $\nu_e$ Appearance (NO$\nu$A)  experiment at Fermilab is a two-detector, long-baseline, neutrino oscillation experiment 
optimised for $\nu_e$ identification. NO$\nu$A uses Fermilab's NuMI beamline having a beam power of 700 KW, as its neutrino
source. The 14 kt far detector is situated at a distance of 812 km, near Ash River, Minnesota and the 0.3 kt near detector is
located at the Fermilab site, near the existing MINOS Near Detector Hall. The 14 mrad off-axis configuration of the NuMI beam gives relatively
narrow band of neutrino energies centered at 2 GeV. Both the NO$\nu$A detectors are highly segmented, highly active tracking calorimeters, filled with liquid scintillator.
NO$\nu$A has started taking data from December 2014 and has given the first physics results recently \cite{nova_recent}. 
More details of \nova\ has been discussed  in \cite{nova}.

\subsection{LBNO}

One of the promising proposals for long-baseline neutrino oscillation experiment, is the 
LAGUNA (Large Apparatus studying Grand Unification and Neutrino Astrophysics) -LBNO (Long Baseline Neutrino Oscillation) project\footnote{
The other goals of the LAGUNA project are the study of proton decay, galactic supernovae, terrestrial and solar neutrinos etc \cite{Nuijten:2011zz}.} in Europe \cite{lbno_eoi}.
The source of neutrinos for this
experiment is likely to be at CERN. Various potential sites for the detector have been identified by LAGUNA, including Boulby (U.K.), Canfranc (Spain), Fr\'{e}jus (France), 
Pyh\"{a}salmi (Finland), Slanic (Romania), SUNLAB (Poland) and Umbria (Italy) \cite{laguna_options}. 
There are three different proposed detectors for the LBNO project: GLACIER (liquid argon), LENA (liquid scintillator) and MEMPHYS (water \cnv). 
In 2011, The LAGUNA collaboration decided to go ahead to investigate three sites in detail: Fr\'{e}jus having 
the shortest baseline of 130 km, Pyh\"{a}salmi with the longest baseline (2300 km) and Slanic having baseline of 1500 km. The corresponding detector for
Pyh\"{a}salmi and Slanic is GLACIER and for Fr\'{e}jus is MEMPHYS. 
To produce the beam for the Fr\'{e}jus (130 km) configuration, the 4 MW, 5 GeV
HP-SPL proton driver was assumed. For the other baselines, the 1.6 MW, 50 GeV HP-PS was
considered. For a operation of 200 days per calendar year, the HP-SPL delivers $10^{21}$ POT and the
HP-PS yields integrated $3 \times 10^{21}$ POT per year. In the past years, study regarding the
beam optimisation, detector simulation and physics potential at the various baselines has been carried out in great detail. 
The details of the physics simulation of LBNO can be found in \cite{Rubbia:2013zqa,lbno2013dec}.

\subsection{LBNE}

The Long Baseline Neutrino Experiment (LBNE) is a proposed long-baseline experiment at Fermilab \cite{lbne_interim2010}. 
The proposed detector is a modular 40 kt Liquid Argon Time Projection Chamber (LArTPC) at Sanford Underground Research Facility (SURF) in
South Dakota at a distance 1300 km from the source.  The first phase of this will
be a 10 kt detector. There will also be a fine-grained ‘near’ neutrino detector.
Performance of LArTPC is unmatched among
massive detectors for precise spatial and energy resolution and for reconstruction of complex neutrino interactions with high efficiency over a broad energy range. 
It thus provides a
``compact, scalable" approach to achieve sensitivity to the oscillation physics goals of LBNE \cite{lbne}.
If the LBNE detector is built underground, it will also be
possible for it to observe atmospheric neutrinos. A detailed
study on atmospheric neutrinos at LBNE is presented in \cite{raj_lbne1,raj_lbne2}.
The beam of LBNE will have
a intense on-axis wide-band profile with a beam power of 1.2 GeV. 
There
are two options being considered for the proton beam –
80 GeV and 120 GeV. For a given beam power, proton
energy varies inversely with the number of protons in the
beam per unit time. Thus a higher proton energy implies a lower flux of neutrinos.

Recently there have been discussions to converge the expertise and technical knowledge of LBNO and LBNE  
into a unified endeavor of a long-baseline experiment named DUNE (Deep Underground Neutrino Experiment) using a Megawatt beam from Fermilab.
One of the major goals of this facility as outlined in \cite{lbnf}
is $3\sigma$ CP sensitivity for 75$\%$ values of $\delta_{CP}$.

\subsection{INO}

The India-based neutrino observatory (INO) project is a multi-institutional effort aimed at building a world-class underground laboratory in India
for studying atmospheric neutrinos in its first phase\footnote{The other physics possibilities of the INO project include
study neutrinoless beta decay, direct dark matter searches etc. \cite{dino}.} \cite{Ahmed:2015jtv}.
It is sensitive to the atmospheric neutrinos in the energy range of 1-10 GeV.
For this experiment, the detector will be 50 kt magnetised iron calorimeter (ICAL) which is sensitive to mainly muon events. 
ICAL will consist of 5.6 m thick iron plates sandwiched between 151 layers of Resistive Plate Chambers (RPCs).
Iron will act as a target and the RPCs will serve as the active detector elements.
Due to the magnetic field of strength 1.5 Tesla, ICAL will be able to distinguish the $\mu^+$ and $\mu^-$ events in the GeV energy range.
This will make the ICAL detector sensitive towards measuring the neutrino mass hierarchy, which is the primary goal of
the ICAL@INO experiment.
Analysis of hierarchy sensitivity of the ICAL detector using only the muon momentum information can be found in \cite{gct}.
It is also capable of reconstructing the hadron energy and average direction of hadron shower.
The improvement in the sensitivity due to the inclusion of hadrons is discussed in \cite{ino3d}.
Apart from determining the neutrino mass hierarchy, the other physics goals of the ICAL@INO experiment include measurement of
octant of $\theta_{23}$, indirect detection of dark matter, searches for magnetic monopoles, non standard interactions, Lorentz and CPT violation etc.
The site for the INO has been identified at Pottipuram in Bodi West hills of Theni District of Tamil Nadu and
the construction is expected to start soon.

%

\subsection{IceCube}

The IceCube Neutrino Observatory is a neutrino telescope constructed at the Amundsen-Scott South Pole Station in Antarctica.
The main aim of IceCube is to study neutrinos from astrophysical sources. 
IceCube consists of 4800
optical sensors installed on 80 strings between 1450 m and 2450 m below the surface. 
Strings are deployed in a triangular grid pattern
with a characteristic spacing of 125 m enclosing an area
of 1 km$^2$ \cite{Achterberg:2006md}. The IceCube telescope was deployed in the summer of 2004 and is taking data since 2005. 
But the evidence of extraterrestrial neutrinos are found just recently. 
IceCube observed 37 neutrino candidate events in the energy range 30 TeV to 2000 TeV \cite{Aartsen:2014gkd}. 
The data in this energy range can be explained
by an $E^{-2}$ neutrino spectrum with a per-flavour normalisation (1:1:1).
On 4th august 2015 IceCube has observed a 2.3 PeV neutrino event \cite{icecubewebsite} which corresponds to the 
highest energy neutrino ever detected.

%% file: cp.tex

\section{Overview}

In this chapter we will study the sensitivity reach of the present/future generation experiments for
determining the remaining unknown neutrino oscillation parameters, namely: (i) the neutrino mass hierarchy, (ii) the octant of the mixing angle $\theta_{23}$ 
and (iii) the leptonic CP phase $\dcp$. 
We will analyse the sensitivities of the current long-baseline experiments T2K and \nova,
future atmospheric experiment ICAL@INO and the proposed long-baseline experiments LBNO and LBNE. 
We will also analyse the recent data of the IceCube experiment in this context. 
This chapter is organised in the following way. In the Section \ref{sec1} we will study the CP sensitivity of the T2K, \nova\ and ICAL experiments.
Taking the projected exposures of T2K and \nova\, we will show that the combined CP sensitivity of T2K and \nova\ is limited due to the presence of parameter
degeneracies and we will demonstrate how ICAL@INO can be used as a remedy of this problem. Next we will explore the CP sensitivity reach of these setups
by taking combinations of different exposures. We will present our results in terms of both discovery of CP violation (CPV) and precision of $\dcp$. 
We will also
study dependence of CP sensitivity on different oscillation parameters and compare the sensitivities of T2K and \nova\ for a given exposure.
As the sensitivity of \nova\ and T2K is limited due to shorter baselines and less statistics, 
it is necessary to study neutrino oscillations at higher baseline with higher statistics.
In Section \ref{sec2} and \ref{sec3} we will study the potential of the proposed long-baseline experiments LBNO and LBNE respectively for determining
all the three above-mentioned unknowns. These experiments have longer baselines as compared to
T2K and \nova\ and due to huge detector volume and high beam power they have higher statistics. 
As the exact configurations of these experiments are not yet decided, it is very important to
find an optimal and economised configuration for LBNO and LBNE. It is also important to remember that when these experiments will be operational, 
the data from T2K, \nova\ and ICAL experiments will
also be available.
Thus, in our analysis, we have calculated the minimum exposures of LBNO and LBNE for
determining hierarchy, octant and CP violation at a given confidence level in conjunction with 
T2K, \nova\ and ICAL. We will show that due to the synergy between T2K, \nova\ and ICAL, the required exposures of the LBNO and LBNE experiments are reduced significantly.
The reduction of the exposure signifies the fact that the same physics sensitivity can be obtained with a lower
beam power, small detector mass and/or less runtime. 
For the LBNE experiment, we will also present results
showing the effect of adding a near detector, the role of the second oscillation maximum and that of antineutrino runs. 
We will show that addition of a near detector reduces the systematic error significantly, whereas the second maxima plays a non-trivial role only in the determination of hierarchy.
While studying the role of the antineutrinos we find that the combination of equal neutrino and antineutrino do not always gives the best result.
In Section \ref{sec4} of this chapter we will study the
CP sensitivity of the IceCube experiment. As we have discussed in the Chapter \ref{chap:intro}, the IceCube experiment at south pole is designed mainly for studying neutrinos from
astrophysical sources. While coming from the extragalactic sources, ultra high energy neutrinos oscillate and due to the very large distance, 
their mass dependent oscillatory terms average out and they have no explicit dependence on the oscillation probabilities. 
Thus in principle it is possible to probe the mixing parameters in the IceCube events irrespective of the value of the neutrino mass. 
In our analysis, we have investigated the possibility of constraining the leptonic phase $\dcp$ from the recent IceCube data. We show that
in the oscillations of the ultra high energy neutrinos
$\dcp$ have a very weak dependence and thus it is not possible to put any constraint on $\dcp$
with significant confidence level. But we found that the results significantly depend on the initial sources of the neutrinos. We will present the results showing
how the properties of different sources can be constrained from this data. Finally we will summarize all the results of this chapter in Section \ref{sec5}.

 
\section{Evidence for Leptonic CP Phase from NO$\nu$A, T2K and ICAL}
\label{sec1}

In this section, we study the potential for measuring CP phase $\delta_{CP}$ in the current generation long-baseline
experiments T2K, \nova\ and the atmospheric neutrino oscillation experiment ICAL@INO. 
In the PMNS matrix, $\dcp$ is associated with 
$\theta_{13}$. Thus a non-zero  $\theta_{13}$ is
required for any measurement of $\dcp$. The
$10\sigma$ signature for non-zero $\theta_{13}$ 
leads naturally to the question to what 
extent CPV discovery is possible by 
the current superbeam experiments T2K and NO$\nu$A and/or with how much precision a true value of $\dcp$ can be measured.
In these experiments, the sensitivity to $\dcp$ comes mainly from the 
$\nu_\mu - \nu_e$ (and $\overline{\nu}_\mu - \overline{\nu}_e$)
oscillation probability, $P_{\mu e}$ ($\bar{P}_{\mu e}$) which are sensitive to $\delta_{CP}$.
But as discussed in Chapter \ref{chap:oscillation}, the measurement of any oscillation parameter is difficult due to the presence of parameter degeneracies
as the correct signal can  also be faked by a 
wrong solutions due to hierarchy-$\dcp$ and octant-$\dcp$ degeneracy. 
In \cite{t2knova}, it was shown that a prior knowledge of the hierarchy
facilitates the measurement of $\dcp$ by 
\nova\ and T2K. However, the determination of the hierarchy by \nova\ and T2K itself 
suffers from being dependent on the `true' value of $\dcp$ in nature. For the favourable combinations 
(\{$\dcp \in [-180^\circ,0^\circ]$, NH\} or \{$\dcp \in [0^\circ,180^\circ]$, IH\}), 
\nova\ and T2K will be able to determine the 
hierarchy at 90\% C.L. with their planned runs. 
But their hierarchy determination ability 
and hence their CP sensitivity will be poor if nature has chosen the
unfavourable combinations \cite{sanjib_glade}.
The octant determination capability of T2K and \nova\ are also affected due to the completely unknown value of $\dcp$ \cite{suprabhoctant}.
For the case of neutrinos, $\dcp \in [-180^\circ,0^\circ]$ is the favourable (unfavourable) half-plane for determination of 
octant if $\theta_{23}$ belongs in HO (LO) and $\dcp \in [0^\circ,180^\circ]$ is the favourable (unfavourable) half-plane for determination of 
octant if $\theta_{23}$ belongs to LO (HO). This is opposite for the case of antineutrinos.
So to have a octant sensitivity for all $\dcp$ values, a balanced neutrino and antineutrino run is required. 
On the other hand, the hierarchy and octant sensitivities of atmospheric neutrino 
experiments, for example ICAL, are independent of $\dcp$ \cite{ushier,usoctant}. 
Hence, a combination of long-baseline (LBL) 
and atmospheric data would be able to enhance the hierarchy and octant sensitivity for the $\dcp$ values which are adverse for the LBL experiments. 
This can substantially improve the ability of the LBL 
experiments to measure $\dcp$ in  the unfavourable regions of.
 
This section is organised as follows. 
First we will give the necessary experimental details of T2K, \nova\ and ICAL, used in our simulations and briefly describe the 
the method of $\chi^2$ analysis\footnote{We will explain the method of calculating $\chi^2$ in detail in the appendix.}. 
Next we present the CP sensitivity of the combined T2K and \nova\ experiments by taking their projected exposures
and show that the CP sensitivity of these experiments are compromised due to the hierarchy-$\dcp$ and the octant-$\dcp$ degeneracy.
Next we  demonstrate that, though ICAL do not have its own CP sensitivity but still the CP 
sensitivity of T2K and \nova\ can be enhanced significantly 
by including ICAL data in the analysis.
Then we expand our discussions by studying synergies between these setups and dependence of different oscillation 
parameters by taking different exposures of T2K, \nova\ and ICAL.

\subsection{Experimental Specification}
 
For our study we simulate \nova\ and T2K using 
the GLoBES package \cite{ globes1,globes2,messier_xsec,paschos_xsec,t2k,twobase1,globes_t2k3,globes_t2k4,globes_t2k5,nova,sanjib_glade}.
For \nova, we have assumed a $14$ kt totally active scintillator detector (TASD) 
and a neutrino beam having power of $0.7$ MW.
We have used a re-optimised \nova\ set-up with 
refined event selection criteria \cite{sanjib_glade,Kyoto2012nova}. 
T2K is assumed to have a $22.5$ kt water \v{C}erenkov detector and a $0.77$ MW beam. 
This beam power correspond to a $10^{21}$ POT per year.
In our analysis we give the exposures of \nova/T2K as a+b where a and b
denote the number of years of neutrino and antineutrino running of the experiments respectively.
For these experiments, we have used the systematic errors and background 
rejection efficiencies as used in Ref. \cite{sanjib_glade,Kyoto2012nova}. 

 
For atmospheric neutrinos we consider 
ICAL@INO with a proposed mass of 50 kt. The detector is 
capable of detecting muon events with charge
identification \cite{Ahmed:2015jtv}.  
For our analysis 
we use  neutrino energy and angular resolutions of (10\%, $10^\circ$)   
unless noted otherwise. 
These are representative values giving similar sensitivity as obtained 
in \cite{ino3d} using the informations of both muons and hadrons in the simulation. 

Now let us discuss briefly about our treatment of systematic errors. For the ICAL experiment
we have considered the following five sources of systematic errors: (i) 20\% flux normalisation error,
(ii) tilt error \cite{pulls_gg}  which includes the effect of
deviation of the atmospheric fluxes from a power
law, (iii) 5\% zenith angle uncertainty, (iv) 10\% cross section error and (v) an overall systematic uncertainty of 5\%.
But for our simulation of the long-baseline experiments we have used an overall normalisation error and an overall tilt error
as used in the references mentioned above. In our analysis we have implemented the systematic errors by the method of pulls \cite{pulls_gg,pull_lisi}.
We will discuss this method in the appendix.

\subsection{Details of the Simulation}

In our analysis, we give the sensitivity of the experiments in terms of $\chi^2$. The statistical $\chi^2$ for a Gaussian distribution\footnote{If the number of events
are very less then we use the Poisson $\chi^2$ formula. See appendix for detail.} is defined as
\be
\chi^2_{{\rm stat}} = \text{min} \frac{(N_{{\rm ex}} - N_{{\rm th}} )^2}{N_{{\rm ex}}},
\label{chisq}
\ee

where $N_{{\rm ex}}$ and $N_{{\rm th}}$ are the number of true (corresponds to data) and test (corresponds to theory) events respectively.
In our calculation we include a
marginalisation over the 
systematic errors by the method of pulls.
The resultant $\chi^2$ 
from the various experiments are then added and finally marginalised 
over the test parameters over their allowed $3\sigma$ range. 
We have added the external (projected) information on $\theta_{13}$ from the reactor experiments 
in the form of a prior on $\theta_{13}$:   
\begin{equation}
\chi^2_{{\rm prior}} = 
\left(\frac{{\sin^2 2\theta_{13}^{{{\rm tr}}}} - \sin^2 2\theta_{13}}
{\sigma(\sin^2 2\theta_{13})}\right)^2,
\end{equation}
where $\sigma(\sin^2 2\theta_{13})$ is the $1 \sigma$ error of $\sin^2 2\theta_{13}$.
 
For the analysis of CP sensitivity we will use two kinds of $\chi^2$ which are defined in the following way\footnote{For simplicity we give the diagnostics in terms of
$\chi^2_{{\rm stat}}$. However in each case we have included the systematic errors and priors in our numerical analysis.}: 
 
(i) The CP violation discovery $\chi^2$

\be
\chi^2_{{\rm stat}} = \text{min} \frac{(N_{{\rm ex}} (\dcp^{{\rm tr}}) - N_{{\rm th}} (\dcp^{{\rm test}}=0,180^\circ))^2}{N_{{\rm ex}}
(\dcp^{{\rm tr}})}
\label{chisq_violation}
\ee
which is the potential of an experiment to differentiate a true value of $\dcp$ from the CP conserving values $0^\circ$ and $180^\circ$.
This is obtained by varying $\dcp$ in the true spectrum and keeping it fixed in $0^\circ$ and $180^\circ$ in the test spectrum. 
As expected, the CP violation discovery potential of the experiments is zero for true 
$\dcp=0\circ$ and $180^\circ$, while it is close to maximum at the maximally CP violating 
values $\dcp=\pm 90^\circ$. 

(ii) The CP precision $\chi^2$

\be
\chi^2_{{\rm stat}} = \text{min} \frac{(N_{{\rm ex}} (\dcp^{{\rm tr}}) - N_{{\rm th}} (\dcp^{{\rm test}}))^2}{N_{{\rm ex}}
(\dcp^{{\rm tr}})}
\label{chisq_precision}
\ee
which describes how well an experiment can exclude the wrong $\dcp$ values other than the true values. 
This is obtained by varying $\dcp$ in the test spectrum over the full range $[-180^\circ, 180^\circ)$ for each value of true $\dcp$.
We will present our CP sensitivity results in the $\dcp({\rm true})-\dcp({\rm test})$ plane.

In our analysis we use the following transformations relating the effective measured values of the 
atmospheric parameters $\Delta_{\mu \mu}$ and $\theta_{\mu \mu}$ to their natural values $\Delta_{31}$ 
and $\theta_{23}$ \cite{degouvea_defn,parke_defn,spuriousth23}:
\begin{equation}
 \sin\theta_{23} = \frac{\sin\theta_{\mu \mu}}{\cos\theta_{13}} \ ~,
 \label{corr_th23}
\end{equation}
\begin{equation}
 \Delta_{31} = \Delta_{\mu \mu} + (\cos^2\theta_{12} - \cos\delta\sin\theta_{13}\sin2\theta_{12}
\tan\theta_{23})\Delta_{21}\ ~.
\label{corr_del31}
\end{equation}
The effective values $\Delta_{\mu \mu}$ and $\theta_{\mu \mu}$ correspond to
parameters measured by muon disappearance experiments. It is 
advocated to use these values in the definitions of priors if 
the prior is taken from muon disappearance measurements. 
The corrected definition of $\tmm$ is significant due to the large measured 
value
of $\theta_{13}$, while for $\dmmm$ the above transformation would be relevant 
even for 
small $\theta_{13}$ values.
In our analysis we do not use any external priors for these parameters 
as the experiments themselves are sensitive to these parameters. 
However it is to be noted that for the effective parameters, 
there is an exact mass hierarchy degeneracy between 
$\dmmm$ and ($-\dmmm$) and an exact intrinsic 
octant degeneracy between $\theta_{\mu\mu}$ and ($90^\circ - \theta_{\mu\mu}$). 
Therefore use of these values in the analysis ensures that one hits the exact 
minima for the wrong hierarchy and wrong octant in the numerical analysis 
for the muon disappearance channel. 
Measurements with the appearance channel and  
the presence of matter effects can break these degeneracies. 
Also, the generalised octant degeneracy occurring between values of 
$\theta_{\mu\mu}$ in opposite octants 
for different values of $\theta_{13}$ and $\dcp$ is still present for the 
effective atmospheric mixing angle. 
For such cases, a fine marginalisation grid has to be 
used in the analysis in order to capture the $\chi^2$ minima occurring in the
wrong hierarchy and wrong octant. 

For our analysis, we have fixed the solar parameters $\theta_{12}$ and $\Delta_{21}$ to their current best-fit values as obtained from the recent global fits
\cite{global_fogli,global_valle,global_nufit}
in both true and test spectrum.
We have taken true value of $\sin^22\theta_{13}=0.1$ throughout our analysis and added a 5\% prior on $\sin^22\theta_{13}$ 
which is the expected precision from the current reactor experiments.
True value of $\Delta_{\mu \mu}$ is taken as $+(-)2.4 \times 10^{-3}$ eV$^2$ for NH (IH).
In the test spectrum, we have marginalised $\theta_{\mu \mu}$ in the range $35^\circ$ to $55^\circ$ and $|\Delta_{\mu \mu}|$ from $2.19 \times 10^{-3}$ eV$^2$ 
to $2.61 \times 10^{-3}$ eV$^2$.
We have also marginalised hierarchy in the test spectrum unless otherwise mentioned.

\subsection{CP Sensitivity of T2K and \nova\ with Their Projected Exposure}

\begin{figure}[ht!]
\begin{center}
\vspace{0.3 in}
\includegraphics[scale=0.9]{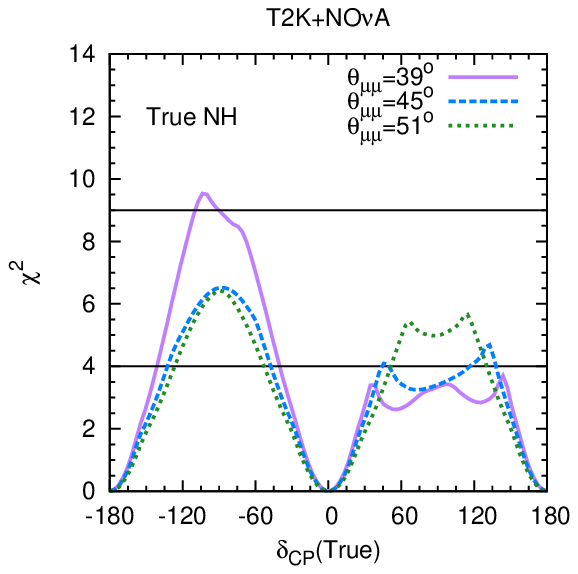}
\hspace{-0.8 in}
\includegraphics[scale=0.9]{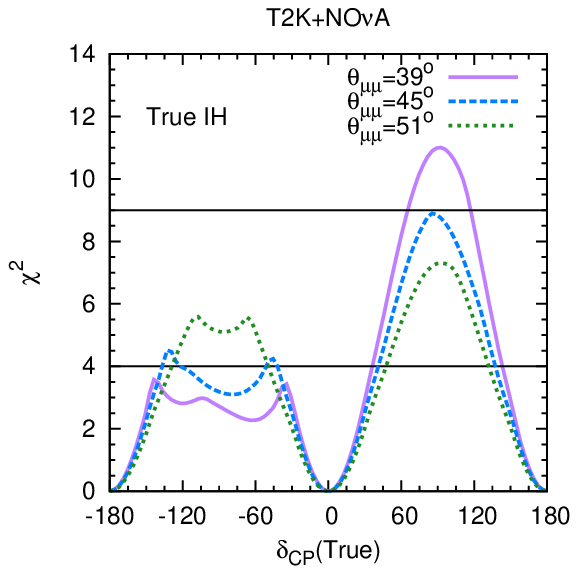}
\caption[CPV discovery vs true $\dcp$ for \nova(3+3)+T2K(8+0) 
taking $\theta_{\mu\mu}^{tr} = 39^\circ$, $45^\circ$ and $51^\circ$ and $\sin^2 2\theta_{13} = 0.1$.]{$\chi^2$ for CPV discovery vs true $\dcp$ for \nova+T2K
for $\sin^2 2\theta_{13} = 0.1$, three values
of $\theta_{\mu \mu}$  and a true normal (left panel) or 
inverted (right panel) mass hierarchy.}
\label{discovery}
\end{center}
\end{figure}

In Fig. \ref{discovery}, 
we plot the combined CPV discovery potential (as defined in Eq. \ref{chisq_violation}), of the 
LBL experiments
\nova\ and T2K\footnote{We have explained the procedure of adding $\chi^2$ of different experiments in the appendix.}.
In these plots we have considered \nova\ running for 3 years in neutrino mode and 3 years in antineutrino mode i.e., \nova(3+3) 
according to the planned run time of \nova.
For T2K we have considered a total exposure of $8 \times 10^{21}$ POT, which is the projected exposure of T2K. We consider T2K running only in neutrino mode.
As our flux corresponds to $10^{21}$ POT per year, this corresponds to 8 years running of T2K i.e., T2K(8+0).
Though currently T2K is also running in the antineutrino mode, in our analysis we have not considered the antineutrino run of T2K because when T2K is combined with
\nova(3+3), the pure neutrino run and equal neutrino-antineutrino run of T2K give almost similar sensitivity.
We will discuss this point in more detail in Section \ref{role_anu}.

The left panel gives the CP discovery for true NH, while 
the right panel depicts the results for true IH. 
From the figure, it may be observed that the CPV discovery of \nova+T2K 
suffers a drop in one of the half-planes 
of $\dcp$ for all the three values of $\theta_{\mu \mu}$. The drop is in the region 
$[0^\circ,180^\circ]$ (upper half-plane: UHP) if it is NH,
and $[-180^\circ,0^\circ]$ (lower half-plane: LHP) if it is IH.
These are the unfavourable half-planes corresponding to the hierarchy-$\dcp$ degeneracy where the $\chi^2$ minima occurs with the wrong hierarchy\footnote{Henceforth the
favourable and unfavourable half-planes that are mentioned throughout this chapter will correspond to that with respect to the hierarchy-$\dcp$ degeneracy.}.
Here we can see that the results depend significantly on the true value of 
$\theta_{\mu \mu}$.
As hierarchy sensitivity increases with $\theta_{\mu \mu}$, the CP sensitivity of the unfavourable half-planes increases
with increasing $\theta_{\mu \mu}$.
But on the other hand CP sensitivity decreases with increasing 
$\theta_{\mu \mu}$ in the favourable half-planes. 
We will discuss the reason for this in Section \ref{dependence}.
There is also a drop around $\dcp=-90^\circ$ for true NH and $\theta_{23}=39^\circ$.
This occurs because of the presence of octant-$\dcp$ degeneracy. As discussed earlier (see Fig. \ref{fig:figure5}), the octant-$\dcp$
degeneracy occurs in LO-LHP for neutrinos and absent in antineutrinos. As in these plots we have considered T2K(8+0) and \nova(3+3), the neutrino run dominates and thus
the wrong octant solution appears at $\dcp=-90^\circ$ for true NH and $\theta_{23}=39^\circ$.

From these plots we observe that with the projected exposure of T2K and \nova, $3\sigma$ CPV sensitivity can be achieved only around $\dcp=-90^\circ (+90^\circ)$ 
if true hierarchy is NH (IH).

\subsection{CP Sensitivity of Atmospheric Neutrinos}
  
Now let us discuss the CP sensitivity of the atmospheric neutrino experiment ICAL.
\begin{figure}[ht!]
\begin{center}
\vspace{0.3 in}
\includegraphics[scale=0.65]{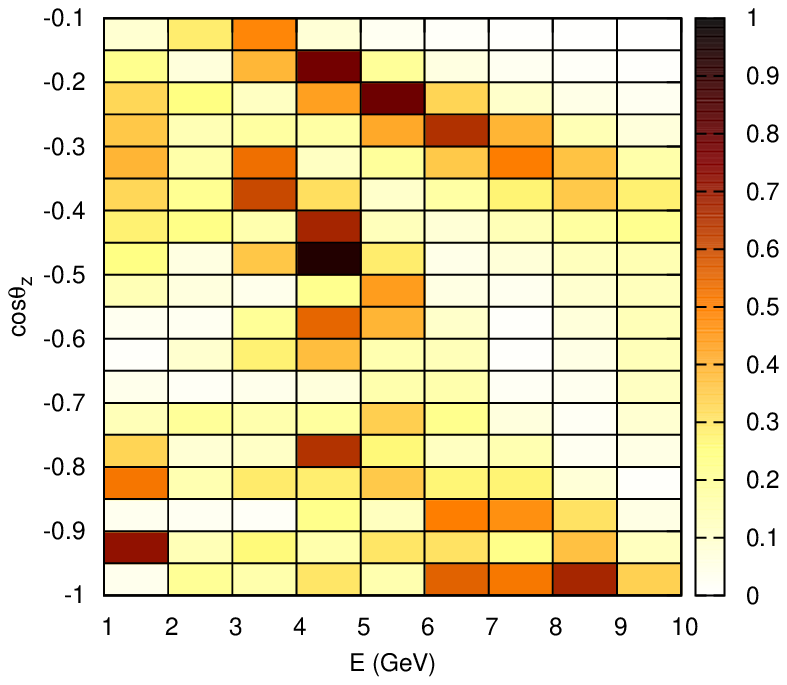}
\hspace{-1.0 in}
\includegraphics[scale=0.65]{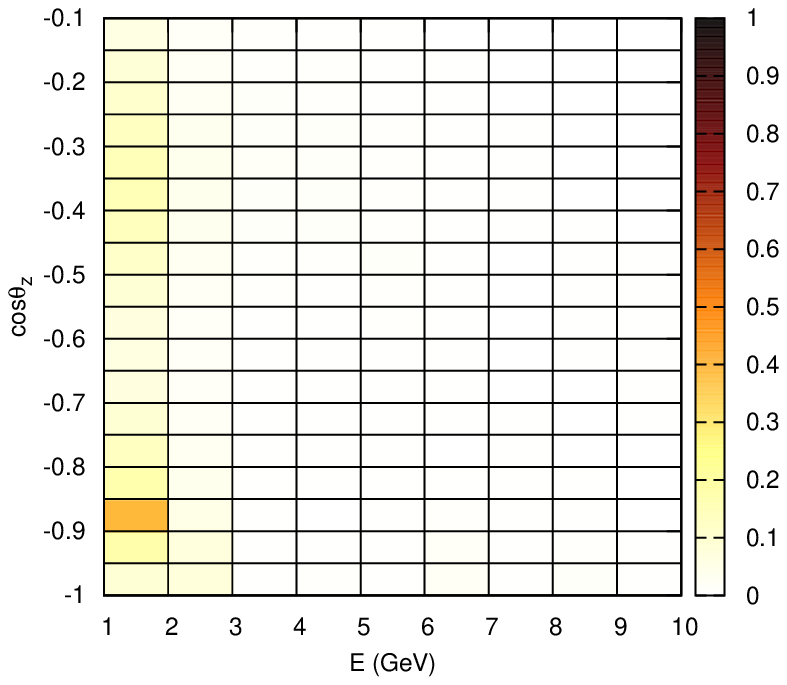}
\caption[CP sensitivity of ICAL experiment.]{$S_\mu + S_{\overline{\mu}}$, a measure of ICAL $\dcp$-sensitivity in the 
$E-\cos {\theta_z}$ plane for 
$\sin^2 2\theta_{13} = 0.1$, $\sin^2 \theta_{\mu \mu} = 0.5$ and NH. 
The grid represents bins in energy and $\cos\theta_z$.
The left panel is with ideal detector resolution and the right panel is with a resolution of $10^\circ$ in angle and $10\%$ in energy.
}
\label{nodcp}
\end{center}
\end{figure} 
The muon events in atmospheric neutrinos get contributions from both  
$P_{\mu \mu}$ and $P_{e \mu}$.
In these probabilities, the $\dcp$-dependent 
term always appears along with a factor of $\cos\Delta$ or $\sin\Delta$ (cf. Eq. \ref{alpha_s13}).
In atmospheric neutrinos, the baseline is associated with the direction of the incoming neutrinos or the zenith angle $\theta_z$.
If we consider even a $10\%$ error range in
the zenith angle and 
energy of the neutrino, the oscillating term varies over an entire cycle 
in this range. 
As a result, the $\dcp$-sensitivity of the channel gets washed out because of smearing. 
In Fig. \ref{nodcp}, we have plotted the quantity $S = S_\mu + S_{\overline{\mu}}$ in 
the $E-\cos\theta_{z}$ plane, which 
is a measure of the $\dcp$-sensitivity of the atmospheric neutrino experiment. Here, 
$S_\mu = (\delta N_\mu)^2 / N_\mu(\textrm{avg})$,
where $\delta N_\mu$ is the maximum difference in the number of events
obtained by varying $\dcp$ 
and $N_\mu(\textrm{avg})$ is the average number of events over all values of $\dcp$ 
(and likewise $S_{\overline{\mu}}$ for $\overline{\mu}$ events). The quantity $S$ is 
thus a measure of the 
maximum possible relative variation in events due to $\dcp$ in each bin. 
In the left panel, we show the results 
for an ideal detector with an exposure of 500 kt yr, with infinite energy and angular 
precision. 
Here we see substantial sensitivity to $\dcp$, with $S$ exceeding $0.5$ 
in some bins \cite{Samanta:2009hd}. 
However, when we introduce realistic resolutions 
($10^\circ$ in angle and $10\%$ in energy), we  
see from the right panel of Fig. \ref{nodcp} that the sensitivity is lost.
To study this point in more detail, we investigate how the intrinsic CP 
sensitivity of atmospheric neutrinos depend on the energy and angular 
resolutions and how much sensitivity can be achieved for an ideal detector.  
In Fig. \ref{res} the CP violation discovery potential of ICAL is plotted as a
function of the energy and angular resolution. 
\begin{figure}[ht!]
\begin{center}
\vspace{0.3 in}
\includegraphics[scale=0.9]{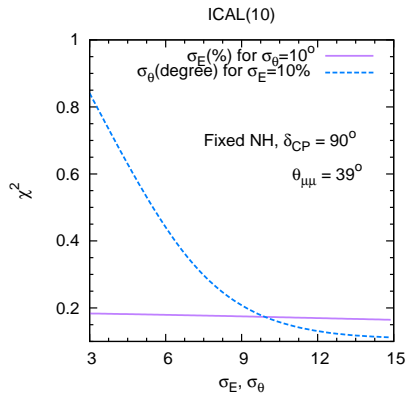}
\caption[CPV discovery potential of ICAL as a function of the detector
energy and angular resolutions.]{CP violation discovery potential of ICAL as a function of the detector
energy and angular resolutions for $\dcp^{tr} = 90^\circ$. $\theta_{\mu\mu}^{tr} =
39^\circ$ and a true NH is assumed.}
\label{res}
\end{center}
\end{figure} 
The curve for angular (energy) resolution is plotted by varying the respective smearing widths  
between $3^\circ - 15^\circ$ ($3\% - 15\%$) while holding the 
energy (angular) resolution fixed at $10\%(10^\circ)$.
The figure illustrates the significant role played by the angular resolution of
an atmospheric neutrino detector in controlling the CP sensitivity.
With present realistic values of detector smearing (15$\%$,15$^\circ$), the CP
sensitivity of such an experiment is  washed out 
by averaging over bins in energy and direction, due to the coupling between
$\dcp$ and $\Delta = \Delta_{31}L/4E$ 
in the term $\cos (\dcp + \Delta)$ in $P_{\mu e}$. With a hypothetical
improved angular resolution of $3^\circ$, the CP violation
discovery $\chi^2$ may reach values close to 1, going up to 5 for an ideal detector with no smearing.

Thus atmospheric neutrino experiments by themselves are not sensitive to 
$\dcp$.
For beam experiments, since the direction of the neutrinos 
is known, angular smearing is not needed and hence 
the sensitivity to $\dcp$ is not
compromised due to this reason.

\subsection{Effect of ICAL on the CP Sensitivity of T2K and \nova}

In this section we discuss the combined CP sensitivity of the T2K, \nova\ and ICAL detectors.
Previously we have seen that CP sensitivity of T2K and \nova\ is compromised due to unknown hierarchy and octant.
An atmospheric neutrino detector like ICAL gives  
hierarchy and octant sensitivity 
which is remarkably
stable over the entire range of $\dcp$, 
even though it does not offer any significant 
CPV discovery potential by itself.
Thus the hierarchy and octant sensitivity of the atmospheric neutrinos 
can exclude the wrong solutions for CPV discovery.
\begin{figure}[ht!]
\begin{center}
\vspace{0.3 in}
\includegraphics[scale=0.9]{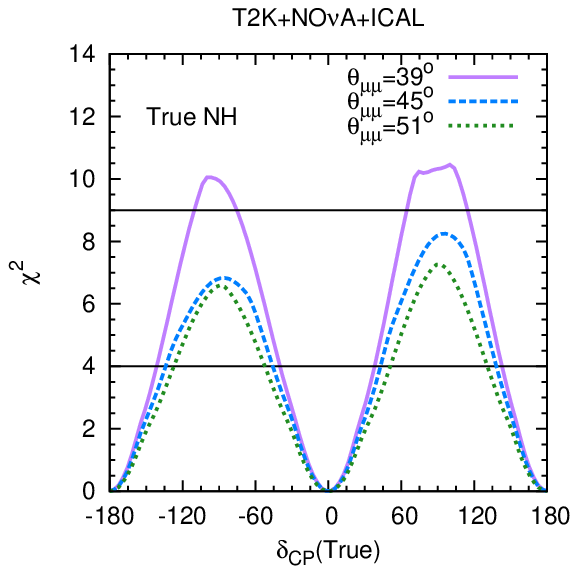}
\hspace{-0.8 in}
\includegraphics[scale=0.9]{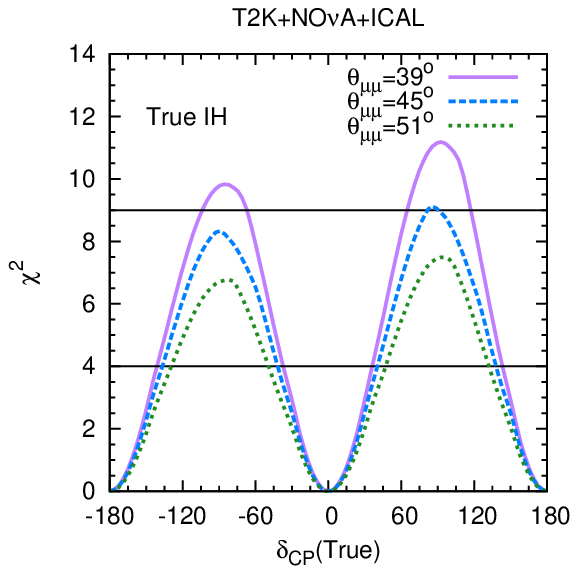}
\caption[CPV discovery vs true $\dcp$ for \nova(3+3)+T2K(8+0)+ICAL(500 kt-yr)
taking $\theta_{\mu\mu}^{tr} = 39^\circ$, $45^\circ$ and $51^\circ$ and $\sin^2 2\theta_{13} = 0.1$.]
{CPV discovery vs true $\dcp$ for \nova+T2K+ICAL 
for $\sin^2 2\theta_{13} = 0.1$, three values
of $\theta_{\mu \mu}$ and a true normal (left panel) or 
inverted (right panel) mass hierarchy.}
\label{discovery_INO}
\end{center}
\end{figure} 

In Fig. \ref{discovery_INO} we have plotted the combined CP sensitivity of \nova+T2K+ICAL.
Left panel is for true NH and right panel is for true IH.
In these figures we see that
when the information of ICAL is added to \nova+T2K, the $\chi^2$ in the unfavourable region increases significantly.
For NH and $\theta_{23}=39^\circ$, the shape of the curve shows that the sensitivity of ICAL is not sufficient to rule out the wrong 
hierarchy minima completely for $\dcp=90^\circ$.
We also observe that after the addition of ICAL data, the $\theta_{\mu \mu}$ dependence in the unfavourable region becomes similar to that of favourable region.
The wrong octant solution for NH, $\theta_{23}=39^\circ$ and $\dcp=-90^\circ$ also vanishes. 
The advantage offered by combining ICAL with the LBL data is most prominent for
$\theta_{23} = 39^\circ$
and progressively diminishes with increasing $\theta_{23}$. 
In general, the atmospheric neutrino contribution to the CPV discovery
potential of \nova+T2K+ICAL
is effective till the wrong solutions are disfavoured and the minimum 
comes with the true hierarchy and octant. Once that is achieved, a further increase in the 
sensitivity of atmospheric neutrinos will not affect the CPV discovery results, 
since atmospheric neutrinos by themselves do not have CPV sensitivity 
for realistic resolutions.
From the figures we can see that for true NH (IH), T2K+\nova\ can  discover CPV at $2\sigma$ for 
$\sim$ 28\%(29\%) fraction of $\dcp$ values for $\theta_{\mu \mu}= 39^\circ$. 
By adding ICAL information, this improves to $\sim$ 58\%. 
For maximal CPV ($\dcp = \pm 90^\circ$), inclusion of 
ICAL gives a $\sim 3\sigma$ signal for both hierarchies. 
Without the ICAL { contribution} this is true only in one of the half-planes 
depending on the hierarchy.

\begin{figure}[ht!]
\begin{center}
\vspace{0.3 in}
\includegraphics[scale=0.9]{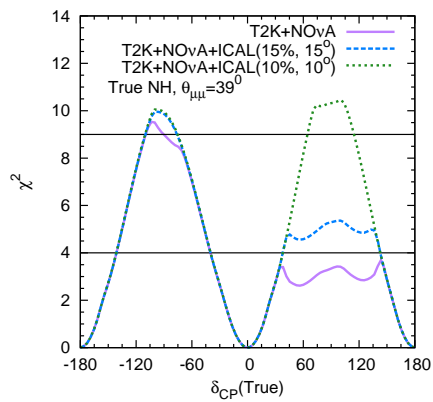}
\caption[CPV discovery vs true $\dcp$ for \nova(3+3)+T2K(8+0) and for
\nova(3+3)+T2K(8+0)+ICAL(500 kt yr) taking two sets of ICAL detector resolutions
for $\theta_{\mu\mu}^{tr} = 39^\circ$ and $\sin^2 2\theta_{13} = 0.1$.]{CPV discovery vs true $\dcp$ for \nova+T2K and 
\nova+T2K+ICAL {{(500 kt yr)}} for two sets of ICAL detector resolutions for $\theta_{\mu \mu}=39^\circ$, 
$\sin^2 2\theta_{13} = 0.1$ and true NH.}
\label{resvary}
\end{center}
\end{figure}

To study the effect of ICAL detector resolutions on the results, we plot in 
Fig. \ref{resvary} the CPV discovery potential of \nova+T2K+ICAL for 
$\theta_{\mu \mu} = 39^\circ$ and true NH assuming two sets of energy and angular 
smearing for ICAL -- 
(15\%,$15^\circ$) and (10\%,$10^\circ$).
In the former case, an indication of  CPV at  
$2\sigma$ is seen to be 
achieved around $\dcp=+90^\circ$ but the improvement in the CP sensitivity is very less. 
For the latter (better) smearing set, though the $\chi^2$ minimum still comes with the wrong hierarchy
but the sensitivity increases significantly in the unfavourable half-plane.
An improvement in the resolution beyond 
(10\%,$10^\circ$) can improve the CP sensitivity even more by excluding the wrong hierarchy solution completely.
For $\dcp=-90^\circ$, the drop due to the wrong octant solution is no longer visible after adding the ICAL data with (15\%,$15^\circ$) resolution.
Since the wrong solution is already removed, there is no further improvement in the CP sensitivity for this value of $\dcp$ when the ICAL resolution improved to (10\%,$10^\circ$).

\begin{figure}[ht!]
\begin{center}
\vspace{0.3 in}
\includegraphics[scale=0.9]{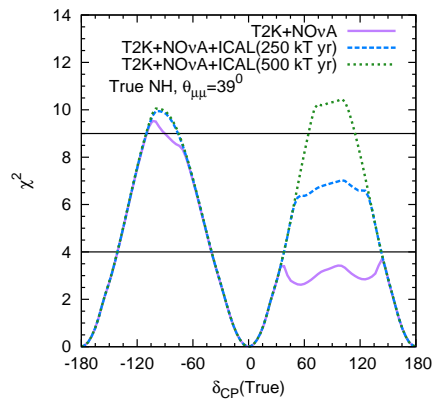}
\caption[CPV discovery vs true $\dcp$ for \nova(3+3)+T2K(8+0) and for 
\nova(3+3)+T2K(8+0)+ICAL taking two sets of ICAL exposure
for $\theta_{\mu\mu}^{tr} = 39^\circ$ and $\sin^2 2\theta_{13} = 0.1$.]
{CPV discovery vs true $\dcp$ for \nova+T2K and 
\nova+T2K+ICAL for two exposures, 250 kt yr and 500 kt yr for $\theta_{\mu \mu}=39^\circ$, 
$\sin^2 2\theta_{13} = 0.1$ and true NH. The ICAL resolutions are assumed to be 
10$\%$ in energy and 10$^\circ$ in angle.}
\label{exposure}
\end{center}
\end{figure}

In order to gauge the contribution from ICAL with a reduced exposure, we plot 
in Fig. \ref{exposure} the CPV discovery as a function of $\dcp$ for \nova+T2K and 
\nova+T2K+ICAL for two ICAL exposures, 250 kt yr and 500 kt yr for $\theta_{\mu \mu}=39^\circ$, 
$\sin^2 2\theta_{13} = 0.1$ and true NH, using the (10\%,$10^\circ$) ICAL resolution set.
The figure shows that
with an ICAL exposure of 250 kt yr, 
a 2.6$\sigma$ hint for CPV is  
achieved for $\dcp=+90^\circ$.

\subsection{A Chronological Study of $\delta_{CP}$ Using T2K, \nova\ and ICAL}

In this section we further explore the CPV discovery potential of T2K and \nova\ using different run times. We also use exposure beyond that projected in order to assess
the capabilities of these experiments with enhanced statistics. 
In addition to CPV discovery, we will present the results for $\dcp$ precision in $\dcp$ (true) vs $\dcp$ (test) plane.
We will also study the effect of ICAL in improving the CP sensitivity of T2K and \nova. 
For our study we will consider the following exposures of T2K, \nova\ and ICAL:

\begin{itemize}
\item 5 year run of T2K:  either (5+0) configuration or  (3+2) configuration and 10 year run of T2K in (5+5) configuration. 
\item For \nova\ we will take either (3+3) or (5+5) configuration. 
\item For ICAL we will consider either 5 year (250 kt yr) or 10 year (500 kt yr) exposure.
\end{itemize}

For detail see \cite{ourlongcp}.

\paragraph*{} 
In Table \ref{tablesummary} we summarize the maximum values of CP violation discovery 
potential, and the percentage of true $\dcp$ values capable 
of giving a CP violation discovery signal at 2$\sigma$ and 3$\sigma$, 
for different combinations of the experiments T2K, \nova\ and ICAL.
\begin{table}[ht!]
\begin{center}
\begin{tabular}{|c || c | c | c | c |} \hline 
       Experiment (exposure) & 
\multicolumn{2}{|c|}{max $\chi^2$ in}
& \multicolumn{2}{|c|}{$\dcp$ fraction for CPV}  \\
\cline{2-5}
    & UVHP & FVHP & $2 \sigma$ & $3 \sigma$ \\ 
  \hline
  \hline 
 T2K(8+0)+\nova (3+3) & $3.5$ & $9.5$ & $28\%$ & $5\%$ \\ \hline
 T2K(8+0)+\nova (3+3)+ICAL & $10.5$ & $10.0$ & $58\%$ & $24\%$ \\ \hline
 \hline
  T2K(3+2) & $0.9$ &$3.3$& $-$ & $-$ \\   \hline
  T2K(5+0) & $1.2$ & 0.8 & $-$ & $-$    \\ \hline 
  T2K(3+2) + \nova(3+3) & $3.1$ & 7.5 & $24\%$ & $-$    \\ \hline
  T2K(5+0) + \nova(3+3)  & $3.3$ & 8.2 & $25\%$ & $-$   \\ \hline
  T2K(5+0) + \nova(5+5)  & $4.8$ & 10.7 & $36\%$ & $11\%$  \\ \hline
  T2K(5+5) + \nova(5+5) & $4.9$ &12.5 & $41\%$ & $17\%$  \\ \hline
  T2K(5+0) + \nova(3+3) + ICAL 5 & $6.4$ & 8.3 & $52\%$ & $-$  \\ \hline
  T2K(5+0) + \nova(5+5) + ICAL 5 & $7.4$ & 10.8 & $60\%$ & $12\%$  \\ \hline
  T2K(5+5) + \nova(5+5) + ICAL 5 & $7.7$ & 12.7 & $62\%$ & $17\%$  \\ \hline
  T2K(5+0) + \nova(5+5) + ICAL 10 & $10.7$ & 11.0 & $60\%$ & $27\%$  \\ \hline
  T2K(5+5) + \nova(5+5) + ICAL 10 & $11.1$ & 12.7 & $62\%$ & $36\%$  \\ \hline  
           \hline
\end{tabular}
\caption[Values of maximal CP violation discovery $\chi^2$ in the favourable and unfavourable half-planes and 
percentage of true $\dcp$ values allowing CP violation discovery at 2$\sigma$/3$\sigma$
for different exposures of T2K, \nova\ and ICAL.]{Values of maximal CP violation discovery $\chi^2$ in the favourable and unfavourable half-planes (FVHP and UVHP) and 
percentage of true $\dcp$ values allowing CP violation discovery at 2$\sigma$/3$\sigma$
for combinations of experiments. Here $\theta_{\mu\mu}^{tr} = 39^\circ$, $\sin^2 2\theta_{13} = 0.1$ and true NH.}
\label{tablesummary}
\end{center}
\end{table}
From the table we see that, with the projected exposure, the combination of T2K and \nova\ gives $2\sigma$ CPV discovery sensitivity for $28\%$ fraction of total $\dcp$ values and 
$3\sigma$ CPV discovery sensitivity for $5\%$ fraction of total $\dcp$ values. 
With the inclusion of ICAL, the sensitivity improves to $58\%$ and $24\%$ for $2\sigma$ and $3\sigma$ respectively. 
If one considers the most optimistic run times for these experiments
i.e., the T2K(5+5) and \nova(5+5) configurations,
then also with the inclusion of 10 year data of ICAL gives a maximum CPV discovery sensitivity 
corresponding to a fraction of $62\%$ for $2\sigma$ and $36\%$ for $3\sigma$ of the total $\dcp$ values. 

From these results we understand that, to establish CPV (as well as CP precision) at higher confidence level, covering larger fraction of $\dcp$ values, one needs
experiments more powerful than T2K and \nova. This is also true for determining hierarchy and octant. In \cite{t2knova,suprabhoctant} it has been shown that
the combination of T2K and \nova\ is not sensitive enough to determine hierarchy and octant with a very high confidence level. To achieve higher sensitivity,
several experiments have been proposed which will have comparatively longer baselines and powerful neutrino beams. The projects 
LBNO and LBNE are the examples of such proposed long-baseline experiments.
LBNO is an European project where neutrinos will be delivered from the CERN accelerator, whereas for the LBNE project, Fermilab will generate the neutrinos.
As the exact design of these experiments are
still under consideration, one needs to carry out a detail analysis to find out the optimal configuration for measuring the oscillation parameter with
greater confidence level.
In the next two sections we will study the physics potential of the LBNO and LBNE experiments and find out their optimal exposure
to determine the unknown neutrino oscillation parameters.

%% file: LBNO.tex

\section{Physics Potential of LBNO in Conjunction with T2K, \nova\ and ICAL}
\label{sec2}

In this section, we will discuss the contributions of \nova, T2K, ICAL@INO and 
LBNO towards determining the mass hierarchy, octant of $\theta_{23}$ as well as for discovery of CP violation. 
As the precise configuration of LBNO is under consideration, our aim is to
determine the configuration for LBNO with 
`adequate' exposure in conjunction with T2K, \nova\ and ICAL,
which can determine the unknown 
oscillation parameters.
The `adequate' configuration is defined as one with the minimal exposure
which would give a 5$\sigma$ discovery potential for hierarchy 
and octant and 3$\sigma$ discovery potential for $\dcp$ 
in the most unfavourable case. 
This configuration can be viewed as 
the first step in a staged approach that has been advocated by previous 
studies \cite{incremental}. 

The plan of this section is as follows. First we give the 
experimental specifications that we have used for the proposed LBNO experiment and other simulation details. 
The next three 
subsections thereafter are devoted to 
the analysis of the experimental reach of the combination of experiments 
for determining the mass hierarchy, octant of $\theta_{23}$ and discovery of CP violation 
respectively.

\subsection{Experimental Specifications and Other Simulation Details}

Out of the various possible options for the LBNO experiment as described in chapter \ref{chap:oscillation} 
, we consider the following three options that are 
prominent in the literature: 
CERN-Pyh\"{a}salmi, CERN-Slanic and CERN-Fr\'{e}jus. 
The specifications that we have used in this study are listed below in 
Table \ref{tab:exp_lbno}. We have used the superbeam fluxes from 
Ref. \cite{lbnoflux}. 

For \nova, T2K and ICAL we consider the same specification as mentioned in section \ref{sec1}.
For \nova\ we have considered an equal 3 year running in both neutrino and antineutrino mode.
For T2K we assumed a total exposure corresponding to $8 \times 10^{21}$ POT running in completely neutrino mode and we have taken 10 year running of ICAL, which corresponds to
a total exposure of 500 kt-yr.

\begin{table}[htb]
\begin{center}
 
     \begin{tabular}{ || l || c | c | c ||}

         \hline
         \hline
         Detector site & Pyh\"{a}salmi & Slanic & Fr\'{e}jus \\
         \hline
         \hline
         Baseline & $2290$ km & $1540$ km & $130$ km \\
         Detector Type & LArTPC & LArTPC & Water \v{C}erenkov \\
	 Proton energy & $50$ GeV & $50$ GeV & $4.5$ GeV \\
         Resolutions, efficiencies & as in Ref. \cite{incremental} &  
			as in Ref. \cite{incremental} &  as in Ref. \cite{newmemphys} \\
	 Signal systematics & 5\% & 5\% & 5\%\\
         Background systematics & 5\% & 5\%  & 10\% \\
         \hline
         \hline
      \end{tabular}
         
\caption{Experimental characteristics of the LBNO options.}

\label{tab:exp_lbno}

\end{center}
\end{table}

We have fixed the `true' values of the parameters close to the values obtained
from global fits of world neutrino data. 
We have taken: 
$\sin^2 \theta_{12} = 0.304$, $|\Delta_{31}| = 2.4 \times 10^{-3}$ eV$^2$,
$\Delta_{21} = 7.65 \times 10^{-5}$ eV$^2$ and $\sin^2 2\theta_{13} = 0.1$. 
Three representative
true values of $\theta_{23}$ have been considered -- $39^\circ$, 
$45^\circ$ and $51^\circ$ (except in the case of octant determination where 
a wider range and more intermediate values have been included). The true 
value of $\dcp$ is varied in its entire allowed range. All our 
results are shown for both NH and IH. The
`test' values of the parameters are allowed to vary in the following ranges -- 
$\theta_{23} \in \left[35^\circ,55^\circ\right]$, 
$\sin^2 2\theta_{13} \in \left[0.085,0.115\right]$, 
$\dcp \in \left[0,360^\circ\right)$. The test hierarchy is also allowed 
to run over both possibilities. 
We have imposed a prior 
on the value of $\sin^2 2\theta_{13}$ with an error 
$\sigma(\sin^2 2\theta_{13}) = 0.005$. 
We have however not imposed any prior on the atmospheric parameters, 
instead allowing the $\nu_\mu$ disappearance channels to restrict 
their range. In all our simulations, we 
have taken into account the three-flavour-corrected definitions of the 
atmospheric parameters \cite{parke_defn,degouvea_defn,spuriousth23} as given in Eqs. \ref{corr_th23} and \ref{corr_del31}. 

The aim of this exercise is to determine the least exposure required from LBNO 
in order to determine hierarchy and octant with a statistical significance corresponding to 
$\chi^2=25$ and that CP violation detected with $\chi^2=9$ for 20\% value of $\dcp$\footnote{Conventionally, these values are taken to correspond to 
$5\sigma$, and  $3\sigma$, respectively. 
However, it was recently pointed out in Refs. \cite{stats_hier1,stats_hier2} 
that for a binary question such as hierarchy, the relation between $\chi^2$ 
and confidence levels is somewhat involved. For more recent discussions on 
statistical interpretation, see Refs. \cite{blennowstats1,blennowstats2,lbno2013dec}.}. 
Therefore, we have plotted the sensitivity 
to hierarchy/octant/CP violation for various different exposures of LBNO, 
combined with \nova, T2K and INO. 
From this, we estimate the adequate amount 
of exposure required by LBNO. We express the exposure in units of POT-kt. 
This is a product of three experimental quantities: 
\begin{equation}
 \textrm{exposure (POT-kt)} = \textrm{beam intensity (POT/yr)} \times 
 \textrm{runtime (yr)} \times \textrm{detector mass (kt)} ~.
\end{equation}
Thus, a given value of exposure can be achieved experimentally by 
adjusting the intensity, runtime and detector mass. 
The advantage of using this 
measure is that while the physics goals are expressed in terms of simply 
one number (the exposure), the experimental implementation of this exposure 
can be attained by various combinations of beam, detector and runtime settings. 
For example, an exposure of $45\times10^{21}$ POT-kt could be achieved with 
a $1.5\times10^{21}$ POT/yr beam running for $3$ years 
with a $10$ kt detector or a $3 \times 10^{21}$ POT/yr beam running for 
$3$ years with a $5$ kt detector. 
According to our terminology,
the exposures given correspond to each 
mode (neutrino and antineutrino). 
Thus, 
a runtime of $n$ years implies $n$ years 
each in neutrino and antineutrino mode totaling to $2n$ years.

\subsection{Determination of Mass Hierarchy}

\begin{figure}[ht!]
\begin{center}
\vspace{0.3 in}
\includegraphics[scale=0.7]{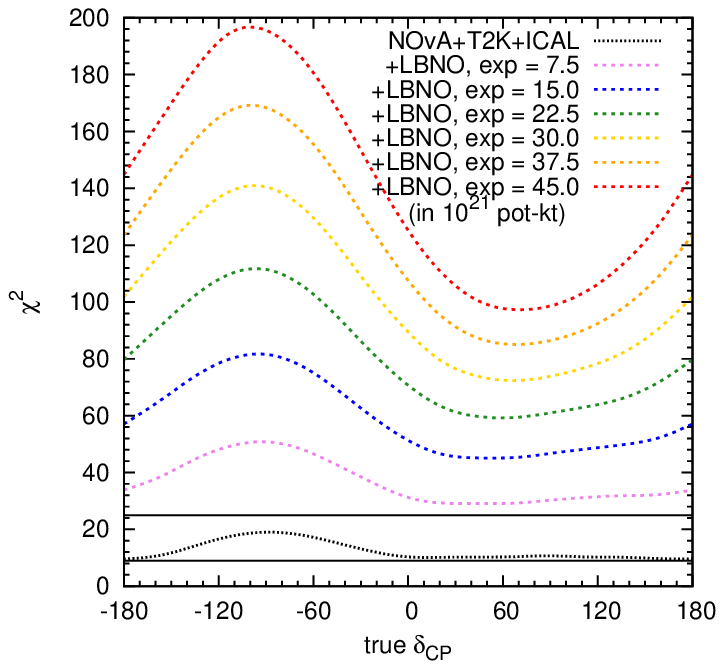}
\hspace{-0.7 in}
\includegraphics[scale=0.7]{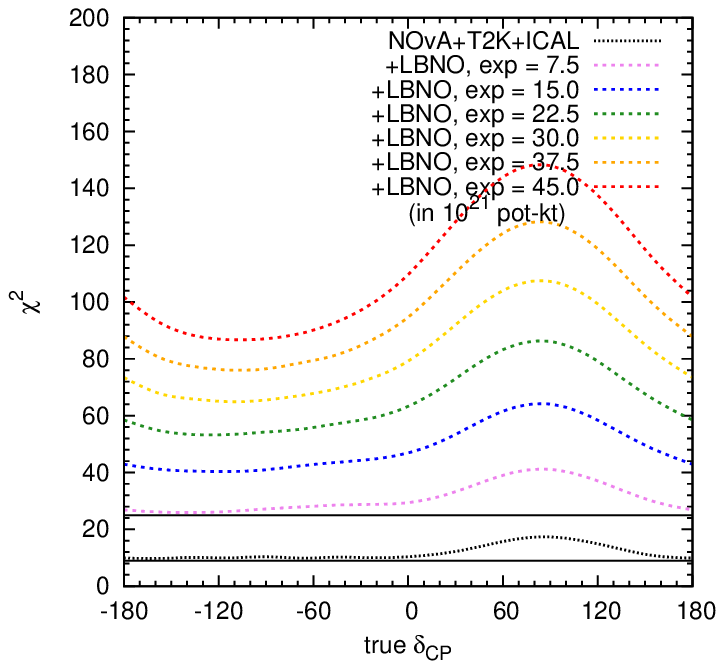} \\
\vspace{0.1 in}
\includegraphics[scale=0.7]{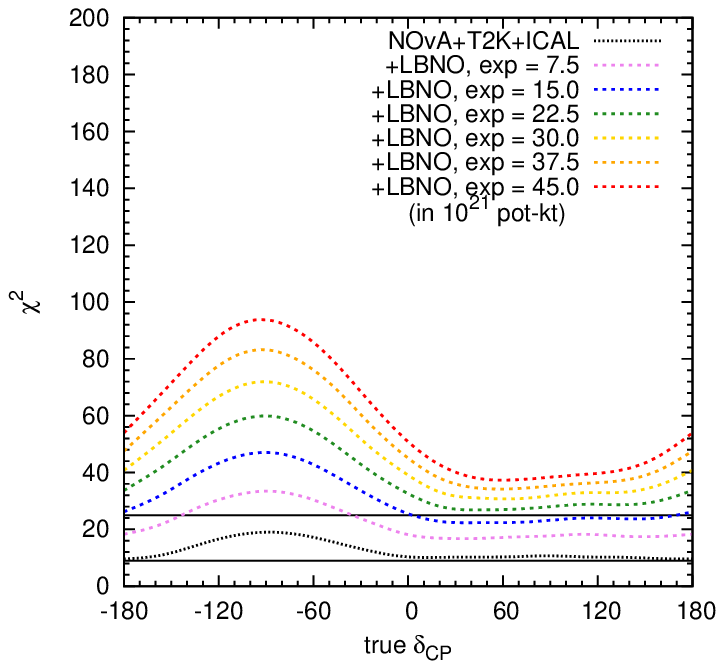}
\hspace{-0.7 in}
\includegraphics[scale=0.7]{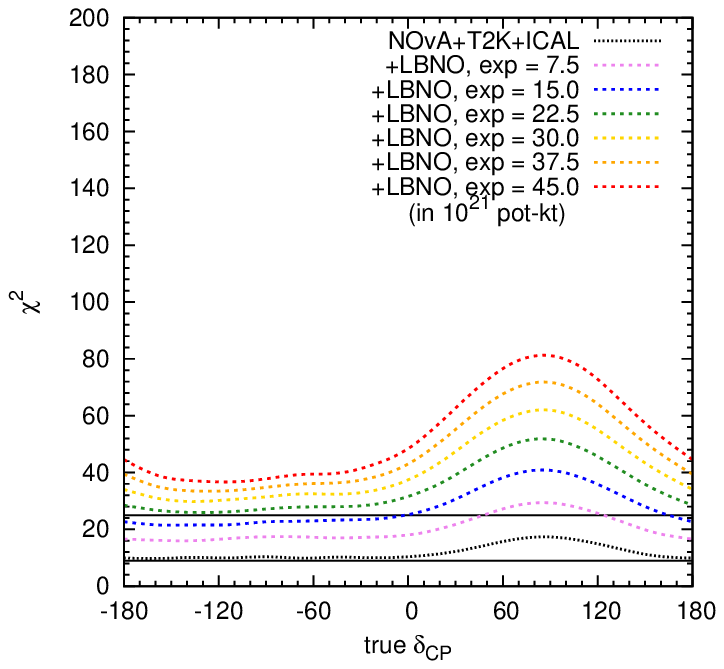} \\
\caption[Hierarchy sensitivity $\chi^2$ of LBNO vs true $\dcp$ for the 2290 km and 1540 km baselines.]{Hierarchy sensitivity $\chi^2$ vs true $\dcp$. 
The top (bottom) panels are for the $2290$ ($1540$) km baseline. The left (right) 
panels are for true NH (IH). In all the panels, the lowermost densely-dotted (black) 
curve is for \nova+T2K+ICAL, while the curves above are for \nova+T2K+ICAL+LBNO, 
for various values of LBNO exposure. All the plotted sensitivities are for the least 
favourable value of true $\theta_{23}$.}
\label{fig:hierall}
\end{center}
\end{figure}

Among the three chosen prospective baselines for LBNO, the 130 km 
set-up has the lowest hierarchy sensitivity due to small matter effects. 
As the baseline increases, the hierarchy sensitivity becomes better because 
of enhanced matter effects. In particular, the 2290 km set-up has the 
unique advantage of being close to satisfying the bimagic conditions 
\cite{bnlhs,bnlhs_long,bimagic}(cf. Section \ref{bimagic}). 
This feature makes the baseline particularly suited for 
hierarchy determination.  The above features are reflected in Fig. \ref{fig:hierall}.
In each of the panels of Fig. \ref{fig:hierall}, 
the lowermost densely-dotted (black) 
curve shows the hierarchy sensitivity of the combination \nova+T2K+ICAL. 
We see that these experiments can collectively give $\chi^2 \approx 9$ 
sensitivity to 
the hierarchy\footnote{The hierarchy $\chi^2$ is calculated by taking the correct hierarchy in the true spectrum and the wrong
hierarchy in the test spectrum in Eq. \ref{chisq}.}. Therefore, in keeping with our aims, we need to determine 
the minimum exposure for LBNO, such that the combination \nova+T2K+ICAL+LBNO 
crosses the threshold of $\chi^2=25$ for all values of $\dcp$. For this, 
we have plotted the combined sensitivity of \nova+T2K+ICAL+LBNO for 
various values of LBNO exposure 
(in units of 
$10^{21}$ POT-kt). The results are shown for two baselines -- $2290$ km and 
$1540$ km, and for both hierarchies. 
We find that our results are consistent with 
those shown in Ref. \cite{lbno_eoi}, for the same beam power and oscillation 
parameters.
For the baseline of $130$ km, 
it is not possible to cross $\chi^2=25$ even with extremely high 
exposure. Therefore we have not shown 
the corresponding plots for this baseline. 
We considered three true values of 
$\theta_{23}$ -- $39^\circ, 45^\circ, 51^\circ$ and chose the $\chi^2$ corresponding to the 
least favourable  of these in generating the figures.  
Thus, our results represent the most conservative case.  
We find that in most cases, the minimum $\chi^2$ for hierarchy 
determination occurs for true $\theta_{23} = 39^\circ$. 

\begin{figure}[ht!]
\begin{center}
\vspace{0.3 in}
\includegraphics[scale=0.8]{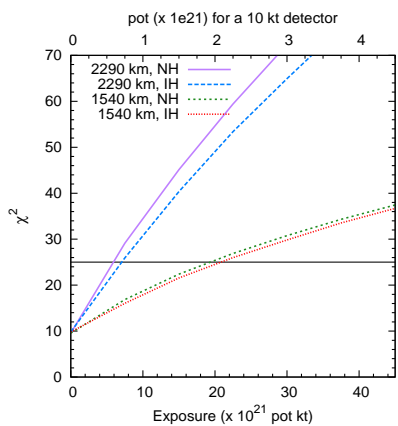}
\caption[Hierarchy sensitivity $\chi^2$ vs LBNO 
exposure for the $2290$ km and $1540$ km baselines.]{Hierarchy sensitivity $\chi^2$ vs LBNO 
exposure, for both baselines and hierarchies under consideration. 
The value of exposure shown here is adequate to exclude the wrong hierarchy 
for all values of $\dcp$. 
The additional 
axis along the upper edge of the graph shows the required total POT assuming a 
detector mass of $10$ kt.}
\label{fig:hierexpo}
\end{center}
\end{figure}

Finally, in Fig. \ref{fig:hierexpo}, we have 
condensed all this 
information into a single plot.
We have shown the sensitivity for the 
experiments as a function of the LBNO exposure. We see that for $2290$ ($1540$) 
km, it is sufficient for LBNO to have an exposure of around 
$7\times10^{21}$ ($21\times10^{21}$) POT-kt in order to get a $\chi^2=25$ 
for all values of $\dcp$. Along the upper edge of the graph, 
we have provided an additional axis, which denotes the total POT required 
if we assume that the detector has a mass of $10$ kt. For $2290$ ($1540$) km, 
we need a total of $0.7\times10^{21}$ ($2.1\times10^{21}$) POT.
To get some idea of the time scale involved we consider for instance the 
beam intensity used in Ref. \cite{incremental} which corresponds to 
$3 \times 10^{21}$ POT/yr delivered by a $50$ GeV proton beam 
from CERN with beam power $1.6$ MW. 
The total POT of $0.7\times10^{21}$ for a 10 kt detector at the $2290$ ($1540$) km 
baseline  would thus need 
less than $1$ ($2$) years (total, inclusive of $\nu$ and $\overline{\nu}$ 
runs) to establish mass hierarchy with $\chi^2=25$. 

\begin{figure}[ht!]
\begin{center}
\vspace{0.3 in}
\includegraphics[scale=0.7]{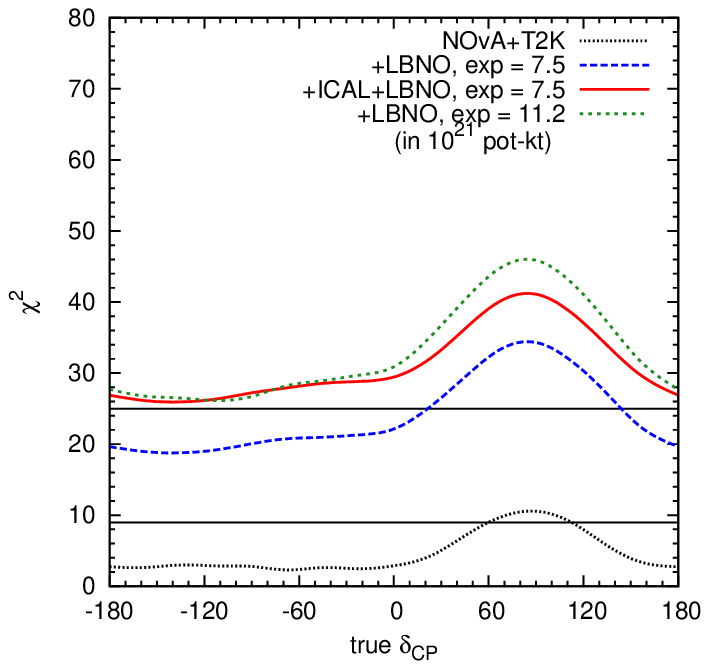}
\hspace{-0.7 in}
\includegraphics[scale=0.7]{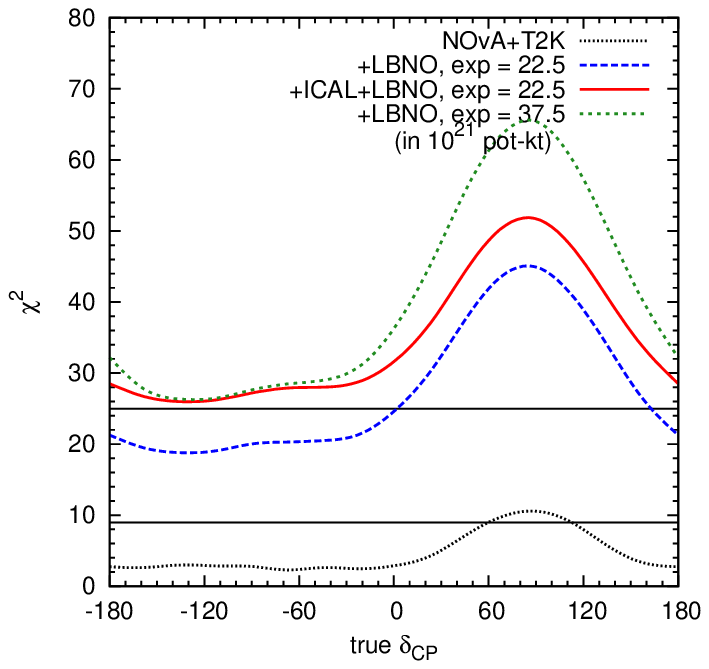} 
\caption[Hierarchy sensitivity $\chi^2$ of LBNO for the 2290 km and 1540 km baselines taking different 
 combinations of experiments, demonstrating the synergy between them.]{Hierarchy sensitivity $\chi^2$ for different 
 combinations of experiments, demonstrating the synergy between them. The left (right) 
 panel is 
 for a LBNO baseline of $2290$ ($1540$) km, assuming IH to be true. With only 
 T2K+\nova+LBNO 
 (dashed, blue), the sensitivity is lower than for T2K+\nova+LBNO+ICAL (red, solid).
 Without ICAL data, the LBNO exposure would have to be increased substantially 
 (dotted, green) in order to get comparable sensitivity. All the plotted sensitivities 
 are for the least favourable value of true $\theta_{23}$.}
\label{fig:hiersyn}
\end{center}
\end{figure}

Fig. \ref{fig:hiersyn}, demonstrates the synergy 
between long-baseline and atmospheric neutrino experiments.
We have chosen 
the $2290$ ($1540$) km baseline as an illustrative case
in the left (right) panels,
with the true hierarchy 
assumed to be IH. The densely-dotted (black) curve at the bottom shows 
the hierarchy sensitivity of \nova+T2K 
without any atmospheric neutrino data included in the analysis.
If the atmospheric information is not included then the combination 
of \nova+T2K+LBNO would need about $11  \times10^{21}$ POT-kt in order to
attain $\chi^2=25$, for the 2290 baseline. 
Assuming a beam intensity of $3 \times 10^{21}$ POT/yr 
this would require less than a year to measure the 
hierarchy with a 10 kt detector. 
Combining these with ICAL reduces the exposure to 
$7 \times10^{21}$ POT-kt. Thus, 
for the same beam intensity one can achieve the same sensitivity
with a 7 kt detector.  
Similar conclusions can be drawn for the 1540 km set-up. 
It should be noted that the numbers in Fig. \ref{fig:hiersyn} are 
sample values at which the simulations are performed. 
The exposure required for each set-up
to attain the `adequate' values can be read off from  
Fig. ~\ref{fig:hierexpo} and is presented in Table
\ref{tab:res}.
 \begin{center}
\begin{table}
 \begin{tabular}{||l||c|c|c||}
  \hline
  \hline
 Adequate exposure for & $2290$ km & $1540$ km & $130$ km \\
  \hline
  \hline 
  Hierarchy exclusion at & $ 7(11) \times 10^{21}$ & $ 21(37) \times 10^{21}$ & $-$ \\
  $\chi^2=25$ &  &  &  \\
  \hline
   Octant exclusion for $39^\circ$ & $ 83(113) \times 10^{21}$ & $ 83(113) \times 10^{21}$ & $ 400(600) \times 10^{21}$ \\
   at $\chi^2=25$ & & & \\
  \hline
  CP violation detection & & &  \\
  at $\chi^2=9$ & $ 240(240) \times 10^{21}$ & $ 170(170) \times 10^{21}$ & $ 35(100) \times 10^{21}$ \\
  for 20\% fraction of $\dcp$ & & & \\
  \hline
  \hline  
 \end{tabular}
 \caption[Adequate exposures of LBNO for determining hierarchy, octant and CP in units of POT-kt.]{Summary of results: `adequate' exposure in POT-kt 
 for three LBNO configurations in conjunction with T2K,\nova\ and ICAL to achieve the physics goals. The numbers given in 
 parentheses indicate the required exposure if atmospheric neutrino data from ICAL 
 is not included.}
 \label{tab:res}
\end{table}
\end{center}

\subsection{Determination of Octant of $\theta_{23}$}

\begin{figure}[ht!]
\begin{center}
\vspace{0.3 in}
\includegraphics[scale=0.7]{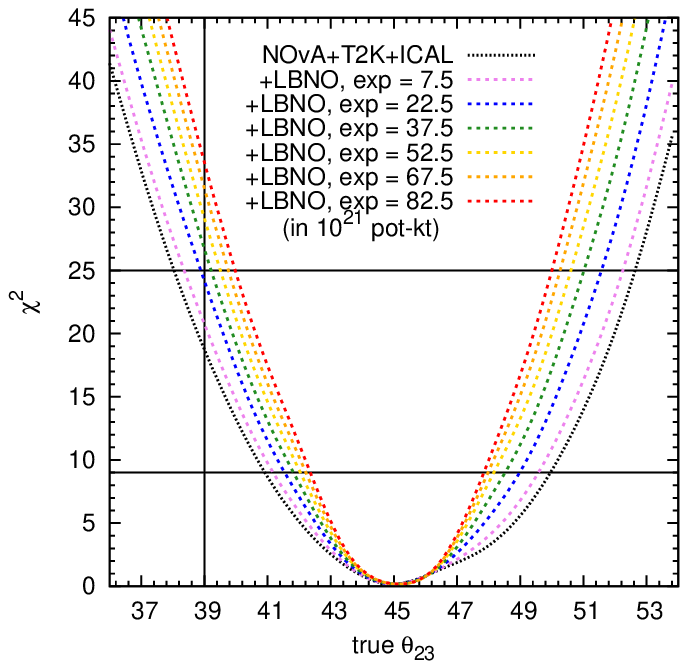}
\hspace{-0.7 in}
\includegraphics[scale=0.7]{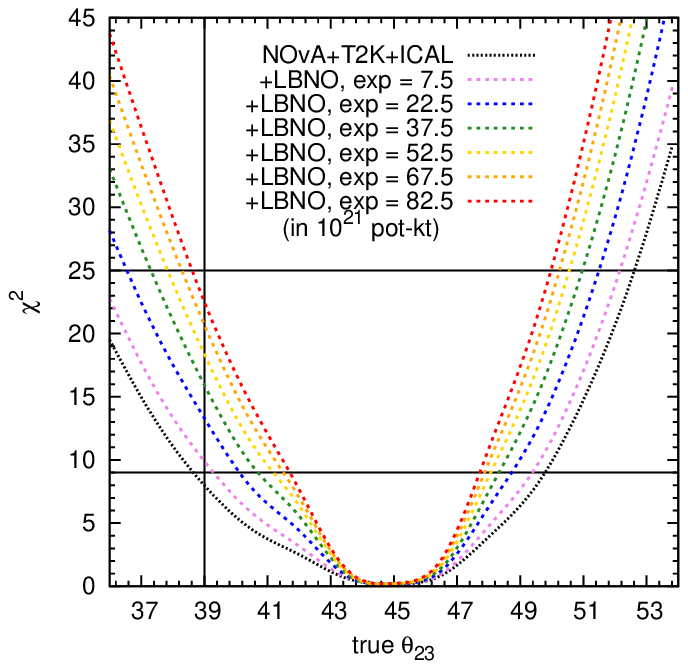} \\
\vspace{0.1 in}
\includegraphics[scale=0.7]{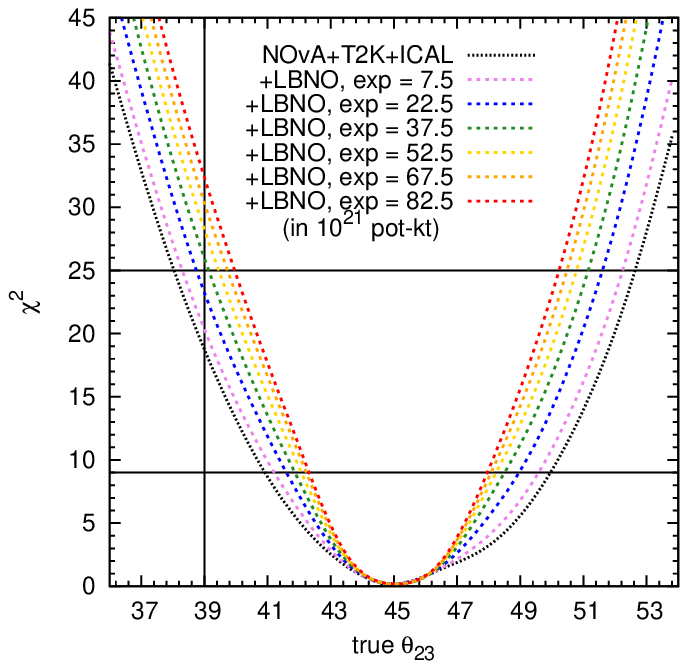}
\hspace{-0.7 in}
\includegraphics[scale=0.7]{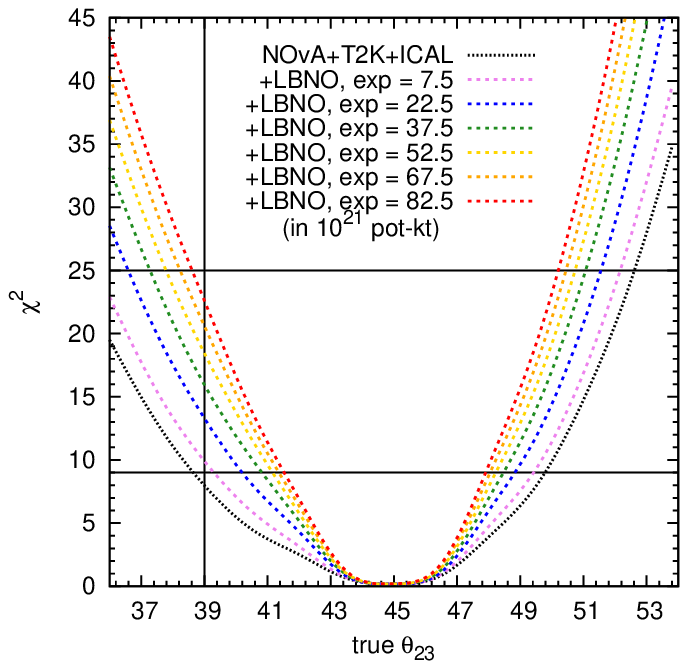} \\
\caption[Octant sensitivity $\chi^2$ of LBNO vs true $\theta_{23}$ for the baselines 2290 km and 1540 km.]{Octant sensitivity $\chi^2$ vs true $\theta_{23}$. 
The top (bottom) panels are for the $2290$ ($1540$) km baseline. The left (right) 
panels are for true NH (IH). In all the panels, the lowermost densely-dotted (black) 
curve is for \nova+T2K+ICAL, while the curves above are for \nova+T2K+ICAL+LBNO, 
for various values of LBNO exposure. All the plotted sensitivities 
 are for the least favourable value of true $\dcp$.}
\label{fig:octall}
\end{center}
\end{figure}

The octant sensitivity of long-baseline experiments has been studied in detail 
recently in \cite{suprabhoctant,suprabhlbnelbno} and also in conjunction 
with atmospheric neutrino experiments \cite{usoctant}. As in the case of 
hierarchy, adding information from various 
experiments enhances the sensitivity. However, the precise knowledge of 
the value of $\theta_{13}$ also plays a very crucial role in enhancing the octant 
sensitivity \cite{Minakata:2002jv}. In Fig. \ref{fig:octall}, the lowermost densely-dotted (black) 
curve denotes the ability of \nova+T2K+ICAL to determine the 
octant\footnote{The octant $\chi^2$ is calculated by taking the correct octant in the true spectrum and the wrong
octant in the test spectrum in Eq. \ref{chisq}.} as 
a function of the true value of $\theta_{23}$ in nature.
Again, the other curves 
denote the combined 
sensitivity of \nova+T2K+ICAL+LBNO(2290 km and 1540 km) for various values of LBNO exposure (in units of 
$10^{21}$ POT-kt). We generated the results for various 
true values of $\dcp$, and the 
results shown in the figure are for the most conservative case. We see that 
only with \nova+T2K+ICAL, the octant can be determined at $>3 \sigma$ C.L. 
when $\theta_{23}=39^\circ$. For values closer to $45^\circ$, the sensitivity 
gets steadily worse. The addition of LBNO data increases the sensitivity. For 
the range of exposures considered, it is possible to get a $\chi^2=25$ 
sensitivity to the octant as long as $\theta_{23}$ deviates from maximality 
by at least $\sim 6^\circ$. 

\begin{figure}[ht!]
\begin{center}
\vspace{0.3 in}
\includegraphics[scale=0.8]{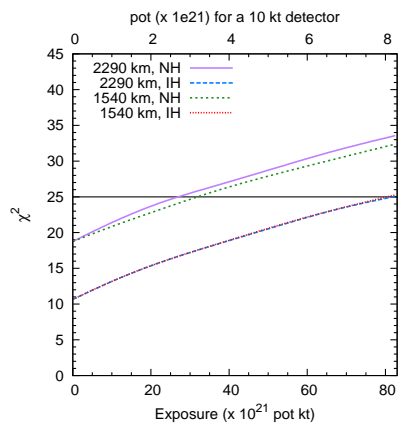}
\caption[Octant sensitivity $\chi^2$ vs LBNO 
exposure, for the $2290$ km and $1540$ km baselines.]
{Octant sensitivity $\chi^2$ vs LBNO 
exposure, for the $2290$ km and $1540$ km baselines and both hierarchies, with 
$\theta_{23}=39^\circ$. The additional axis along 
the upper edge of the graph shows the required total POT assuming a detector 
mass of $10$ kt.}
\label{fig:octexpo}
\end{center}
\end{figure}

In  Fig. \ref{fig:octexpo}, we have shown how the octant sensitivity of these experiments 
increases as the exposure for LBNO is increased. 
For this, we have chosen the 
true value of $\theta_{23}$ to be $39^\circ$. 
Because of the better performance of \nova+T2K+ICAL when NH is true, the adequate 
exposure for LBNO is higher when IH is true. Given our current state of ignorance 
about the true hierarchy in nature, we give the $\chi^2$ for the worst case
i.e.,we present the octant sensitivity results (cf. Table \ref{tab:res}) for the hierarchy that requires a higher exposure.
The fig shows that it is sufficient to have an exposure of around 
$83\times10^{21}$ POT-kt to reach $\chi^2=25$ for both the baselines. 
The upper axis shows the total POT required, with a $10$ kt 
detector. For instance, we see that $8.3\times10^{21}$ POT is sufficient if we 
have a $10$ kt detector. This translates to a runtime of a little under 
$3$ years in 
each $\nu$ and $\overline{\nu}$ mode, given an intensity of $3\times10^{21}$ POT/yr. 

\begin{figure}[ht!]
\begin{center}
\vspace{0.3 in}
\includegraphics[scale=0.7]{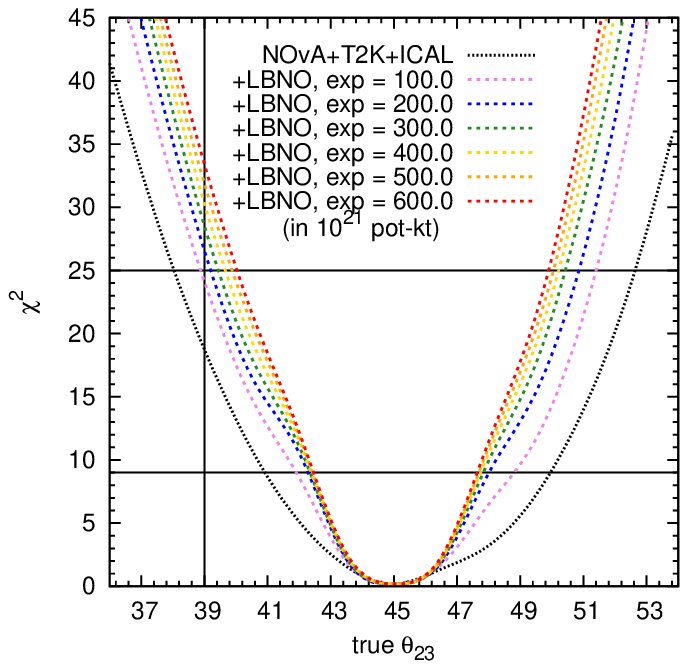}
\hspace{-0.7 in}
\includegraphics[scale=0.7]{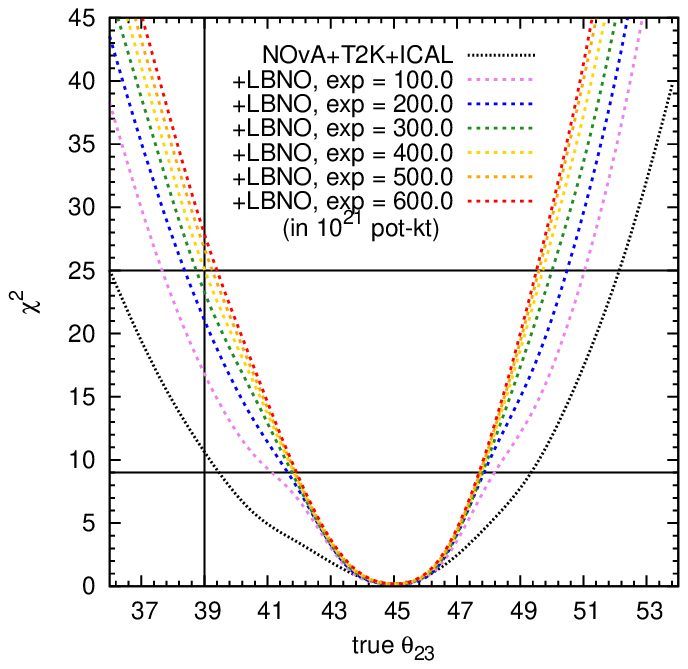} 
\caption[Octant sensitivity $\chi^2$ of LBNO vs true $\theta_{23}$  
for the $130$ km baseline.]{Octant sensitivity $\chi^2$ vs true $\theta_{23}$ 
for the $130$ km baseline. The left (right) 
panel is for true NH (IH). In both panels, the lowermost densely-dotted (black) 
curve is for \nova+T2K+ICAL, while the curves above are for \nova+T2K+ICAL+LBNO, 
for various values of LBNO exposure. All the plotted sensitivities 
 are for the least favourable value of true $\dcp$.}
\label{fig:oct130}
\end{center}
\end{figure}

\begin{figure}[ht!]
\begin{center}
\vspace{0.3 in}
\includegraphics[scale=0.8]{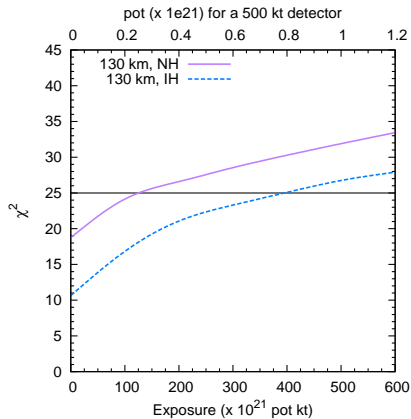}
\caption[Octant sensitivity $\chi^2$ vs LBNO 
exposure, for the $130$ km baseline.]{Octant sensitivity $\chi^2$ vs LBNO 
exposure, for the $130$ km baseline and both hierarchies, with 
$\theta_{23}=39^\circ$. The additional 
axis along the upper edge of the graph shows the required total POT assuming a 
detector mass of $500$ kt.}
\label{fig:oct130expo}
\end{center}
\end{figure}

Fig. \ref{fig:oct130} is the same as Fig. \ref{fig:octall}, but for the $130$ km 
baseline.
As expected, because of smaller matter effects, the exposure required 
to determine the octant is much higher than that for the other two baselines. 
However, for a large mass detector like MEMPHYS that is being planned for the 
Fr\'{e}jus site, this exposure is not difficult to attain. The sensitivity as 
a function of LBNO exposure for this baseline is shown in Fig. \ref{fig:oct130expo}.
We need an exposure of around $400\times10^{21}$ POT-kt in this case. For this 
graph, the upper axis shows the required POT if we consider a $500$ kt detector, 
as proposed for MEMPHYS \cite{newmemphys}. We see that for such a large mass detector, 
only around $0.8\times10^{21}$ POT is adequate to exclude the octant for 
$\theta_{23} = 39^\circ$. 

\begin{figure}[ht!]
\begin{center}
\vspace{0.3 in}
\includegraphics[scale=0.7]{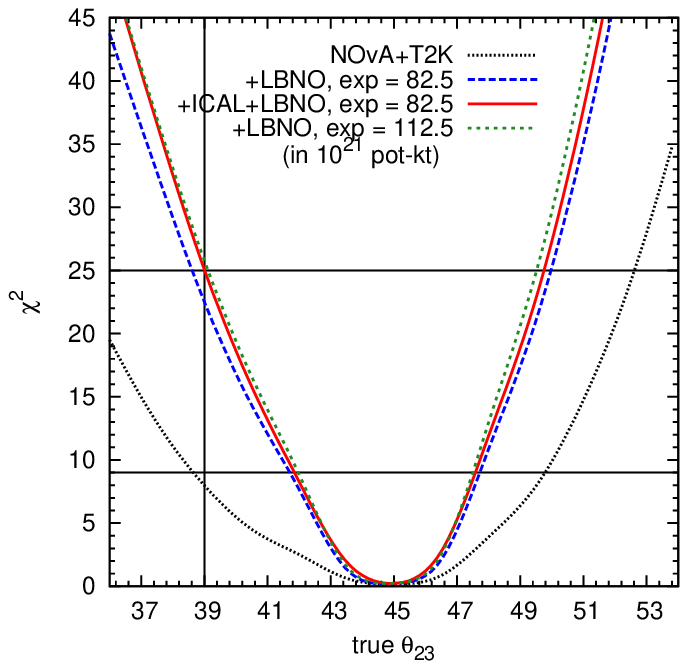}
\hspace{-0.7 in}
\includegraphics[scale=0.7]{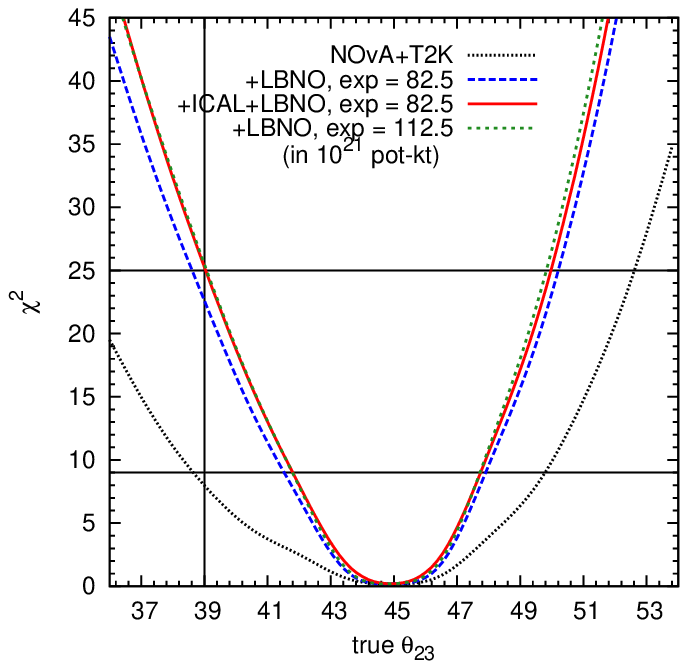} 
\caption[Octant sensitivity $\chi^2$ of LBNO for the 2290 km and 1540 km baselines taking different 
 combinations of experiments, demonstrating the synergy between them.]{ Octant sensitivity $\chi^2$ for different 
 combinations of experiments, demonstrating the synergy between them. The left (right) 
 panel is  for a LBNO baseline of $2290$ ($1540$) km, assuming IH to be true. 
 With only T2K+\nova+LBNO 
 (dashed, blue), the sensitivity is lower than for T2K+\nova+LBNO+ICAL (red, solid).
 Without ICAL data, the LBNO exposure would have to be increased substantially 
 (dotted, green) in order to get comparable sensitivity. All the plotted sensitivities 
 are for the least favourable value of true $\dcp$.}
\label{fig:octsyn}
\end{center}
\end{figure}

Fig. \ref{fig:octsyn} shows the synergy 
between LBL experiments and ICAL. 
In the left (right) 
panel, we have chosen the LBNO baseline of $2290$ ($1540$) km to illustrate this 
point. IH is assumed to be the true hierarchy. The sensitivity of 
T2K+\nova\ alone (densely-dotted, black curve) is enhanced by adding data 
from ICAL and LBNO. The solid (red) curve in the left panel shows that an exposure 
of $82.5\times10^{21}$ POT-kt is enough to determine the octant with $\chi^2=25$ 
at $39^\circ$. But without ICAL data (dashed, blue curve), the sensitivity 
would be lower. The dotted (green) curve shows that only with $112.5\times10^{21}$ POT-kt
 (more than 35\% higher than the 
adequate amount), can we attain $\chi^2=25$ without ICAL. For $1540$ km 
(right panel) also, 
similar features are observed. This demonstrates the 
advantage of adding atmospheric neutrino data.

\subsection{Evidence for CP Violation}

\begin{figure}[ht!]
 \begin{center}
 \vspace{0.3 in}
 \includegraphics[scale=0.7]{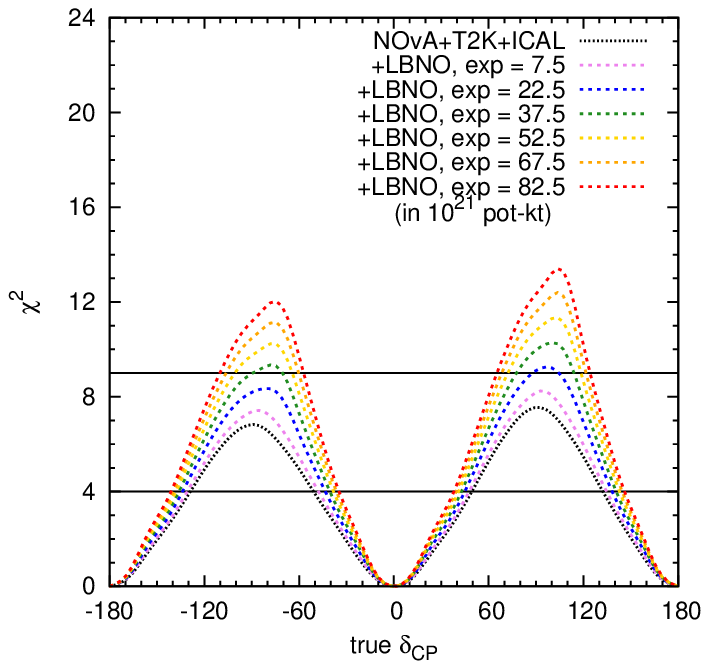}
 \hspace{-0.7 in}
 \includegraphics[scale=0.7]{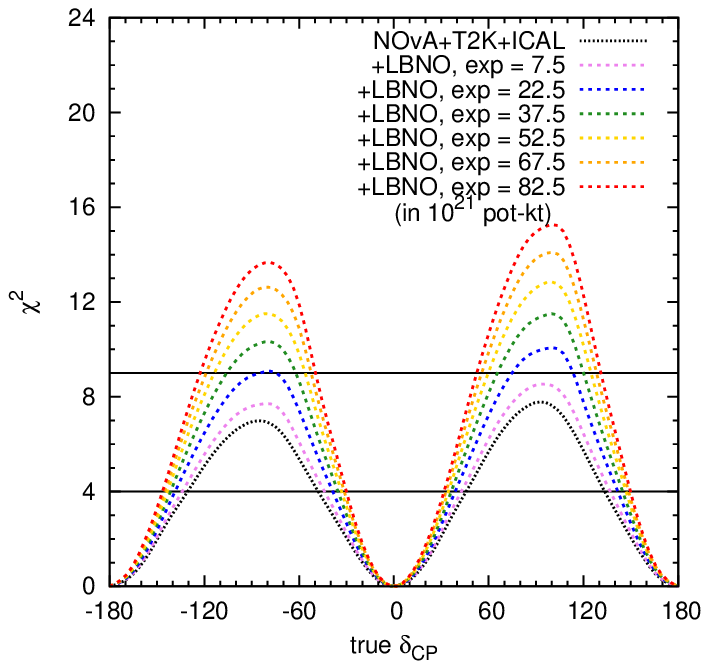} \\
 \vspace{0.1 in}
 \includegraphics[scale=0.7]{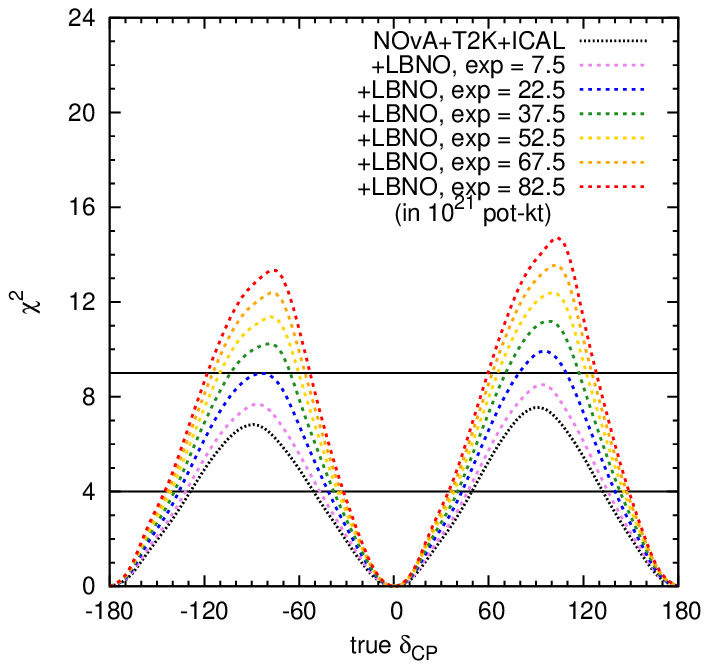}
 \hspace{-0.7 in}
 \includegraphics[scale=0.7]{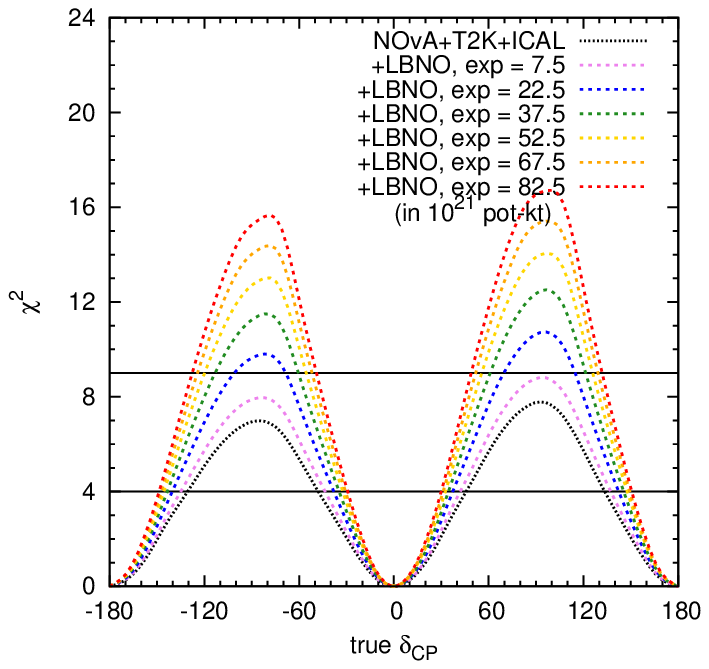} \\
 \caption[CP violation detection $\chi^2$ of LBNO vs true $\dcp$ for the 2290 km and 1540 km baselines.]{ CP violation detection $\chi^2$ vs true $\dcp$. 
 The top (bottom) panels are for the $2290$ ($1540$) km baseline. The left (right) 
 panels are for true NH (IH). In all the panels, the lowermost densely-dotted (black) 
 curve is for \nova+T2K+ICAL, while the curves above are for \nova+T2K+ICAL+LBNO, 
 for various values of LBNO exposure. All the plotted sensitivities 
 are for the least favourable value of true $\theta_{23}$.}
\label{fig:cpdiscall}
\end{center}
\end{figure}

 Now, we discuss the detection
 of CP violation.
 We show our results as a function of true $\dcp$ in 
 Fig. \ref{fig:cpdiscall}.
As in the case of hierarchy exclusion, we have 
minimised over three different true 
values of $\theta_{23}$ and have chosen the most 
conservative case among these for each true $\dcp$.
 
\begin{figure}[ht!]
 \begin{center}
 \vspace{0.3 in}
 \includegraphics[scale=0.8]{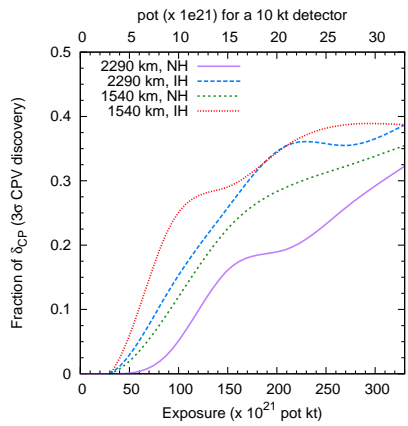}
 \caption[Fraction of the full $\dcp$ range for which it is 
 possible to detect CP violation at $3\sigma$ vs LBNO 
 exposure, for the $2290$ km and $1540$ km baselines.]{Fraction of the full $\dcp$ range for which it is 
 possible to detect CP violation (exclude $\dcp=0,180^\circ$) at $3\sigma$ vs LBNO 
 exposure, for the $2290$ km and $1540$ km baselines and both hierarchies. The 
 additional 
 axis along the upper edge of the graph shows the required total POT assuming a 
 detector mass of $10$ kt.}
 \label{fig:cpdiscexpo}
 \end{center}
 \end{figure}

 We see in Fig. \ref{fig:cpdiscall} that with \nova+T2K+ICAL, only around 
 $\chi^2=4$ can be attained, for a small range of $\dcp$ values 
 around $\pm 90^\circ$. Adding LBNO (2290 km and 1540 km) data with increasing exposure can enhance 
 this and can even help to achieve $\chi^2=9$ for CP detection for some range of 
 $\dcp$. In Fig. \ref{fig:cpdiscexpo}, we have plotted the fraction of $\dcp$ 
 for which CP violation can be detected with $\chi^2=9$, as a 
 function of the LBNO exposure.
 As an example, if we aim to detect CP 
 violation for at least 20\% of $\dcp$ values, then we require around 
 $240\times10^{21}$ ($170\times10^{21}$) POT-kt exposure from LBNO with a 
 baseline of $2290$ ($1540$) km. 
 It can also be seen from the figure that with $350\times10^{21}$ POT-kt exposure,
 the maximum CP fraction for which a $3\sigma$ sensitivity is achievable 
 ranges from 30\% to 40\%. 
 The upper axis shows that these values correspond 
 to $24\times10^{21}$ ($17\times10^{21}$) POT, if we consider a $10$ kt detector. 
 
\begin{figure}[ht!]
 \begin{center}
 \vspace{0.3 in}
 \includegraphics[scale=0.7]{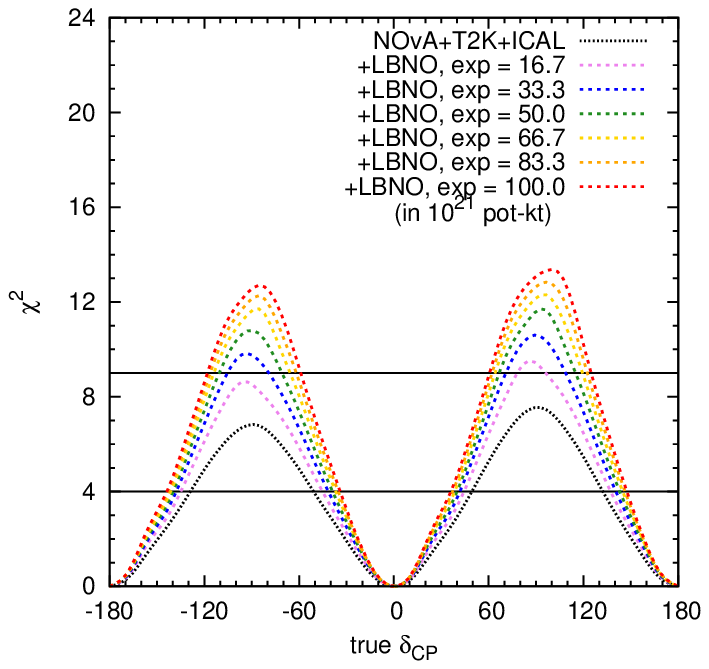}
 \hspace{-0.7 in}
 \includegraphics[scale=0.7]{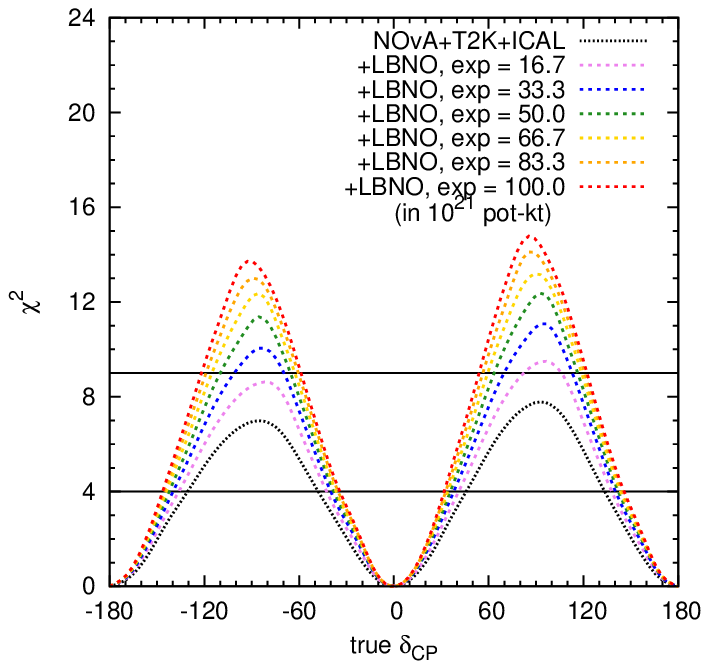} 
 \caption[CP violation detection $\chi^2$ of LBNO vs true $\dcp$ 
 for the $130$ km baseline.]{CP violation detection $\chi^2$ vs true $\dcp$ 
 for the $130$ km baseline. The left (right) 
 panel is for true NH (IH). In both panels, the lowermost densely-dotted (black) 
 curve is for \nova+T2K+ICAL, while the curves above are for \nova+T2K+ICAL+LBNO, 
 for various values of LBNO exposure. All the plotted sensitivities 
  are for the least favourable value of true $\theta_{23}$.}
 \label{fig:cpdisc130}
 \end{center}
 \end{figure}

\begin{figure}[ht!]
\begin{center}
\vspace{0.3 in}
\includegraphics[scale=0.8]{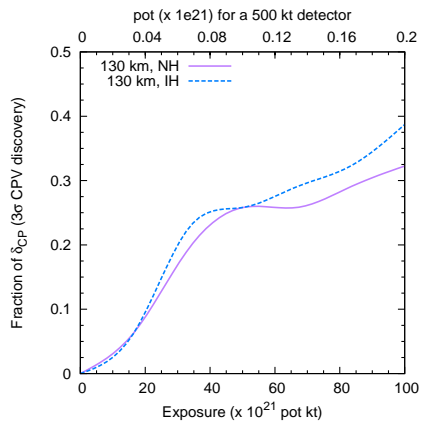}
\caption[Fraction of the full $\dcp$ range for which it is 
possible to detect CP violation at $3\sigma$ vs LBNO 
exposure, for the $130$ km baseline and both hierarchies.]{Fraction of the full $\dcp$ range for which it is 
possible to detect CP violation at $3\sigma$ vs LBNO 
exposure, for the $130$ km baseline and both hierarchies. The additional 
axis along the upper edge of the graph shows the required total POT assuming a 
detector mass of $500$ kt.}
\label{fig:cpdisc130expo}
\end{center}
\end{figure}

 Figs. \ref{fig:cpdisc130} and \ref{fig:cpdisc130expo} show the results for 
 the $130$ km option.
Once again, we see that an exposure much higher than 
the longer baselines is required. In this case, CP detection for 20\% $\dcp$ 
values requires an exposure of around $35\times10^{21}$ POT-kt. This 
is not difficult to achieve with a large MEMPHYS-like detector. In fact, 
the total POT required by a $500$ kt detector at $130$ km is only around 
$0.07\times10^{21}$ POT. 
Moreover, an underground megaton scale detector like MEMPHYS can 
also be used to collect atmospheric neutrino data \cite{campagne}, which can further 
enhance the sensitivity because of the ability of the atmospheric neutrinos to rule out the wrong hierarchy solutions. 

\begin{figure}[ht!]
\begin{center}
\vspace{0.3 in}
\includegraphics[scale=0.8]{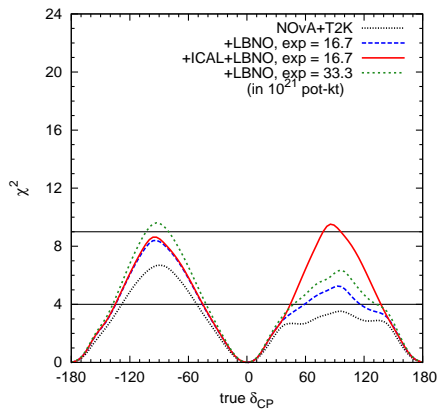}
\caption[CP detection sensitivity $\chi^2$ of LBNO for the 130 km baseline taking different 
 combinations of experiments, demonstrating the synergy between them.]{CP detection sensitivity $\chi^2$ for different 
 combinations of experiments, demonstrating the synergy between them. This plot is 
 for a LBNO baseline of $130$ km, assuming NH to be true. With only T2K+\nova+LBNO 
 (dashed, blue), the sensitivity is lower than for T2K+\nova+LBNO+ICAL (red, solid).
 Without ICAL data, the LBNO exposure would have to be increased substantially 
 (dotted, green) in order to get comparable sensitivity. All the plotted sensitivities 
 are for the least favourable value of true $\theta_{23}$.}
\label{fig:cpdiscsyn}
\end{center}
\end{figure}

In Fig. \ref{fig:cpdiscsyn}, we have demonstrated the synergy between 
atmospheric and long-baseline experiments for the baseline of $130$ km 
and with NH.
We see that with only T2K+\nova\ 
(densely-dotted, black curve), one suffers from the hierarchy-$\dcp$ 
degeneracy in the unfavourable region of $\dcp$. This degeneracy is lifted by 
adding information from other experiments. On adding data from ICAL and 
$16.7\times10^{21}$ POT-kt of LBNO (solid, red curve), we just reach 
$\chi^2=9$ sensitivity. 
With the same LBNO exposure, absence of ICAL data reduces the detection 
reach, as seen from the dashed (blue) curve. Reaching $\chi^2=9$ without 
ICAL will require the LBNO exposure to be doubled, as the 
dotted (green) curve shows. Thus ICAL plays a significant role in $\dcp$ measurement for this baseline.
For the two longer baselines, LBNO even with very low exposure in 
conjunction with T2K and \nova\ can break the hierarchy-$\dcp$ degeneracy 
by excluding the wrong hierarchy solution.
Therefore, the contribution of ICAL towards detecting CP violation 
becomes redundant in this case.

%% file: LBNE.tex

\section{Physics Potential of LBNE in Conjunction with T2K, \nova\ and ICAL}
\label{sec3}
In this section, we explore the sensitivity reach of the LBNE experiment
in determining the remaining unknowns in neutrino oscillation, in combination with the
other experiments T2K, \nova\ and ICAL.
In our study we carry out a similar analysis as that of LBNO i.e.,
we determine the most conservative specifications that this experiment needs, in order to
measure the remaining unknown parameters to a specified level of precision. 
Addition to that, we will also study the effect of near detector
in constraining systematic error, the role of second oscillation maxima. We will also present an optimisation study of neutrino and antineutrino run of LBNE.

\subsection{Experimental Specifications and Other Simulation Details}

As mentioned in chapter \ref{sec2} for LBNE, there are two options being considered
for the proton beam - 80 GeV and 120 GeV.
In our study, we have chosen the 120 GeV proton beam.
As the neutrino flux decreases with proton energy, this
gives us a lower flux of neutrinos and hence a conservative estimate of our results.

The specifications for the liquid argon detector have been taken from 
Ref. \cite{lbne_interim2010}.  In this study we use the flux corresponding to 
1.2 MW beam power \cite{dancherdack}.  However we give our results 
in terms of MW-kt-yr.  This will enable one to interpret the results in
terms of varying detector volume, timescale and beam power. 
Note that although we use the flux corresponding to 1.2 MW beam power, if the 
accelerator geometry remains the same, then the change in the value of the 
beam power will proportionally change the flux. 
Therefore, the flux for a different value of beam power can be obtained by 
simply scaling the `standard' flux file by the appropriate factor.

The specifications for T2K, \nova\ and ICAL is same as that of LBNO analysis.

In the analyses that follow, we have evaluated the $\chi^2$ for determining the 
mass hierarchy, octant of $\theta_{23}$ and CP violation using a 
combination of LBNE and the current/upcoming experiments T2K, \nova\ and ICAL. 
For each set of `true' values assumed, we evaluate the $\chi^2$ marginalised 
over the `test' parameters. 
For true $\theta_{23}$, we have considered three values -- $39^\circ$, $45^\circ$ 
and $51^\circ$ which are  within
the  current $3\sigma$ allowed range.
The systematic uncertainties are parametrised in terms of four nuisance 
parameters -- signal normalisation error (2.5\%), signal tilt error (2.5\%), 
background normalisation error (10\%) and background tilt error (2.5\%).

 \begin{figure*}[ht!]
 \begin{tabular}{rcl}
 \epsfig{file=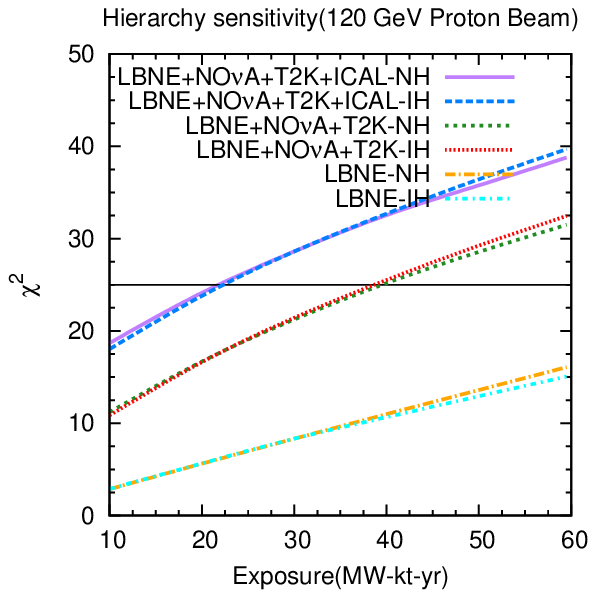, width=0.33\textwidth, bbllx=89, bblly=50, bburx=260, bbury=255,clip=}
 \epsfig{file=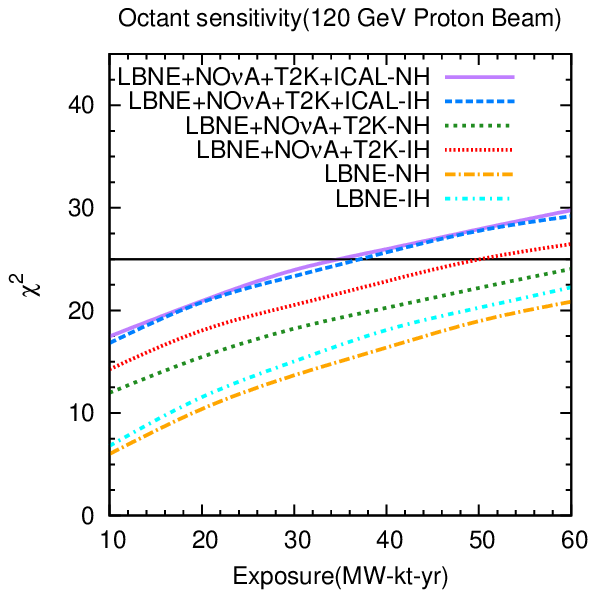, width=0.33\textwidth, bbllx=89, bblly=50, bburx=260, bbury=255,clip=}
 \epsfig{file=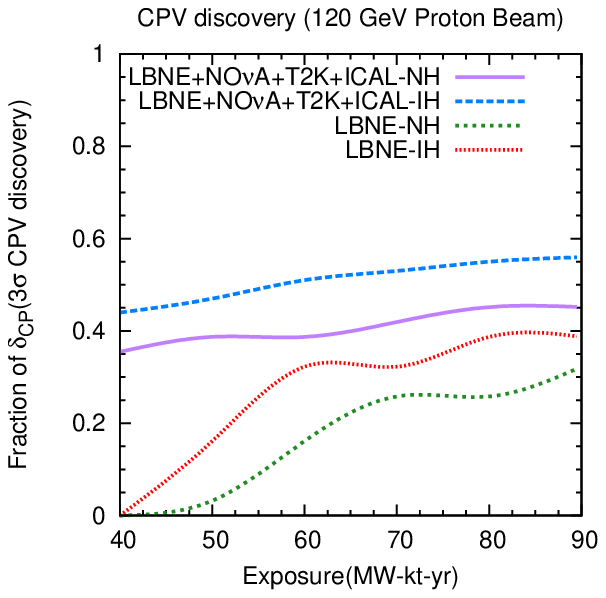, width=0.33\textwidth, bbllx=86, bblly=50, bburx=260, bbury=255,clip=}
 \end{tabular}
 \caption[Hierarchy/Octant/CPV discovery $\chi^2$ vs LBNE 
 exposure, for both hierarchies]{Hierarchy (Octant) sensitivity $\chi^2$ vs LBNE 
 exposure, for both hierarchies in the left (middle) panel. 
 The value of exposure shown here is adequate to exclude the wrong hierarchy 
 for all values of $\dcp$. 
 Two additional sets of curves are shown to show the fall in $\chi^2$ without data
 from ICAL, and the hierarchy sensitivity of LBNE alone. 
 The right panel shows the fraction of $\dcp$ range for which it is possible 
 to exclude the CP conserving cases of $0$ and $180^\circ$, at the 
 $\chi^2=9$ level.
 An additional set of curves is shown to show the the CP sensitivity of LBNE alone.
 }\label{fig:expo}
 \end{figure*}
To economise the configuration of LBNE with 
the help of the current generation of experiments, We 
evaluate the `adequate' exposure for LBNE. The qualifier `adequate', as 
defined in the context of LBNO, means the exposure 
required from the experiment to determine the hierarchy and octant with 
$\chi^2 = 25$, and to detect CP violation with $\chi^2 = 9$. To do so, 
we have varied the exposure of LBNE, and determined the combined 
sensitivity of LBNE along with T2K, \nova\ and ICAL. The variation of 
total sensitivity with LBNE exposure tells us what the adequate exposure 
should be. In this study, we have quantified the exposure for LBNE in units of 
MW-kt-yr. This is a product of the beam power (in MW), the runtime of 
the experiment (in years)
{\footnote{A runtime of $n$ years is to be interpreted as $n$ years each in 
neutrino and antineutrino mode. In this study, we have always considered equal 
runs in both modes for LBNE unless otherwise mentioned.}}
and the detector mass (in kilotons). As a 
phenomenological study, we will only specify the total exposure. 
This may be interpreted experimentally as different combinations of 
beam power, runtime and detector mass whose product quantifies the  
exposure. For example, an exposure of 20 MW-kt-yr could be achieved by 
using a 10 kt detector for 2 years (in each, $\nu$ and $\overline{\nu}$ mode), 
with a 1 MW beam.
We use events in the energy range 0.5 - 8 GeV for LBNE which covers both 
first and second oscillation maxima.

\subsection{Adequate Exposure for LBNE}

\subsubsection{Hierarchy Sensitivity}

In the left panel of Fig. \ref{fig:expo}, we have shown the combined sensitivity of LBNE, 
\nova, T2K and ICAL for determining the mass hierarchy, as the exposure for 
LBNE is varied.
Note that the hierarchy sensitivity depends very strongly on 
the true value of $\dcp$ and $\theta_{23}$. In this study, as in LBNO, we 
are interested in finding out the least exposure needed for LBNE, 
irrespective of the true values of the parameters in nature. Therefore, we 
have evaluated the $\chi^2$ for various true values of these parameters, and taken the most conservative 
case out of them. Thus, the exposure plotted here is for the most 
unfavourable values of true $\dcp$ and $\theta_{23}$. Since hierarchy 
sensitivity of the $P_{\mu e}$ channel increases with $\theta_{23}$, the 
worst case is usually found at the lowest value considered which is 
$\theta_{23} = 39^\circ$. The most unfavourable of $\dcp$ is around 
$+(-)90^\circ$ for NH (IH) \cite{t2knova} (cf. Section \ref{sec1}.). 
Separate curves are shown for both hierarchies, but the results 
are almost the same in both the cases. 
We find that 
the adequate exposure for LBNE including T2K, \nova\ and ICAL data 
is around 22 MW-kt-yr 
for both NH and IH. 
This is shown by the upper curves. 
The two intermediate curves show the same sensitivity, but without including 
ICAL data in the analysis. In this case, the adequate exposure is around 
39 MW-kt-yr. 
Thus, in the absence of ICAL data, LBNE would have to increase 
its exposure by over 75\% to achieve the same results. 
For the benchmark values of 1.2 MW power and 10 kt 
detector, the exposure of 22 MW-kt-yr  implies 
under 2 years of running in each mode
whereas 
the adequate exposure of
39 MW-kt-yr 
corresponds to about 3 years exposure in neutrino and 3 years in antineutrino mode.  

Finally, we show 
the sensitivity from LBNE alone, in the lowermost curves of Fig. \ref{fig:expo}. For the range 
of exposures considered, LBNE can achieve hierarchy sensitivity up to 
the $\chi^2=16$ level.    
The first row of Table \ref{tab:adequate} shows the adequate exposure 
required for hierarchy sensitivity reaching $\chi^2 = 25$ for 
only LBNE and also  
after adding the data from T2K, \nova\ and ICAL. 
The numbers in the parentheses correspond to IH. With only LBNE, the 
exposure required to reach $\chi^2=25$ for hierarchy sensitivity is seen to be 
much higher .  

\begin{table*}[htb]
\begin{center}
\begin{tabular}{|c|c|c|c|}
\hline
Sensitivity   & LBNE+NO$\nu$A+T2K+ICAL  & LBNE+NO$\nu$A+T2K & LBNE   \\          
\hline
Hierarchy($\chi^2=25$)         & 22(22)    &  39(39)  & 95(106)    \\
Octant($\chi^2=25$)            & 22(37)     &  65(50)  & 84(76)    \\
CP(40$\%$ at $\chi^2=9$)        & 65(36)     &  65(36)  & 114(90)    \\
\hline

\hline
\end{tabular}
\end{center}
\caption{Adequate exposures of LBNE for determining hierarchy, octant and CP in units of MW-kt-yr for NH (IH).}
\label{tab:adequate} 
\end{table*}

\begin{figure*}
\begin{tabular}{rcl}
\epsfig{file=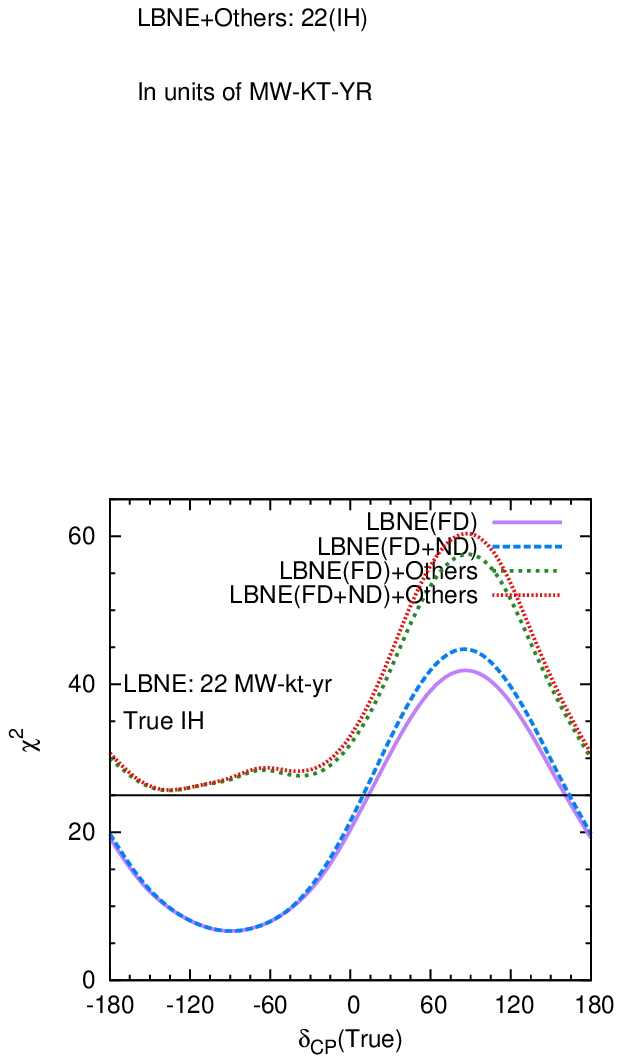, width=0.33\textwidth, bbllx=89, bblly=50, bburx=260, bbury=255,clip=}
\epsfig{file=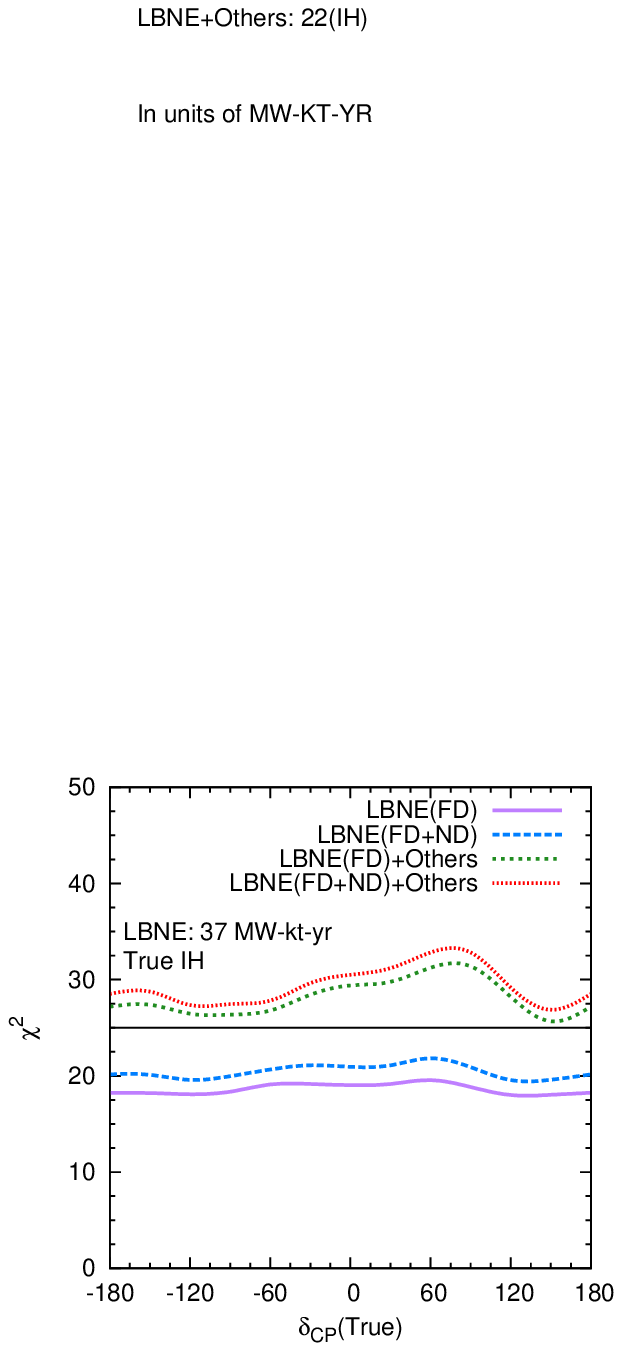, width=0.33\textwidth, bbllx=89, bblly=50, bburx=260, bbury=255,clip=}
\epsfig{file=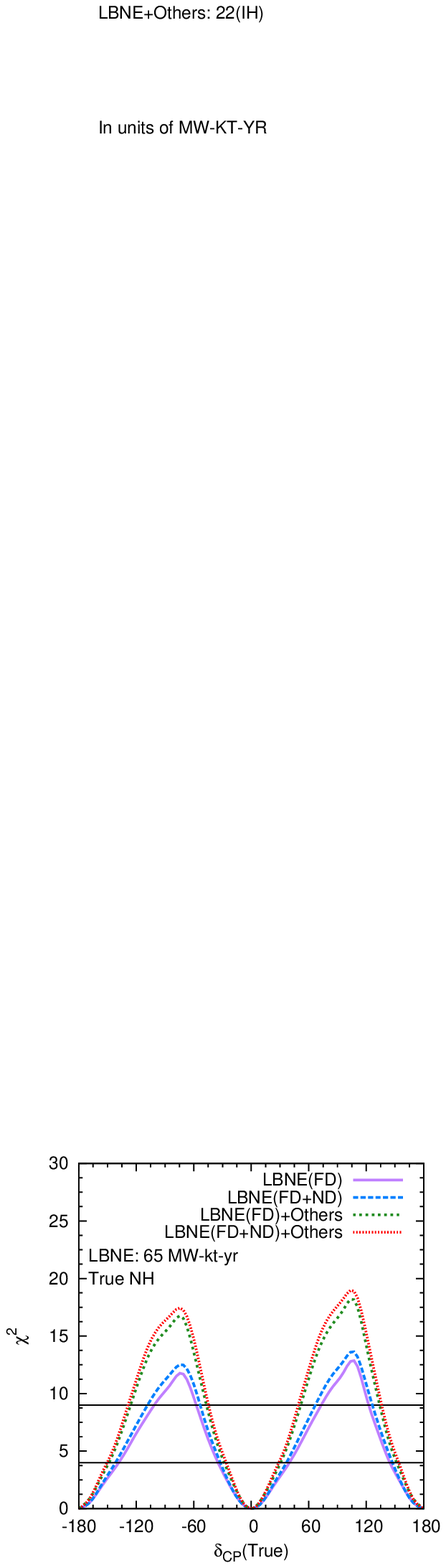, width=0.33\textwidth, bbllx=86, bblly=50, bburx=260, bbury=255,clip=}
\end{tabular}
\caption[Hierarchy/Octant/CPV discovery 
$\chi^2$ of LBNE vs true $\dcp$ showing effect of including a near detector.]{Hierarchy/Octant/CP violation discovery sensitivity 
$\chi^2$ vs true $\dcp$ in the left/middle/right panel. The various curves show 
the effect of including a near detector on the sensitivity of 
LBNE alone and LBNE combined with the other experiments. 
}\label{fig:syst}
\end{figure*}

\subsubsection{Octant Sensitivity}

The  mass hierarchy as well as the values of 
$\dcp$ and $\theta_{23}$ in nature affect  the octant sensitivity
of experiments significantly. In our analysis, we have considered various 
true values of $\dcp$ across its full range, and two representative 
true values 
of $\theta_{23}$ -- $39^\circ$ and $51^\circ$. 
Having evaluated the 
minimum $\chi^2$ for each of these cases, we have chosen the lower 
value to get a more conservative estimate. Thus, we have ensured that the adequate exposure
shown here holds, irrespective of the true octant of $\theta_{23}$. 
Note that octant sensitivity reduces as we go more towards $\theta_{23}= 45^\circ$. 
Thus the above choice of true $\theta_{23}$ does not correspond to the most 
conservative case.

The middle panel of Fig. \ref{fig:expo} shows the combined octant 
sensitivity of the experiments, as a function of LBNE exposure. 
Around  35 (37) MW-kt-yr for NH (IH) is the required exposure 
for LBNE, to  measure the octant 
with 
 \nova, T2K and ICAL.
This implies a runtime of around 3 years in 
each mode for the `standard' configuration of LBNE. 
Without 
information from ICAL however, LBNE would have to increase its exposure to 
around 65 (50) MW-kt-yr for NH (IH) to measure the octant with $\chi^2=25$.
For a 1.2 MW beam and a 10 kt detector this implies about 
5 (4)  years  for NH (IH) in each mode.  
While only LBNE would need a higher exposure of 84 (76) MW-kt-yr 
for NH (IH) corresponding to about 7 (6) years in each mode. 
Thus including ICAL data reduces the exposure required from LBNE. 
This is summarized in the 2nd row of Table \ref{tab:adequate}. 

\subsubsection{Detecting CP Violation}

Here, we have tried to 
determine the fraction of the entire $\dcp$ range for which the 
setups can detect CP violation with at least $\chi^2=9$. We have always 
chosen the smallest fraction over the true values of $\theta_{23}$ considered
($39^\circ$, $45^\circ$ and $51^\circ$), 
so as to get a conservative estimate. 

We find in the right panel of Fig. \ref{fig:expo} that for the range of 
exposures considered, 
the fraction of $\dcp$ is between 0.35 and 0.55. While the exposure 
increases by a factor of two, the increase in the fraction of $\dcp$ is 
very slow. This combination of experiments can detect CP violation over 
40\% of the $\dcp$ range with an exposure of about 65 MW-kt-yr at LBNE for NH 
(i.e. a runtime of around 5.5 years for LBNE). 
Without including T2K and \nova\ information the exposure required will 
be 114 MW-kt-yr for 40\% coverage for discovery of $\dcp$. 
In this context, we want to remind that, one of the mandates of LBNE/DUNE is
$3\sigma$ CP coverage for 75\% values of $\dcp$ \cite{lbnf}. 
We find that 
an exposure of 300 MW-kt-yr in neutrinos and 300 MW-kt-yr in antineutrinos
gives 69\% (73\%) CP coverage at
$3 \sigma$ for $\theta_{23}=39^\circ$ and 60\% (65\%) for $51^\circ$ in NH (IH). We also find that
addition of NO$\nu$A and T2K data does not help much for such
high values of exposure. The results
are summarized in Table \ref{tab:lbnf_frac}. 

\begin{table*}
\begin{center}
\begin{tabular}{|c|c|c|}
\hline
3$\sigma$ CPV coverage for $\theta_{23}$    &  LBNE   & LBNE+NO$\nu$A+T2K   \\          
\hline
$39^o$         & 69(73)    &  71(74)    \\
$51^o$            & 60(65)     &  63(67)    \\
\hline

\hline
\end{tabular}
\end{center}
\caption{CPV coverage fraction of LBNE at 3$\sigma$ for total 600 MW-Kt-yr exposure.}
\label{tab:lbnf_frac} 
\end{table*}
 
In the following subsections, we fix the exposure in each case to be the 
adequate exposure as listed in Table \ref{tab:adequate}, for the 
most conservative parameter values. 

\subsection{Role of the Near Detector in Reducing Systematics}
\label{sec:syst}

The role of the near detector (ND) in long-baseline neutrino experiments 
has been well discussed in the literature, see for example 
Refs. \cite{nd_t2k1,nd_t2k2,nd_minos}. The measurement of events at the 
near and far detector (FD) reduces the uncertainty associated with the 
flux and cross-section of neutrinos. Thus the role of the near 
detector is to reduce systematic errors in an 
oscillation experiment. It has recently been seen that the near detector for 
the T2K experiment can bring about a spectacular reduction of systematic 
errors \cite{nd_t2k3}. 

In this study, we have tried to quantify the improvement in results, once 
the near detector is included. Instead of putting in reduced systematics 
by hand, we have explicitly simulated the events at the near detector using 
GLoBES. The design for the near detector is still being planned. For our 
simulations, we assume that the near detector has a mass of 5 tons and 
is placed 459 meters from the source. The flux at the near detector site 
has been provided by the LBNE collaboration \cite{dancherdack}. The 
detector characteristics for the near detector are as follows \cite{nd_lbneindia}. 
The muon (electron) detection efficiency is taken to be 95\% (50\%). The NC 
background can be rejected with an efficiency of 20\%. The energy resolution 
for electrons is $6\%/ \sqrt{E \rm{(GeV)} }$, while 
that for muons is $37$ MeV across the entire energy range of interest. 
Therefore, for the neutrinos, we use a (somewhat conservative) energy 
resolution of $20\%/ \sqrt{E \rm{(GeV)} }$. The systematic errors that the 
near detector setup suffers from are assumed to be the same as those of 
the far detector. 

In order to have equal runtime for both FD and ND, we fix
the far detector volume as 10 kt and consider both the detectors to 
receive neutrinos from 1.2 MW beam. This
fixes the runtime of FD which is then also used in the simulation for ND. 
The run times corresponding to
the exposures which are taken in this section are following:
1.8 year for hierarchy sensitivity, 3.1 year for octant sensitivity and 
5.4 year for CPV discovery sensitivity.

In order to simulate the ND+FD setup for LBNE, we use GLoBES to generate 
events at both detectors, treating them as separate experiments. 
We then use these two data sets to perform a correlated systematics analysis using 
the method of pulls \cite{pulls_gg}. This gives us the combined sensitivity 
of LBNE using 
both near and far detectors. Thereafter, the procedure of combining results 
with other experiments and marginalising over oscillation parameters 
continues in the usual manner. The results are shown in 
Fig. \ref{fig:syst}. The effect of reduced systematic errors is felt most 
significantly in regions where the results are best. This is because 
for those values of $\dcp$, the experiment typically has high enough 
statistics for systematic errors to play an important role. 

\begin{figure}
\begin{center}
\epsfig{file=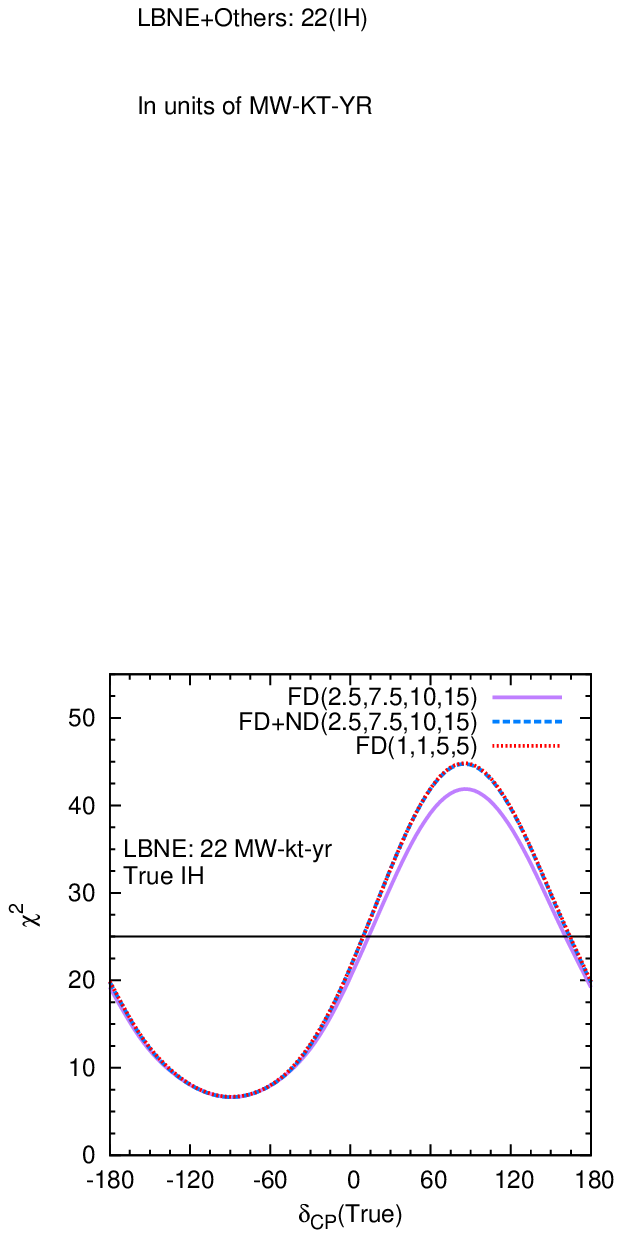, width=0.5\textwidth, bbllx=80, bblly=50, bburx=275, bbury=220,clip=}
\caption[$\chi^2$ vs $\dcp$ for LBNE, showing the effect of including the near detector.]
{$\chi^2$ vs $\dcp$ showing the improvement in systematics due to inclusion of the 
near detector. The numbers in brackets denote $\nu_e$ appearance signal 
norm error, $\nu_\mu$ disappearance signal norm error, $\nu_e$ appearance 
background norm error and $\nu_\mu$ disappearance background norm error.
}
\label{fig:syst2}
\end{center}
\end{figure}

Next, we have tried to quantify the reduction in systematic errors seen 
by the experiment, when the near detector is included. To be more 
specific, if the systematic errors seen by the far detector setup are
denoted by 
$\vec{\pi}$, then what is the effective set of errors 
$\vec{\pi}_{\rm eff}$ for the far detector setup, once the 
near detector is also included. In other words, for given systematic 
errors $\vec{\pi}$, we have found the value of $\vec{\pi}_{\rm eff}$ 
that satisfies the relation
\begin{equation}
 \chi^2 ({\rm{FD}} (\vec{\pi}_{\rm eff})) \equiv 
 \chi^2 ({\rm{FD}} (\vec{\pi}) + {\rm{ND}} (\vec{\pi}) ) ~,
\end{equation}
where the right-hand side denotes the correlated combination. The result of the computation is shown in 
Fig. \ref{fig:syst2}, for the case of hierarchy determination. We have 
chosen typical values of systematic errors for the detector: 
$\nu_e$ appearance signal norm error of 2.5\%, $\nu_\mu$ disappearance signal 
norm error of 7.5\%, $\nu_e$ appearance background norm error of 10\% and 
$\nu_\mu$ disappearance background norm error of 15\%.
The tilt error is taken as 2.5\% in both appearance and 
disappearance channels.
The first four numbers 
constitute $\vec{\pi}$, as labeled in the 
figure.
We have not varied the tilt errors in this particular 
analysis because their effect on overall results is quite small.
The sensitivity 
of FD+ND obtained using these numbers, are matched by an FD setup with 
effective errors as follows: $\nu_e$ appearance signal norm error of 1\%, 
$\nu_\mu$ disappearance signal 
norm error of 1\%, $\nu_e$ appearance background norm error of 5\% and 
$\nu_\mu$ disappearance background norm error of 5\%. Similar results 
are obtained in the case of octant and CP sensitivity also. 
These results are summarized in Table \ref{tab:systresults}. 

\begin{table}
\begin{center}
  \begin{tabular}{|l|c|c|}
  \hline
  Systematic error & only FD & FD+ND \\
  \hline
   $\nu_e$ app signal norm error & 2.5\% & 1\% \\
   $\nu_\mu$ disapp signal norm error & 7.5\% & 1\% \\
   $\nu_e$ app background norm error & 10\% & 5\% \\ 
$\nu_\mu$ disapp background norm error & 15\% & 5\% \\
\hline
  \end{tabular}
\caption{Reduction in systematic errors in LBNE with the addition 
of a near detector.}
\label{tab:systresults}
\end{center}
\end{table}

 \subsection{Significance of the Second Oscillation Maximum}
\label{sec:secondmax}

For a baseline of 1300 km, the oscillation probability $P_{\mu e}$ has its 
first oscillation maximum around 2-2.5 GeV. This is easy to explain from 
the formula 
\[
 \frac{\mlj^{(m)} L}{4 E} = \frac{\pi}{2} ~,
\]
where $\mlj^{(m)}$ is the matter-modified atmospheric mass-squared difference. 
In the limit $\mkj \to 0$, it is given by 
\[
 \mlj^{(m)} = \mlj \sqrt{ (1-\hat{A})^2 + \sin^2 2\theta_{13} } ~.
\]
The second oscillation maximum, for which the oscillating term takes the 
value $3\pi/2$, occurs at an energy of around 0.6-1.0 GeV. Studies have 
discussed the advantages of using the second oscillation maximum to 
get information on the oscillation 
parameters \cite{wbb_vs_oa,kopphuber_2base}. 
In fact, one of the main aims of the proposed ESSnuSB 
project \cite{Baussan:2013zcy,suprabh_ess} is to study neutrino oscillations 
at the second oscillation maximum. 

\begin{figure}
\begin{center}
\epsfig{file=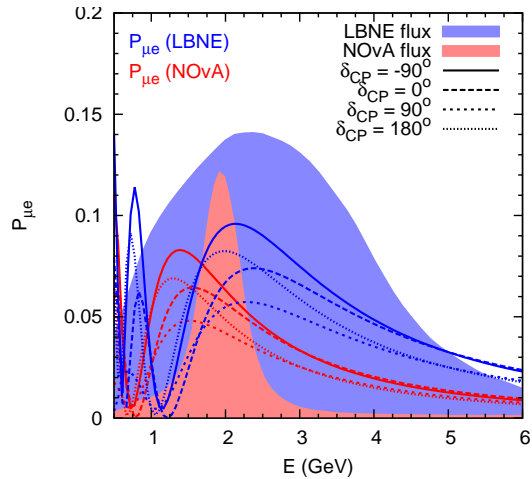, width=0.5\textwidth, bbllx=80, bblly=50, bburx=275, bbury=220,clip=}
\end{center}
\caption[$\nu_\mu$ flux and neutrino oscillation probability $P_{\mu e}$ for 
various representative values of $\dcp$ and normal hierarchy, 
for the \nova\ and LBNE baselines.]{Neutrino oscillation probability $P_{\mu e}$ for 
various representative values of $\dcp$ and normal hierarchy, 
for the \nova\ and LBNE baselines. 
Also shown as shaded profiles in the background are the $\nu_\mu$ flux
for both these experiments (on independent, arbitrary scales).
}
\label{fig:beamprofile}
\end{figure}

The neutrino 
flux that LBNE will use has a wide-band profile, which can extract 
physics from both, the first and second maxima. Fig. \ref{fig:beamprofile} 
shows $P_{\mu e}$ for the LBNE baseline, superimposed on the $\nu_\mu$ flux. 
This is in contrast with \nova, which uses a narrow-band 
off-axis beam concentrating on its first oscillation maximum, in order 
to reduce the $\pi^0$ background at higher energies.

\begin{figure*}
\begin{tabular}{rcl}
\epsfig{file=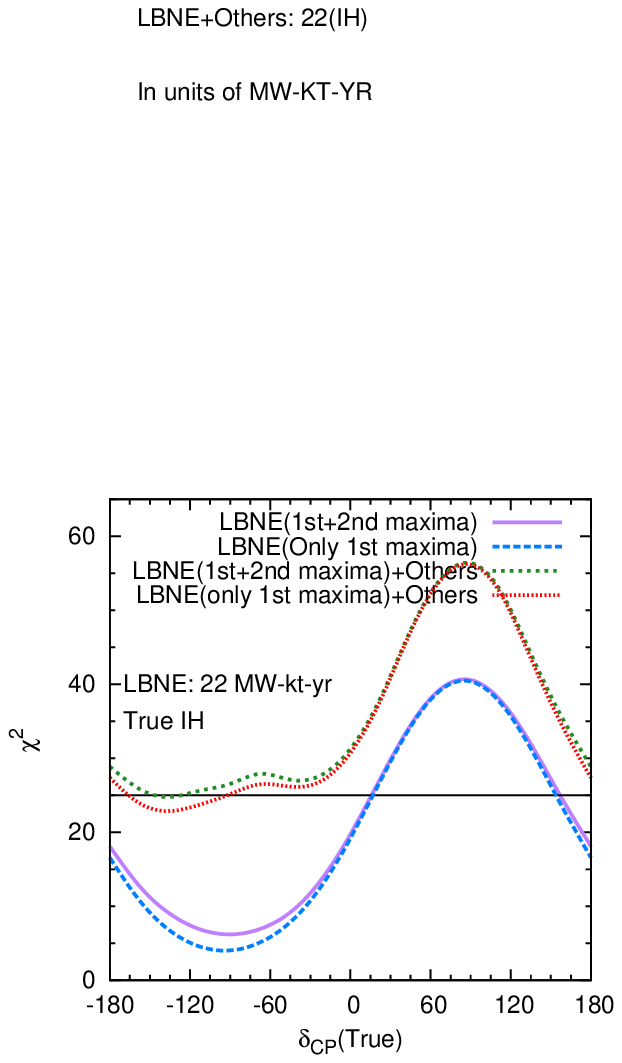, width=0.33\textwidth, bbllx=89, bblly=50, bburx=260, bbury=255,clip=}
\epsfig{file=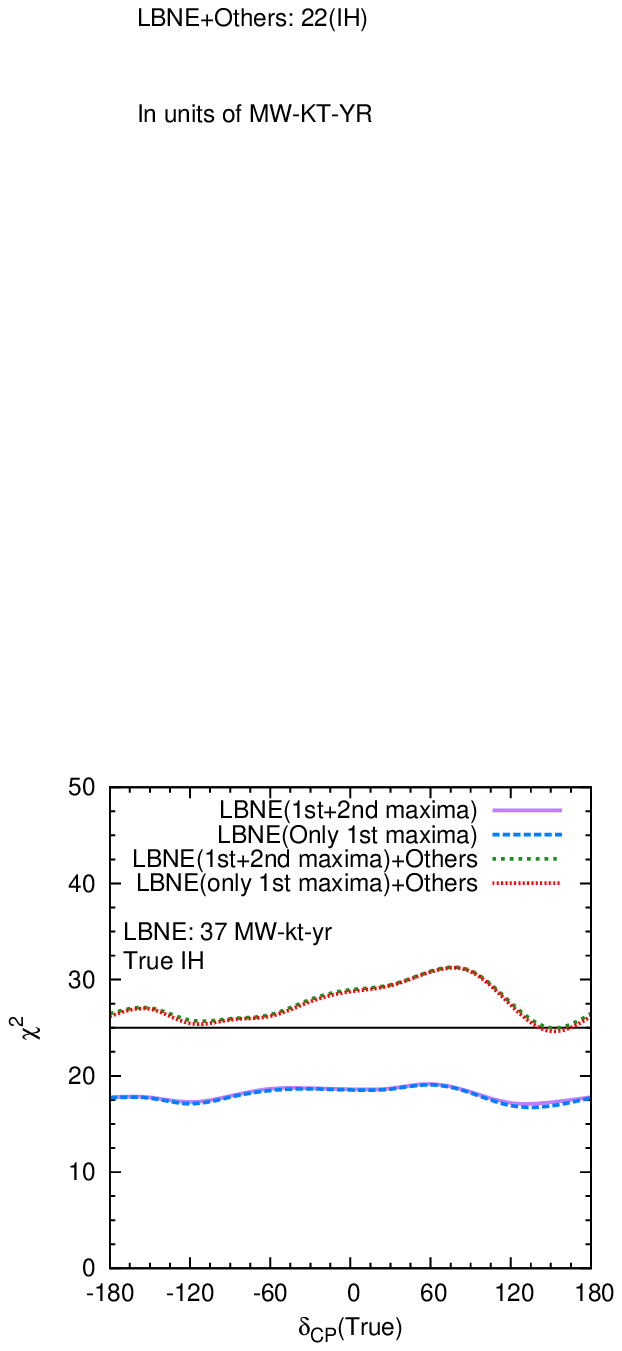, width=0.33\textwidth, bbllx=89, bblly=50, bburx=260, bbury=255,clip=}
\epsfig{file=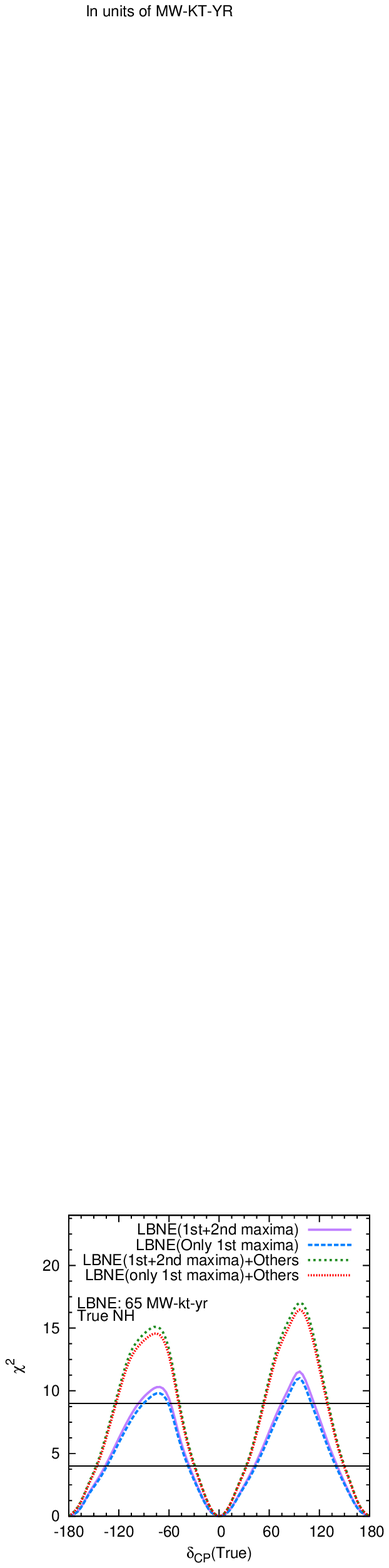, width=0.33\textwidth, bbllx=86, bblly=50, bburx=260, bbury=255,clip=}
\end{tabular}
\caption[Hierarchy/Octant/CPV discovery 
$\chi^2$ of LBNE vs true $\dcp$ showing effect of the second oscillation maximum.]{Hierarchy/Octant/CP violation discovery sensitivity 
$\chi^2$ vs true $\dcp$ in the left/middle/right panels respectively. The various curves show 
the effect of data from the second oscillation maximum on the sensitivity of 
LBNE alone and LBNE combined with the other experiments. 
}\label{fig:secondmax}
\end{figure*}

In order to understand the impact of the second oscillation maximum, 
we have considered two different energy ranges. Above 1.1 GeV, only 
the first oscillation cycle is relevant. However, if we also include 
the energy range from 0.5 to 1.1 GeV, we get information from 
the second oscillation maximum. Fig. \ref{fig:secondmax} compares the 
sensitivity to the hierarchy, octant and CP violation only 
from the first oscillation cycle and from both the oscillation cycles assuming the adequate exposures
obtained previously. We see 
that inclusion of data from the second oscillation maximum only increases 
the $\chi^2$ by a small amount. This increase is visible only for hierarchy sensitivity.
As expected, the effect is pronounced in the region $\dcp=-90^\circ$.
The results for all three performance indicators are given in Table \ref{tab:secondmax}.
It is seen that the effect of including the second oscillation maxima is more for
only LBNE for which a higher exposure is needed and this effect is significant for hierarchy sensitivity.

\begin{table*}
\begin{center}
\begin{tabular}{|c|c|c|c|c|}
\hline
  Sensitivity   & \multicolumn{2}{|c|}{LBNE+NO$\nu$A+T2K+ICAL} & \multicolumn{2}{|c|}{Only LBNE} \\
\cline{2-5}
                & 1st + 2nd & Only 1st & 1st + 2nd & Only 1st \\
\hline
Hierarchy ($\chi^2 = 25$)                  & 22       &  28   & 106  &  218  \\
Octant ($\chi^2 = 25$)                     & 37        &  39  & 84   &  95   \\
CP ($40\%$ coverage                        & 65        &  70  & 114  &  128  \\
at $\chi^2 = 9$)                           &           &      &      &        \\
\hline
\end{tabular}
\caption[Effect of the second oscillation maximum on the sensitivity of LBNE.]{Effect of the second oscillation maximum on the sensitivity of LBNE. The 
numbers indicate the adequate exposure (in MW-kt-yr) required by LBNE for determining 
the oscillation parameters, with and without the contribution from the second 
oscillation maximum. For each of the three unknowns, the true parameters 
(including hierarchy) are taken 
to be ones for which we get the most conservative sensitivity.}
\label{tab:secondmax}
\end{center}
\end{table*}

\subsection{Optimising the Neutrino-Antineutrino Runs}
\label{sec:nunubar}

One of the main questions while planning any beam-based neutrino experiment 
is the ratio of neutrino to antineutrino run. Since the 
the neutrino and antineutrino  
probabilities are different due to $\dcp \rightarrow -\dcp$, an antineutrino run can provide a different set of 
data which may be useful in determination of the parameters. However, the 
interaction cross-section for antineutrinos in the
detectors is smaller by a factor 
of 2.5-3 than the neutrino cross-sections. Therefore, an antineutrino run typically 
has lower statistics. Thus, the choice of neutrino-antineutrino ratio is often 
a compromise between new information and lower statistics. 

\begin{figure*}
\begin{tabular}{cc}

 \epsfig{file=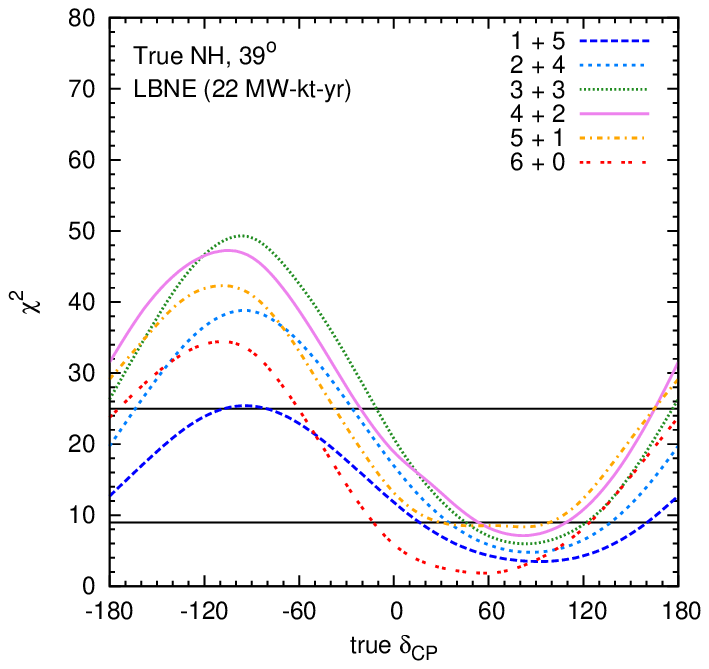, width=0.4\textwidth, bbllx=89, bblly=50, bburx=300, bbury=255,clip=} &
 \epsfig{file=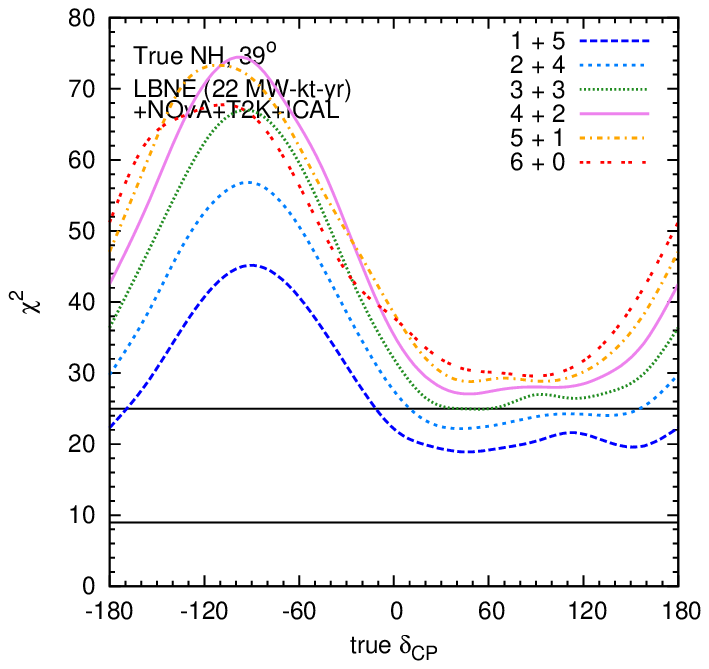, width=0.4\textwidth, bbllx=89, bblly=50, bburx=300, bbury=255,clip=} 
 \\
 
\epsfig{file=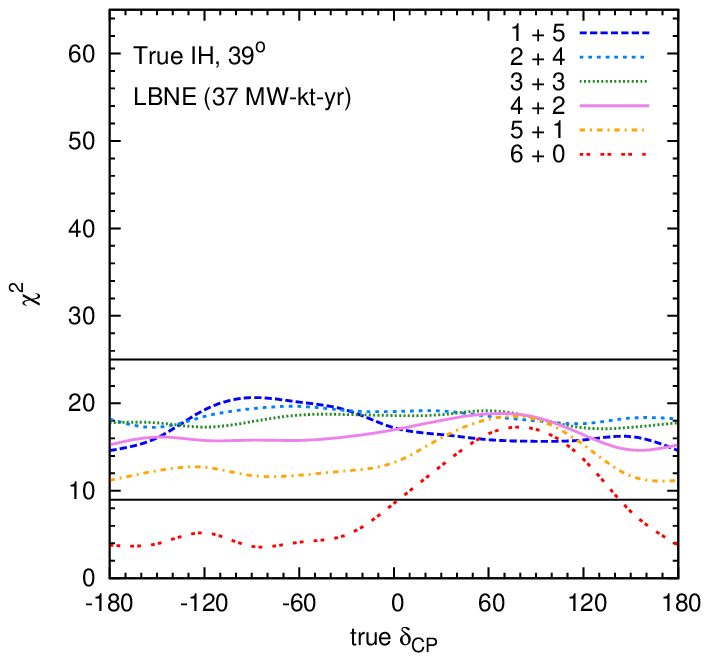, width=0.4\textwidth, bbllx=89, bblly=50, bburx=300, bbury=255,clip=} &
 \epsfig{file=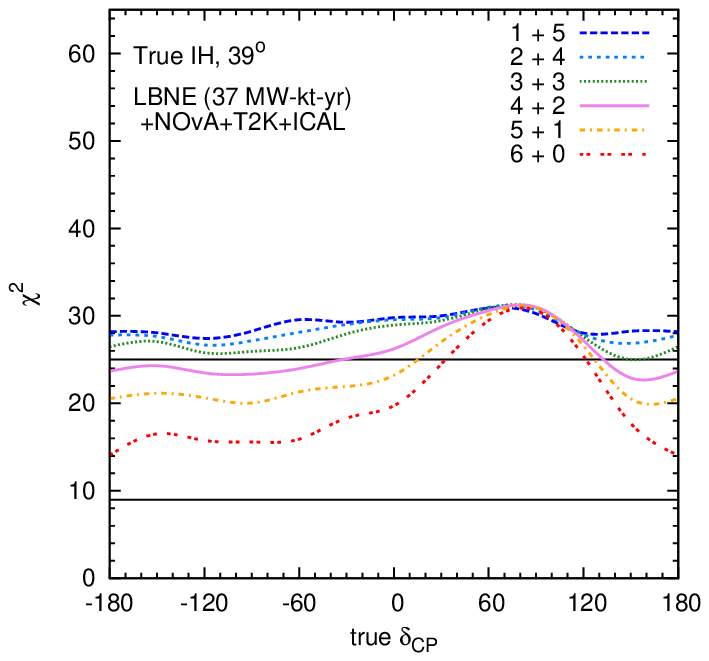, width=0.4\textwidth, bbllx=89, bblly=50, bburx=300, bbury=255,clip=} 
 \\
 
\epsfig{file=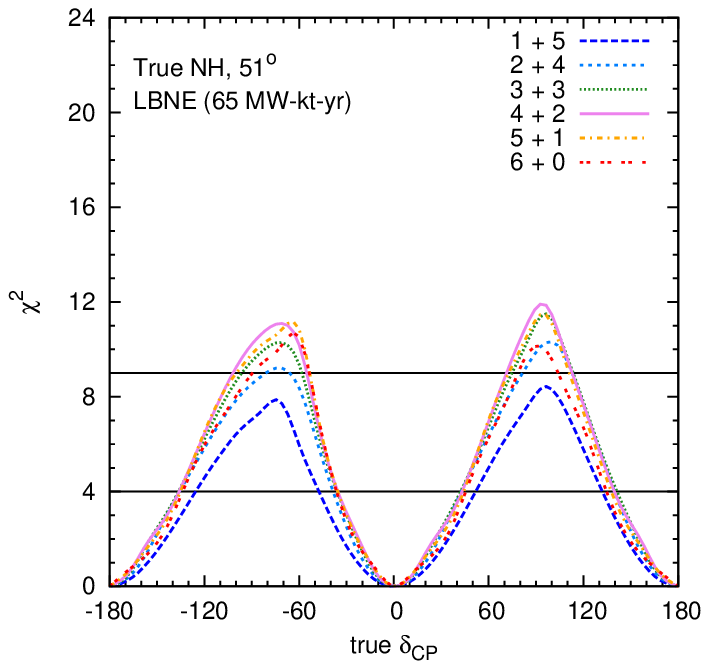, width=0.4\textwidth, bbllx=89, bblly=50, bburx=300, bbury=255,clip=} &
 \epsfig{file=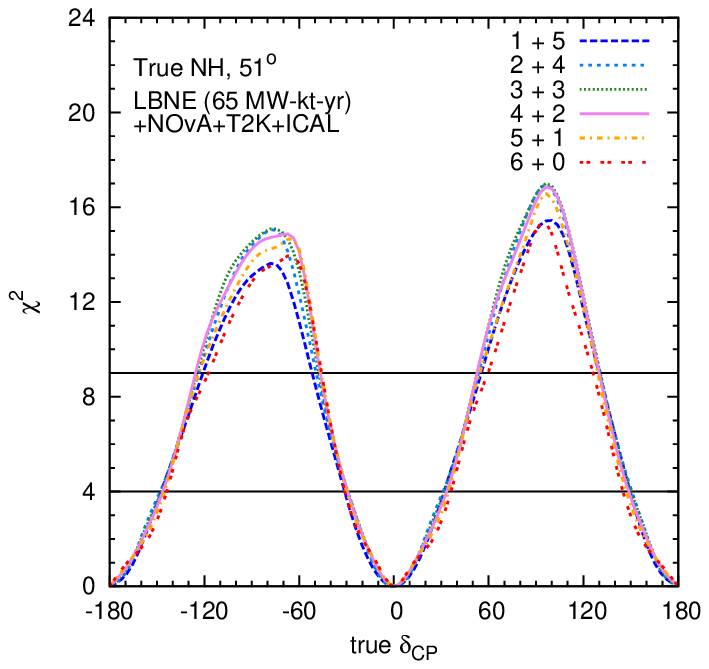, width=0.4\textwidth, bbllx=89, bblly=50, bburx=300, bbury=255,clip=} 
 
\end{tabular}

\caption[Sensitivity of LBNE for various 
 combinations of neutrino and antineutrino run by itself and in 
 conjunction with T2K, \nova\ and ICAL.]{Sensitivity of LBNE for various 
 combinations of neutrino and antineutrino run by itself (left panel) and in 
 conjunction with T2K, \nova\ and ICAL (right panel). The top/middle/bottom 
 row shows the sensitivity to hierarchy/octant/CP violation detection. 
 The total exposure has been divided into 6 equal parts and distributed
between neutrinos and antineutrinos. For example,
for hierarchy sensitivity, 6+0 corresponds to 44 MW-kt-yr in only neutrino; 
3+3 correspond to 22 MW-kt-yr in both neutrino and antineutrino mode. 
}
\label{fig:nuopt}
\end{figure*}

It is well known that neutrino and antineutrino oscillation 
probabilities suffer from the same form of hierarchy-$\dcp$ degeneracy \cite{t2knova}. 
However, the octant-$\dcp$ degeneracy has the opposite form for neutrinos and 
antineutrinos \cite{suprabhoctant}.
Thus, inclusion of an 
antineutrino run helps in lifting 
this degeneracy
for most of the values of $\dcp$ \cite{suprabhlbnelbno}.
For  measurement of $\dcp$,
it has been  shown, for T2K, that   
the antineutrino run is required only for those true hierarchy-octant-$\dcp$
combination for which octant degeneracy is present \cite{Ghosh:2014zea}.   
Once this degeneracy is lifted by including 
some amount of antineutrino data, further antineutrino run does not 
help much in CP discovery; 
in fact it is then better to run with neutrinos to gain in 
statistics \cite{Ghosh:2014zea}.   
But this conclusion may change for a different baseline and matter effect. 
From Fig. \ref{fig:beamprofile} we see that for \nova\
the oscillation peak does not coincide with the flux peak. 
Around the energy where the flux peaks,  
the probability spectra with $\dcp = \pm0, 180^\circ$ are not equidistant from the 
$\dcp = \pm 90^\circ$ spectra. For antineutrino mode the curves for 
$\pm 90^\circ$ switch position. Hence for neutrinos 
$\dcp = 0^\circ$ is closer to $\dcp = -90^\circ$ and $\dcp = 180^\circ$ is 
closer to $\dcp = 90^\circ$, while the opposite is true for 
antineutrinos. This gives a synergy and hence running in both neutrino 
an antineutrino modes can be helpful. 
For T2K the  energy where the flux peak occurs coincides with the oscillation peak. 
At this point the curves for $\dcp = 0, 180^\circ$ are equidistant from 
$\dcp = \pm 90^\circ$ and hence this synergy is not present. Thus, the role 
of antineutrino run is only to lift the octant degeneracy.
In what follows we have varied the proportion  
of neutrino and antineutrino runs to
ascertain what is the 
optimal combination. 
The adequate exposure is split into various combinations of neutrinos 
and antineutrinos -- 1/6$\ \nu$ + 5/6$\ \nubar$, 2/6$\ \nu$ + 4/6$\ \nubar$, ... 
6/6$\ \nu$ + 0/6$\ \nubar$. The intermediate configuration 3/6$\ \nu$ + 3/6$\ \nubar$ 
corresponds to the equal-run configuration used in the other sections. For 
convenience of notation, these configurations are referred to simply as 
1+5, etc., i.e. without appending the `/6'.
The results are shown in Fig. \ref{fig:nuopt}. 

The top row of Fig. \ref{fig:nuopt} shows the hierarchy sensitivity of 
LBNE for various 
combinations of neutrino and antineutrino run. Normal hierarchy and 
$\theta_{23}=39^\circ$ have been assumed as the true parameters.
For LBNE, we have chosen an 
exposure of 22 MW-kt-yr which was found to be the adequate exposure assuming equal neutrino and antineutrino runs. 
In the left panel, we see the results for LBNE alone. 
the figure shows that in the favourable region of $\dcp \in [-180^\circ,0]$ 
the best sensitivity comes from the combination 3+3 or 4+2. 
Although the statistics is more for neutrinos, the antineutrino run 
is required to remove the wrong-octant regions.
For normal hierarchy, $\dcp \in [0,180^\circ]$ is the unfavourable region 
for hierarchy determination \cite{t2knova}, as is evident from the figure. 
In this region, we see that the results are worst for pure neutrino
run.   The best sensitivity comes for the case 5+1. 
This amount of antineutrino run is required to remove the octant degeneracy.
The higher proportion of neutrino run ensures better statistics.  
In the right panel, along with LBNE we have also combined data from \nova, 
T2K and ICAL. With the inclusion of these data the hierarchy sensitivity 
increases further and even in the unfavourable region $\chi^2=25$ sensitivity 
is possible with only neutrino run from LBNE. 
This is because \nova, which will run in antineutrino mode 
 for 3 years and the antineutrino component in the atmospheric 
neutrino flux at ICAL, will provide the necessary amount of information to 
 lift the parameter degeneracies that reduce hierarchy sensitivity. 
 Therefore, the best option for LBNE is to run only in neutrino mode, which 
 will have the added advantage of increased statistics. 
In the favourable region also the sensitivity 
is now better for 6+0 and 5+1  i.e. less amount of antineutrinos 
from LBNE is required because of the antineutrino information 
coming from \nova. Note that overall, the amount of antineutrino run 
depends on the value of $\dcp$. However combining information 
from all the experiments 4+2 seems to be the best option over the 
largest fraction of $\dcp$ values. 

In the middle row of Fig. \ref{fig:nuopt}, we have shown the octant 
sensitivity of LBNE 
 alone (left panel) and in combination with the current experiments (right 
 panel). For LBNE we have used an exposure of 37 MW-kt-yr.
We have fixed the true hierarchy to be inverted, and 
 $\theta_{23}=39^\circ$ i.e in the lower octant.
For this case the probability for neutrinos is maximum for 
$\dcp \sim -90^\circ$  and overlaps with the higher octant probabilities. 
Thus the octant sensitivity in neutrino channel is very poor. Thus the 
worst results for these values of $\dcp$ come from only neutrino runs. 
For antineutrino channel because of the flip in $\dcp$ the probability
for $\dcp = -90^\circ$ is well separated from those for HO. Therefore the 
octant sensitivity comes mainly from antineutrino channel \cite{suprabhoctant}. 
Thus, addition of antineutrino runs help in enhancing octant sensitivity.   
Therefore at $-90^\circ$ the best sensitivity is from 1+5 i.e $1/6^{th}$  
neutrino + $5/6^{th}$ antineutrino combination. 
On the other hand the neutrino probability is minimum 
for $\dcp = +90^\circ$ and  LO and therefore there is octant sensitivity 
in the neutrino channel. However since we are considering IH the 
antineutrino probabilities are enhanced due to matter effect and
for a broadband beam 
some sensitivity comes from the antineutrino channel also. 
Therefore there is slight increase in octant sensitivity 
by adding antineutrino data as can be seen. 
Overall, the best compromise is  seen to be reached for 
2+4 i.e $1/3^{rd}$ neutrino and 
$2/3^{rd}$ antineutrino combination, which gives the best results over 
the widest range of $\dcp$ values. 
Addition of \nova, T2K 
and ICAL data increases the octant sensitivity. The octant sensitivity 
is best for combinations having more antineutrinos. For $\dcp \sim +90^\circ$
all combinations give almost the same sensitivity.  

The left and right panels of the bottom row in Fig. \ref{fig:nuopt} show the 
ability  of LBNE (by itself, and in conjunction with the current generation 
of experiment, respectively) to detect CP violation.
Here the true hierarchy is NH and true $\theta_{23}$ is $51^\circ$. 
Although this true combination does not suffer from any octant degeneracy, 
we see in the left panel that 6+0 is not the best combination. 
This is due to the  synergy between neutrino-antineutrino runs
for larger baselines 
as discussed earlier. 
 In both cases, we find that 
 the best option is to run LBNE with antineutrinos for around a third of the 
 total exposure. 
On adding information from the other experiments, 
 we find great improvement in the CP sensitivity. From the right panel, we 
 see that the 
 range of $\dcp$ for which $\chi^2 = 9$ detection of CP is possible is almost 
 the same for most combinations of neutrino and antineutrino run. Therefore, 
 as in the case of octant determination, the exact choice of combination 
 is not very important.

%% file: chap3_sens.tex
 
 \section{Constraining $\dcp$ using First Three Years of IceCube Data}
\label{sec4}

In this section, we will try to find constraint on $\dcp$ by analysing data from ultra high energy neutrinos (UHE) coming from various astrophysical sources.
The study of cosmic particles, and through them the study of astrophysical 
phenomena has gradually moved up the energy scale over the last few decades.
The first data set announced by the IceCube collaboration consists of 28 events above 25 TeV, 
detected over a period of 662 days of live time (May 2011 -- May 2012 with 79 strings, 
and May 2012 -- May 2013 with 86 strings). 
7 out of these 28 events are tracks signifying 
($\nu_\mu+\bar{\nu}_{\mu}$) charged-current (CC) events; while the other 21 are 
showers indicating 
either ($\nu_e+\bar{\nu}_{e}$) or ($\nu_\tau+\bar{\nu}_{\tau}$), or ($\nu_\mu+\bar{\nu}_{\mu}$) 
neutral-current (NC) events \cite{Aartsen:2013bka}. 
This $4\sigma$ detection marked the first discovery of UHE neutrinos. Further data was collected for next one year. For the full 988 days IceCube 
collected 37 events, adding 1 track, 7 shower events and 1 was produced by a coincident pair of background muons from 
unrelated air showers that cannot be reconstructed with
a single direction and energy. 

In this section, we will analyse the IceCube neutrino data to 
measure $\dcp$ and determine the source of astrophysical neutrinos.
In Ref. \cite{Winter:2006ce}, the author discussed in detail the complementary 
nature of astrophysical 
and terrestrial neutrino experiments in studies regarding the detections of $\dcp$. In this paper (and more recently 
in Ref. \cite{Meloni:2012nk}), data in the form of 
flavour ratios of observed neutrinos was used. In this study, we have analysed data from 
IceCube using a similar approach to get a hint about the value of $\dcp$. 

\subsection{Astrophysical Sources}

The data recorded by the IceCube telescope is the first evidence of extra-terrestrial events 
in the UHE range. These neutrinos can have their origin in extragalactic astrophysical 
sources like low power Gamma-Ray Burst (GRB) jets in stars \cite{Murase:2013ffa} or 
Active Galactic Nuclei (AGN) cores \cite{Stecker:2013fxa}.
The energy of the 37 detected neutrino events are in the range $25 - 2000$ TeV. 
By tracing the hadronic origin \cite{Razzaque:2002kb} of these events, 
one can estimate the proton energies at their 
sources to be within $0.5-40$ PeV. Supernova Remnants (SNRs), AGNs, 
GRBs and other astrophysical sources can accelerate protons 
to very high energies. 
The interactions of these protons with soft photons or matter from 
the source
can give UHE neutrinos through the following process: $p\gamma,pp
\rightarrow{\pi^{\pm}}X,\,~{\pi^{\pm}} \, \rightarrow \,{\mu^{\pm}}\nu_{\mu}(\bar\nu_{\mu}), 
\,~{\mu^{\pm}} \, \rightarrow \, {e^{\pm}}\bar\nu_{\mu}(\nu_{\mu})\nu_e(\bar\nu_e)$ \cite{Waxman:1998yy,Halzen:2008vz} 
with a flux ratio of $\phi_{\nu_{e}}:\phi_{\nu_{\mu}}:\phi_{\nu_{\tau}} =
1:2:0$ (known as \piS\ process). Some of the muons, due to their light mass, can get cooled in the 
magnetic field quickly resulting in a 
neutrino flux ratio of $0:1:0$ (\muDS\ process). K-mesons, produced from $p\gamma$ interactions with a 
cross-section two orders of magnitude less than pions, will cool in the magnetic field of the source at 
higher energies compared to the pions. $K^{+}\rightarrow{\mu^{+}\bar{\nu_{\mu}}}\,$ is 
the dominant channel of neutrino production from cooled pions, with a branching fraction of 63\%, 
and with the same flux ratio as the pion decay \cite{Hummer:2011ms}.  
The $p\gamma$ interaction also produces high energy neutrons which would decay as 
$n\,\rightarrow p +e^-+\overline{\nu}_e$ to antineutrinos \cite{Moharana:2010su} with the flux 
ratio of $1:0:0$ (\nS\ process). The relative contribution of each channel depends on different parameters of the 
astrophysical source like the magnetic field, the strength of 
the shock wave and density of photon background \cite{Moharana:2011hh}. 
Apart from the neutrinos these processes also produce high energy photons inside the source. 
Correlation of high energy photons with the UHE neutrinos can be considered as a signature 
of hadron production inside the source. For example, a TeV neutrino can have an accompanying TeV photon at the source. 
However due to attenuation in the background radiation during propagation, PeV photons will have typical mean free path 
$\sim 10$ kpc \cite{Protheroe:1996si}. Thus, the associated photons of TeV neutrinos from extragalactic sources cannot reach earth. 

\subsection{Analysis}

The main sources of astrophysical neutrinos
are the \piS, \muDS\ and \nS\ channels. 
However, the exact fraction of events in the detector from each of these sources 
is not known. Therefore, we have introduced relative fractions $k_1$, $k_2$ and $k_3$ 
for these three sources respectively, which are treated as free parameters in the problem 
subject to the normalisation constraint $\sum k_i = 1$. In this study, we have 
not considered any other sub-dominant mode of neutrino production. 

Neutrinos oscillate during propagation and
given that the value of $L/E$ for such neutrinos 
is very large, we can only observe the average oscillation probability. 
Therefore, the probabilities take the simple form:
\be
P(\nu_\alpha \to \nu_\beta) \equiv P_{\alpha\beta} = \sum_i |U_{\alpha i}|^2 |U_{\beta i}|^2 ~.
\ee
It is worth emphasizing that this oscillation probability depends only on 
the mixing angles and CP phase, but not on the mass-squared differences. 
Therefore, unlike in beam-based experiments where knowledge of the mass hierarchy 
is crucial for CP sensitivity \cite{t2knova}, in this case we can 
(at least in principle) detect CP violation without suffering from the hierarchy 
degeneracy. 
Also 
note that $P_{\alpha\beta} = P_{\beta\alpha}$, therefore the probability 
can only be an even function of $\dcp$. As a consequence, we can treat neutrino 
and antineutrino oscillations on an equal footing. Another consequence of this is 
that every value of $\dcp$ allowed by the data will be accompanied by a degenerate 
solution ($-\dcp$).

The distinction between tracks (which we 
assume to be $\nu_\mu$ CC events) and showers (which we assume to be 
$\nu_e$ or $\nu_\tau$ or $\nu_\mu$ NC events) is quite clear in the IceCube detector. 
We have folded the relative initial fluxes with the oscillation probabilities to 
get the relative number of events at the detector. Separation of muon events 
into CC and NC has been done using the ratio of the cross-sections at the 
relevant energy \cite{Gandhi:1995tf}.
We have done a simple analysis 
using the total events, instead of binning the data in energy and angle. Since the 
probability is 
almost independent of energy, this simplification is not expected to affect the 
analysis. This also allows us to neglect the effect of energy resolution. 
In Ref. \cite{Aartsen:2013jdh}, the number of background events 
in the IceCube data set is estimated to be $10.6^{+5.0}_{-3.6}$. 
Of these, $6.0 \pm 3.4$ are expected to be veto penetrating atmospheric muons 
and $4.6^{+3.7}_{-1.2}$ are from the atmospheric neutrino background above energy 10 TeV.
The background assumed by IceCube could be an overestimation \cite{Mena:2014sja}, 
since (a) it has been estimated by extrapolating data, and (b) for atmospheric 
neutrinos the background has been calculated from 10 TeV while the events have been detected 
with lowest energy nearly 28 TeV. Therefore, we have used an estimate of 3 
background atmospheric muon tracks and 3.4 (the lower limit) background 
atmospheric neutrinos. IceCube have predicted a total of $8.4\pm4.2$ muon events and 
$6.6_{-1.6}^{+5.9}$ atmospheric neutrinos \cite{Aartsen:2014gkd} including the next set of neutrino events detected for the period of 988 days. Using 
the same analysis method we have taken the lowest limit of the backgrounds for our calculation. We have separated the background atmospheric neutrinos 
into tracks and showers using the same cross-sections as mentioned earlier.
These background events are subtracted from 
the data set in our analysis.

In Refs. \cite{Serpico:2005sz,Winter:2006ce,Meloni:2012nk,Rodejohann:2006qq}, the authors have 
proposed the use of the variable $R = {N_\mu}/{(N_e+N_\tau)} $
for the study of CP violation with astrophysical sources, where 
$N_\alpha$ is the flux of $\nu_\alpha+\bar{\nu}_{\alpha}$ at the detector. 
This variable helps by 
eliminating the overall source and detector-dependent normalisation. Moreover, as 
studies of the up/down ratio as well as data/MC ratio in atmospheric 
neutrinos have shown, taking 
ratios of event rates can also reduce the effect of systematics \cite{Fukuda:1998mi,Foot:1997rz}. 
For our study, we have constructed a similar quantity $ \rho = {N_{track}}/{N_{shower}}$, 
with the flavour compositions of the track and shower events as mentioned before.

We have constructed the quantity $\rho^{data}$ using the IceCube data, and calculated 
$\rho^{theory}$ for a certain value of $\dcp$ as described above. Background events 
are subtracted from the data, as mentioned above.
The statistical $\chi^2$ is then computed using the Gaussian definition
\be
\chi^2(\dcp) = \left( \frac{ \rho^{data}-\rho^{theory}}{\sigma_\rho} \right)^2 ~,
\ee
where $\sigma_\rho$ is the corresponding error .
We have incorporated systematic effects using the method of pulls, with a systematic error 
of 5\%. Note that, we have marginalised the $\Delta \chi^2$ over the mixing angles 
($\theta_{23}$, $\theta_{13}$, $\theta_{12}$) within the ranges 
$\theta_{23}$ = 35$^\circ$ to 55$^\circ$, $\sin^{2} 2\theta_{13}$= 0.085 to 0.115 and 
$\theta_{12}$= 30$^\circ$ to 36$^\circ$ respectively.
Here we take priors on all the three mixing angles.
The priors added are $\sigma(\sin^{2} 2\theta_{13})$ = 0.01, 
$\sigma(\sin^2 2\theta_{23})$ = 0.1 and $\sigma(\sin^2 \theta_{12})$ = 0.0155.


\subsection{Results}

\begin{figure}[ht!]
\begin{center}
\includegraphics[scale=0.9]{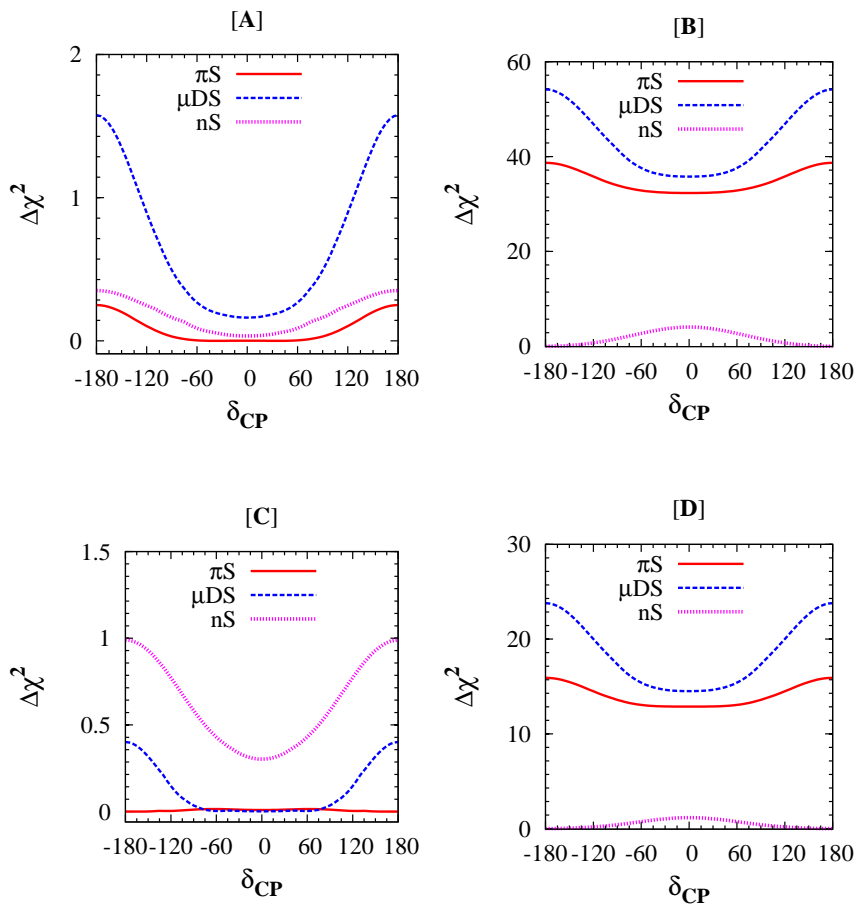}
\caption[Fit of the IceCube data to $\dcp$ considering a single source i.e., 
\piS\ ($k_1=1, k_2=0, k_3=0$), 
\muDS\ ($k_1=0, k_2=1, k_3=0$), \nS\ ($k_1=0, k_2=0, k_3=1$).]
{Fit to $\dcp$ considering a single source i.e., 
\piS\ ($k_1=1, k_2=0, k_3=0$, solid red), 
\muDS\ ($k_1=0, k_2=1, k_3=0$, dashed blue), \nS\ ($k_1=0, k_2=0, k_3=1$, 
dotted magenta), considering all the events being from astrophysical environment. 
Panel [A]: Three-year data, without background; 
Panel [B]: Three-year data, with background; 
Panel [C]: Two-year data, without background; 
Panel [D]: Two-year data, with background. }
\label{sing_wbg-source}
\end{center}
\end{figure}

To demonstrate the impact of the origin of these astrophysical 
neutrinos on the precision of $\dcp$, we start with various possibilities, 
like, single, double or a combination of three sources as the origin.
First we show the fit to the data as a function of $\dcp$ for the 
single source assumption, in Fig. \ref{sing_wbg-source}.
The upper row shows the results of our analysis of the full 
three-year data set. We have also included the results from analysing data from 
only the first two years (lower row) to show the improvement in results from 
additional data.

In the left panels we assume 
that all the events seen at IceCube are purely of astrophysical origin 
whereas in the right panels we include the effect of 
backgrounds. The latter is the realistic assumption. 
 From these figures 
we can see that, in case of no background the \piS\
source is 
favoured by the data as compared to the \nS\ and \muDS\ source (though the sensitivity 
is quite small, as $\Delta \chi^2$ is always $\textless$ 1.5). However, when we 
include the background, the scenario changes completely.
Pure \piS\ and pure \muDS\ sources are ruled out by the data 
at $> 3\sigma$, while the pure \nS\ source is favoured by data, though 
it is not sufficient to put any significant constraint 
on the value of $\dcp$. This has also been pointed 
out recently in Ref. \cite{Mena:2014sja}.
This result can be  understood  qualitatively in the following way. 
In the 2nd column of Table \ref{tab:werner} we have listed the 
theoretically calculated values  
for track by shower ratio for all the three sources keeping 
the oscillation parameters fixed at their tri-bimaximal (TBM) values 
($\theta_{23}=45^o$, $\theta_{13}=0^o$, $\sin^2\theta_{12}=
\frac{1}{3}$)\footnote{Due to the present non-zero value of 
$\theta_{13}$, there will be deviations from the TBM values but 
as shown in Ref. \cite{Meloni:2012nk}, this deviation is quite small.}
whereas the third column contains the experimental values of the track 
by shower ratio without and with backgrounds. 
 We can clearly see that 
for a pure signal, the track to shower ratio
for \piS\ is closest to the data. But the difference 
becomes quite high when backgrounds 
are taken under consideration, resulting in a very high $\Delta \chi^2$. A comparison of the upper and lower panels shows a marked 
increase in $\Delta \chi^2$. This shows the importance of additional data 
in both, excluding certain combinations of sources as well as constraining 
the value of $\dcp$.

\begin{figure}[ht!]
\begin{center}
\includegraphics[scale=0.9]{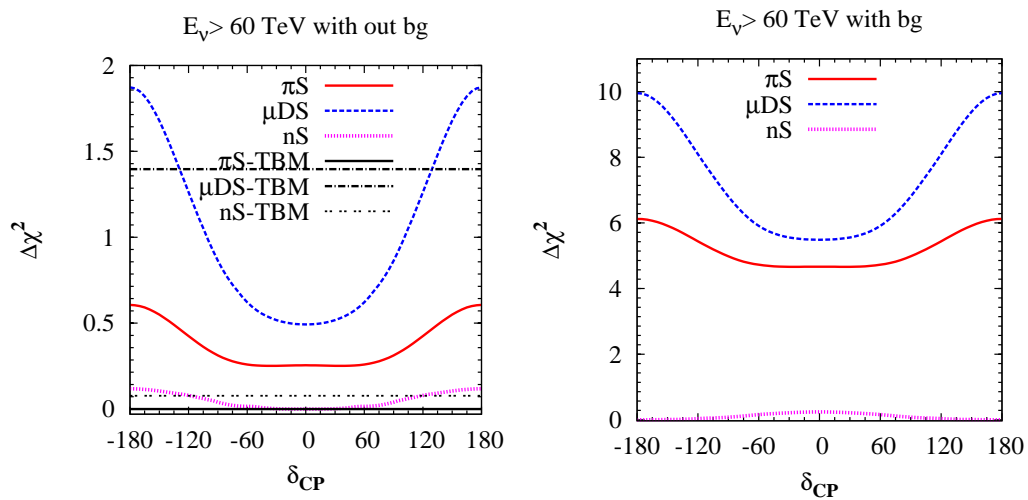}
\caption{Similar plot as that of Fig. \ref{sing_wbg-source} but for neutrinos within energy 60 TeV to 3 PeV.}
\label{with60}
\end{center}
\end{figure}

 We have also done an analysis of the events in the energy range 60 TeV $< E <$ 3 PeV considering the 3 years of IceCube data. 
This is motivated by the fact that, this energy interval contains
the atmospheric muon background less than one. In this energy range there are 4 track events and 16 shower events with an atmospheric muon background of 0.435 and atmospheric
neutrino background of 2.365 \cite{Aartsen:2014gkd}. The result is plotted in Fig. \ref{with60}. 
In the left panel there is no background and in the right panel 
background has been considered. From the right panel we can see that we are still getting \nS\ as the favoured source whereas \piS\ and \muDS\ sources are excluded at more than
$2 \sigma$. This is due to the fact that though the atmospheric muon background is less than one in this energy range but due to the presence of atmospheric neutrino
background \nS\ is getting preferred over \piS\ source.  This can bee seen from the left panel where no background is considered. There we can note that
the data agrees with the final flavor ratio 1:1:1; i.e., it favours the \piS\
source over \nS\ source marginally
when TBM mixing is assumed. But when we vary the oscillation parameters in their allowed $3 \sigma$ range then due to the deviation from TBM, \nS\ is getting slightly
preferred over \piS.

\begin{figure}[ht!]
\begin{center}
\includegraphics[height=6cm,width=7.cm]{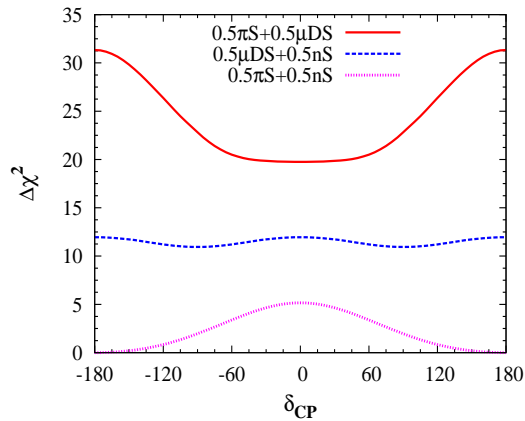}
\caption[Fit of the IceCube data to $\dcp$ considering contribution from two 
sources at a time, in equal proportion i.e., 
$k_1=k_2=0.5, k_3=0$, $k_1=0, k_2= k_3=0.5$ and 
$k_1=k_3=0.5, k_2=0$.]{Fit to $\dcp$ considering contribution from two 
sources at a time, in equal proportion i.e., 
$k_1=k_2=0.5, k_3=0$ (solid, red), $k_1=0, k_2= k_3=0.5$ (dashed, blue), 
$k_1=k_3=0.5, k_2=0$ 
(dotted, magenta).}
\label{doub-source}
\end{center}
\end{figure}

\begin{figure}[ht!]
\begin{center}
\includegraphics[height=6cm,width=8.cm]{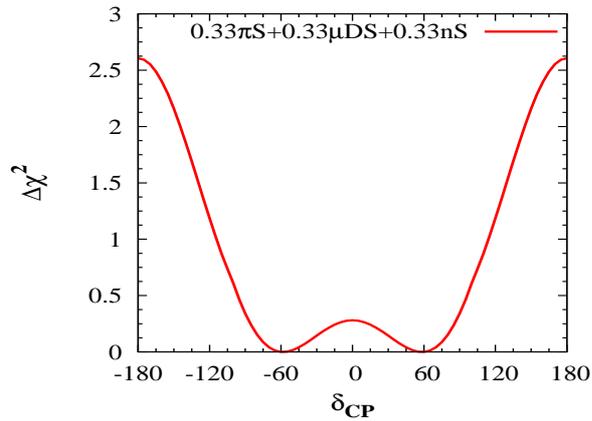}
\caption[Fit of the IceCube data to $\dcp$ considering equal contribution from all the 
sources, i.e., $k_1= k_2= k_3=0.33$.]{Fit to $\dcp$ considering equal contribution from all the 
sources, i.e., $k_1= k_2= k_3=0.33$.}
\label{equal-source}
\end{center}
\end{figure}

In Fig. \ref{doub-source} and Fig. \ref{equal-source} we show the fit to the data 
when neutrinos are coming from two/all the three sources 
respectively, with equal contributions. 
These results are for the full data set, and 
backgrounds have been included in generating these plots. 
In Fig. \ref{doub-source}, we find only the combination of \piS\ and \nS\ 
neutrinos are allowed at $3\sigma$ level. We also see that the CP dependence is 
maximum if the neutrinos come from the combination of \piS\ and 
\muDS\ modes. The data may also rule out one-third of $\dcp$ values 
(approximately $-60^\circ$ to $60^\circ$) at $\sim 2\sigma$. The poor sensitivity 
from \nS\ neutrinos is the reason why the combination 
of \piS+\muDS\ in Fig. \ref{doub-source} 
has a higher $\chi^2$ than the combinations involving \nS.
When we 
consider equal contributions from all these channels (Fig. \ref{equal-source}), 
we find that the data favours the first and fourth quadrants of $\dcp$ at $1\sigma$.


\begin{table}
\begin{center}
\renewcommand{\arraystretch}{1.5}
 \begin{tabular}{|c|c||c|}
  \hline
  Source & $\frac{N_{track}}{N_{shower}}$(Calculated) & $\frac{N_{track}}{N_{shower}}$(Data) \\
  \hline
  \piS & 0.30 &  \\
       &      &  8/28=0.287(Without background) \\
  \muDS & 0.38 &  \\
        &      & 0.05(With background)   \\
  \nS & 0.18 & \\
  \hline
 \end{tabular}
\renewcommand{\arraystretch}{1}
\caption[Theoretical values of track by shower ratio 
for \piS, \muDS\ and \nS\ sources along with value calculated from IceCube data.]{Theoretical values of track by shower ratio 
for all the three sources along with experimental values with and without background.}
\label{tab:werner}
\end{center}
\end{table}

\begin{figure}[ht!]
\begin{center}
\includegraphics[height=6.cm,width=10.cm]{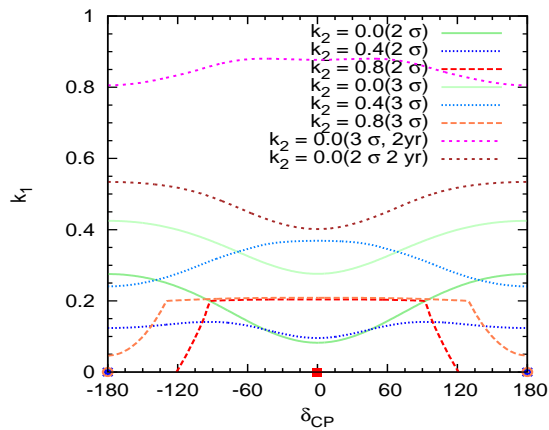}
\caption[Contour plots for allowed region in the $k_1-\dcp$ plane, 
for three representative values of $k_2$.]{Contour plots for allowed region in the $k_1-\dcp$ plane, 
for three representative values of $k_2$. The points marked in the respective colours 
indicate the best-fit point with new IceCube data.}
\label{cont_k1}
\end{center}
\end{figure}

\begin{figure}[ht!]
\begin{center}
\includegraphics[height=6.cm,width=10.cm]{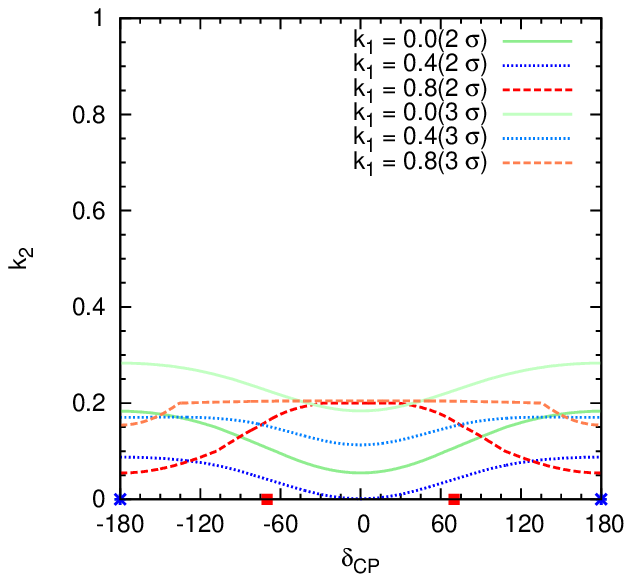}
\caption[Contour plots for allowed region in the $k_2-\dcp$ plane, 
for three representative values of $k_1$.]{Contour plots for allowed region in the $k_2-\dcp$ plane, 
for three representative values of $k_1$. The points marked in the respective colours 
indicate the best-fit point.}
\label{cont_k2}
\end{center}
\end{figure}

\begin{figure}[ht!]
\begin{center}
\includegraphics[height=6.cm,width=10.cm]{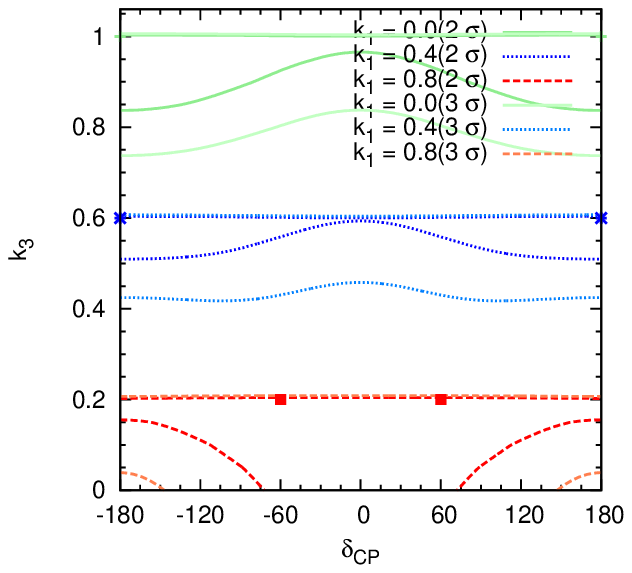}
\caption[Contour plots for allowed region in the $k_3-\dcp$ plane, 
for three representative values of $k_1$.]{Contour plots for allowed region in the $k_3-\dcp$ plane, 
for three representative values of $k_1$. The points marked in the respective colours 
indicate the best-fit point.}
\label{cont_k3}
\end{center}
\end{figure}

We have then performed a check to constrain the astrophysical 
parameters $k_i$ vs $\dcp$ using the IceCube 
data, by plotting the allowed contours in the $k_i-\dcp$ 
plane. In Fig. \ref{cont_k1}, 
we have showed the $2\sigma$ (light)
and $3\sigma$ (dark) contours in the $k_1-\dcp$ plane for 
three fixed values of $k_2$. 
The best-fit point indicated by the data has been 
marked with a red dot. 
We see that the data favours a smaller value of $k_1$ and 
larger values of $k_2$ and $k_3$. 
Similarly, Fig. \ref{cont_k2} shows that for a given value of $k_1$, the 
data disfavours the \muDS\ process (small value of $k_2$) but favours the \nS\ process 
(large value of $k_3$).
Likewise, Fig. \ref{cont_k3} shows the data favouring the
largest possible value of $k_3$ allowed by the normalisation condition. 
These features 
can be understood from Fig. \ref{sing_wbg-source}, where we see that the data prefers the 
\nS\ source. From these contours, we may draw certain constraints on the 
astrophysical sources most favoured. In particular if we obtain a good prior 
on $\dcp$ from other experiments, then the most favoured ratio of $k_1$, $k_2$ and 
$k_3$ may be obtained. Alternately, if we obtain a better picture of the 
sources of the IceCube events, a refined and constrained range on  $\dcp$ would 
be predicted.


 To show the statistical improvement of the 3 year data over 2 year data, 
in Fig. \ref{cont_k1} we have also plotted the $2\sigma$ and $3\sigma$ contours 
for the latter for $k_2=0$. Here we can clearly see that for $\delta_{CP}=0$, 
3 year data can exclude $73\%$ ($91.5\%$) of $k_1$ values at $3\sigma$ ($2\sigma$) 
where as the 2 year data can only rule 
out $13\%$ ($61\%$) of $k_1$ values at $3\sigma$ ($2\sigma$). For $\delta_{CP}=\pi$ 
the exclusion percentages are $58\%$ ($73\%$) at $3\sigma$ ($2\sigma$) for 3 years
 and $20\%$ ($48\%$) at $3\sigma$ ($2\sigma$) for 2 years. One can understand this 
 qualitatively from the \piS\ curve of Fig. \ref{sing_wbg-source} showing a significant 
 improvement in the 
$\Delta\chi^2$ with 3 years of data compared to 2 years.

\section{Summary}
\label{sec5}

In this chapter we have studied the capabilities of the current/future generation neutrino experiments to
constrain the remaining unknowns of neutrino oscillation parameters.
In Section \ref{sec1}, we have studied the CP sensitivity of the T2K, \nova\ and ICAL@INO experiments.
We emphasized the critical impact that 
atmospheric neutrinos can have in obtaining the first hint of  CPV   
from the LBL experiments T2K/\nova. This is achieved by the  
ability of the atmospheric neutrino data to exclude the 
degenerate solutions.
Taking the projected run times of T2K and \nova, we show that  
adding ICAL data  
can provide a signature of  
CPV at $2\sigma$ confidence level for $\sim 58\%$ $\dcp$ values. 
We also analyse the synergies
between these setups which may aid in CP violation discovery and a precision
measurement of $\dcp$. This has been studied for different combinations of these
experiments. We find that, 
while the CP sensitivity principally arises from the appearance channel of
\nova/T2K, the appearance and disappearance channels are synergistic due to
their different dependences on $\dcp$.  $P_{\mu e}$ depends on $\dcp$ through
the quantity $\cos(\Delta+\dcp)$, while $P_{\mu \mu}$ only has a $\cos \dcp$ 
dependence.
Thus their combination gives a CP sensitivity significantly higher than the sum
of sensitivities of the two channels. 
We also note that for smaller values of $\theta_{13}$, the CP-discovery 
$\chi^2 \propto \theta_{13}$ and hence increases with $\theta_{13}$. 
On the other hand, for larger values of $\theta_{13}$ the CP-discovery $\chi^2
\propto (90^\circ - \theta_{13})^2$ which 
decreases with $\theta_{13}$. The discovery $\chi^2$ 
attains its highest value in the range $\sin^2 2\theta_{13}
\sim 0.08 - 0.2$. This tells us that the range of
$\theta_{13}$ provided by nature lies in an optimal region which is favourable
for CP sensitivity with such experiments.

In Section \ref{sec2} we have studied the physics potential of the LBNO setup
to determine neutrino mass hierarchy, octant of $\theta_{23}$ and CP violation in leptonic sector
in conjunction with T2K, \nova\ and ICAL.
We have explored the minimum exposure needed
for such a set-up and quantified  the
`adequate' configuration that can
exclude the wrong hierarchy
($\chi^2=25$), exclude the wrong octant ($\chi^2=25$) and detect CP violation
($\chi^2=9$). We have determined the adequate exposure required
for LBNO in units of POT-kt and for the least favourable 
true hierarchy, $\theta_{23}$ and $\delta_{CP}$. 
In determining the requisite exposure
we fully exploit the possible synergies between the
existing LBL experiment T2K and \nova\, 
and the  atmospheric neutrino
experiment ICAL@INO which is likely to commence data taking
in five years time.
For the prospective LBNO configuration we consider
three options:
CERN-Pyh\"{a}salmi ($2290$ km) baseline
with a LArTPC, CERN-Slanic
($1500$ km) with a LArTPC
and CERN-Fr\'{e}jus ($130$ km) with a Water \v{C}erenkov detector.
The `adequate' exposure needed is summarized
in Table \ref{tab:res} where we give the results for T2K+\nova+LBNO
with and without ICAL. 
Inclusion of the atmospheric data from ICAL can play a significant role in
reducing the exposure required for hierarchy and octant determination
for the 2290 and 1540 km setups and for octant and CP detection for the
130 km set-up. Of the two longer baselines, we find that $2290$ km is best 
suited to determine the mass
hierarchy, while $1540$ km is better for detecting CP violation.
However, $130$ km is the best candidate for CP violation physics.

In Section \ref{sec3}, we have carried out a similar analysis for the LBNE experiment.
We have evaluated the adequate exposure for LBNE (in units of MW-kt-yr), 
i.e., the minimum exposure for LBNE to determine the unknown parameters 
in combination with T2K, \nova\ and ICAL, for all values of the oscillation 
parameters. The threshold for determination is taken to be $\chi^2=25$ for the 
mass hierarchy and octant, and $\chi^2=9$ for detecting CP violation.
The results are summarized in Table \ref{tab:adequate}.
We  find that adding information from \nova\ and T2K helps in reducing the 
exposure required by only LBNE for determination of all the three unknowns--
hierarchy, octant and $\dcp$. Adding ICAL data to this combination further   
help in achieving the same level of sensitivity with a reduction in exposure 
of LBNE (apart from $\dcp$). 
Thus the synergy between various experiments can be helpful in 
economising the LBNE configuration. 
We have also  probed the role of the near detector in improving the results 
by reducing systematic errors. We have simulated events at the near and far 
detectors and performed a correlated systematics analysis of both sets of events. 
We find an improvement in the physics reach of LBNE when the near detector is 
included. We have also evaluated the drop in systematics because of the near
detector. Our results are shown in Table \ref{tab:systresults}.

Further  we have checked the role of information from the lowest energy bins 
which 
are affected by the second oscillation maximum of the probability. 
We find that for the combined study of LBNE and the other experiments, the second oscillation maxima
do not play much role for the adequate exposure. However for only LBNE, 
with a higher adequate exposure, the
second maxima has a significant role in the hierarchy sensitivity.
 
Finally, we have done an optimisation study of the 
neutrino-antineutrino run for LBNE.  The amount of antineutrino run 
required depends on the true value of $\dcp$. 
It helps in achieving two objectives -- (i) reduction in 
octant degeneracy and (ii) synergy between neutrino and antineutrino data
for  octant and CP sensitivity. 
For hierarchy determination using a total exposure 
of 44 MW-kt-yr
the optimal combination for only LBNE is (3+3) which corresponds to
22 MW-kt-yr in 
neutrino and antineutrino mode each, for $\dcp$ in the 
lower half-plane $[-180^\circ,0]$ and true NH-LO.   
For $\dcp$ in the upper half-plane ($[0,180^\circ]$) the  optimal ratio is 
$5/6^{th}$ of the total exposure 
in neutrinos and + $1/6^{th}$ of the total exposure in antineutrinos.   
Adding information from T2K, \nova\ and ICAL 
the best combination for LBNE is $2/3^{rd}$ neutrino + $1/3^{rd}$ antineutrino 
for $\dcp$ in the lower half-plane. 
In the upper half-plane, pure neutrino run gives the best sensitivity.  
In the latter case, the 
antineutrino component coming from \nova\ and 
ICAL helps in reducing the required antineutrino run from LBNE. 
For octant sensitivity the best result from the combined experiments 
comes from the proportion $(1/6^{th}+5/6^{th})$ except for 
$\dcp = +90^\circ$ where
all combinations
give almost the same sensitivity. For CPV discovery, all combinations give 
similar results when the data are added together, with equal neutrino and
antineutrino or $2/3^{rd}$ neutrino + $1/3^{rd}$ antineutrino combination 
faring slightly better. 
 
Finally in Section \ref{sec4} we have explored the possibility to constrain the leptonic phase $\dcp$ from the IceCube data.
we have analysed the first IceCube data on TeV-PeV scale neutrinos. 
We have used the flux ratios 
of the three neutrino flavours to put constraints on $\dcp$. We find that the results 
depend strongly on the source of the neutrinos. 
After taking into account the effect of backgrounds, we find that the \nS\ source 
of neutrinos is favoured by the data. Depending on the particular combination of sources for these neutrinos, 
current data can only hint at the allowed region of the $\dcp$ range. However, 
we have shown that additional data gives a remarkable improvement in results, which 
underlines the importance of future data from IceCube.
We have also put constraints on the astrophysical parameters $k_1$, $k_2$ and $k_3$ 
that determine which of the 
modes of neutrino production is more close to the data.

%% file: chap4_intro.tex
\section{Overview}

In this chapter we discuss the structure of the low energy neutrino mass matrix in the presence of one extra light sterile neutrino. 
As mentioned in the introduction, the neutrino mass matrix in flavour basis is not diagonal and can be written as
\begin{eqnarray}
M_{\nu}&=&V^*M_{\nu}^{diag}V^{\dagger} 
\end{eqnarray}
where $V$ is the leptonic mixing matrix which contains the neutrino oscillation parameters, 
in a basis where charged lepton mass matrix is diagonal and $M_{\nu}^{diag}$ is the diagonal mass matrix.
From the above equation it is clear that the neutrino oscillation parameters can determine the elements of the low energy neutrino mass matrix $M_\nu$.
One of the popular themes to study the structure of the low energy mass matrix is in terms of texture zeros\footnote{For a recent review see \cite{Adhikary:2012zx}.}.
Texture zero means one or more elements of the mass matrix are relatively small compared to the others. 
Such studies help in understanding the underlying parameter space
and the nature of the mass spectrum involved and often predict
correlations between various parameters which can be
experimentally tested.
At the fundamental level the structure of the mass matrix
is determined by the Yukawa couplings which
are essentially free parameters in most models.
Knowing the form of the low energy neutrino mass matrix may help to constrain 
the high scale structures including zero textures in the Yukawa matrix itself \cite{Goswami:2008rt,Goswami:2009bd,Choubey:2008tb,Adhikary:2012zx}.
The origin of texture zero could be due to $U(1)$ symmetry like Froggatt-Neilsen \cite{Froggatt:1978nt} or other flavour symmetries \cite{Grimus:2004hf}, discreet or continuous.
Texture zeros in the low energy mass matrices in the context of three
generations have been extensively explored both in the
quark and lepton sector \cite{Kang:2000fn,Frampton:2002yf,Dev:2006qe,Xing:2002ta,Xing:2002ap,Desai:2002sz,Dev:2007fs,Dev:2006xu,Kumar:2011vf,Fritzsch:2011qv,Meloni:2012sx,
Ludl:2011vv,Grimus:2012zm}.
In particular for three generations of neutrinos, a very remarkable result was
obtained in \cite{Frampton:2002yf} that there can be at the most two zeros in the low energy neutrino mass
matrix in the flavour basis. 

As mentioned in the Introduction, data from LSND and MiniBooNE experiments can not be accommodated in the three neutrino scenario.
These experiments have reported oscillations which can only be explained by the inclusion of one or more sterile neutrino having mass in the eV scale.
The recently observed Gallium and reactor anomaly also provide additional support to the sterile neutrino hypothesis. 

These evidences of sterile neutrino motivated us to study the texture zero properties of the neutrino mass matrix in the presence of 
the light sterile neutrinos and compare the results with that of
three generation case. In our study we considered the one-zero and two-zero textures in 3+1 scenario.
We find that the allowed textures and correlations between different parameters differ significantly as compared to the results of the three generation case
\footnote{Unlike three generation case,
in the context of the 4-neutrinos, more than two zeros can be allowed.
The results for three texture zero in 3+1 scheme can be found in \cite{Zhang:2013mb}.}.

This chapter is organised in the following way. In Section \ref{3gen} we will first review and update the texture zero results of the 3 generation case
and in section \ref{3+1_mixing} we will describe the mass and mixing pattern in the 3+1 scheme. In Section \ref{two_zero} and \ref{one_zero} we
present the results for two-zero and one-zero textures of $M_\nu$ in 3+1 scenario. We summarize our results in Section \ref{sum}.

\section{Texture Zero Results for 3 Generation}
\label{3gen}

In three generation picture, the low energy Majorana neutrino mass matrix $M_\nu$ is a $3 \times 3$ complex symmetric matrix having six independent elements given by
\begin{eqnarray}
M_{\nu}&=&
       \begin{pmatrix}
        m_{ee} & m_{e\mu} & m_{e\tau} \\
m_{e \mu}& m_{ \mu \mu} & m_{\mu \tau} \\
m_{e \tau }& m_{\mu \tau} & m_{\tau \tau}\\
       \end{pmatrix}.
\end{eqnarray}
Thus for three generations there are 15 possible two-zero textures and 6 possible one-zero textures. The two-zero textures are categorised
in different classes as shown in Table \ref{man3}.
\begin{table}
\begin{center}
\begin{small}
\begin{tiny}
\begin{small}
\begin{tabular}{|c|c|c|c|}
\hline $ A_1$& $A_2$ &  &   \\
\hline $\left(
\begin{array}{ccc}
0 & 0 & \times \\
0 & \times & \times \\ 
\times & \times & \times
\end{array}
\right)$ & $\left(
\begin{array}{ccc}
0 & \times & 0\\ 
\times & \times & \times \\
0 & \times & \times
\end{array}
\right)$  & & \\
\hline $ B_1$ & $B_2$ & $B_3$ & $B_4$  \\
\hline
 $\left(
\begin{array}{ccc}
\times & \times & 0 \\  \times &0 & \times \\ 0 & \times & \times
\end{array}
\right)$  & $\left(
\begin{array}{ccc}
\times& 0 & \times \\  0 &\times & \times \\ \times & \times & 0 
\end{array}
\right)$& $\left(
\begin{array}{ccc}
\times & 0 & \times \\  0 & 0 & \times \\ \times & \times & \times
\end{array}
\right)$  & $\left(
\begin{array}{ccc}
\times & \times& 0 \\  \times & \times & \times \\0 & \times & 0
\end{array}
\right)$ \\
\hline $C$& & &  \\
\hline
 $\left(
\begin{array}{ccc}
\times & \times & \times \\ \times & 0 & \times \\ \times & \times & 0
\end{array}
\right)$ & & & \\
\hline $D_1$& $D_2$ & & \\
\hline
$\left(
\begin{array}{ccc}
\times &\times & \times  \\  \times & 0 & 0 \\ \times & 0 & \times
\end{array}
\right)$ & $\left(
\begin{array}{ccc}
\times & \times & \times \\  \times & \times &0 \\ \times & 0 &0 
\end{array}
\right)$& & \\
\hline $E_1$ & $E_2$ & $E_3$ & \\
\hline
$\left(
\begin{array}{ccc}
0 & \times & \times \\ \times & 0 & \times \\ \times & \times
\end{array}
\right)$& $\left(
\begin{array}{ccc}
0 & \times & \times \\  \times & \times & \times \\ \times & \times & 0
\end{array}
\right)$&  $\left(
\begin{array}{ccc}
0 &\times & \times \\ \times & \times & 0 \\ \times & 0 & \times
\end{array}
\right)$& \\
\hline $F_1$& $F_2$ & $F_3$ & \\
\hline
$\left(
\begin{array}{ccc}
\times & 0 & 0 \\  0 & \times & \times \\0 & \times & \times
\end{array}
\right)$&  $\left(
\begin{array}{ccc}
\times& 0 & \times \\  0 & \times & 0 \\ \times & 0 & \times
\end{array}
\right)$ & $\left(
\begin{array}{ccc}
\times &\times & 0 \\ \times & \times &0 \\ 0 & 0 & \times
\end{array}
\right)$&\\
\hline
\end{tabular}
\end{small}
\end{tiny}
\caption{Possible two-zero textures in the three generation scenario.}
\label{man3}
\end{small}
\end{center}
\end{table}
The first analysis of the two-zero textures in the three generation has been done in \cite{Frampton:2002yf} and shown that 
among the 15 possible textures, only 7 textures are phenomenologically allowed.
A detailed analysis regarding the parameter space of the allowed textures were presented in \cite{Dev:2006qe,Xing:2002ta}. 
At that time $\theta_{13}$ was unknown and those analysis were performed taking the CHOOZ upper bound of $\theta_{13}$.
After the precise measurement of $\theta_{13}$, the same analysis has been done by many groups in view of the current data \cite{Fritzsch:2011qv,Liao:2013saa,Grimus:2012zm}.
The main conclusion of the all the analysis are almost same but with the precise measurements of the oscillation parameters 
the allowed parameter space of the viable textures have been reduced significantly.
Below we briefly discuss the main results of the two-zero textures\footnote{We have also done the analysis
regarding the two-zero textures in three generation case. We will discuss our results in Section \ref{3_active}.}.
Textures belonging to $A$ class
are allowed in only normal hierarchy. 
The textures in $B$ class are allowed in both normal and inverted hierarchy. 
The classes $B1$ and $B4$ predict negative values of $\cos\dcp$ whereas the classes $B2$ and $B3$ predict positive values of $\cos\dcp$.
The textures belonging to the $B$ class can also predict the octant of $\theta_{23}$. The textures $B1$ and $B3$ predict $\theta_{23}$ in the lower octant and
the textures $B2$ and $B4$ predict $\theta_{23}$ in the upper octant for normal 
hierarchy. The predictions are opposite for the inverted hierarchy. 
The texture belonging to $C$ class is allowed mainly in the inverted hierarchy. 
This class is marginally allowed in the normal hierarchy when $\theta_{23}$ is close to $45^\circ$.
For the viability of this class, when $\theta_{23} < 45^\circ$, one must have $-90^\circ < \dcp < 90^\circ$ and when
$\theta_{23} > 45^\circ$, one must have $90^\circ < \dcp < 270^\circ$.
The textures in the classes $D$, $E$ and $F$ are forbidden by the data.

Analysis of the one-zero textures in 3 generations has been done in Ref. \cite{Merle:2006du,Lashin:2011dn}. 
Regarding the matrix element $m_{ee}$
which is the effective mass governing the neutrinoless double beta decay, it is well known that it
can vanish only for normal hierarchy. The element $m_{e \mu}$ can be zero in both normal hierarchy and inverted hierarchy. For the case of inverted hierarchy
this predicts $\sin\alpha = 0$ where $\alpha$ being the Majorana phase. The predictions for $m_{e \tau}=0$ is similar as that of $m_{e \mu}$.
The elements $m_{\mu \mu}$ and $m_{\tau \tau}$ can vanish in both the hierarchies. For
quasi-degenerate masses i.e., $m_1 \approx m_2 \approx m_3$, $\theta_{23}$ needs to be below $45^\circ$ if $m_{\mu \mu}$ is zero and
above $45^\circ$ if $m_{\tau \tau}$ is zero. They also predict $\sin\alpha = 0$. The element $m_{\mu \tau}$ can only vanish for
quasi-degenerate masses and predict\footnote{ While discussing the one-zero textures in 3+1 scenario, we will also present 
the status of the one-zero textures in three generation in view of the current experimental data.} $\sin\alpha = 0$.

%% file: two_zero.tex
\section{Masses and Mixing in the 3+1 Scheme}
\label{3+1_mixing}

Addition of one extra sterile neutrino to the standard three
generation picture gives rise to two possible mass patterns --
the 2+2 and 3+1 scenarios \cite{GomezCadenas:1995sj,Goswami:1995yq,Okada:1996kw}. Of these,
the 2+2 schemes are strongly disfavored
by the solar and atmospheric neutrino oscillation data \cite{Maltoni:2002ni}.
The 3+1 picture also suffers from some tension
between observation of oscillations in antineutrino channel by
LSND and MiniBooNE and non-observation of oscillations in the
neutrino channels as well as in disappearance measurements.
However, it was shown recently in \cite{Giunti:2011cp} that a reasonable
goodness-of-fit can still be obtained.
Although introduction of more than one sterile neutrinos
may provide a better fit to the
neutrino oscillation data \cite{Kopp:2011qd,Conrad:2012qt}, the 3+1 scheme
is considered to be minimal and to be more consistent with the
cosmological data \cite{Mangano:2011ar}. Very recently combined analysis of
cosmological and short baseline (SBL) data in the context of additional sterile
neutrinos have been performed in \cite{Giunti:2011gz,Joudaki:2012uk}.
The analysis in \cite{Giunti:2011gz} found a preference of
the 3+1 scenario over 3+2
while the analysis in \cite{Joudaki:2012uk} shows that the status of the
3+2 scenario depends on the cosmological data set
used and the fitting procedure and no conclusive statement can be made
regarding whether it is favored or disallowed.

Theoretically, sterile neutrinos are naturally included in
Type-I seesaw model
\cite{Minkowski:1977sc,Mohapatra:1979ia}. But their mass scale is usually very high to
account for the small mass of the neutrinos.
Light sub-eV sterile neutrinos as suggested by the data
can arise in many models
\cite{Abazajian:2012ys}.

Irrespective of  the mechanism for generation of neutrino masses
the low energy Majorana mass matrix  in presence of an extra sterile
neutrino will be of dimension $4\times4$
with  ten independent entries and is given as,

\begin{eqnarray}
M_{\nu}&=&
       \begin{pmatrix}
        m_{ee} & m_{e\mu} & m_{e\tau} & m_{es} \\
m_{e \mu}& m_{ \mu \mu} & m_{\mu \tau}& m_{\mu s} \\
m_{e \tau }& m_{\mu \tau} & m_{\tau \tau}& m_{\tau s}\\
m_{es} & m_{ \mu s} & m_{\tau s} & m_{ss} \\
       \end{pmatrix}.
\end{eqnarray}

There are two ways in which one can add a
predominantly sterile state separated by $\sim$ eV$^2$ from the
standard 3 neutrino mass states.
In the first case the additional
neutrino  can be of higher mass than the other three
while in the second case the
the fourth neutrino is  the lightest state.
The later turns out to be incompatible with cosmology
since  in this case
three active neutrinos, each with mass $\sim$  eV results in an
enhanced cosmological energy density.
Thus it suffices to consider only the first case which admits
two possibilities displayed in Fig. \ref{fig1}. These are:

\begin{figure}
 \begin{center}
 \includegraphics[scale=0.5,angle=0]{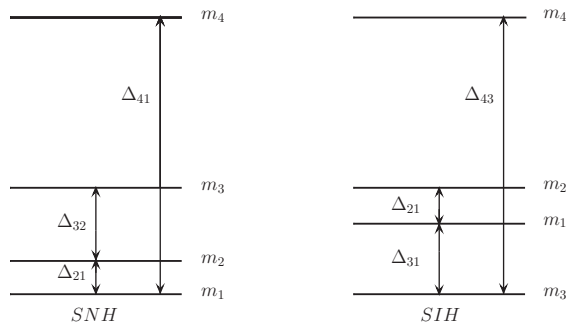}
 \end{center}
\caption{The allowed 3+1 mass ordering.}
\label{fig1}
 \end{figure}

\begin{itemize}
\item[(i)] SNH: in this
$m_1 \approx m_2 < m_3 < m_4$  corresponding to a normal hierarchy (NH)
among the active neutrinos which implies,
\\
$m_2=\sqrt{m_1^2+\Delta_{21}}~~ ,
m_3= \sqrt{m_1^2+\Delta_{21}+\Delta_{32}}~~ ,
m_4=\sqrt{m_1^2+\Delta_{41}}.
$

\item[(ii)] SIH : this corresponds to
$m_3 < m_2 \approx m_1 < m_4$ implying an inverted ordering
among the active neutrinos with masses expressed as,
\\
$m_1= \sqrt{m_3^2+\Delta_{31}}~~,
m_2=\sqrt{m_3^2+\Delta_{31}+\Delta_{21}}~~,
m_4=\sqrt{m_3+\Delta_{43}},
$
\end{itemize}
with $\Delta_{ij} = m_i^2 - m_j^2$. These cases correspond to complete hierarchy among the active neutrinos.
The active neutrino mass spectrum can also be quasi-degenerate (QD) where the three active
neutrinos have approximately equal masses i.e., 
SQD: where
$ |m_4|\gg|m_1|\approx|m_2|\approx|m_3|\approx m_0 $.

In the 3+1 scenario,
the neutrino mixing matrix, $V$ in the flavor basis
will be a $4 \times 4$ unitary matrix. As described in the introduction, 
in general a N $\times$ N unitary mixing matrix contains $\frac{N(N-1)}{2}$ mixing angles and $\frac{1}{2}(N-1)(N-2)$ Dirac type CP violating phases.
It will also have
(N-1) number of additional Majorana phases if neutrinos are Majorana particles.
So in our case V can be parametrized in
terms of sixteen parameters. In addition to the three mixing angles between the active flavors,
($\theta_{13}$, $\theta_{12}$, $\theta_{23}$) we now have three more mixing angles from sterile
and active mixing, ($\theta_{14}$, $\theta_{24}$, $\theta_{34}$). There are six CP violating phases,
three Dirac ($\delta_{13}$, $\delta_{14}$, $\delta_{24}$) and three additional Majorana phases as
($\alpha$, $\beta$, $\gamma$) as neutrinos here are considered to be Majorana particles. Then,
there are four masses of neutrino $m_1$, $m_2$, $m_3$ corresponding to three active states and $m_4$
which is predominantly the mass of heavy sterile neutrino.

The mixing matrix $V$ can be expressed as
$V=U.P$ \cite{Goswami:2005ng} where
\begin{equation}
U={R_{34}}\tilde R_{24}\tilde R_{14}R_{23}\tilde R_{13}R_{12},
\end{equation}
where $R_{ij}$ denotes rotation matrices in the \textit{ij} generation space
and is expressed as,
\begin{center}
$R_{34}$=$\left(
\begin{array}{cccc}
1~ &~0 & 0 & 0 \\  0~ &~ 1 & 0 & 0 \\ 0~ & ~0 & c_{34}& s_{34} \\0 ~& ~0 & -s_{34} & c_{34}
\end{array}
\right)$ , $\tilde R_{14}$=$\left(
\begin{array}{cccc}
c_{14}~ & ~0 &~ ~0 &~s_{14}e^{-i \delta_{14}} \\ 0 ~ & ~ 1&~~ 0 & 0 \\ 0 ~& ~0 &~~ 1 & 0 \\-s_{14}e^{i \delta_{14}}  & ~ 0& ~~0 &c_{14}
\end{array}
\right)$. \\
\end{center}
Here we use the abbreviations $s_{ij}=\sin\theta_{ij}$ and $c_{ij}=\cos\theta_{ij}$. The  phase matrix is diagonal and is expressed as,
\begin{center}
$P=diag(1,e^{i \alpha}, e^{i {(\beta+\delta_{13})}},e^{i {(\gamma+\delta_{14})}})$.
\end{center}

%
%

\section{Analysis of Two-Zero Textures}
\label{two_zero}

\subsection{Formalism}

The two-zero textures in the neutrino mass matrix give two complex
equations viz.
 \begin{eqnarray} \label{mnuzero}
M_{\nu (ab)}&=&0, \\  \nonumber 
M_{\nu (pq)}&=&0.
\end{eqnarray}
where \textit{a, b, p} and \textit{q} can take the values $e$, $\mu$, $\tau$ and $s$.
The above Eq. \ref{mnuzero} can be written as
\begin{equation}
U_{a1} U_{b1} + \frac{1}{x} U_{a2} U_{b2} + \frac{1}{y} U_{a3} U_{b3} e^{2 i \delta_{13}} + z U_{a4} U_{b4}=0,
\label{M11}
\end{equation}
\begin{equation}
U_{p1} U_{q1} + \frac{1}{x} U_{p2} U_{q2} + \frac{1}{y} U_{p3} U_{q3} e^{2 i \delta_{13}} + z U_{p4} U_{q4}=0,
\label{M22}
\end{equation}
where 
\be
x  =  \frac{m_1}{m_2}e^{i\alpha},~~
y =  \frac{m_1}{m_3}e^{i\beta},~~
z  =  \frac{m_4}{m_1}e^{-2i(\gamma/2-\delta_{14}).}
\label{xyz}
\ee

Solving Eqs. (\ref{M11}) and (\ref{M22}) simultaneously we get the two mass ratios as
\begin{equation}
x=\frac{U_{a3}U_{b3}U_{p2}U_{q2}-U_{a2}U_{b2}U_{p3}U_{q3}}{U_{a1}U_{b1}U_{p3}U_{q3}-U_{a3}U_{b3}U_{p1}U_{q1}+z(U_{a4}U_{b4}U_{p3}U_{q3}-U_{a3}U_{b3}U_{p4}U_{q4})},
\end{equation}
\begin{equation}
y=-\frac{U_{a3}U_{b3}U_{p2}U_{q2}+U_{a2}U_{b2}U_{p3}U_{q3}}{U_{a1}U_{b1}U_{p2}U_{q2}-U_{a2}U_{b2}U_{p1}U_{q1}+z(U_{a4}U_{b4}U_{p2}U_{q2}-U_{a3}U_{b3}U_{p4}U_{q4})}e^{2 i \delta_{13}}.
\end{equation}
The modulus of these quantities gives the magnitudes
$x_m$, $y_m$ while the argument determines
the Majorana phases $\alpha$ and $\beta$.
\be
x_m = \left|x\right|,~~
y_m  =  \left|y\right|
\ee
\be
\alpha=arg\left(x\right),~~
\beta=arg\left(y\right).
\ee
Thus, the number of free parameters is five, the lowest
mass $m_1$ (NH) or $m_3$ (IH), three Dirac and one Majorana type CP phases.
We can check for the two mass spectra in terms of the magnitude of the 
mass ratios $x_m$, $y_m$ and $z_m = |z|$  as,
\begin{itemize}
\item
SNH which corresponds to  $x_m<1$ , $y_m<1$ and $z_m>1$.
\item
SIH which implies
$x_m<1$ , $y_m>1$ and $z_m>1$.
\end{itemize}

Thus, it is $y_m$ which determines if the hierarchy among the
three light neutrinos is normal or inverted. Note that if
the three light neutrinos are quasi-degenerate then we will have $x_m \approx y_m \approx 1$.
Unlike the three generation case,
the lowest mass cannot be determined in the four neutrino analysis
in terms of $x_m$ and $y_m$ since these ratios also depend on $m_1$
through $z$. Thus, we keep the lowest mass as a free parameter.
To find out the allowed two-zero textures we adopt the following procedure.\\
We vary the lowest mass randomly from 0 to  0.5 eV. 
The upper limit chosen by us is guided by the cosmological upper bound on neutrino masses. 
All the five mixing angles
(apart from $\theta_{34}$) 
and the three mass-squared differences
are distributed normally
about the best-fit values with their corresponding $1\sigma$ errors as given in Table \ref{Table:parameters}\footnote{The extraction of the sterile mixing parameters $\theta_{14}$,
$\theta_{24}$ and $\theta_{34}$ from the global analysis of the SBL experiment data is given in the appendix.}.
\begin{table}[ht!]
\begin{center}
\begin{tabular}{lccc}
\hline
\hline
Parameter & Best fit & $1\sigma$ range & $3\sigma$ range \\
\hline
$\Delta_{21}/10^{-5}~\mathrm{eV}^2 $ (NH or IH) & 7.54 & 7.32 -- 7.80 & 6.99 -- 8.18 \\
\hline
$\sin^2 \theta_{12}/10^{-1}$ (NH or IH) & 3.07 & 2.91 -- 3.25  & 2.59 -- 3.59 \\
\hline
$\Delta_{32}/10^{-3}~\mathrm{eV}^2 $ (NH) & 2.43 & 2.33 -- 2.49  & 2.19 -- 2.62 \\
$\Delta_{31}/10^{-3}~\mathrm{eV}^2 $ (IH) & 2.42 & 2.31 -- 2.49  & 2.17 -- 2.61 \\
\hline
$\sin^2 \theta_{13}/10^{-2}$ (NH) & 2.41 & 2.16 -- 2.66 & 1.69 -- 3.13 \\
$\sin^2 \theta_{13}/10^{-2}$ (IH) & 2.44 & 2.19 -- 2.67 & 1.71 -- 3.15 \\
\hline
$\sin^2 \theta_{23}/10^{-1}$ (NH) & 3.86 & 3.65 -- 4.10  &                       3.31 -- 6.37 \\
$\sin^2 \theta_{23}/10^{-1}$ (IH) & 3.92 & 3.70 -- 4.31  & 3.35 -- 6.63 \\
\hline
$ \Delta_{LSND}(\Delta_{41}^2 or \Delta_{43}^2) ~\mathrm{eV}^2$ & 0.89 & 0.80 -- 1.00 & 0.6 -- 2 \\
\hline
$ \sin^2\theta_{14} $ & 0.025 & 0.018 -- 0.033 & 0.01 -- 0.05 \\
\hline
$ \sin^2\theta_{24} $ & 0.023 & 0.017 -- 0.037 & 0.005 -- 0.076 \\
\hline
$ \sin^2\theta_{34} $ & -- &  --  & $ < 0.16 $ \\
\hline
\end{tabular}
\caption[Oscillation parameters \cite{global_fogli_old,Giunti:2011gz,schwetz} used in two-zero texture analysis in 3+1 scenario.]
{The experimental constraints on neutrino oscillation parameters \cite{global_fogli_old}. The constraints on
sterile parameters involving the fourth neutrino are from \cite{Giunti:2011gz,schwetz}.}
\label{Table:parameters} 
\end{center}
\end{table}
The three Dirac and one Majorana type CP phase as well as
the remaining mixing angle $\theta_{34}$ are randomly generated.
Then, we use the above conditions to find out which
mass spectrum is consistent with the
particular texture zero structure under consideration.
We also calculate the three mass-squared difference ratios
\begin{eqnarray}
R_\nu&=&\frac{\Delta_{21}}{|\Delta_{32}|}=\frac{1-x_m^2}{ |(x_m^2/y_m^2)-1|}, \nonumber \\
R_{\nu1}&=& \frac{|\Delta_{31}|}{\Delta_{41}}=\frac{|1-y_m^2|}{y_m^2(z_m^2-1)}, \nonumber \\
R_{\nu2}&=&\frac{\Delta_{21}}{\Delta_{41}}=\frac{1-x_m^2}{ x_m^2(z_m^2-1)}.
\end{eqnarray}
The  $3 \sigma$ ranges of these three ratios calculated from the experimental data are
\begin{eqnarray}
R_{\nu}
&=& (0.02-0.04), \nonumber \\
R_{\nu1} &=& (1.98 \times 10^{-3}-3.3 \times 10^{-3}),  \nonumber \\
R_{\nu2} &=& (0.63 \times 10^{-4}-1.023 \times 10^{-4}).
\label{Reqn}
\end{eqnarray}
The allowed textures are selected by checking that they
give the ratios  within the
above range.

\subsection{Results and Discussions}

\begin{table}
\begin{center}
\begin{small}
\begin{tiny}
\begin{small}
\begin{tabular}{|c|c|c|c|}
\hline $ A_1$& $A_2$ &  &   \\
\hline $\left(
\begin{array}{cccc}
0 & 0 & \times &\times \\  0 & \times & \times & \times \\ \times & \times & \times & \times \\\times & \times & \times & \times
\end{array}
\right)$ & $\left(
\begin{array}{cccc}
0 & \times & 0 &\times \\  \times & \times & \times & \times \\ 0 & \times & \times & \times \\\times & \times & \times & \times
\end{array}
\right)$  & & \\
\hline $ B_1$ & $B_2$ & $B_3$ & $B_4$  \\
\hline
 $\left(
\begin{array}{cccc}
\times & \times & 0 &\times \\  \times &0 & \times & \times \\ 0 & \times & \times & \times \\\times & \times & \times & \times
\end{array}
\right)$  & $\left(
\begin{array}{cccc}
\times& 0 & \times &\times \\  0 &\times & \times & \times \\ \times & \times & 0& \times \\\times & \times & \times & \times
\end{array}
\right)$& $\left(
\begin{array}{cccc}
\times & 0 & \times &\times \\  0 & 0 & \times & \times \\ \times & \times & \times & \times \\\times & \times & \times & \times
\end{array}
\right)$  & $\left(
\begin{array}{cccc}
\times & \times& 0 &\times \\  \times & \times & \times & \times \\0 & \times & 0 & \times \\\times & \times & \times & \times
\end{array}
\right)$ \\
\hline $C$& & &  \\
\hline
 $\left(
\begin{array}{cccc}
\times & \times & \times &\times \\ \times & 0 & \times & \times \\ \times & \times & 0 & \times \\\times & \times & \times & \times
\end{array}
\right)$ & & & \\
\hline $D_1$& $D_2$ & & \\
\hline
$\left(
\begin{array}{cccc}
\times &\times & \times &\times \\  \times & 0 & 0 & \times \\ \times & 0 & \times & \times \\\times & \times & \times & \times
\end{array}
\right)$ & $\left(
\begin{array}{cccc}
\times & \times & \times &\times \\  \times & \times &0 & \times \\ \times & 0 &0 & \times \\\times & \times & \times & \times
\end{array}
\right)$& & \\
\hline $E_1$ & $E_2$ & $E_3$ & \\
\hline
$\left(
\begin{array}{cccc}
0 & \times & \times &\times \\ \times & 0 & \times & \times \\ \times & \times & \times & \times \\\times & \times & \times & \times
\end{array}
\right)$& $\left(
\begin{array}{cccc}
0 & \times & \times &\times \\  \times & \times & \times & \times \\ \times & \times & 0 & \times \\\times & \times & \times & \times
\end{array}
\right)$&  $\left(
\begin{array}{cccc}
0 &\times & \times &\times \\ \times & \times & 0& \times \\ \times & 0 & \times & \times \\\times & \times & \times & \times
\end{array}
\right)$& \\
\hline $F_1$& $F_2$ & $F_3$ & \\
\hline
$\left(
\begin{array}{cccc}
\times & 0 & 0 &\times \\  0 & \times & \times & \times \\0 & \times & \times & \times \\\times & \times & \times & \times
\end{array}
\right)$&  $\left(
\begin{array}{cccc}
\times& 0 & \times &\times \\  0 & \times & 0 & \times \\ \times & 0 & \times & \times \\\times & \times & \times & \times
\end{array}
\right)$ & $\left(
\begin{array}{cccc}
\times &\times & 0 &\times \\ \times & \times &0 & \times \\ 0 & 0 & \times & \times \\\times & \times & \times & \times
\end{array}
\right)$&\\
\hline
\end{tabular}
\end{small}
\end{tiny}
\caption[Allowed two-zero textures in the 3+1 scenario.]{Allowed two-zero textures in the 3+1 scenario. The 15 possible
two-zero textures of three active neutrinos are same as these after omitting 
the 4th row and column.  }
\label{man}
\end{small}
\end{center}
\end{table}

\begin{table}
\begin{center}
\begin{tabular}{lccc}
\hline
\hline
Class & 3 gen(Random) & 3 gen(Gaussian) & 3+1 gen(Gaussian)   \\
\hline
 A & NH & NH & NH \\ 
 B & NH, IH, QD & NH($B_1$, $B_3$), IH($B_2$, $B_4$) & NH, IH, QD \\ 
 C & NH, IH, QD & IH & NH, IH, QD \\ 
 D & - &  - & NH, IH \\ 
 E & - & - & NH, IH \\ 
 F & - & - & NH, IH,QD \\
\hline
\end{tabular}
\begin{center}
\caption[The allowed mass spectra in 3 and 3+1  
scenarios.]{The allowed mass spectra in 3 and 3+1  
scenarios. The last column gives the allowed spectrum for the 3+1 case
assuming normal distribution. For random distribution similar mass spectra get allowed although the parameter space 
is reduced in size. See text for details.}
\label{2results}
 \end{center}
 \end{center}
\end{table}

In this section we present the results of our analysis. 
First we briefly discuss the results that we obtain for the
two-zero textures of the $3 \times 3$ mass matrices.
Next we present the results that we obtain for the 3+1 scenario i.e., $4 \times4$ mass matrices. 

\subsubsection{Results for 3 Active Neutrino Mass Matrix}
\label{3_active}

For the 3 neutrino case, the lowest mass and the two Majorana phases can be determined 
from the mass ratios. Hence, the only unknown parameter 
is the Dirac type CP phase ($\delta_{13}$) which is generated randomly. All the other oscillation parameters 
are distributed normally, peaked at the best-fit and taking their 
one sigma error as width.
 
In our analysis, we find that all 7 textures which were allowed previously (as discussed in section \ref{3gen}) remain so. 
However, the textures belonging to A class allow NH whereas for the B class, 
$B_1$ and $B_3$ admit NH and $B_{2}$ and $B_4$  allow IH solutions. 
Class C gets allowed only for IH. 
The D, E and F classes remain disallowed. 
This is summarized in the second column of Table \ref{2results}. We also display the 
results that we obtain for the two-zero neutrino mass matrices 
with three active neutrinos using 
random distribution of the oscillation parameters. 
The results obtained in this case are somewhat different from
that obtained using normal distribution of oscillation parameters.  
The reason for the difference stems from  
the  different range of values of the atmospheric mixing angle $\theta_{23}$ 
used by these methods. If we assume a Gaussian 
distribution for $\sin^2\theta_{23}$ around its best-fit then there is very less probability of
getting the  3$\sigma$ range in the higher octant as these values lie 
near the tail of the Gaussian distribution. This disallow 
$B_2$ and $B_4$ for NH and $B_1$ and $B_3$ for IH \cite{Fritzsch:2011qv}.
 Similarly QD solutions for B class requires $\theta_{23} \sim \pi/4$ 
\cite{Frampton:2002yf} and for a normal distribution of $\theta_{23}$ with 
the peak at present best-fit the 3$\sigma$ range extends up to $\sim 44^o$ and there is very little 
probability of getting values close to $45^\circ$. Similarly for the C class NH and QD solutions are 
allowed only for $\theta_{23}$ values close to $45^\circ$ and hence is not admissible when Gaussian distribution of 
oscillation parameters about the best-fit value is assumed. 

\subsubsection{Results for 3+1 Scenario}

Adding one sterile neutrino, there exist in total forty five texture structures of the neutrino
mass matrix which can have two zeros.

\begin{enumerate}

\item [(i)] Among these the 9 cases with $|m_{ss}|$= 0 eV are
disallowed as the mass matrix element
$m_{ss}$ contains the term $m_4U_{s4}^2$ which is large
from the current data and suppresses the other terms. Hence, $|m_{ss}|$ cannot vanish.

\item
[(ii)] There are 21 cases where one has at least one zero involving
the mass matrix element of the sterile part i.e.,  $|m_{ks}|$ =0 eV
where $k=e,\mu,\tau$.
This element is of the form,
\bea
m_{ks}  & = &
m_1U_{k1}U_{s1}+m_2U_{k2}U_{s2}e^{-i\alpha} \\
\nonumber
& + &
m_3 U_{k3}U_{s3}e^{i(2\delta_{13}-\beta)}+m_4 U_{k4}U_{s4}e^{i(2\delta_{14}-\gamma)}.
\label{mks}
\eea
The last term in this expression contains the product $m_4 U_{s4}$ which is mostly large  as compared
to first three terms and thus, there can
be no cancellations. However, it is possible that in the regime where the active neutrinos 
are quasi-degenerate, their contribution can match the contribution 
from the sterile part. But we have checked that in any case, it is not possible to obtain a two-zero texture with these elements.
Thus we can exclude these 21 cases from the allowed two-zero textures. 

We will elaborate these points further in the next section at the time of discussing one zero texture.

\item
[(iii)] The remaining cases are the 15 two-zero cases
for which none of the sterile components are zero. Thus, these also belong to the two-zero textures of the three generation
mass matrix.
A general element in this category can be expressed as,
\bea \label{mkl}
m_{kl}  & = &
m_1U_{k1}U_{l1}+m_2U_{k2}U_{l2}e^{-i\alpha} \\
\nonumber
& + &
m_3 U_{k3}U_{l3}e^{i(2\delta_{13}-\beta)}+m_4 U_{k4}U_{l4}e^{i(2\delta_{14}-\gamma)}.
\eea
here, $k,l=e,\mu,\tau$.
We find all these 15 textures, presented in Table \ref{man} 
get allowed with the inclusion of the sterile neutrino.
This can be attributed to additional cancellations that the 
last term in Eq. \ref{mkl} induces. 
Table \ref{2results} displays the nature of the mass spectra that are 
admissible in the allowed textures. 


\begin{figure} [ht!]
\begin{center}
\includegraphics[width=0.38\textwidth,angle=270]{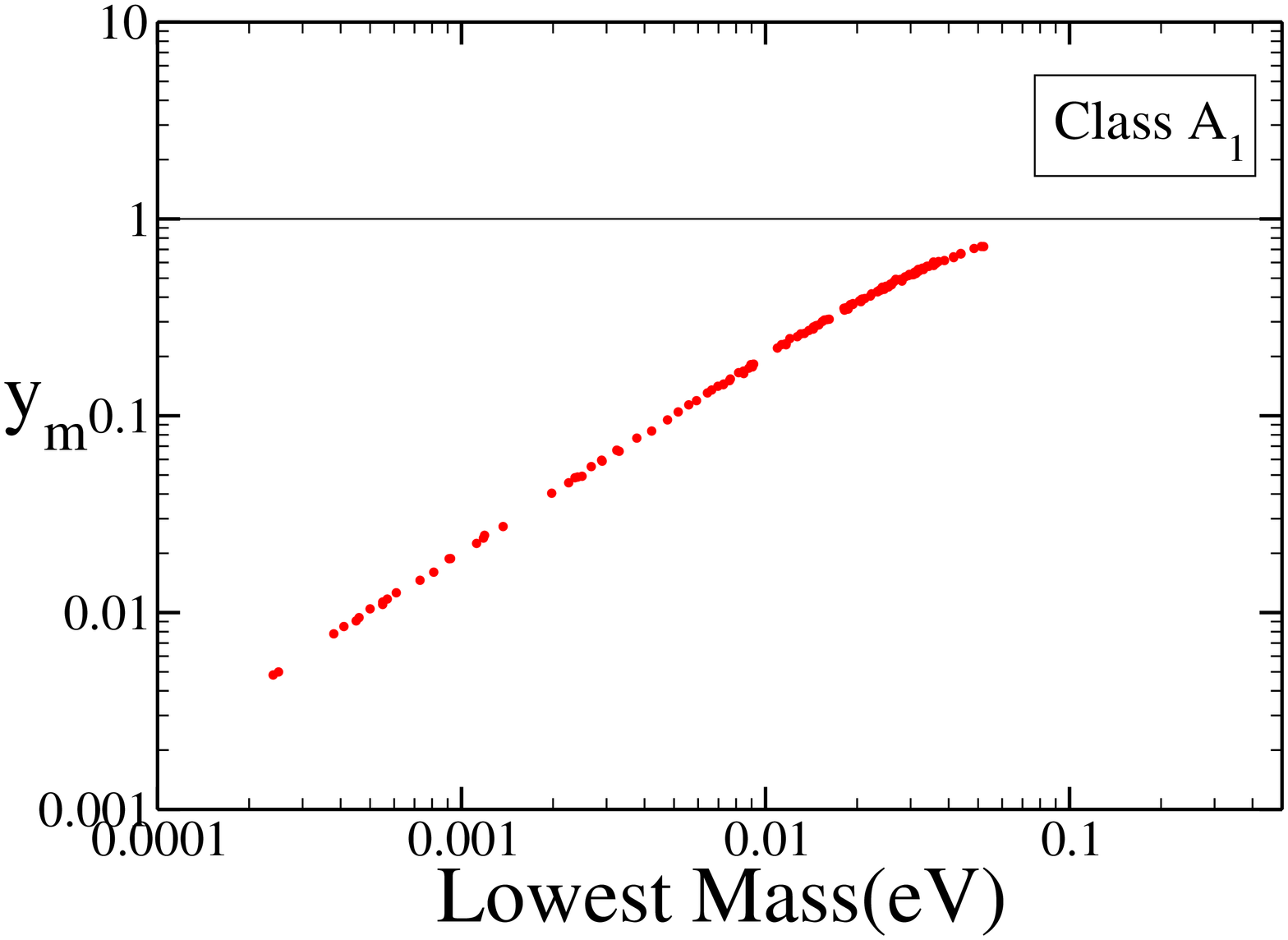}
\includegraphics[width=0.38\textwidth,angle=270]{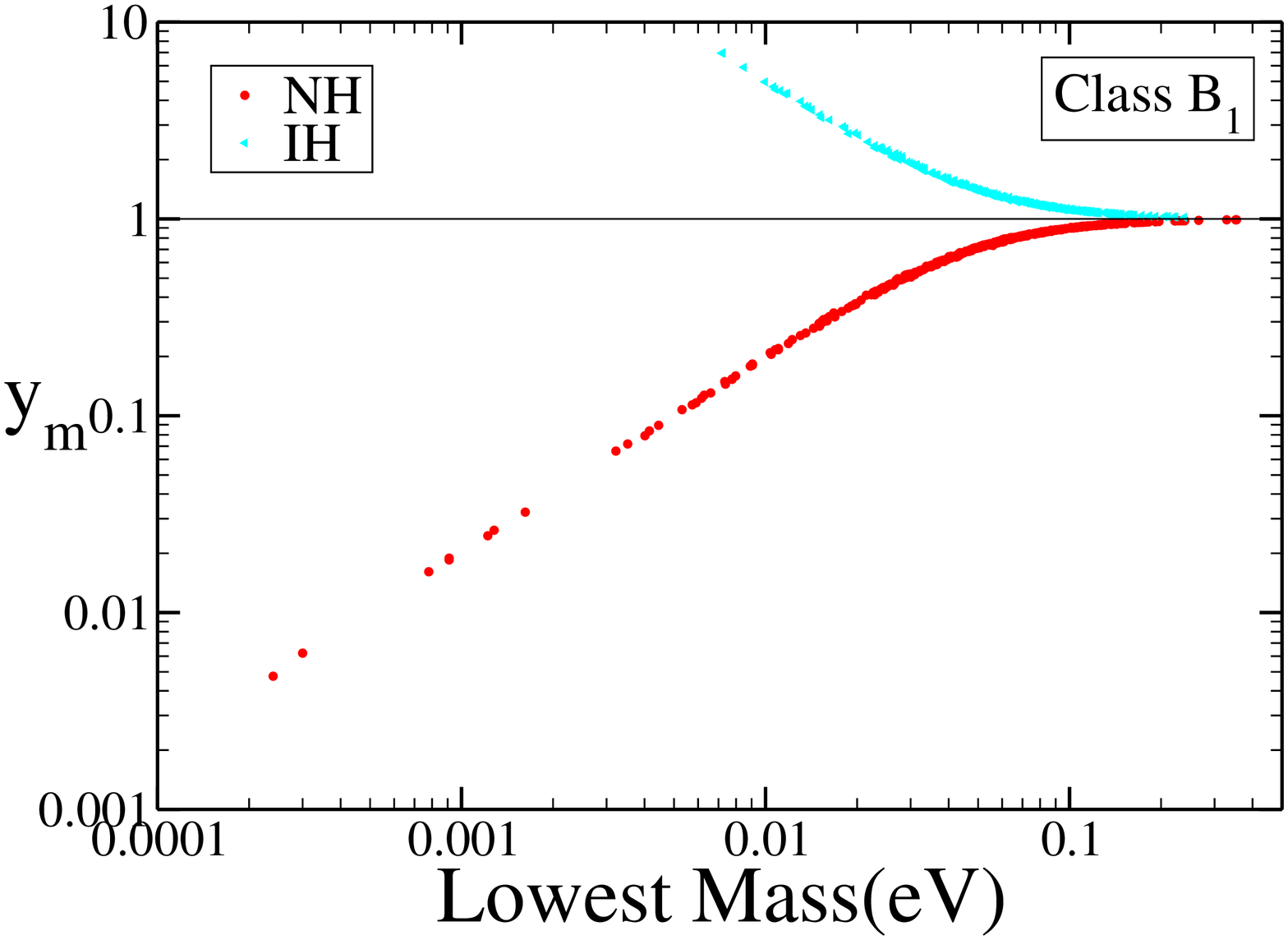} \\
\includegraphics[width=0.38\textwidth,angle=270]{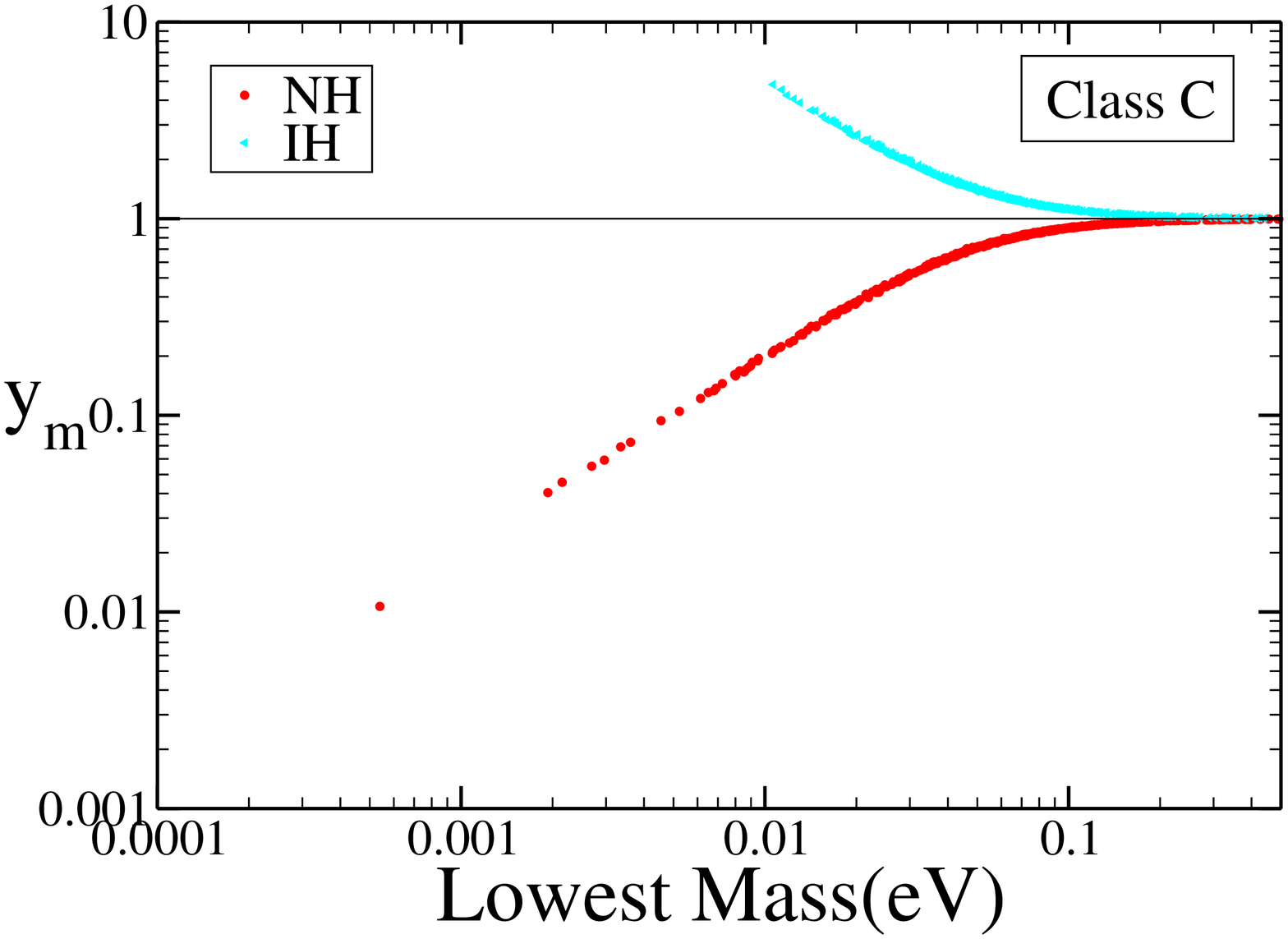}
\includegraphics[width=0.38\textwidth,angle=270]{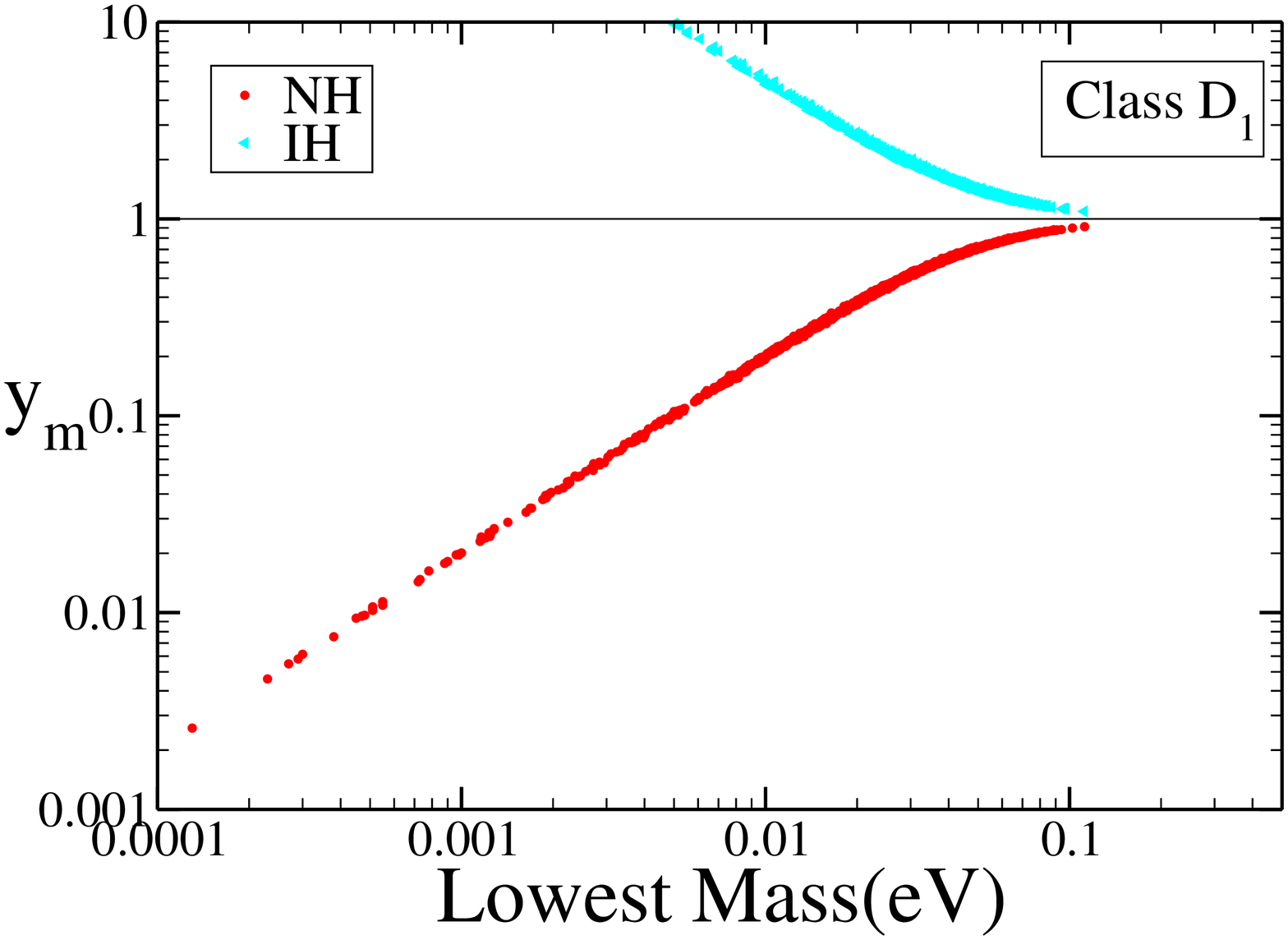} \\
\includegraphics[width=0.38\textwidth,angle=270]{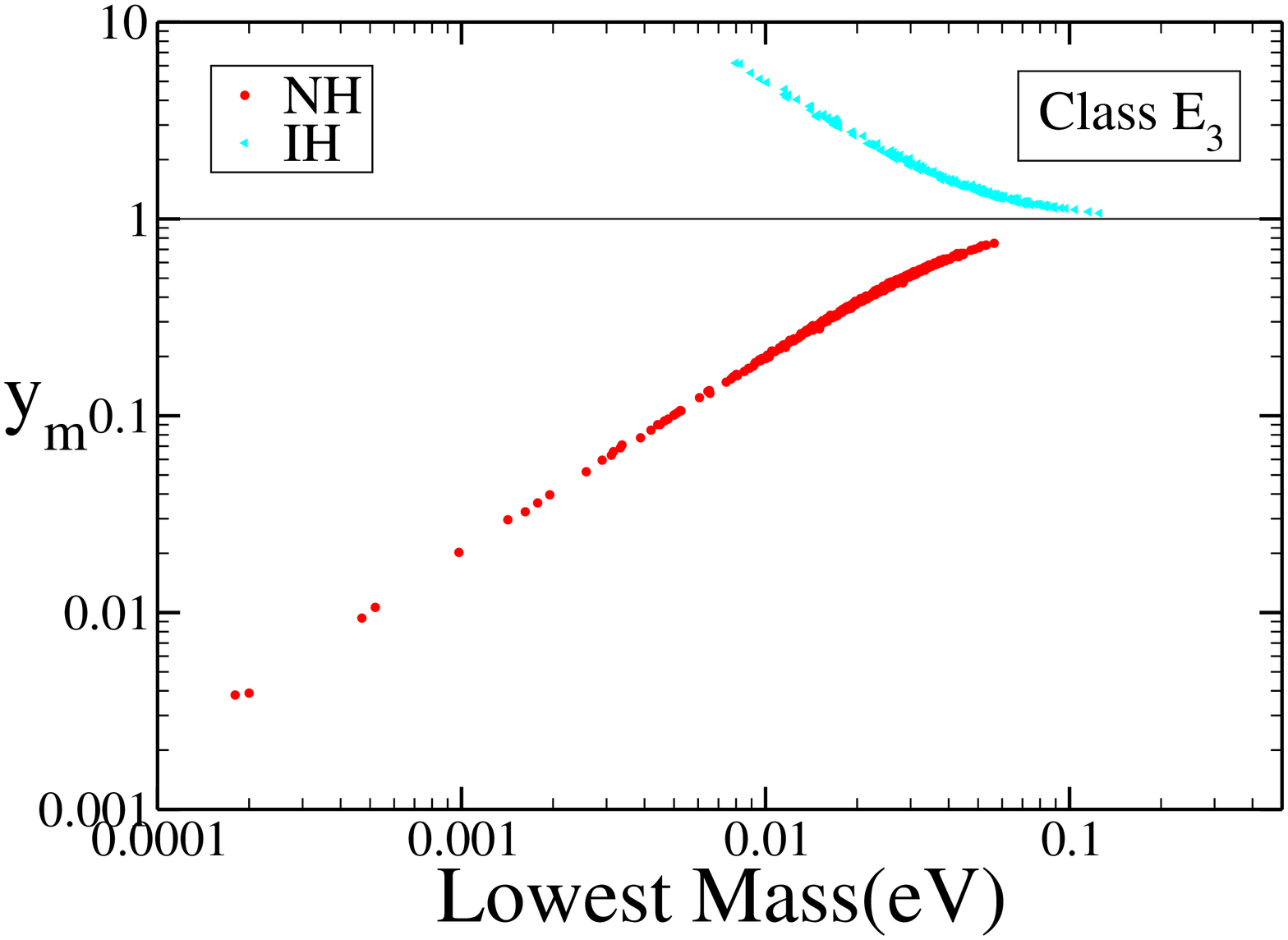}
\includegraphics[width=0.38\textwidth,angle=270]{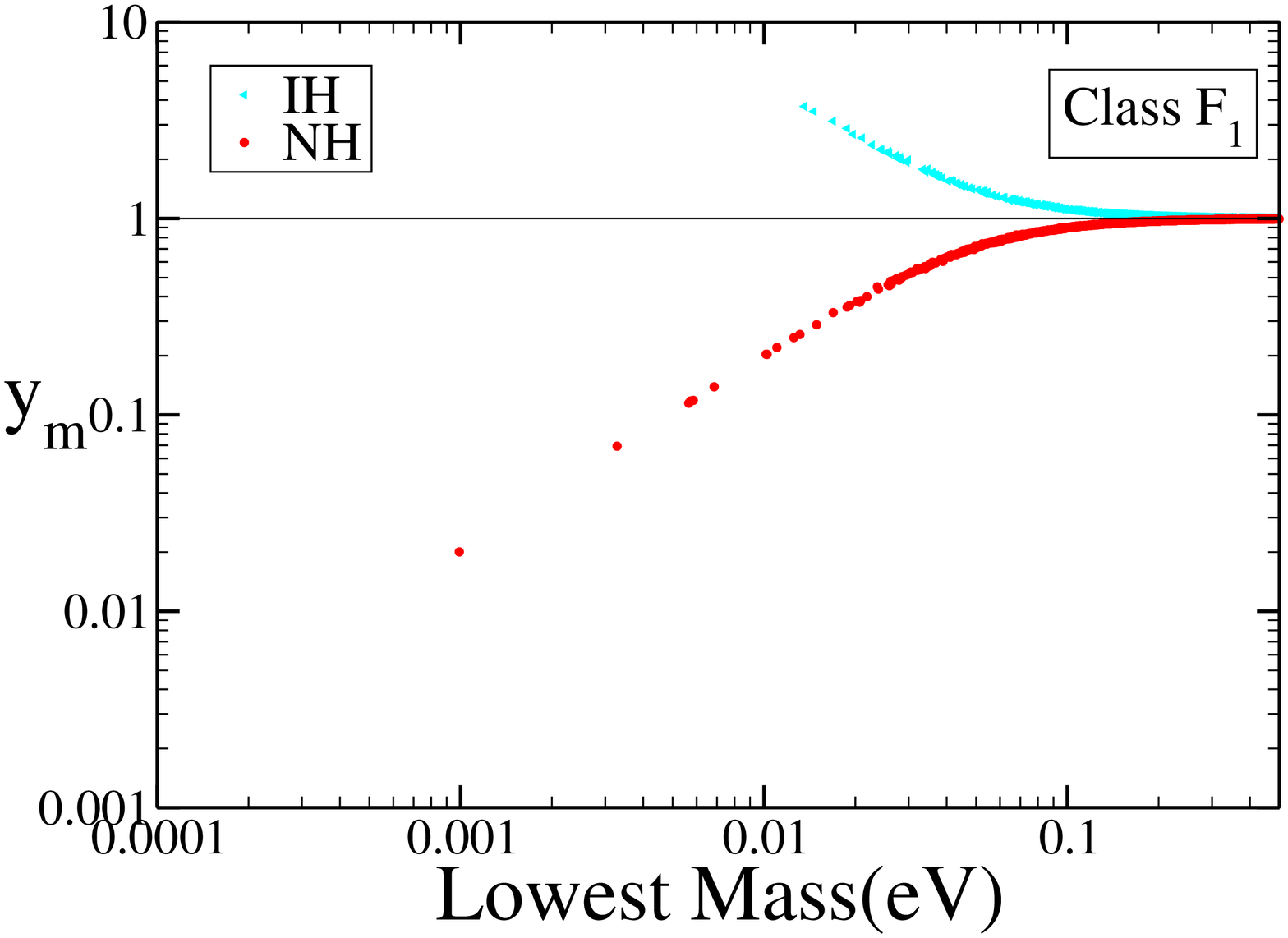}
\caption{The values of $y_m$($=\frac{m_1}{m_3}$) as a function of
the lowest mass ($m_1$ or $m_3$) for the 3+1 case
when the known oscillation parameters are varied 
in Gaussian distributions peaked at their respective best-fit values.}
\label{ym}
\end{center}
\end{figure}

In Fig. \ref{ym} we present the values of $y_m$ vs the lowest mass
for representative textures $A_1$, $B_1$, $C$, $D_1$, $E_3$ and $F_1$
from the 6 classes. The other textures belonging to each class give similar results.
This figure shows that for textures belonging to the A class
$y_m$ remains $<$ 1. Thus, it admits only NH solutions. The textures 
belonging to the D and  E class allow NH and IH 
while the B, C, and F classes allow NH, IH and QD mass spectra.

The textures $A_1-A_2$, $B_1-B_2$, $B_3-B_4$, $D_1-D_2$,
$E_1-E_2$ and $F_2-F_3$ are related by $S_{\mu\tau}$ symmetry where
for the four neutrino framework  $S_{\mu\tau}$ can be expressed as,
\begin{center}
$
S_{\mu\tau}=\left(
\begin{array}{cccc}
 1& 0 & 0 & 0 \\ 0& 0 &1& 0 \\ 0& 1 &0 & 0\\ 0 & 0 & 0 &1
\end{array}
\right)$
\end{center}
in such a way that
\begin{center}$A_2=S_{\mu\tau}^T A_1 S_{\mu\tau} . $ \end{center}

Note that for 3 generation case the angle $\theta_{23}$ in the partner
textures linked by $\mu-\tau$ symmetry was related as
$\bar{\theta}_{23}=(\frac{\pi}{2} - \theta_{23})$.
However, for the 3+1 case no such simple relations are obtained for the mixing
angle $\theta_{23}$. The angles $\theta_{24}$ and $\theta_{34}$ in the two
textures related by $\mu-\tau$ symmetry are also different.
For this case, the mixing angles
for two textures linked  by
$S_{\mu \tau}$ symmetry
are related as
\begin{eqnarray}
\bar{\theta}_{12} &=& \theta_{12},
~~~ \bar{\theta}_{13} = \theta_{13},
~~~ \bar{\theta}_{14} = \theta_{14},  \label{th24} \\
\sin{\bar\theta_{24}} &=&  \sin\theta_{34} \cos{\theta_{24}}, \\
\sin {\bar\theta_{23}} 
&=&\frac{\cos{\theta_{23}}\cos{\theta_{34}}-\sin{\theta_{23}}\sin{\theta_{34}}\sin{\theta_{24}}}{\sqrt{1-\cos{\theta_{24}^2}\sin{\theta_{34}^2}}}, \label{th34} \\
\sin {\bar\theta_{34}}
&=&\frac{\sin{\theta_{24}}}{\sqrt{1-\cos{\theta_{24}^2}\sin{\theta_{34}^2}}}.
\end{eqnarray}


\end{enumerate}

The texture zero conditions together with the constraints imposed by
the experimental data allow us to obtain correlations between various
parameters specially the mixing angles of the 4$^{\mathrm th}$ neutrino
with the other three for the A and E classes. For the B, C, D and F classes one gets constraints
on the effective mass governing $0\nu\beta\beta$.

In order to gain some analytic insight into the results
it is important to understand
the mass scales involved in the problem. The solar mass scale is 
$\sqrt{\Delta_{21}} \approx 0.009$ eV
whereas the atmospheric mass scale is $\sqrt{\Delta_{31}} \approx 0.05$ eV.
Normal hierarchy among the active neutrinos implies
$m_1 << m_2 << m_3$ corresponding to $m_1 \lsim 0.009$ eV.
It is also possible that $m_1 \approx m_2 << m_3$ implying 
$m_1 \sim$ (0.009-0.1) eV. We call this partial normal
hierarchy. IH corresponds to $m_3 << m_1 \approx m_2$.
If on the other hand $m_1 >$ 0.1 eV then
$m_1 \approx m_2 \approx m_3$ which corresponds to quasi-degenerate neutrinos.

\begin{itemize}

\item \underline{\textbf{$A$ and $E$ Class}}

 \begin{figure}[ht!]
 \begin{center}
 \includegraphics[width=0.38\textwidth,angle=270]{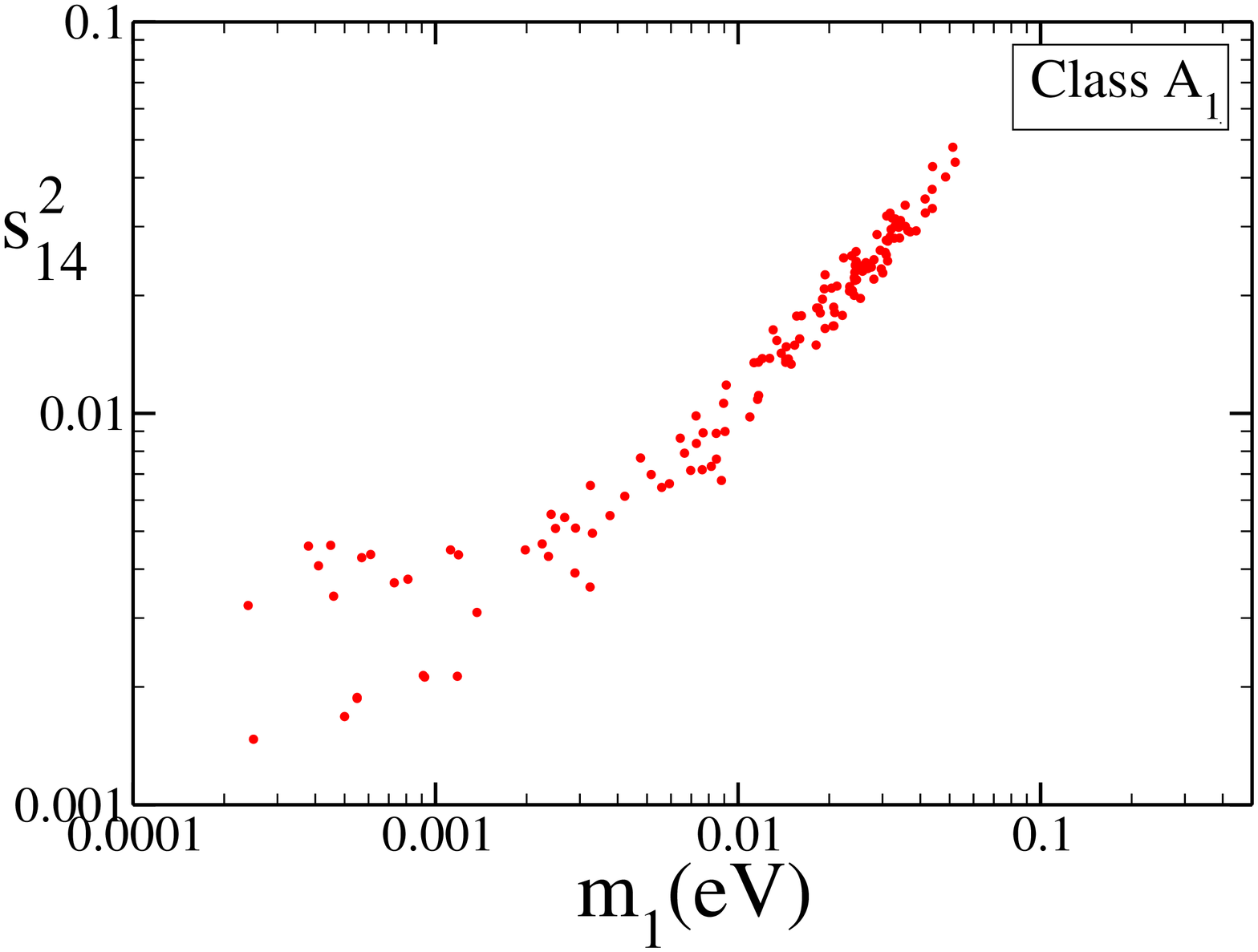}
\includegraphics[width=0.38\textwidth,angle=270]{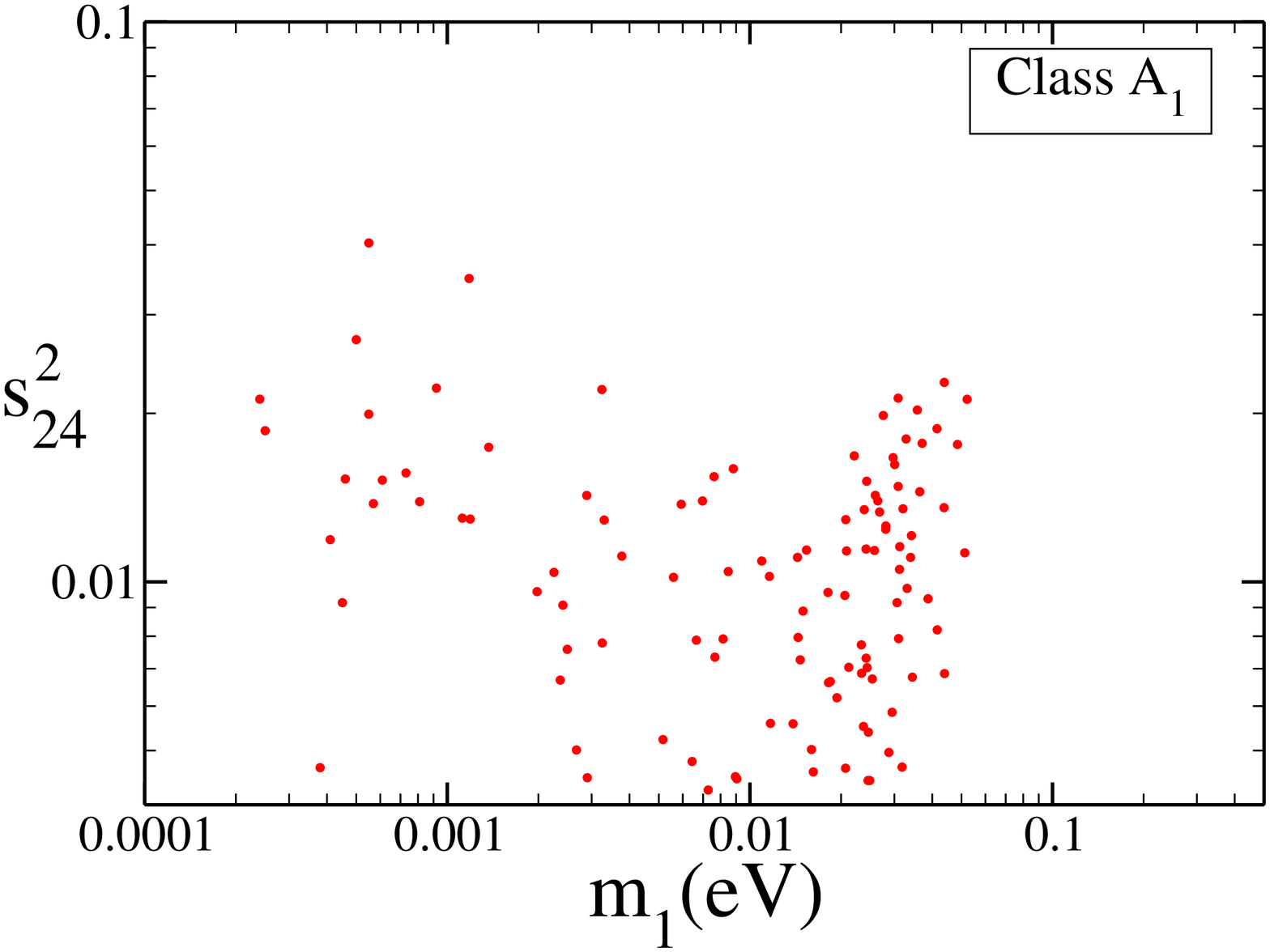} \\
\includegraphics[width=0.38\textwidth,angle=270]{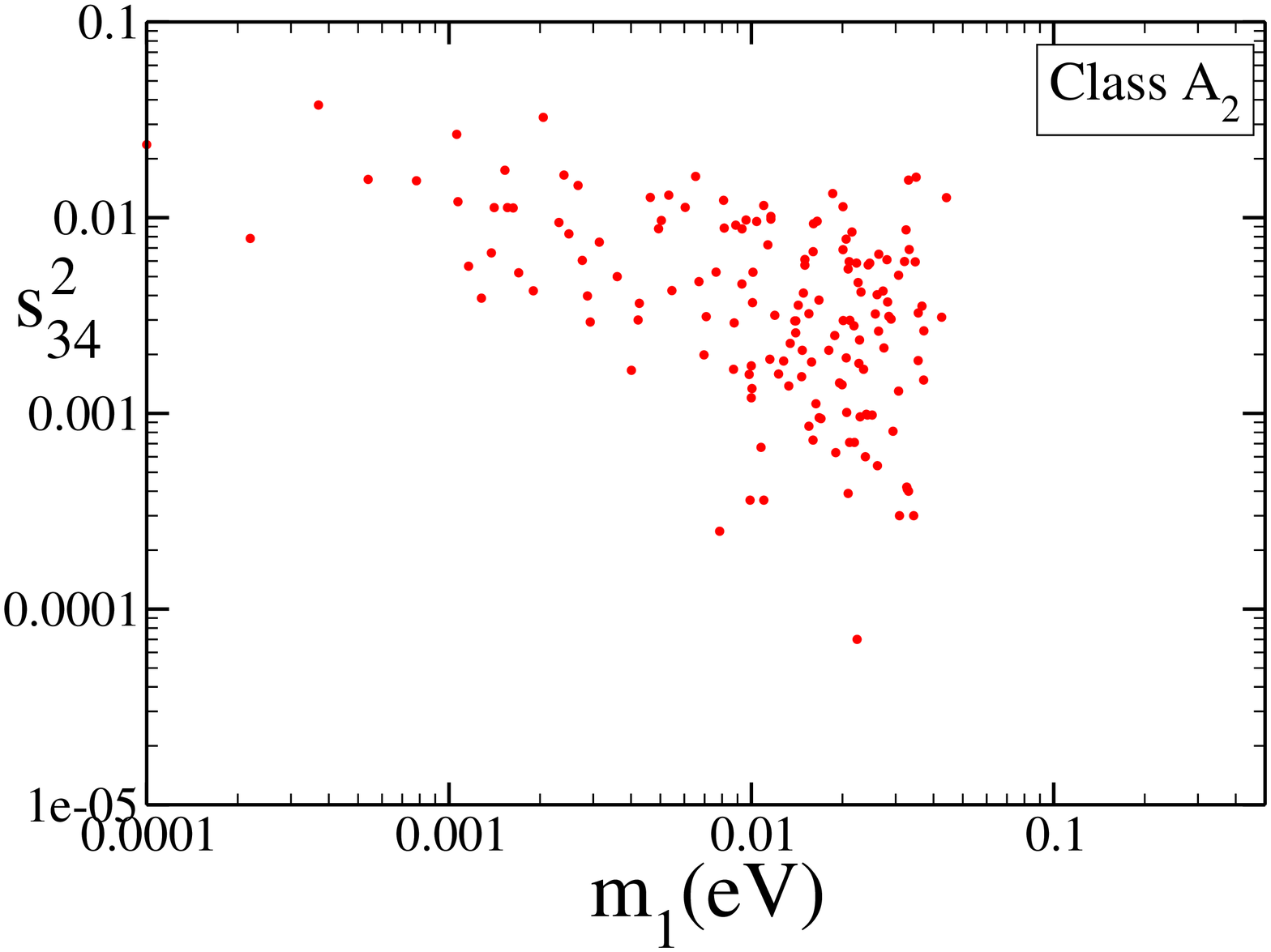} 
\includegraphics[width=0.38\textwidth,angle=270]{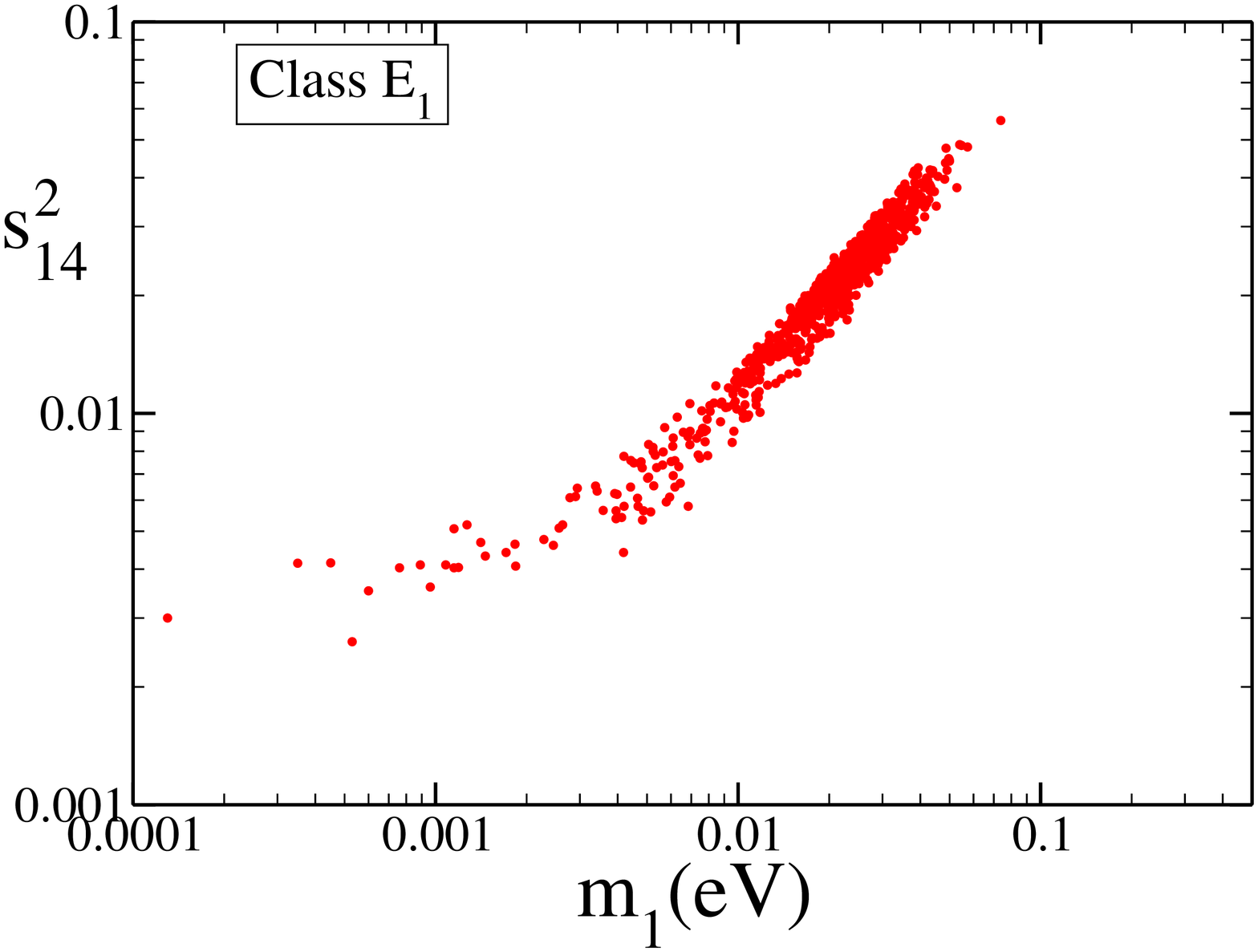} \\
\includegraphics[width=0.38\textwidth,angle=270]{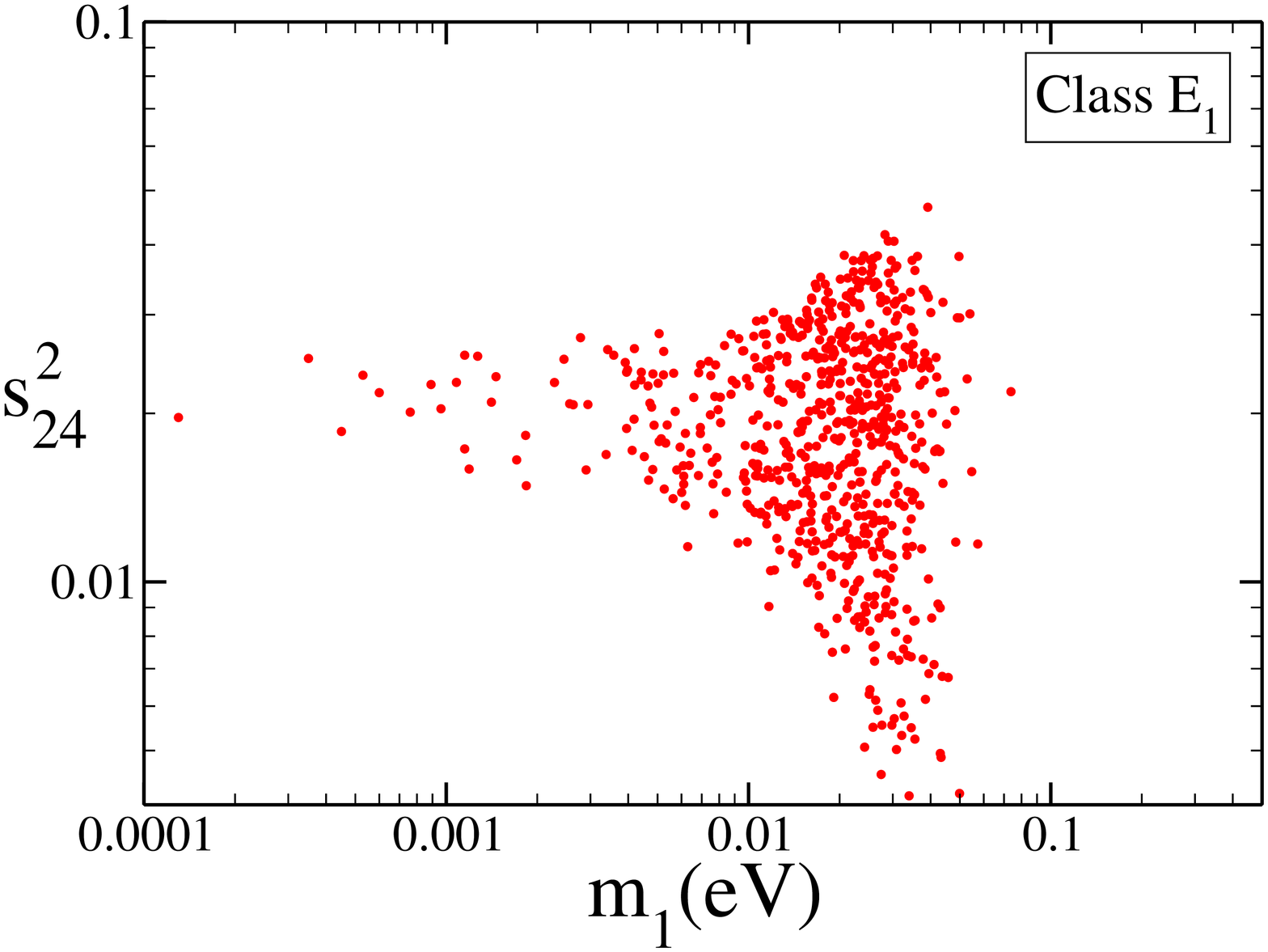}
\includegraphics[width=0.38\textwidth,angle=270]{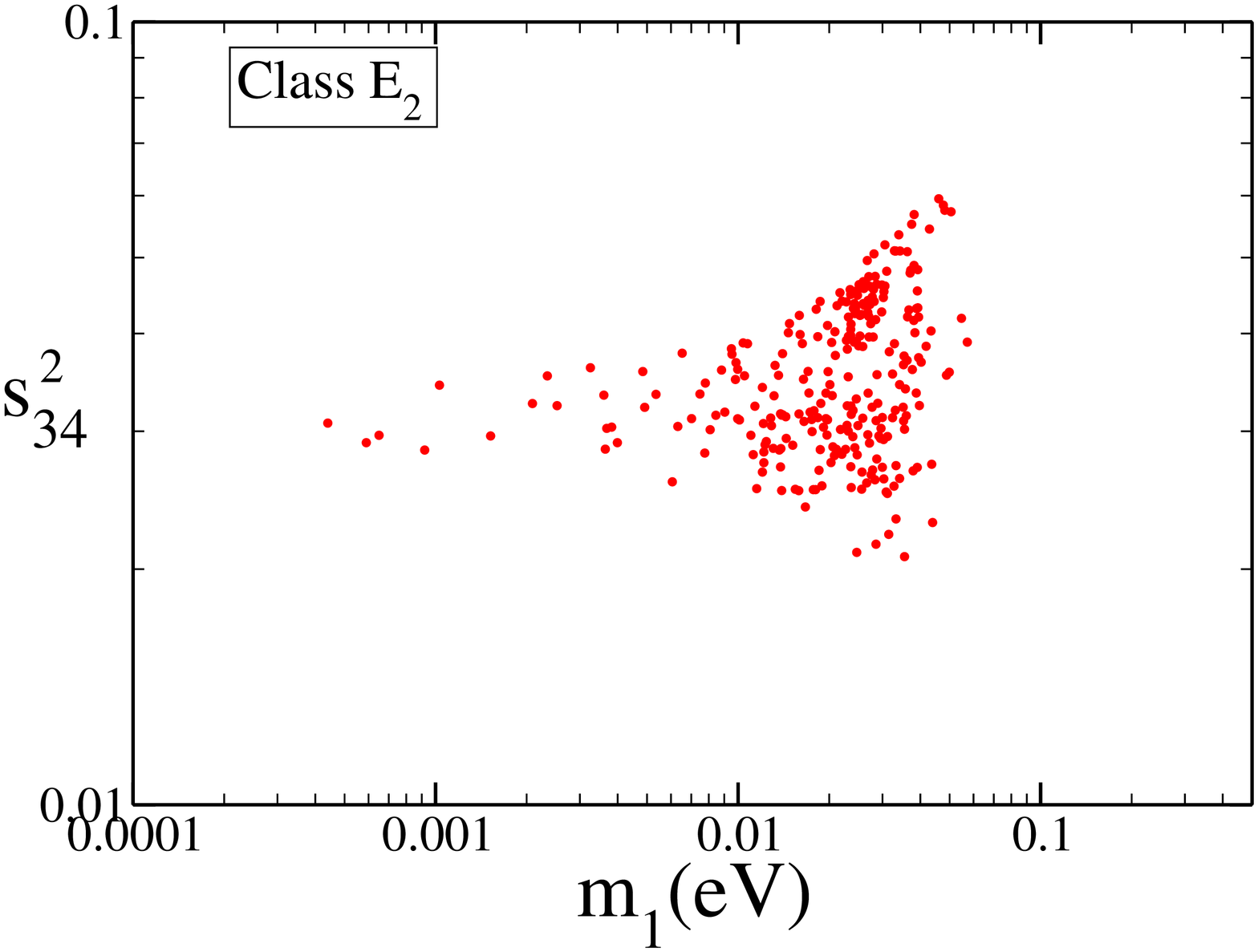}
\caption{Correlation plots for $A$ and $E$ class.}
 \label{Aclass}
 \end{center}
 \end{figure}

For these classes we find $y_m$ to be mainly in the range $>$ 0.0001 eV
extending up to $\sim 0.1$ eV.
Thus, these classes allow normal hierarchy (full or partial)
among the 3 active neutrinos.
These classes are characterised by the condition $|m_{ee}|=0$. $m_{ee}$ for the four neutrino framework can be expressed as,
\bea
m_{ee} & = & c_{12}^2 c_{13}^2 c_{14}^2 m_1 +
c_{13}^2 c_{14}^2 e^{-i \alpha } {m_2} s_{12}^2 \\ \nonumber
&+& c_{14}^2
e^{- i \beta } m_3 s_{13}^2 + e^{- i \gamma } {m_4} s_{14}^2.
\label{mee1}
\eea
For smaller values of $m_1$ and NH,
the dominant contribution to the magnitude of the above term
is expected to  come from the last term
$s_{14}^2 \sqrt{\Delta_{41}} \sim 0.022$. Therefore, very small values of $m_1$ is less likely to give
vanishing $m_{ee}$ for normal hierarchy.
However, we get some allowed points in the small $m_1$ regime which
implies smaller values of $s_{14}^2$.
$m_{ee}$ can be approximated in the small $m_1$ limit as,
\be
m_{ee} \approx  e^{-i \alpha } {m_2} s_{12}^2
+
e^{- i \beta } m_3 s_{13}^2 + e^{- i \gamma } {m_4} s_{14}^2.
\ee
The maximum magnitude of the first two terms  is $\sim 0.003$.
Then using typical values of $m_4$ ($\sim$ 0.9 eV) from the 3$\sigma$ range,
we obtain
$s_{14}^2 \sim (0.003- 0.004$) in the small $m_1$ limit.
This is true for all the textures in the A and E class.
For the $A_1$ class
we also simultaneously need vanishing $|m_{e \mu}|$.
In the small $m_1$ limit approximate expression for $m_{e \mu}$ is
\bea
m_{e \mu} & \approx &
e^{i(\delta _{14}-\delta _{24}- \gamma)}
s_{14} s_{24} m_4  \\ \nonumber
&+& e^{i(\delta _{13}-\beta)}  s_{13} s_{23} m_3
+ e^{-i\alpha} c_{12} c_{23}
s_{12} m_2,
\label{memu}
\eea
and the first term i.e., $m_4 s_{14} s_{24} \sim (0.05 - 0.06) s_{24} $. 
While the other terms are of the order (0.006 - 0.007) which implies
$s_{24}^2 \sim (0.01 - 0.02)$. This is reflected
in the first and second panels of Fig. \ref{Aclass} where the correlation of
$s_{14}^2$ and $s_{24}^2$  with $m_1$ is depicted.
As $m_1$ increases the contribution from the
first three terms in $m_{ee}$ increases  and $s_{14}^2$ becomes larger
for cancellation to occur. For $m_{e \mu}$,
this increase in $s_{14}$ helps to achieve cancellation for higher
values of $m_1$ and  therefore $s_{24}^2$ stays almost the same.
Similar argument also apply to the $E_1$ class which has vanishing $m_{\mu \mu}$.

For $A_{2}$ class, in addition we have vanishing $m_{e \tau}$.
In the limit of small $m_1$, $m_{e \tau}$ can be approximated as,
\be
m_{e \tau} \approx
e^{i(\delta _{14}- \delta _{24}-\gamma)}
s_{14} s_{34} m_4  +  e^{i(\delta_{13} - \beta)} s_{13}^2
m_3 - m_2 e^{-i \alpha}s_{12} c_{12} s_{23}.
\ee
As discussed earlier vanishing $m_{ee}$ implies small $s_{14}^2 \sim (0.002 -0.005$) in
the limit of small $m_1$. Thus, the contribution from the $m_4$ term is $\sim  (0.04 - 0.06) s_{34}$.
The typical contribution from the last two terms is $\sim 0.008$. 
This implies $s_{34}^2$ to be in the range (0.02 -0.04) for smaller values of $m_1$. This is reflected in the third panels of 
Fig. \ref{Aclass}  where we have plotted the correlation of $s_{34}^2$ with $m_1$.
Since with increasing $m_1$, $s_{14}$ increases
to make $|m_{ee}|=0$,
$s_{34}^2$
does not increase further.
Similar bounds on $s_{34}^2$ are also obtained for $E_2$ class.

As one approaches the QD regime then the terms containing
the active neutrino masses starts contributing more.
So for higher values of $m_1$ complete cancellation leading to vanishing $m_{ee}$
even at the highest value of $s_{14}^2$ is not possible. This feature restricts $m_1$ to be $<$ 0.1 eV in A class.

Textures belonging to A and E class contain banishing $m_{ee}$ which is not possible
for IH in the three generation case since the solar mixing angle is not
maximal. In the 3+1 scenario a cancellation leading to vanishing
$m_{ee}$ is possible for IH. In this case, the condition $m_{ee}=0$ 
constraints the majorana phases as $ 80^o < \alpha < 280^o $, $ 160^o < \gamma < 220^o $ and
the 1-4 mixing angle as  $s_{14}^2$
as $0.017 < s_{14}^2 < 0.042$. These conditions are not compatible with the 
condition of vanishing $m_{e \mu}$ or $m_{e \tau}$ for IH. 
Therefore, class A which simultaneously requires one of these elements to 
vanish along with $m_{ee}$ is disfavoured for IH. 
On the other hand, the elements i.e., $m_{\mu \mu}$, $m_{\mu \tau}$ or $m_{\tau \tau}$ 
can assume zero value satisfying these constraints. Thus E class which 
has $m_{ee}$ together with one of these elements as vanishing can have inverted hierarchical spectrum.     


For vanishing $m_{ee}$, 
the effective mass (\textit{$m_{eff}$} = $|m_{ee}|$) governing 
the neutrinoless double beta decay ($0\nu\beta\beta$)
is also vanishing for these classes.


\begin{figure}[ht!]
\begin{center}
\includegraphics[width=0.38\textwidth,angle=270]{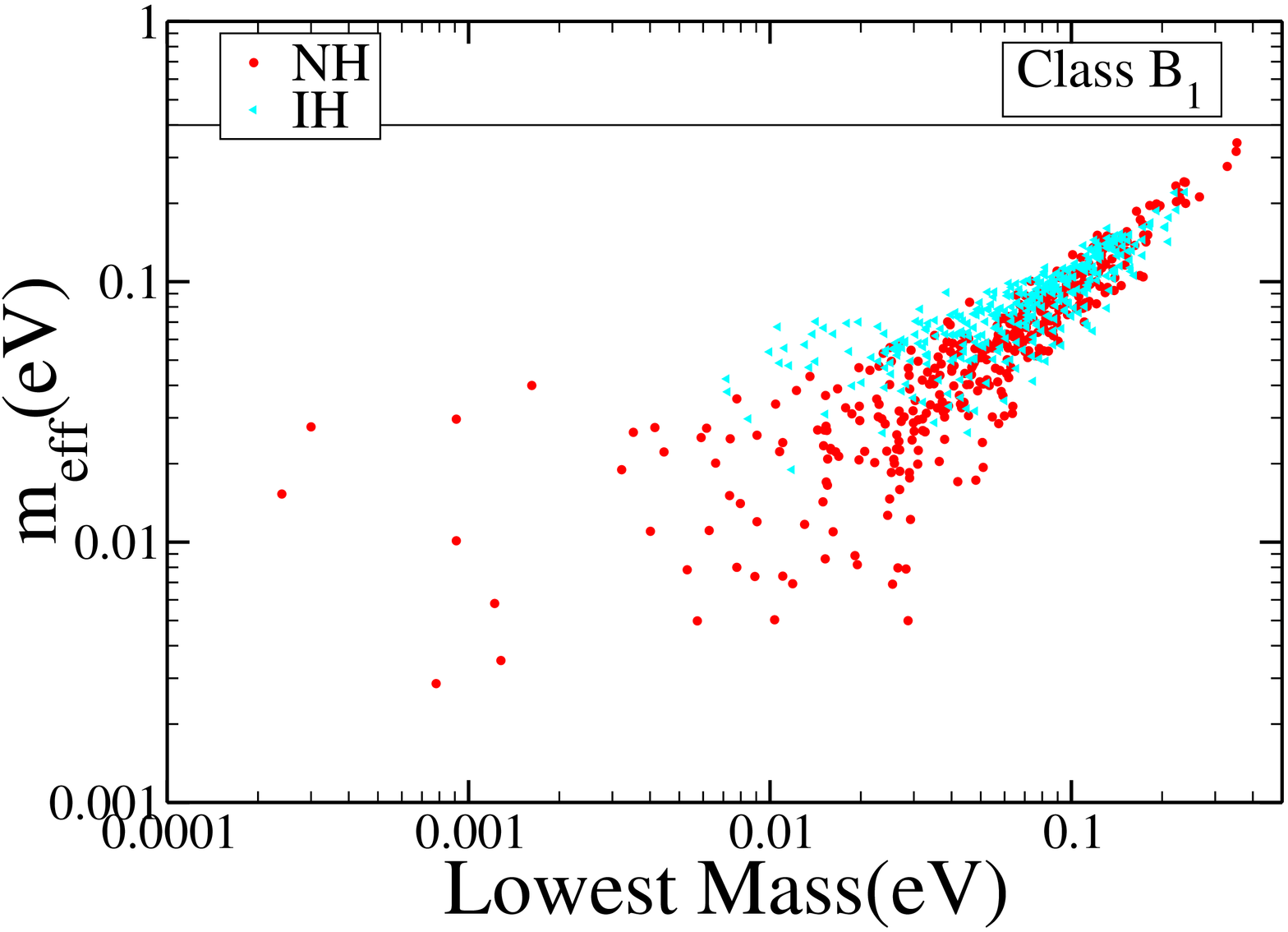}
\includegraphics[width=0.38\textwidth,angle=270]{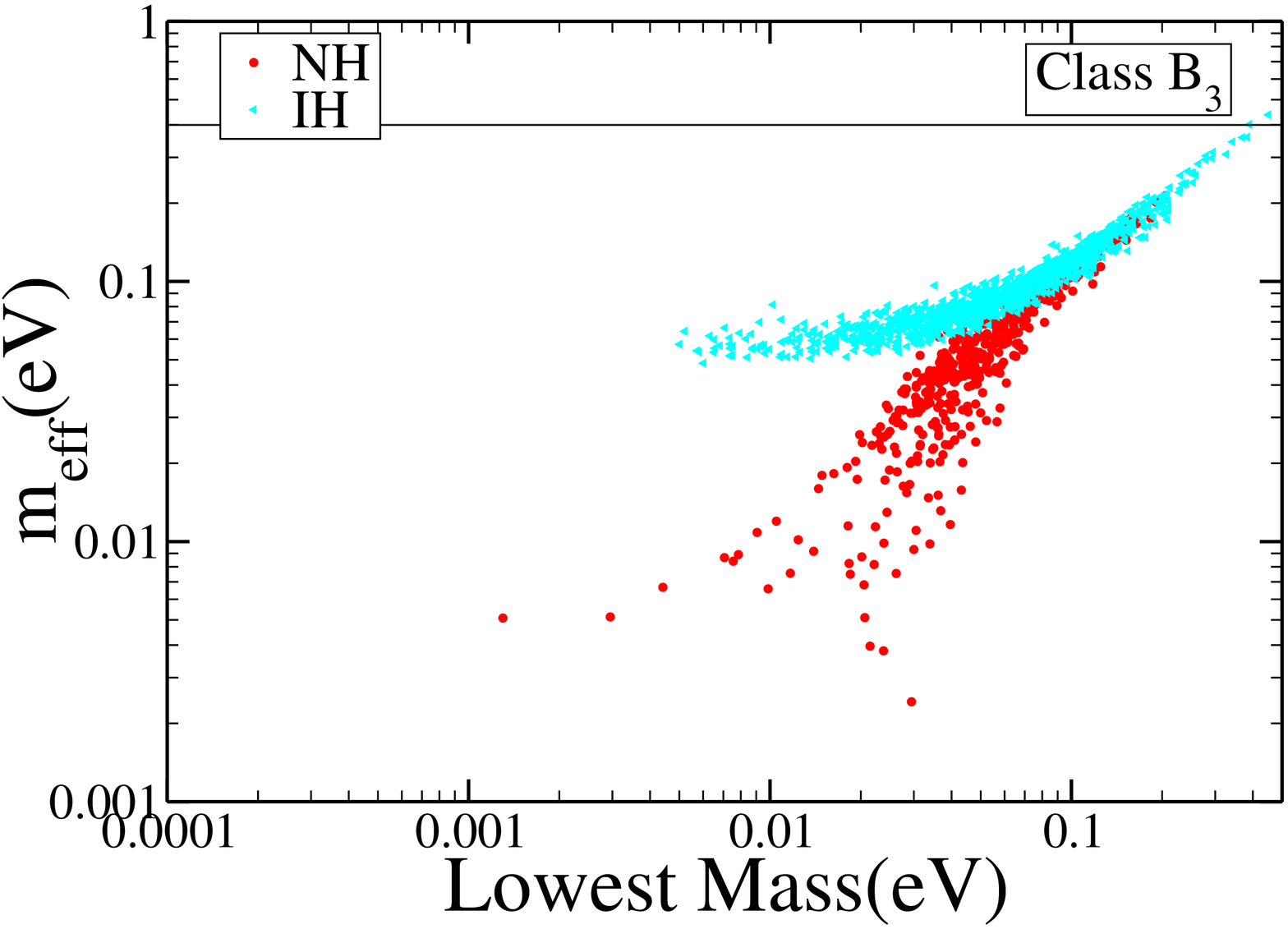} \\
\includegraphics[width=0.38\textwidth,angle=270]{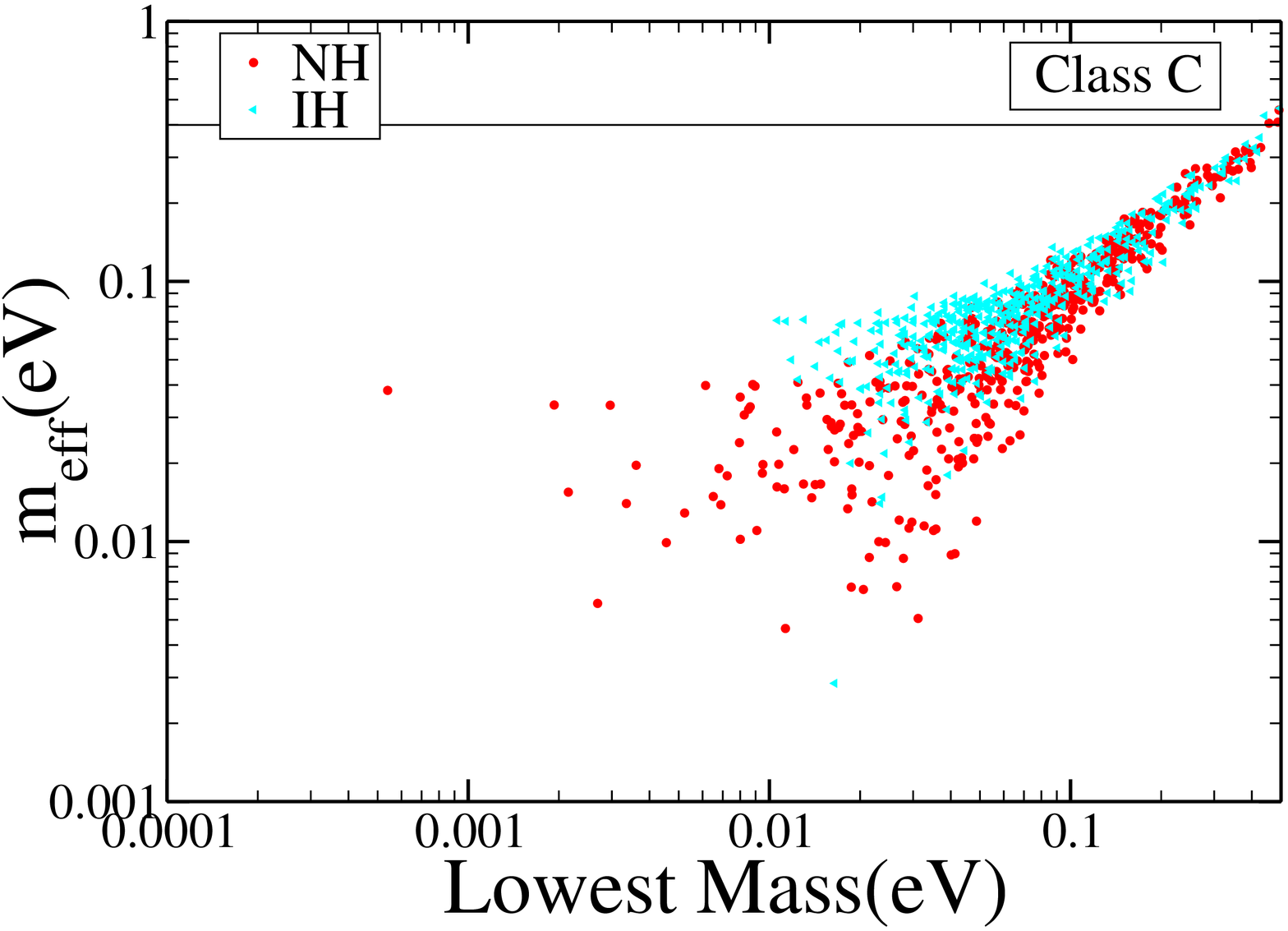} 
\includegraphics[width=0.38\textwidth,angle=270]{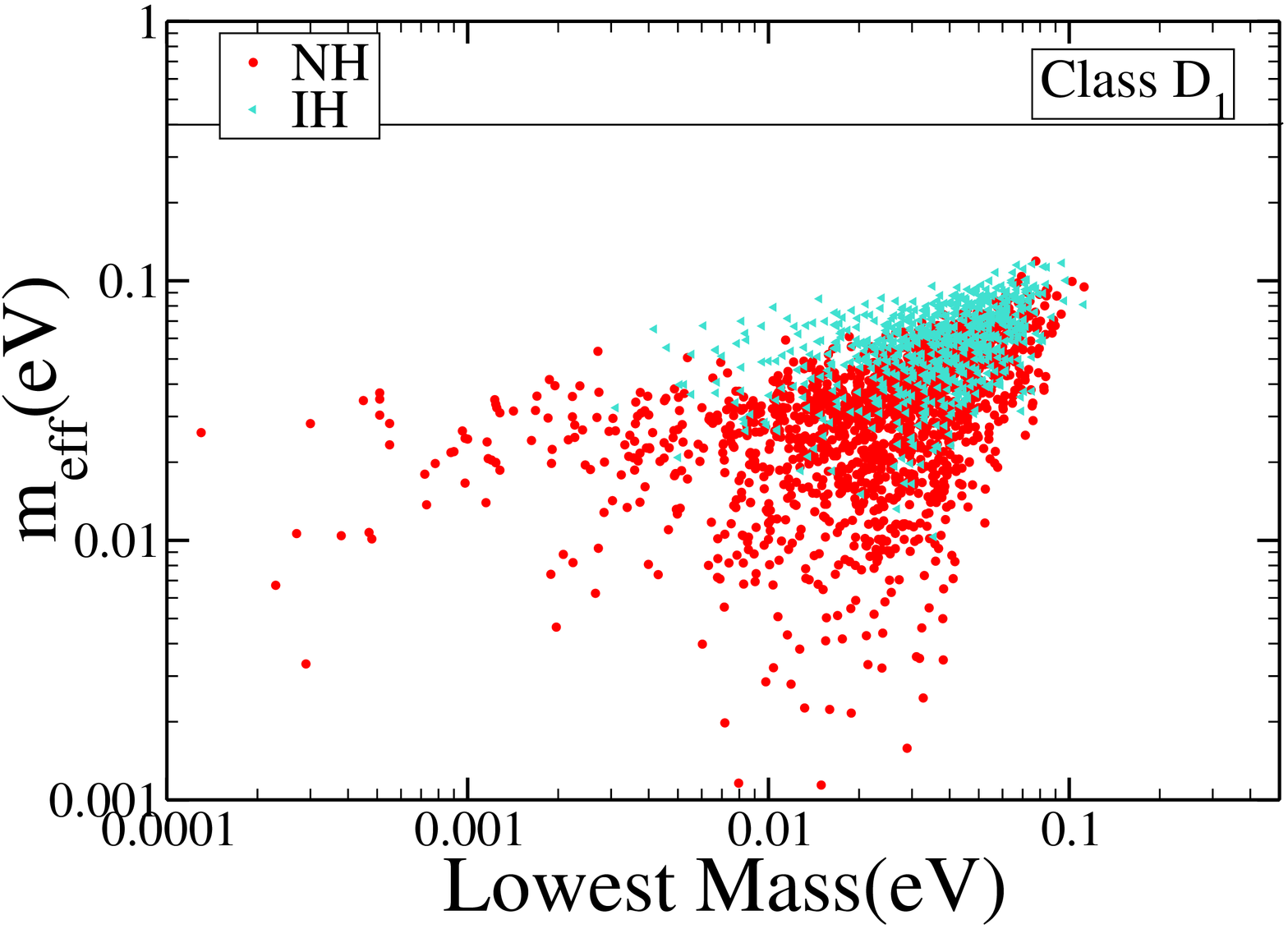} \\
\includegraphics[width=0.38\textwidth,angle=270]{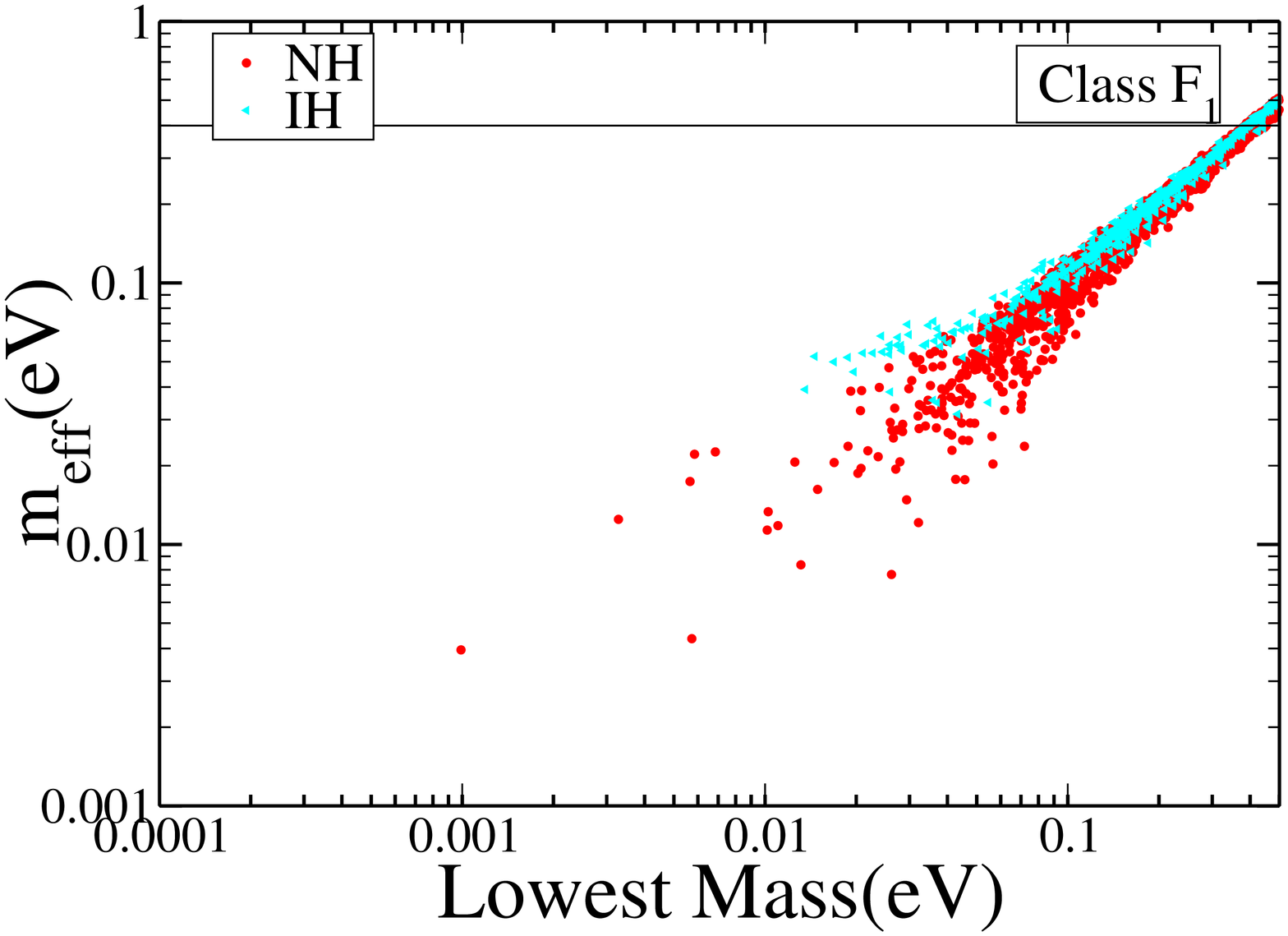} 
\includegraphics[width=0.38\textwidth,angle=270]{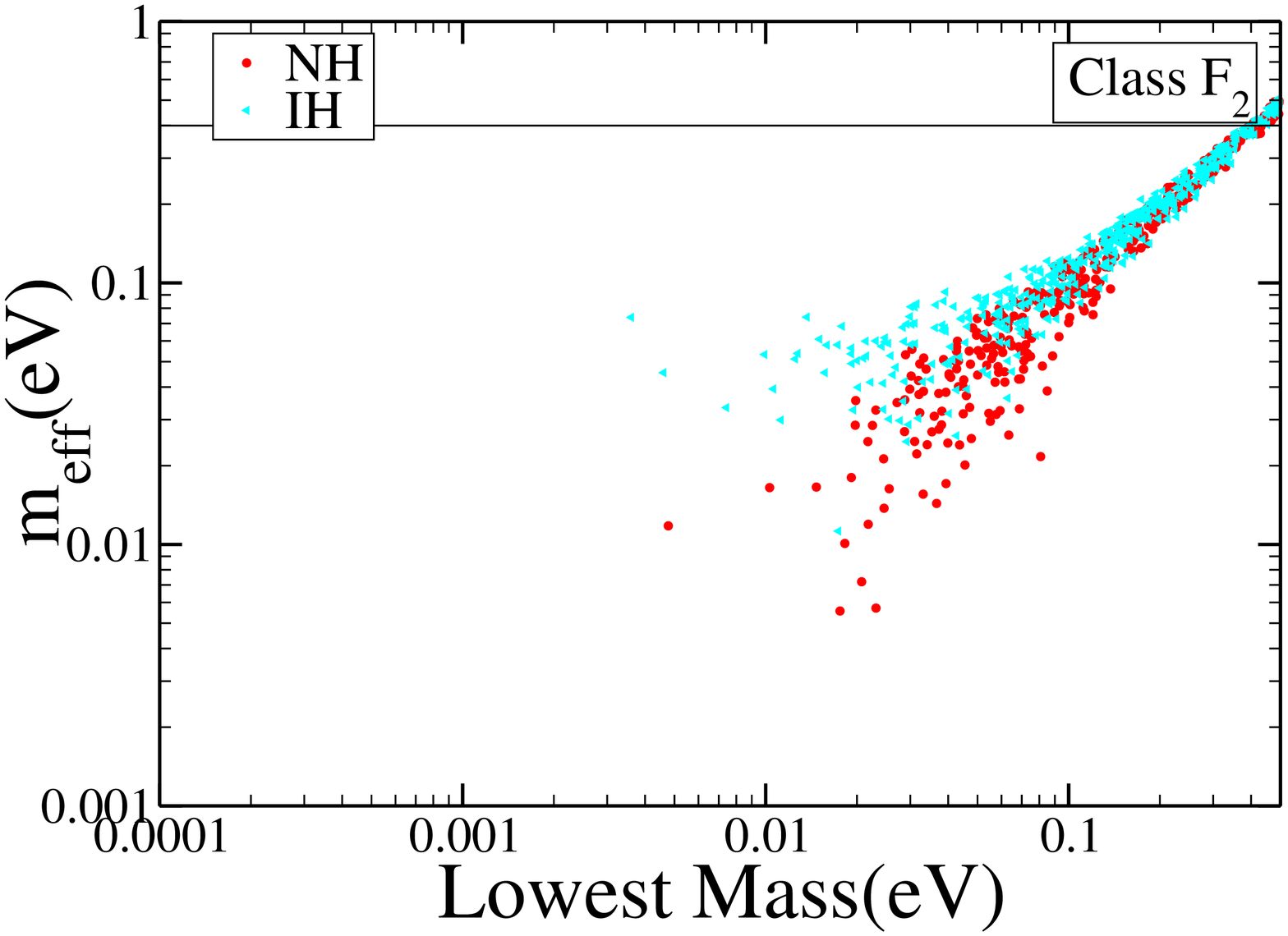}
\caption[The effective mass governing $0\nu\beta\beta$ as a function of 
the lowest mass.]{The effective mass governing $0\nu\beta\beta$ as a function of 
the lowest mass. The red (dark) points correspond to NH while the cyan (light) 
points correspond to IH. The horizontal line is the current
bound from neutrinoless double-beta decay experiments.}
\label{fig:effm}
\end{center}
\end{figure}

\item{\underline{\textbf{$B, C$  Classes}}}

In the 3+1 scenario, B and C classes
allow all three mass spectra -- NH, IH and 
QD  assuming the known oscillation parameters to be normally distributed 
(cf. Table \ref{2results}).
In this case since
$\theta_{23}$ in the textures related by 
$\mu-\tau$  
symmetry is not correlated in a simple 
way, the value of this angle not being in the higher octant does not play 
a significant role as in the 3 generation case. 
Among these only $B_1$ allow few points for smaller values of 
$m_1$ for NH. For the IH solution, larger number of points are obtained 
corresponding to the lowest mass $>$ 0.01 eV as is seen from Fig. \ref{ym}. 
In these textures, for higher values of the lowest mass the 
active neutrino contribution  
to the matrix elements are larger and  it is easier to obtain cancellations.
Hence, textures belonging to these classes 
show a preference for QD solutions.
For these textures the  effective mass
governing  $0\nu\beta\beta$ is non-zero.
In the first row of Fig. \ref{fig:effm} we 
present the effective
mass as a function of the lowest mass for the textures 
$B_1$, $B_3$ and C 
for both NH and IH. These two merge at higher values of the lowest mass
corresponding to the QD solution. The effective mass in these textures is $>0.002$ eV for NH 
and $>$ 0.02 eV for IH. If no signal is
seen in future $0\nu\beta\beta$ experiments then large part of the parameter space belonging to these
textures can be disfavoured.

\item{\underline{\textbf{$D, F$ Classes}}}

These two textures are  disallowed in the three generation  case. 
However for the 3+1 scenario they get allowed.
NH is admissible in all the textures belonging to these classes.
The reason for this is the following. \\
In the three active neutrino scenario, the neutrino mass matrix 
in a $\mu-\tau$ block has the elements of the order of $\sqrt{\Delta_{32}}\approx {0.01}eV$ for normal hierarchy. 
Thus, in general these elements are quite large and cannot vanish
\cite{Frampton:2002yf}. However, in the 3+1 case when there is one additional
sterile neutrino, the neutrino mass matrix elements get contribution from the sterile part of the form $m_4 U_{k4} U_{l4}$ where $k=e,\mu,\tau$.
This term is almost of the same order of magnitude and thus can cancel
the active part, resulting into the possibility of vanishing elements
in the $\mu-\tau$ block. Thus, 
the zero textures which were 
disallowed for NH  are now allowed by the inclusion of sterile neutrino (3+1 case). 
In the case of IH, $m_{\mu \tau}$ element for three active neutrinos is always of the order of $\sqrt{\Delta_{32}}\approx {0.01}eV$ and
thus the textures $D_1$, $D_2$, $F_2$, $F_3$  which 
requires $m_{\mu \tau}$ to vanish, were not allowed.
However, for the 3+1 scenario the
extra term coming due to the fourth state
helps in additional cancellations and IH gets allowed 
in these (cf. Table \ref{2results}). For $F_1$ class, IH for three active neutrino 
is disfavored because of phase correlations. 
However, with the additional sterile neutrino this can be evaded making it 
allowed. 
In the bottom row of Fig. \ref{fig:effm} we present the effective mass governing
$0\nu\beta\beta$ for the textures $D_1$, $F_1$ and $F_2$ 
as a function of the lowest mass. The texture
$D_1$ allows lower values of 
$m_1$ for NH while for IH the lowest mass is largely $\gsim 0.01$ eV. 
QD solution is not allowed in D class . 
For F class more points are obtained 
in the QD regime. 
Future experiments on $0\nu\beta\beta$ would be able to probe these regions 
of parameter space. 

\end{itemize}

\begin{figure}[ht!]
\begin{center}
\includegraphics[width=0.38\textwidth,angle=270]{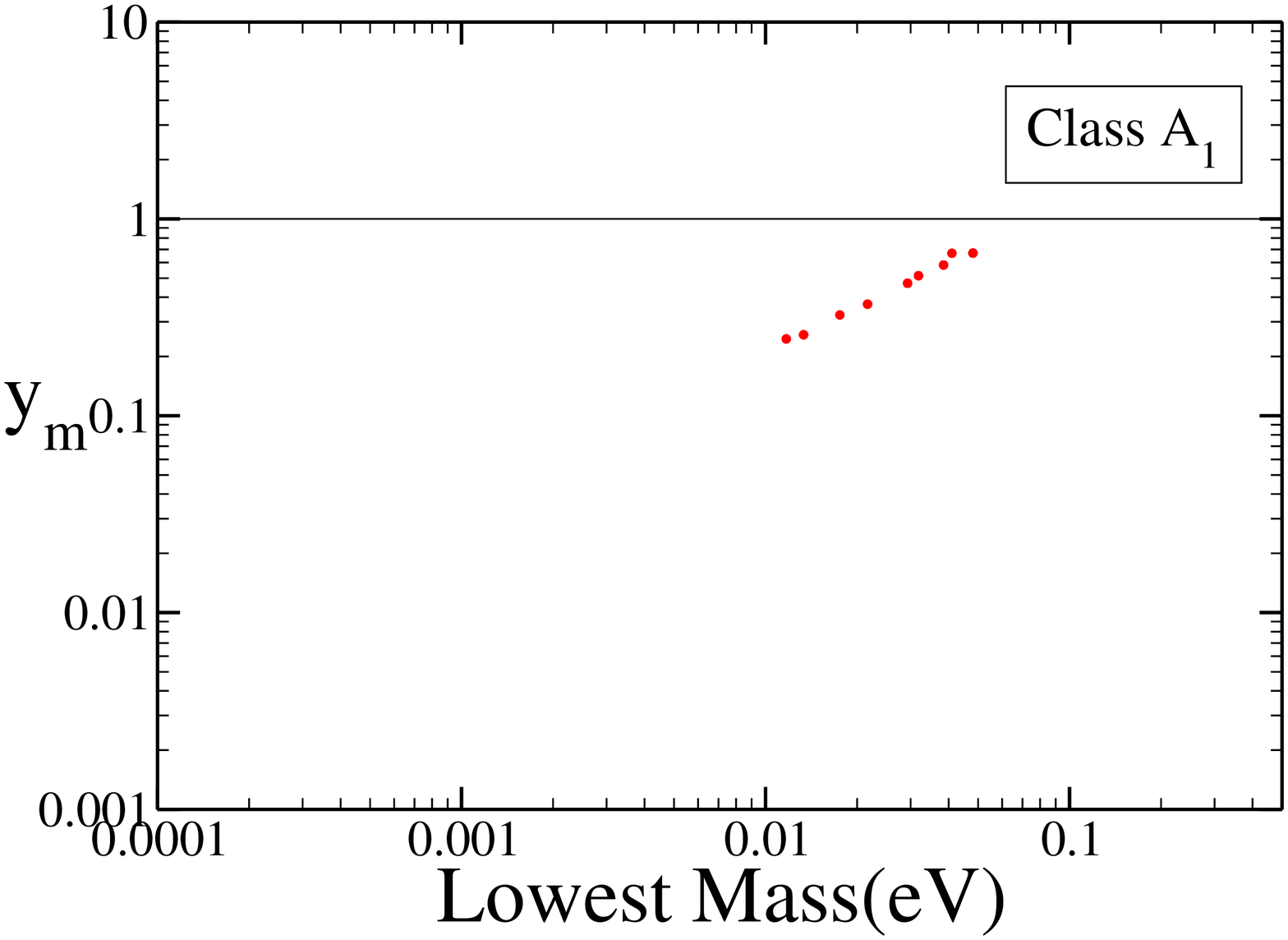}
\includegraphics[width=0.38\textwidth,angle=270]{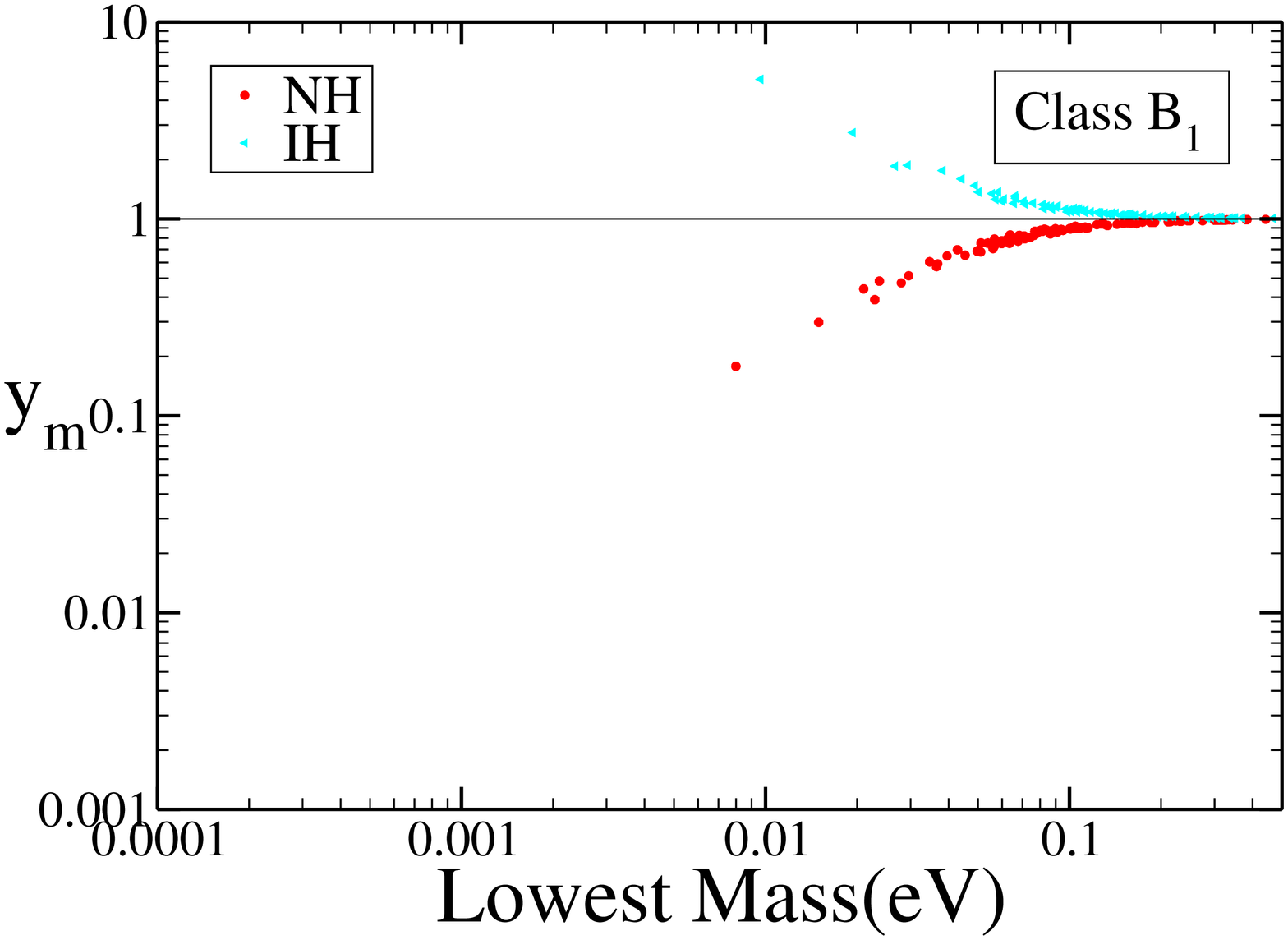} \\
\includegraphics[width=0.38\textwidth,angle=270]{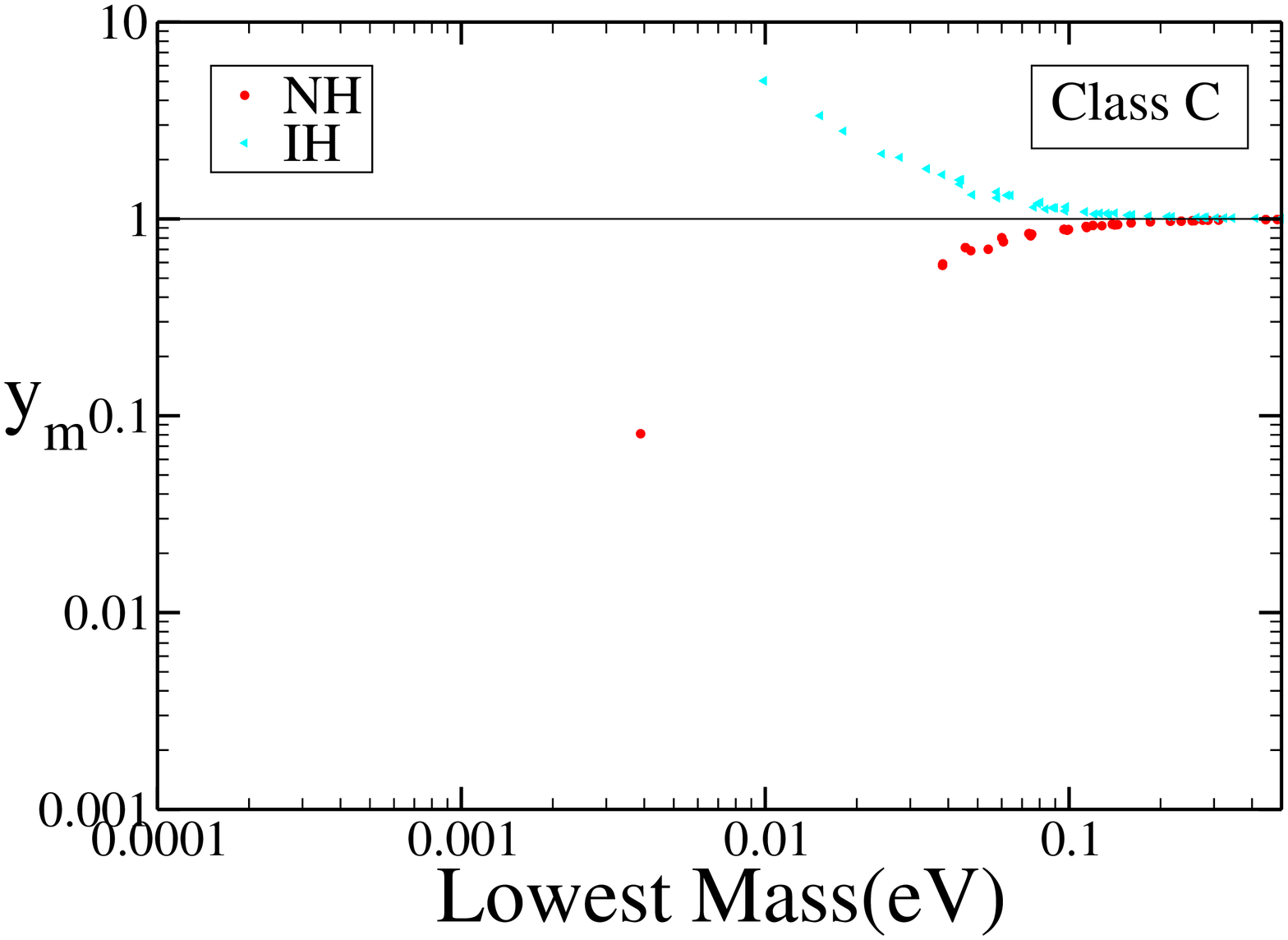}
\includegraphics[width=0.38\textwidth,angle=270]{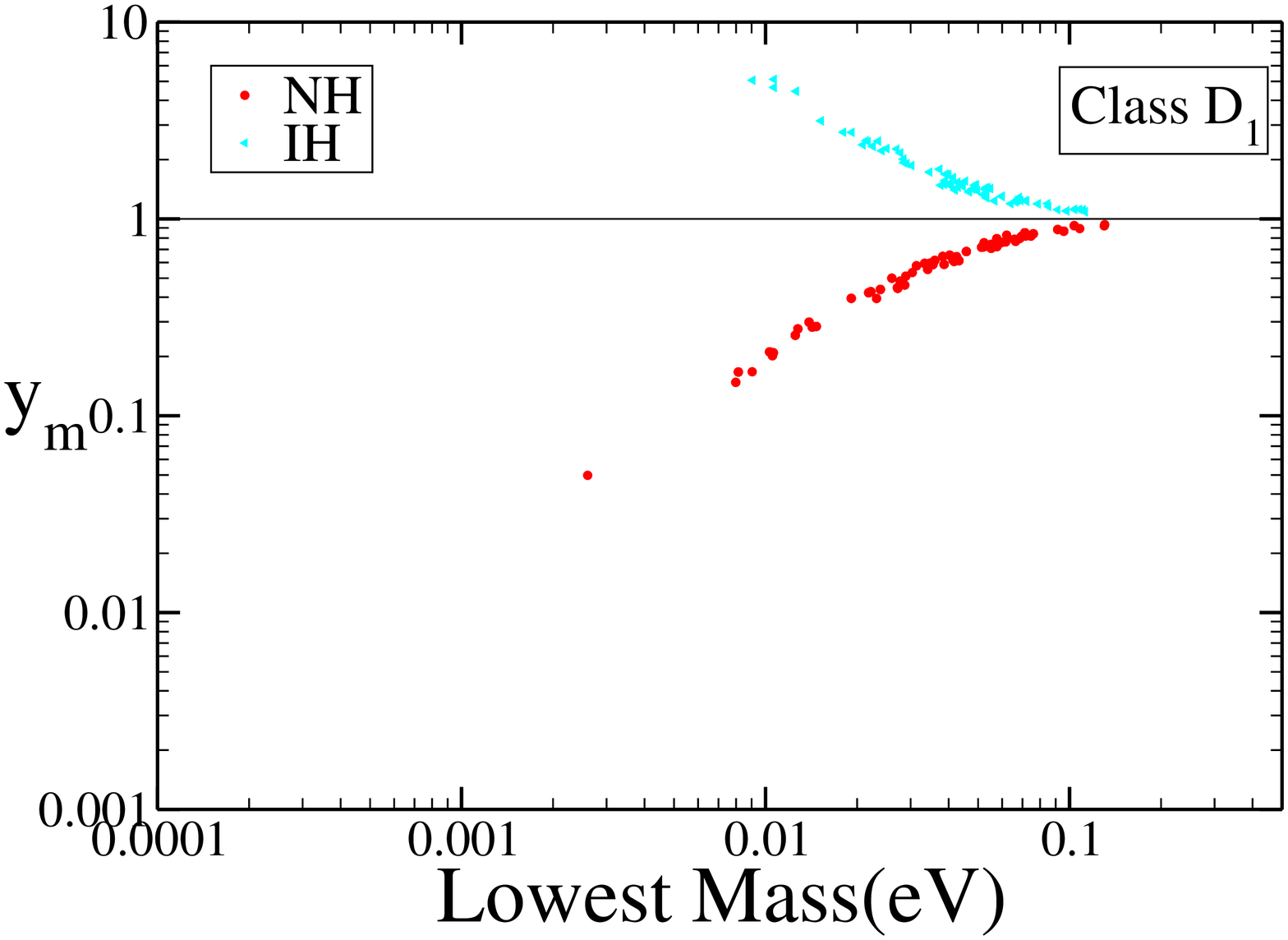} \\
\includegraphics[width=0.38\textwidth,angle=270]{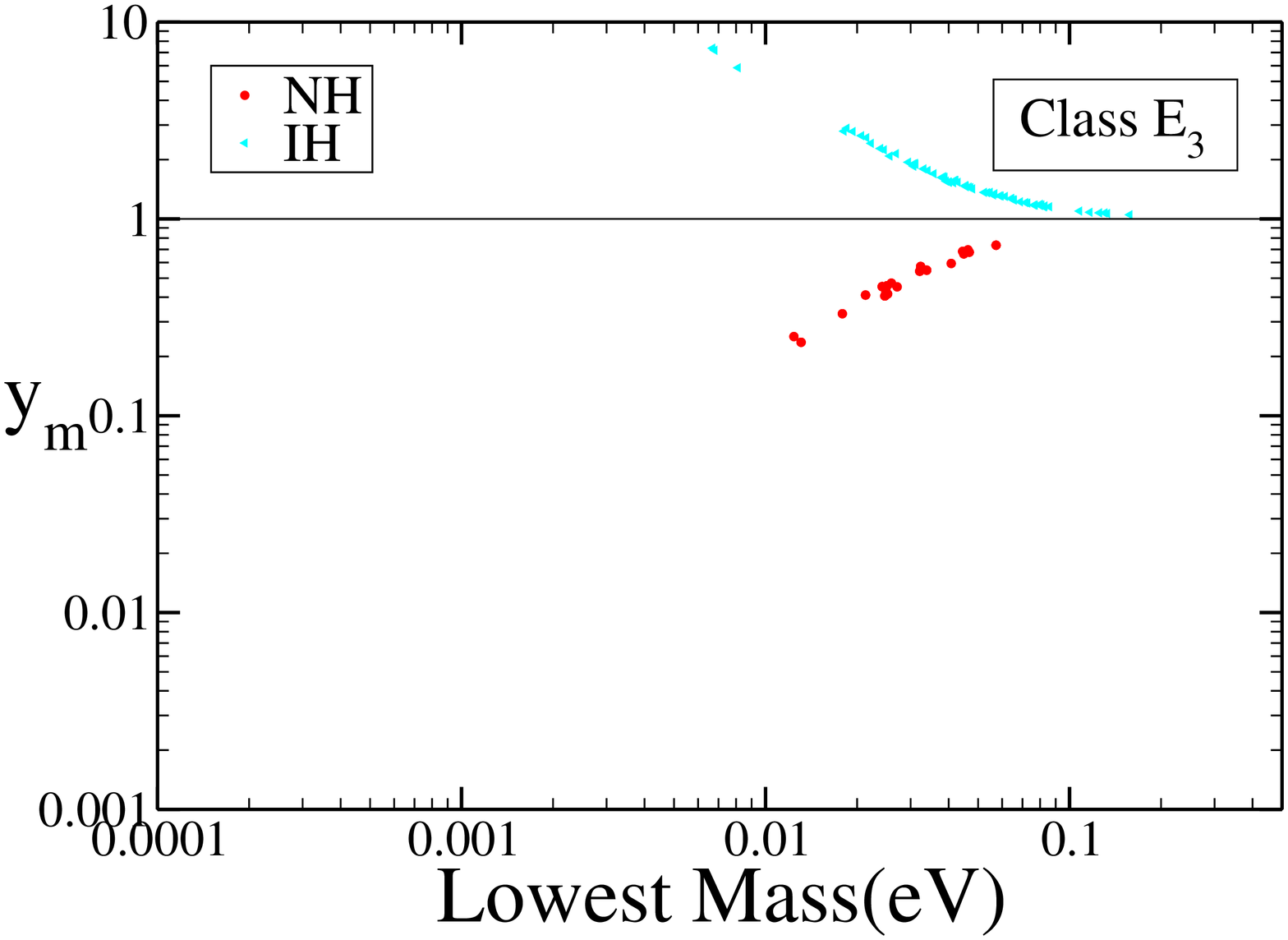}
\includegraphics[width=0.38\textwidth,angle=270]{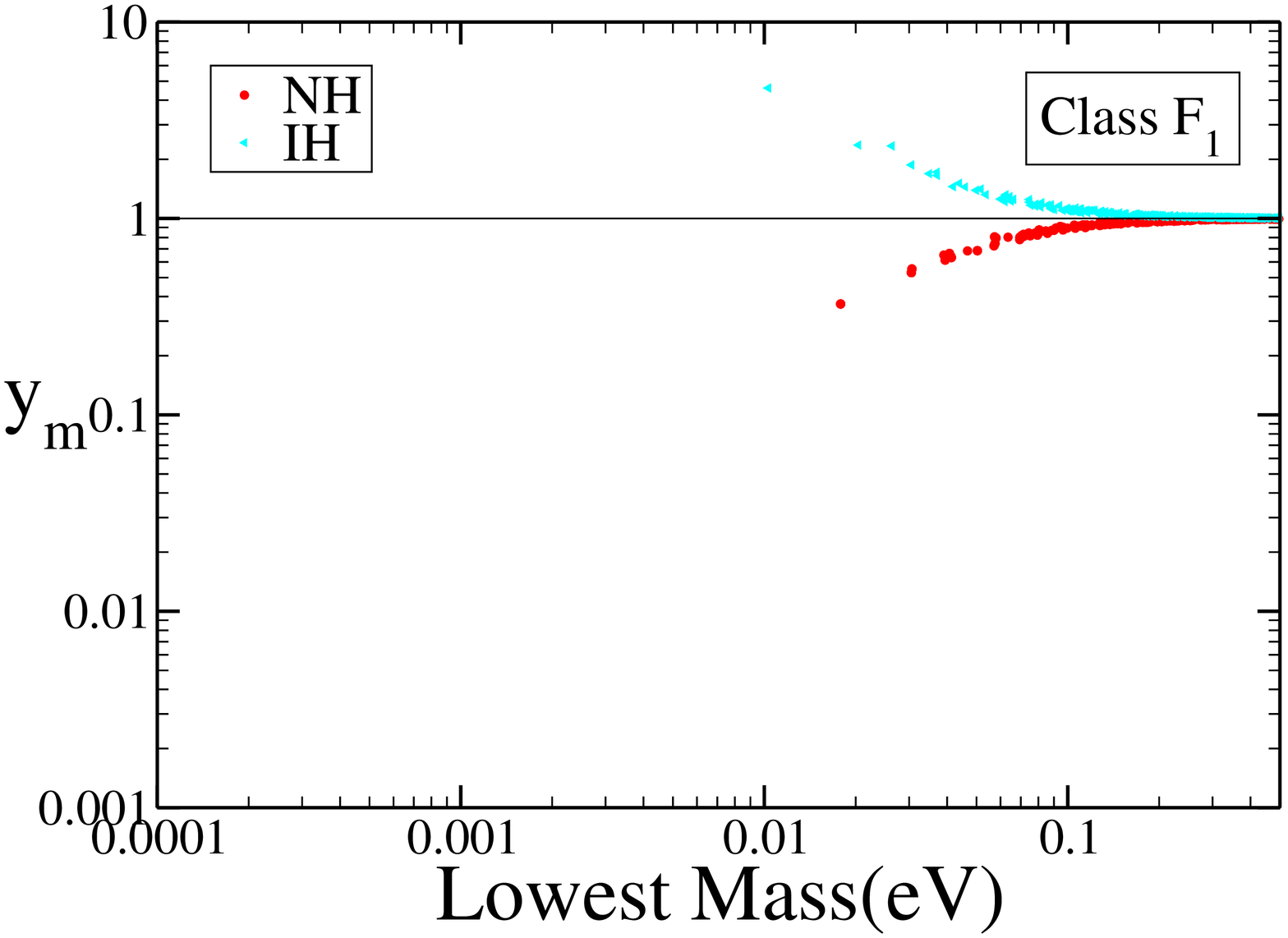}
\caption[The values of $y_m$($=\frac{m_1}{m_3}$) as a function of
the lowest mass ($m_1$ or $m_3$) for the 3+1 case
when parameters are varied randomly.]{The values of $y_m$ as a function of
the lowest mass when parameters are varied randomly.}
\label{yrandom} 
\end{center}
\end{figure}

The results presented above are obtained
by varying the known oscillation parameters 
distributed normally around their best-fit and with a
width given by the 1$\sigma$ range of the parameters.
There is a finite probability of getting the points in the
3$\sigma$ range of this Gaussian distribution although more
points are selected near the best-fit values.
However, note that  for some of the parameters
the 3$\sigma$ range obtained in this procedure
is different from the $3\sigma$ range of the global fits. 
Thus, our results may change if we vary the parameters randomly in their 
3$\sigma$ range as we have seen in the 3 generation case. 
In Fig. \ref{yrandom} we show the allowed values of $y_m$ as a function 
of the lowest mass for the case where all the parameters are varied 
randomly in their 3$\sigma$ range. 
We find that lower values of the smallest mass get disfavored by this method. 
The main reason for this is that if we use the Gaussian method then the 
allowed 3$\sigma$ range of the mixing angle $s_{14}^2$ is 
from (0.002 - 0.048)  while that of $s_{24}^2$ is from (0.001 - 0.06). 
Thus, smaller values of $s_{14}^2$ and $s_{24}^2$  are possible which
helps in achieving cancellation conditions for smaller values of
$m_1$ or $m_3$.
But if the parameters are varied randomly in the 3$\sigma$ range 
then such smaller values 
of the angles  are not allowed and consequently 
no allowed points are obtained for smaller values of masses. 
In particular, we obtain 
$m_1$ (NH) or $m_3$ (IH) $>$ 0.01 eV  in all the textures. 
However, 
main conclusions presented in Table \ref{2results}
regarding the nature of the allowed mass spectrum for the 3+1 scenario remain
unchanged though the allowed parameter space gets reduced.
Specially for the 
A, E and C classes very few points get allowed.
Fully hierarchical neutrinos ($m_1 < m_2 < m_3$) are not 
possible in any of the textures. Textures belonging to 
the B and F classes give more points in the 
QD regime. D class allows partial NH or IH. 

%% file: chap4_matrix.tex
\section{Analysis of One-Zero Textures}
\label{one_zero}

\subsection{Formalism}

For the analysis of the one-zero textures of $M_{\nu}$, we define the ratio of the mass squared differences $\xi$ and $\zeta$ as
\begin{eqnarray}
 \xi= \frac{\Delta_{41}}{\Delta_{32}}~~({\rm NH}) ~~ \text{or} ~~ \frac{\Delta_{43}}{\Delta_{31}}~~({\rm IH}), ~~
\label{xi}
\end{eqnarray}
\begin{eqnarray}
 \zeta =\frac{\Delta_{21}}{\Delta_{32}}~~({\rm NH}) ~~ {\rm or} ~~
\frac{\Delta_{21}}{\Delta_{31}}~~({\rm IH}).
\label{zeta}
\end{eqnarray}
In the extreme cases and using $\zeta \ll 1$, these masses can be written in terms of $\xi$ and $\zeta$ as \\
\be
SNH: |m_4| \approx \sqrt{ \Delta_{32} \xi} \gg |m_3|\approx \sqrt{(1 + \zeta)\Delta_{32}} \approx \sqrt{\Delta_{32}} \gg |m_2| \approx \sqrt{\Delta_{32} \zeta} \gg|m_1|,
\label{xnh}
\ee

\be
SIH: |m_4|\approx \sqrt{\Delta_{31} \xi} \gg|m_2|\approx \sqrt{(1 + \zeta)\Delta_{31}} \approx \sqrt{\Delta_{31}} \approx |m_1| \gg|m_3|,
\label{xih}
\ee

\be
SQD:~~|m_4|\gg|m_1|\approx|m_2|\approx|m_3|\approx m_0.
\label{qd}
\ee

\subsection{Neutrino Mass Matrix Elements}
 
In this section we study the implication of the condition of vanishing
$m_{\alpha \beta}$ for the 3+1 scenario,
where $\alpha, \beta = e, \mu,\tau, s$.
Since $m_{\alpha \beta}$ is complex the above condition implies
both real and imaginary parts are zero.
Therefore to study the one-zero textures we consider $|m_{\alpha \beta}|=0$.  
The best-fit values and the 3$\sigma$ ranges of the sterile neutrino parameters which we have used in this part our analysis
are given in Table \ref{parameters}
where we also
present the mass ratios $\zeta$ and  $\xi$
which would be useful in our analysis.
For the oscillation parameters involving the active neutrinos we use the global analysis results of Ref. \cite{global_fogli_old} 
which we also used in our analysis of two-zero textures (presented in Table \ref{Table:parameters}).
\begin{table}[ht!]
\begin{center}
\begin{tabular}{lccc}
\hline
\hline
Parameter &  Best Fit values & $3\sigma$ range \\
\hline
$ \Delta_{{\rm LSND}}~\mathrm{eV}^2$  & 1.62 & 0.7 -- 2.5 \\
\hline
$ \sin^2\theta_{14} $ & 0.03 & 0.01 -- 0.06 \\
\hline
$ \sin^2\theta_{24} $  & 0.01 & 0.002 -- 0.04 \\
\hline
$ \sin^2\theta_{34} $ &  --  & $ <$ 0.18  \\
\hline
$\zeta /10^{-2}$ (NH) & --   & 2.7 -- 3.7 \\
$\zeta /10^{-2}$ (IH) & --   & 2.7 -- 3.8 \\
\hline
$\xi /10^3$ (NH) & --   &  0.27--1.14 \\
$\xi /10^3$ (IH) & --   &  0.27-- 1.15 \\
\hline
\end{tabular}
\caption[Oscillation parameters \cite{global_fogli_old,Archidiacono:2012ri,schwetz} used in one-zero texture analysis in 3+1 scenario.]
{The constraints on
sterile neutrino parameters are from \cite{Archidiacono:2012ri,schwetz},
 where $\Delta_{{\rm LSND}} = \Delta_{41}({\rm NH})$ or $\Delta_{43}({\rm IH})$.
Also given are the $3\sigma$ ranges of the mass ratios
$\zeta $ and  $\xi$.
}
\label{parameters}
\end{center}
\end{table}
In this part of the analysis we have varied all the parameters randomly between their allowed ranges,
three Dirac phases in the range 0 to $2 \pi$ and the three Majorana phases from 0 to $\pi$.

\subsubsection{The Mass Matrix Element $m_{ee}$ }

The matrix element $m_{ee}$ in the 3+1 scenario is given as,
\be
m_{ee}=m_1 c_{14}^2 c_{13}^2c_{12}^2+m_2 s_{12}^2c_{14}^2c_{13}^2e^{2i\alpha}+m_3 s_{13}^2 c_{14}^2
e^{2i\beta}+m_4s_{14}^2 e^{2i\gamma}.
\label{mee}
\ee
This is of the form
\be \label{meenh}
m_{ee}= c_{14}^2 (m_{ee})_{3\nu}+e^{2i\gamma}s_{14}^2 m_4,
\ee
where $(m_{ee})_{3\nu}$  corresponds to the matrix element in the
3 active neutrino case.
The  contribution of the sterile neutrino to the element $m_{ee}$
depends on the mass $m_4$ and  the
active-sterile mixing angle $\theta_{14}$.
Of all the mass matrix element $m_{ee}$
has the simplest form because of the chosen parametrization and
can be understood quite well.
Using approximation in Eq. \ref{xnh} for the case of extreme hierarchy
one can write this for NH as,
\be \label{meenh}
m_{ee} \approx c_{14}^2 (m_{ee})_{3\nu}+e^{2i\gamma}s_{14}^2\sqrt{\Delta_{41}},
\ee
where $(m_{ee})_{3\nu}\approx \sqrt{\Delta_{32}}(e^{2i\alpha}c_{13}^2s_{12}^2 \sqrt{\zeta}+s_{13}^2e^{2i\beta})$ and $\zeta$ is defined in
Eq. \ref{zeta}.
\begin{figure}[h] 
\begin{center}
\includegraphics[width=0.33\textwidth,angle=270]{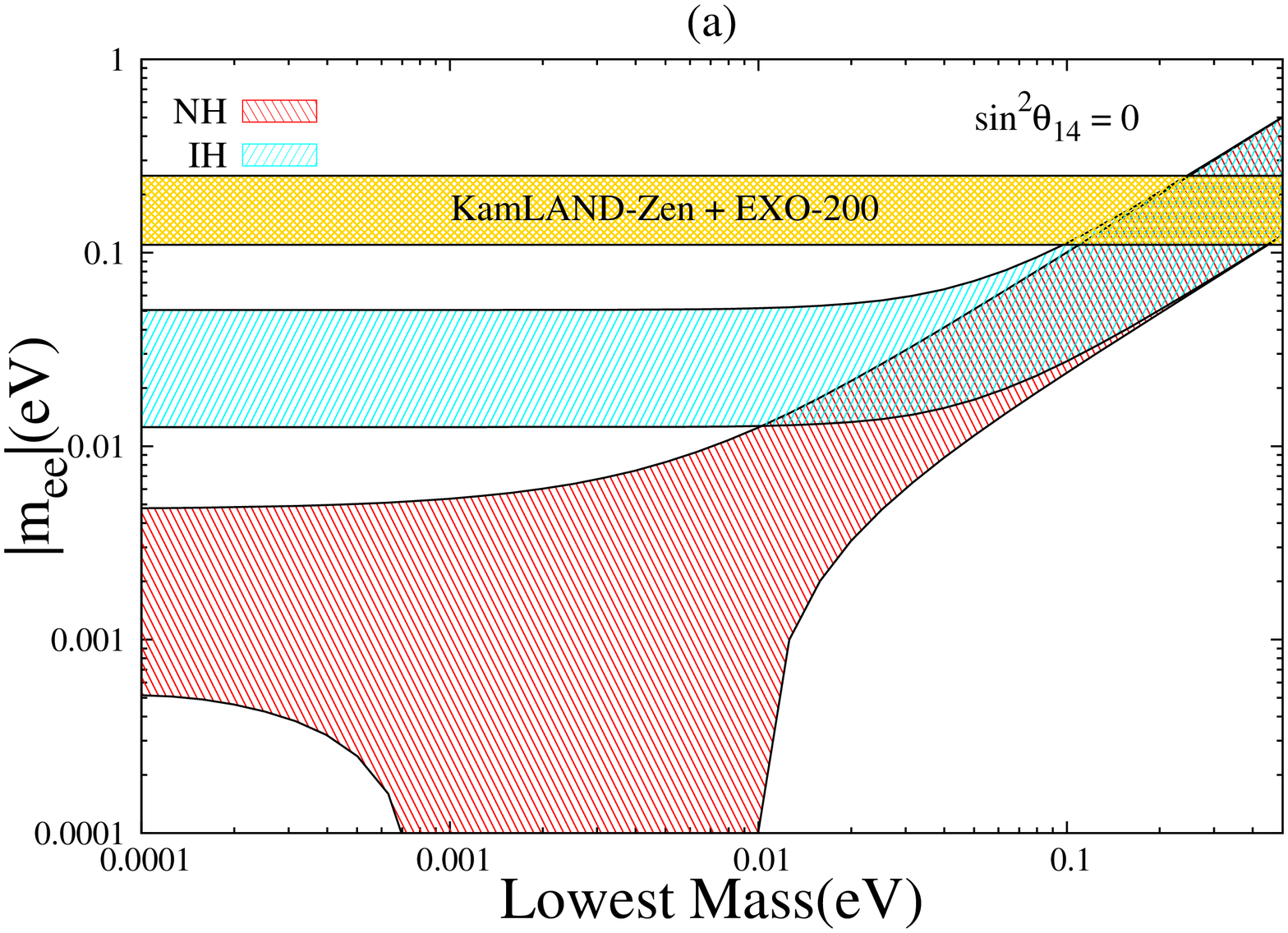}
\includegraphics[width=0.33\textwidth,angle=270]{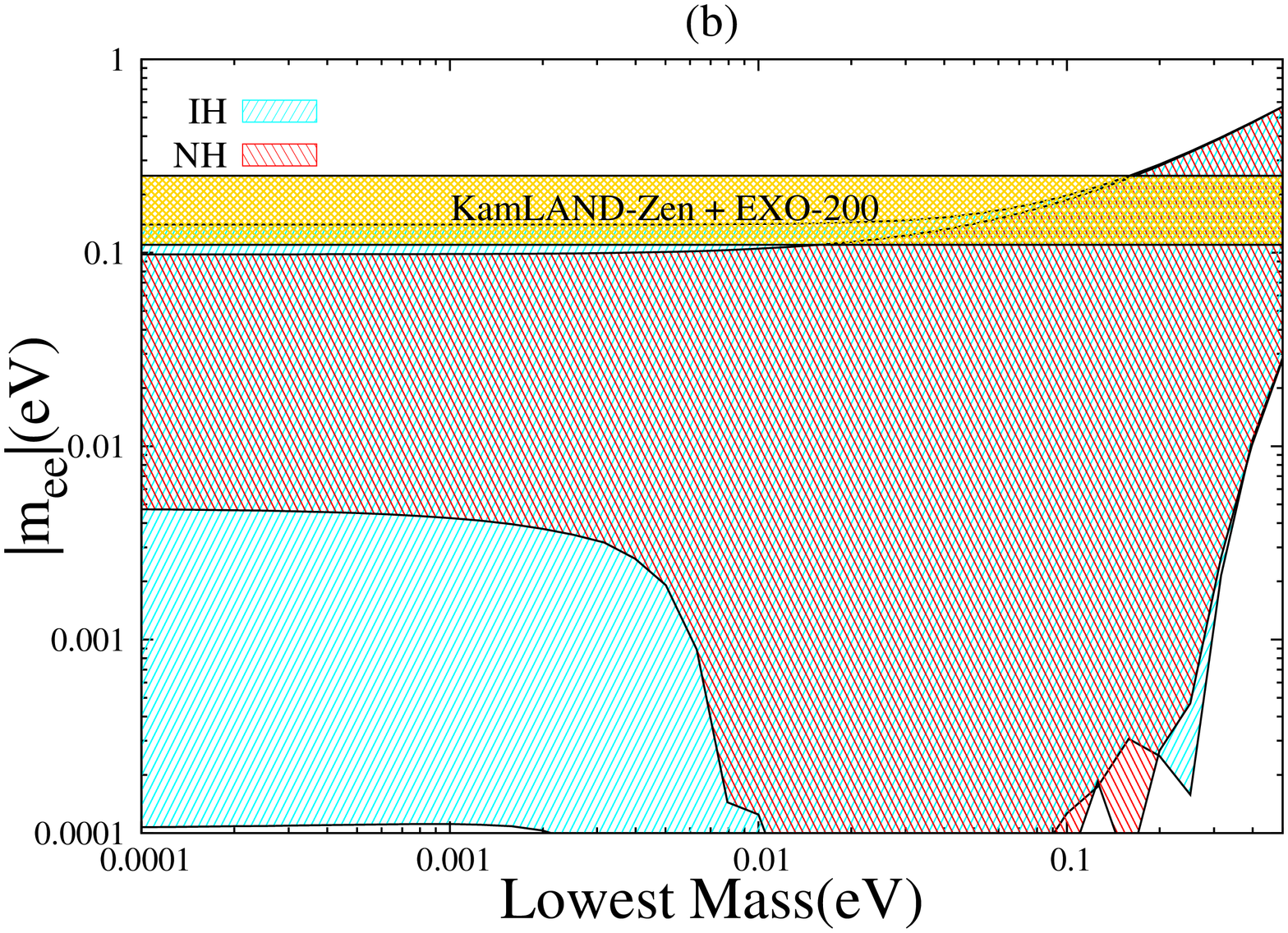} \\
\includegraphics[width=0.33\textwidth,angle=270]{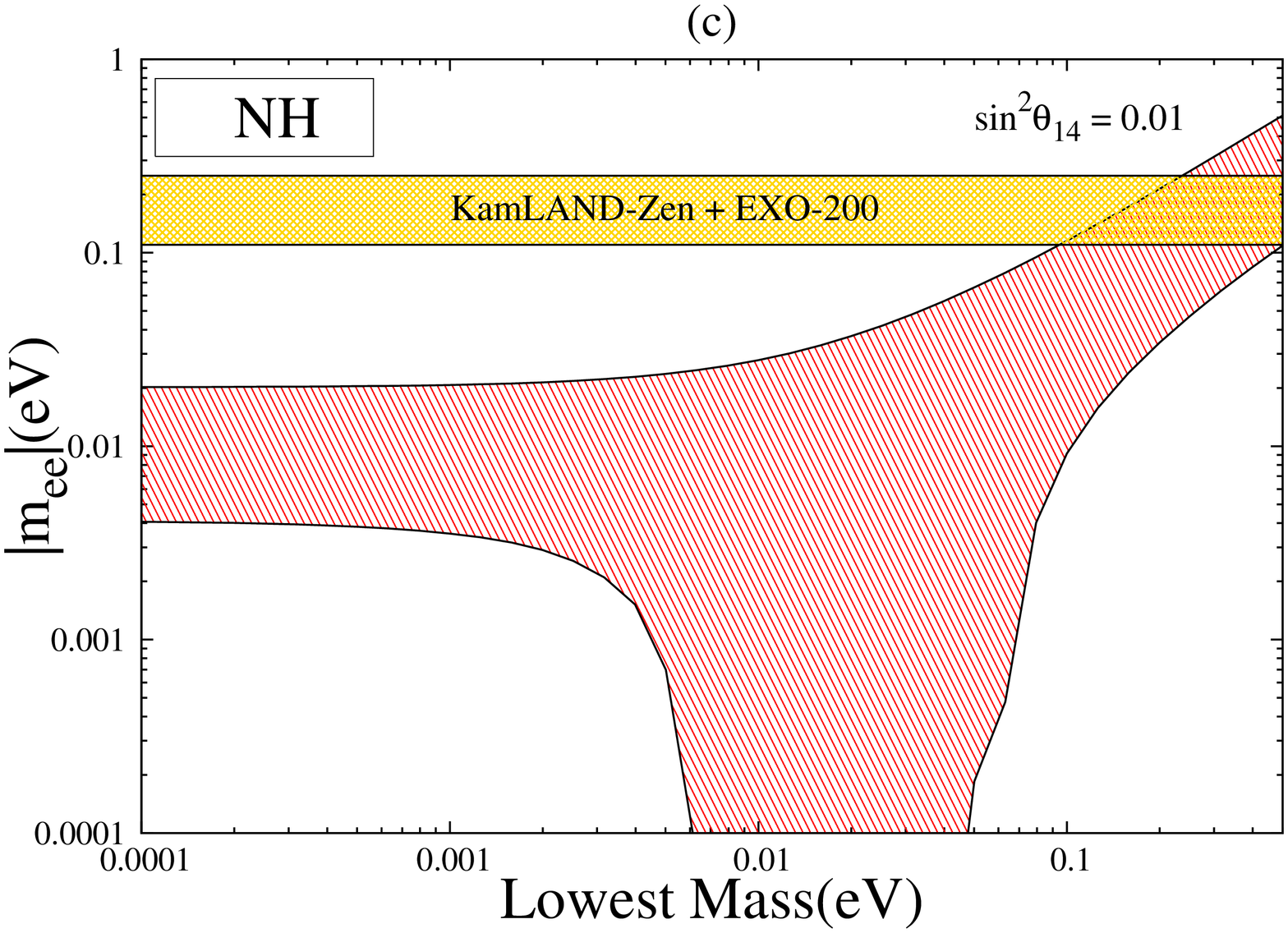}
\includegraphics[width=0.33\textwidth,angle=270]{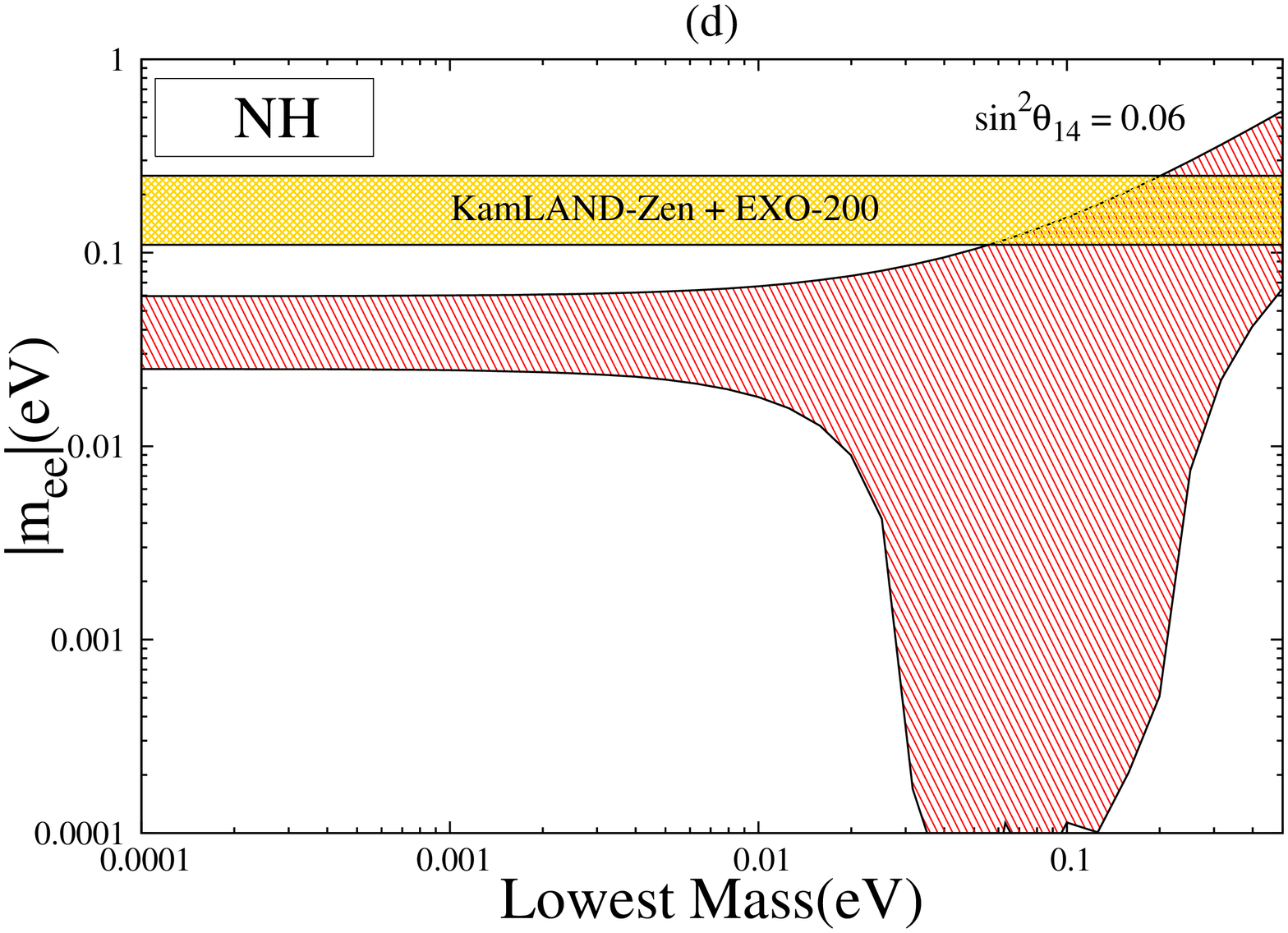}
\caption[Plots of $|m_{ee}|$ versus the lowest mass.]{Plot of $|m_{ee}|$ versus the lowest mass. The panel (a)
corresponds to the three generation case while panel (b) is for 3+1 case. In panel (b)
all the mixing angles are varied in their 3$\sigma$ range and the Majorana CP violating phases are varied in their full range (0-$\pi$).
The panel (c) and (d) are for specific values of
$\theta_{14}$ with all other parameters covering their full range.}
\label{meefig}
\end{center}
\end{figure}
The modulus of $m_{ee}$ is the effective mass that can be extracted from
half-life measurements in neutrinoless double beta decay.
In Fig. \ref{meefig} we plot the effective mass  as a function of the smallest
mass by varying $\theta_{14}$ in its
complete 3$\sigma$ range as well as
for specific values of the mixing
angle $\theta_{14}$.
The Majorana phases are varied randomly
in the range  0 to $\pi$ in all the plots.
The first panel is for $\theta_{14}=0$ i.e., the three generation case.
It is seen that for present values of the oscillation parameters
the cancellation condition is not satisfied for $m_1 \rightarrow 0$ for NH.
However, as one increases $m_1$, complete cancellation can be achieved.
For IH the complete cancellation is never possible.
These results change when we include the sterile contribution as is
evident from the panel (b) in  Fig. \ref{meefig} which shows the effective mass
for NH and IH by varying all the parameters in their full 3$\sigma$
allowed range.
The behaviour can be understood from the expressions of $|m_{ee}|$ in
various limiting cases.
For NH, in the hierarchical limit of  $m_1 \rightarrow 0$
the major contributor will be the additional term due to
the sterile neutrinos because of higher value of $m_4$.
Complete cancellation is only possible for
smaller values of $\theta_{14}$ so that this contribution
is suppressed.
The typical value of $\theta_{14}$ required for cancellation can be
obtained by putting
$\alpha =\beta=0$ (which would maximize the three neutrino contribution)
and $\gamma = \pi/2$, as
\be
\tan^2\theta_{14} \approx \frac{(\sqrt{\zeta} c_{13}^2s_{12}^2+s_{13}^2)}{\sqrt{\xi}} \approx 10^{-3},
\ee
which lies outside the allowed range of $\theta_{14}$.
As we increase $m_1$, $(m_{ee})_{3\nu}$
increases and can be of the same order of magnitude of the
sterile term.  Hence one can get cancellation regions.
The cancellation is mainly controlled by the value of $\theta_{14}$. For higher values of $s_{14}^2$
one needs a higher value of $m_1$ for cancellation to occur. This correlation between $m_1$ and $\theta_{14}$ is brought
out by the panels (c) and (d) in Fig. \ref{meefig}.

For IH case, in the limit of vanishing $m_3$ using approximation in Eq. \ref{xih},
$m_{ee}$ in a 3+1 scenario can be written as
\be
|m_{ee}| \approx |c_{14}^2 c_{13}^2 \sqrt{\Delta_{31}}(c_{12}^2+s_{12}^2 e^{2i\alpha})+\sqrt{\Delta_{43}}s_{14}^2 e^{2i\gamma}|.
\ee
The maximum value of this  is achieved for $\alpha=\gamma=0$
which is slightly lower than that of NH in this limit.
The element vanishes in the limit $m_3\approx$ 0 eV when
$\alpha=0$  and $\gamma = \pi/2$ provided
\be \label{meeih}
\tan^2\theta_{14}  \approx \frac{c_{13}^2}{\sqrt{\xi}} \approx 0.05.
\ee
This
is well within the allowed range.
This behaviour is in stark contrast to that in the 3 neutrino
case \cite{Barry:2011wb} .
There is no significant change in this behaviour as the
smallest mass $m_3$ is increased since this contribution
is suppressed by the $s_{13}^2$ term and the dominant
contribution to  $(m_{ee})_{3\nu}$ comes from the first two
terms in Eq. \ref{mee}. Therefore in this case we do not
observe any correlation between $m_3$ and $s_{14}^2$.

While moving towards the quasi-degenerate regime of
$m_1 \approx m_2 \approx m_3$ we find that effective mass can still be zero.
However, when the
lightest mass approaches a larger value $\sim 0.3$ eV we need very large values of active sterile
mixing angle $\theta_{14}$, outside the allowed range, for cancellation.
Hence the effective mass cannot vanish for such
values of masses.

Also shown is the current limit on effective mass from combined
KamLAND-Zen and  EXO 200
results on the half-life of $0\nu\beta\beta$ in
$^{136}$Xe \cite{Gando:2012zm,Auger:2012ar}.
When translated in terms of effective mass this corresponds to the bound
$|m_{ee}| < 0.11 - 0.24$ eV including nuclear matrix element uncertainties.
For the three generation case, the hierarchical neutrinos
cannot saturate this bound.
But in the 3+1 scenario this bound can be reached even for
very small values of $m_3$ for IH and for some parameter
values it can even exceed the current limit. Thus from the present limits
on neutrinoless double beta decay searches a part of the parameter space
for smaller values of $m_3$ can be disfavoured for IH.
For NH,  the KamLAND-Zen + EXO 200
combined bound is reached for $m_1 = 0.02$ eV
and again  some part of the parameter space can be disfavoured by this bound.

\subsubsection{The Mass Matrix Element $m_{e\mu}$ }
The mass matrix element $m_{e\mu}$ in the presence of extra sterile neutrino is given as
\bea
m_{e \mu}&=&c_{14}(e^{i (\delta _{14}-\delta _{24}+2 \gamma)}m_4s_{14}s_{24}+e^{i(\delta_{13}+2 \beta)} m_3s_{13}(c_{13}c_{24}s_{23}-e^{i
(\delta_{14}-\delta_{13}-\delta_{24})}\\ \nonumber && s_{13}s_{14}s_{24})
+c_{12}c_{13}m_1(-c_{23}c_{24}s_{12}+c_{12}(-e^{i \delta_{13}}c_{24}s_{13}s_{23}-e^{i (\delta_{14}-\delta_{24})}c_{13}s_{14}s_{24})) \\
\nonumber &+& e^{2i \alpha}m_2c_{13}s_{12} (c_{12}c_{23}c_{24}+s_{12}(-e^{i \delta_{13}}c_{24}s_{13}s_{23}-e^{i (\delta_{14}-\delta_{24})}c_{13}s_{14}s_{24}))).
\eea
Unlike $m_{ee}$ here the expression is complicated and an analytic understanding is difficult from the full expression.
The expression for $m_{e \mu}$ in the limit of vanishing active sterile mixing angle $\theta_{24}$ becomes
\bea
m_{e \mu}= c_{14}(m_{e \mu})_{3 \nu}. \nonumber
\eea
Since the active sterile mixing is small, in order to simplify
these expressions we introduce a quantity $\lambda\equiv$0.2
and define these small angles to be of the form $a\lambda$. Thus a
systematic expansion in terms of $\lambda$ can be done.
For sterile mixing angle
\bea \nonumber \label{chi1}
{\mathrm{sin}} \theta_{14} \approx \theta_{14} \equiv \chi_{14}\lambda, \\
{\mathrm{sin}} \theta_{24} \approx \theta_{24} \equiv \chi_{24}\lambda,
\eea
and the reactor mixing angle as
\be \label{chi2}
\sin \theta_{13} \approx \theta_{13} \equiv \chi_{13}\lambda.
\ee
Here $\chi_{ij}$ are parameters of $\mathcal{O}$(1) and their $3 \sigma$ range
from the current constraint on the mixing angles is given by
\begin{eqnarray}
\chi_{13} &=& 0.65 - 0.9, \\ \nonumber
\chi_{14} &=& 0.5 - 1.2, \\ \nonumber
\chi_{24} &=& 0.25 - 1.
\end{eqnarray}
 Note that for the sterile mixing angle $\theta_{34}$ we do not adopt the above approximation because this angle can be large compared to other two sterile mixing angles
and hence the small parameter approximation will not be valid.

\begin{figure}
\begin{center}
\includegraphics[width=0.33\textwidth,angle=270]{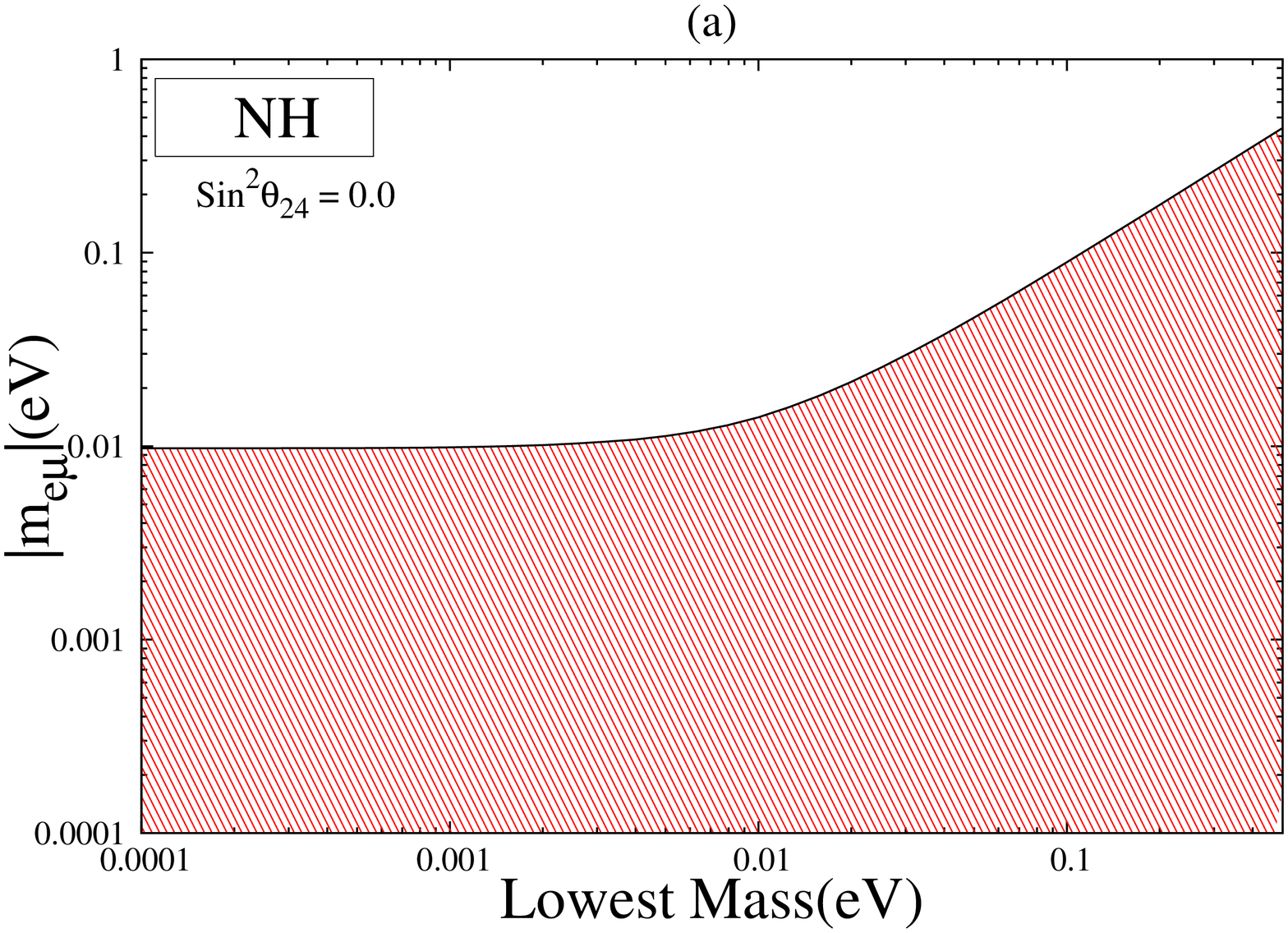}
\includegraphics[width=0.33\textwidth,angle=270]{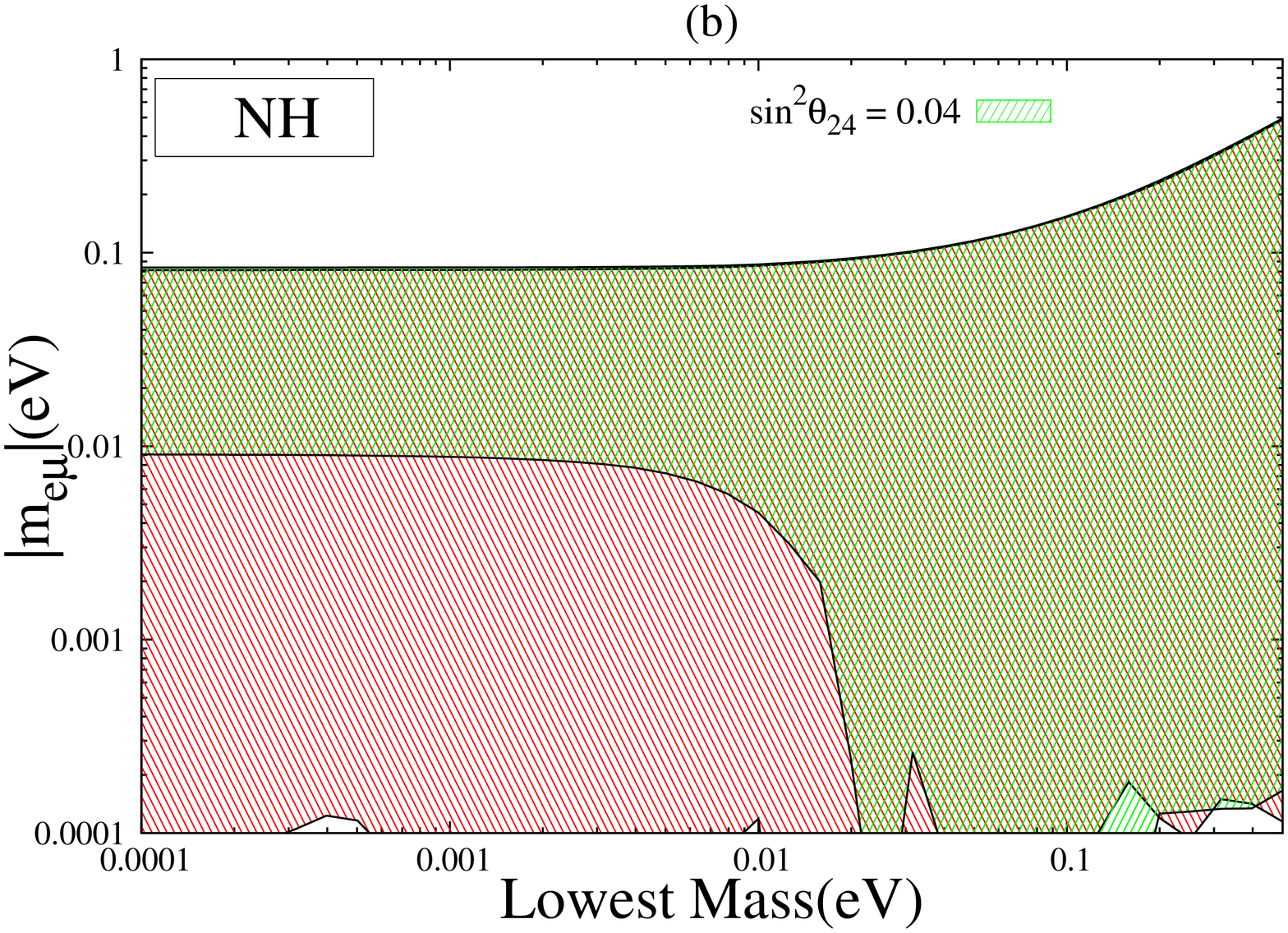}
\caption[Plots of $|m_{e\mu}|$ vs $m_1$ for normal hierarchy.]{Plots of $|m_{e\mu}|$ as a function of
the lowest mass $m_1$ for NH. Panel (a) correspond to the three generation case while (b) (red/light region) is for 3+1 case
and also for $s_{24}^2 = 0.04$ (green/dark region). All the parameters
are varied in their full 3$\sigma$ allowed range, the CP violating Dirac phases are varied from 0 to $2\pi$ and the Majorana phases are varied from 0 to $\pi$ unless
otherwise stated.}
\label{fig4}
\end{center}
\end{figure}

Using the approximation in Eqs. \ref{xnh}, \ref{chi1} and \ref{chi2} we get the expression for $|m_{e\mu}|$ for normal hierarchy as
\bea
|m_{e\mu}| &\approx& |\sqrt{\Delta_{32}}\{\sqrt{\zeta}s_{12}c_{12}c_{23}e^{2i\alpha}+e^{i\delta_{13}}(e^{2i \beta}-e^{2i \alpha}\sqrt{\zeta}s_{12}^2)s_{23} \lambda \chi_{13}\\ \nonumber
&+&\lambda^2
e^{i(\delta_{14}-\delta_{24})}(e^{2i\gamma}\sqrt{\xi} -e^{2i\alpha}\sqrt{\zeta} s_{12}^2)\chi_{14}\chi_{24}\}|.
\eea
To see the order of magnitude of the different
terms we choose vanishing Majorana phases while Dirac CP phases are taken
as $\pi$. The mass matrix element  $m_{e \mu}$ vanishes when

\be \label{mem}
 \sqrt{\zeta}s_{12}c_{12}c_{23}-(1-\sqrt{\zeta}s_{12}^2)s_{23}\lambda \chi_{13}+ \lambda^2(\sqrt{\xi}-\sqrt{\zeta}s_{12}^2)\chi_{14}\chi_{24}=0.\\
\ee
The three generation limit is recovered for $s_{24}^2 = 0$ and in panel (a) of Fig. \ref{fig4} we show $|m_{e\mu}|$ as a function of  $m_1$ of this case, for NH.
Panel (b) (red/light region) of Fig. 4 shows  $|m_{e\mu}|$ for the 3+1 case, with all parameters
varied randomly within their $3 \sigma$ range.
The figures show that $|m_{e\mu}|=0$ can be achieved over the
whole range of the smallest mass for both 3 and 3+1 cases. However, we find that in the
hierarchical limit cancellation is not achieved for
large values of $\theta_{24}$, since
in that case the third term of Eq. \ref{mem} will be of the $\mathcal{O}$ (10$^{-1}$) compared to the leading order term which is of the $\mathcal{O}$ (10$^{-2}$)
and hence there will be no cancellation of these terms.
This can be seen from panel (b) (green/dark region) of Fig. \ref{fig4} for  $s_{24}^2=0.04$.
In the QD limit the contribution from the active terms are large enough to
cancel the sterile contribution and thus $|m_{e\mu}| = 0$ can be achieved.


\begin{figure}[ht!]
\begin{center}
\includegraphics[width=0.33\textwidth,angle=270]{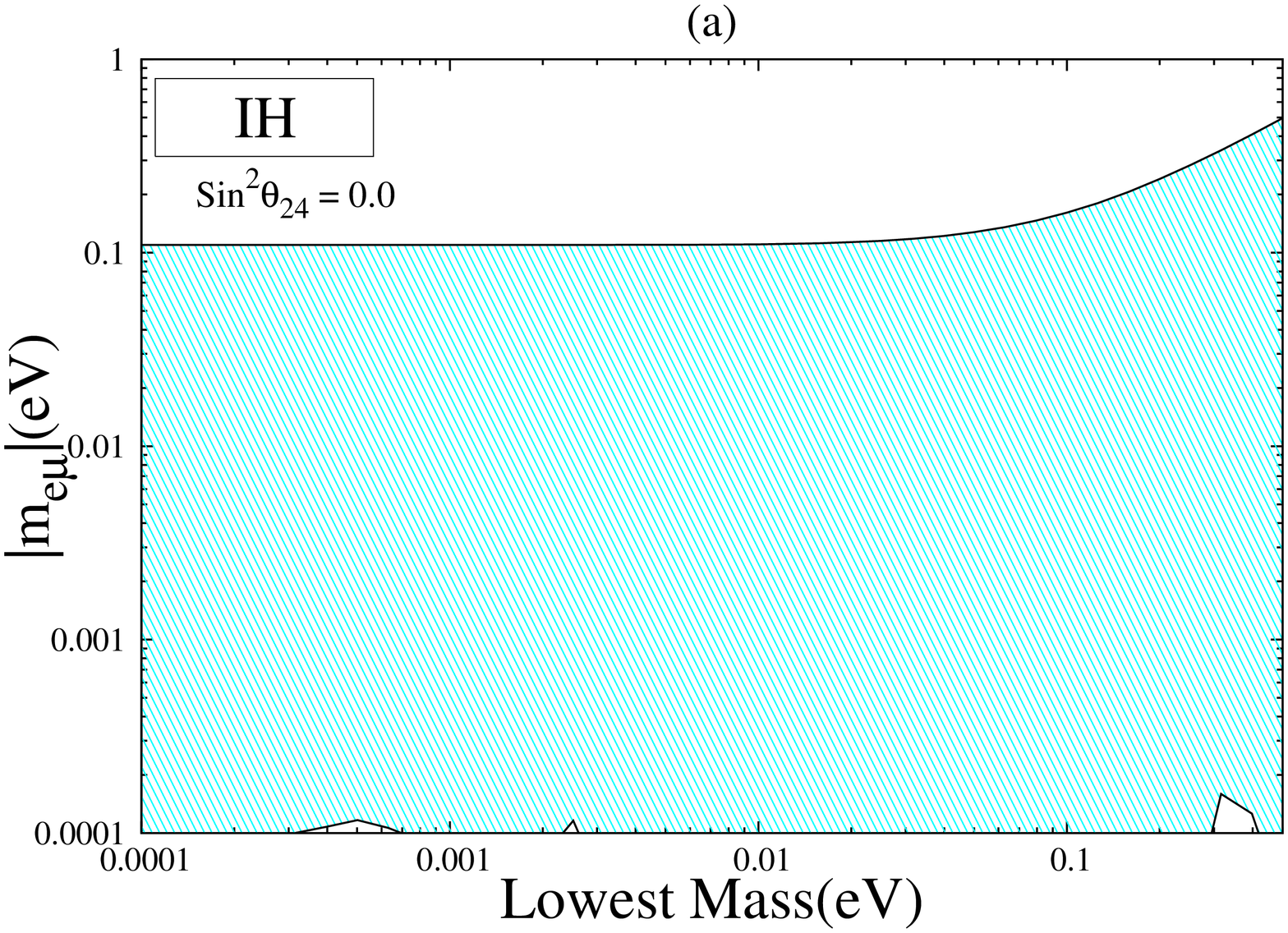}
\includegraphics[width=0.33\textwidth,angle=270]{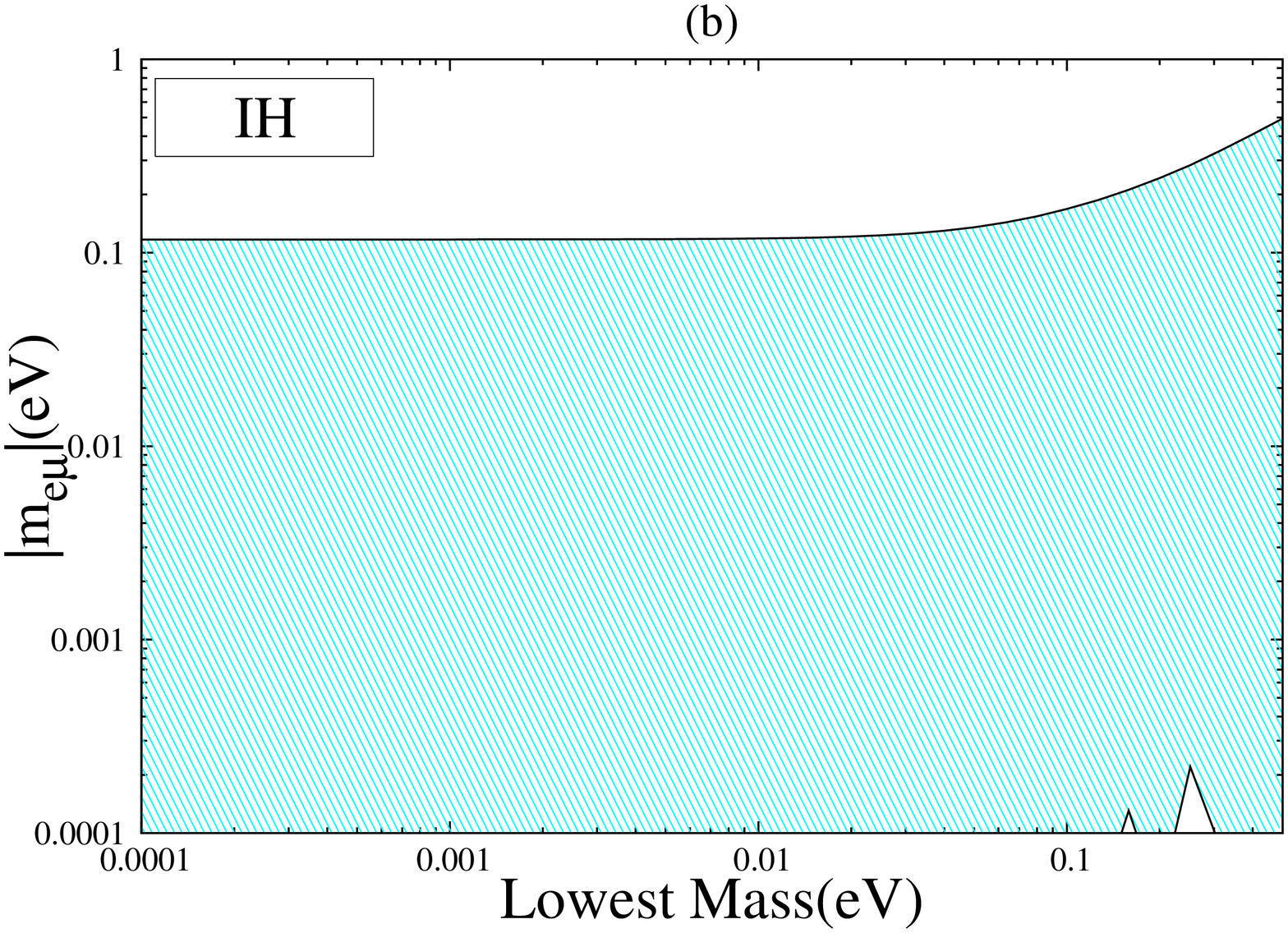}\\
\includegraphics[width=0.33\textwidth,angle=270]{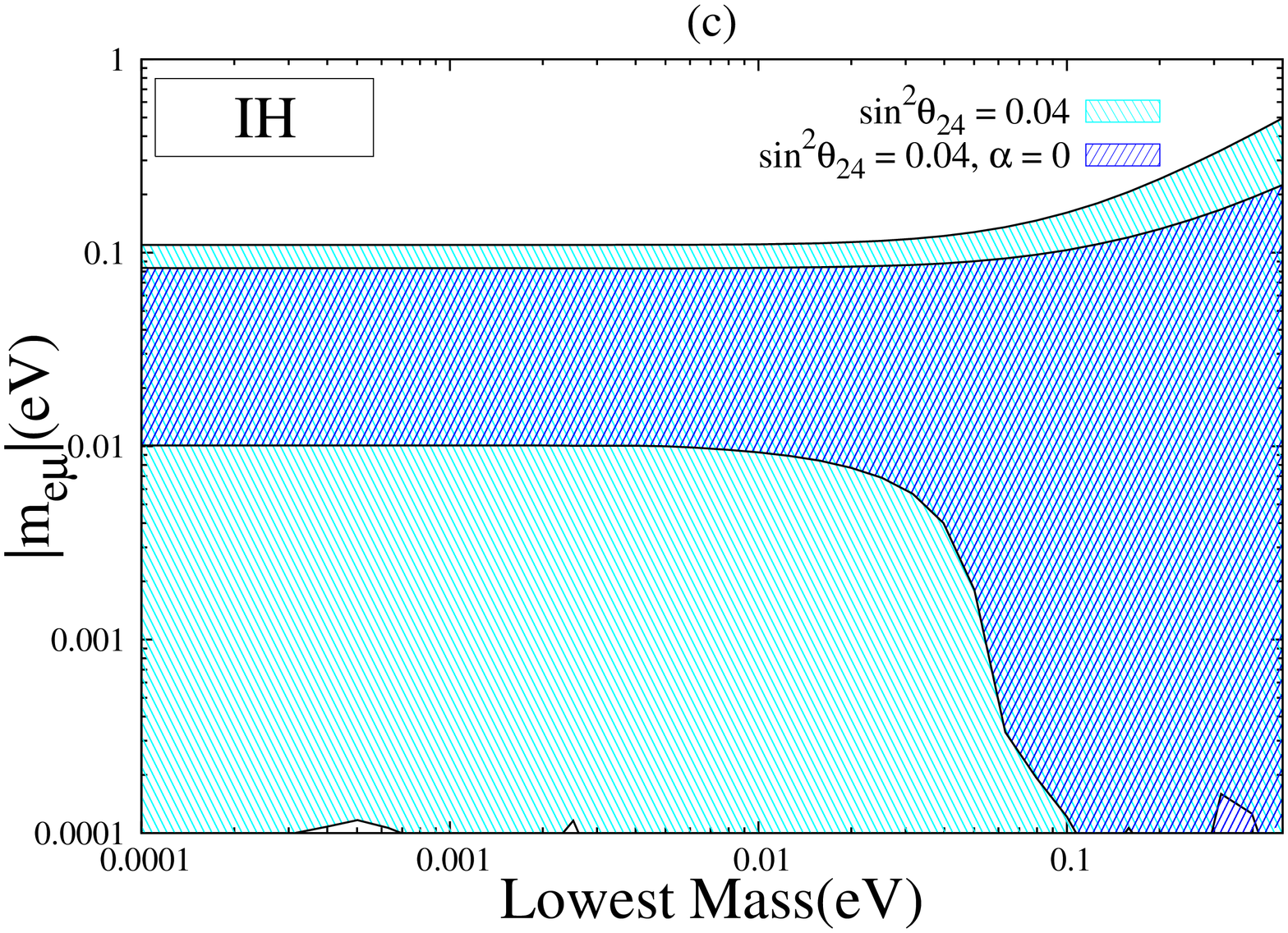}
\includegraphics[width=0.33\textwidth,angle=270]{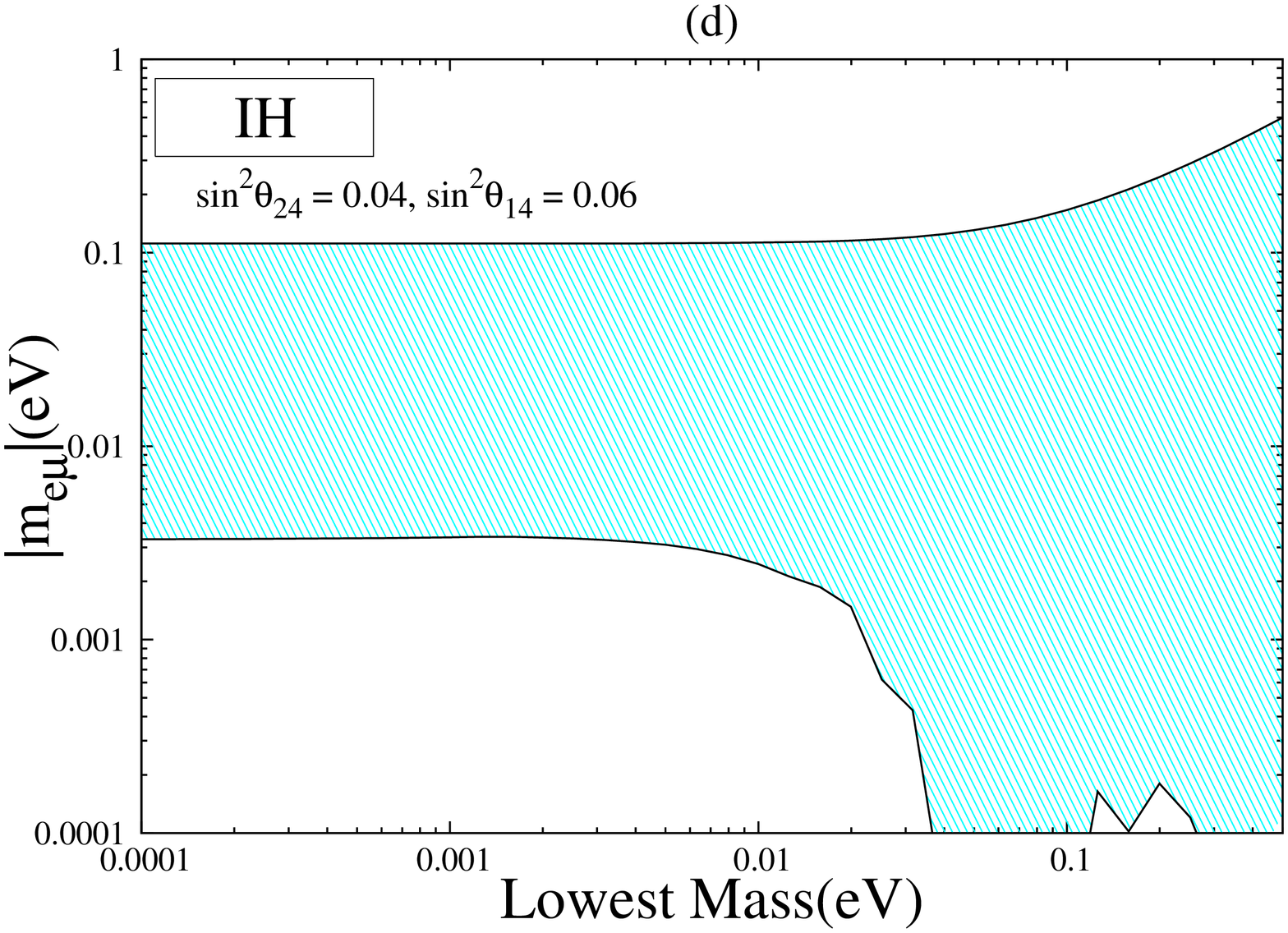}\\
\includegraphics[width=0.33\textwidth,angle=270]{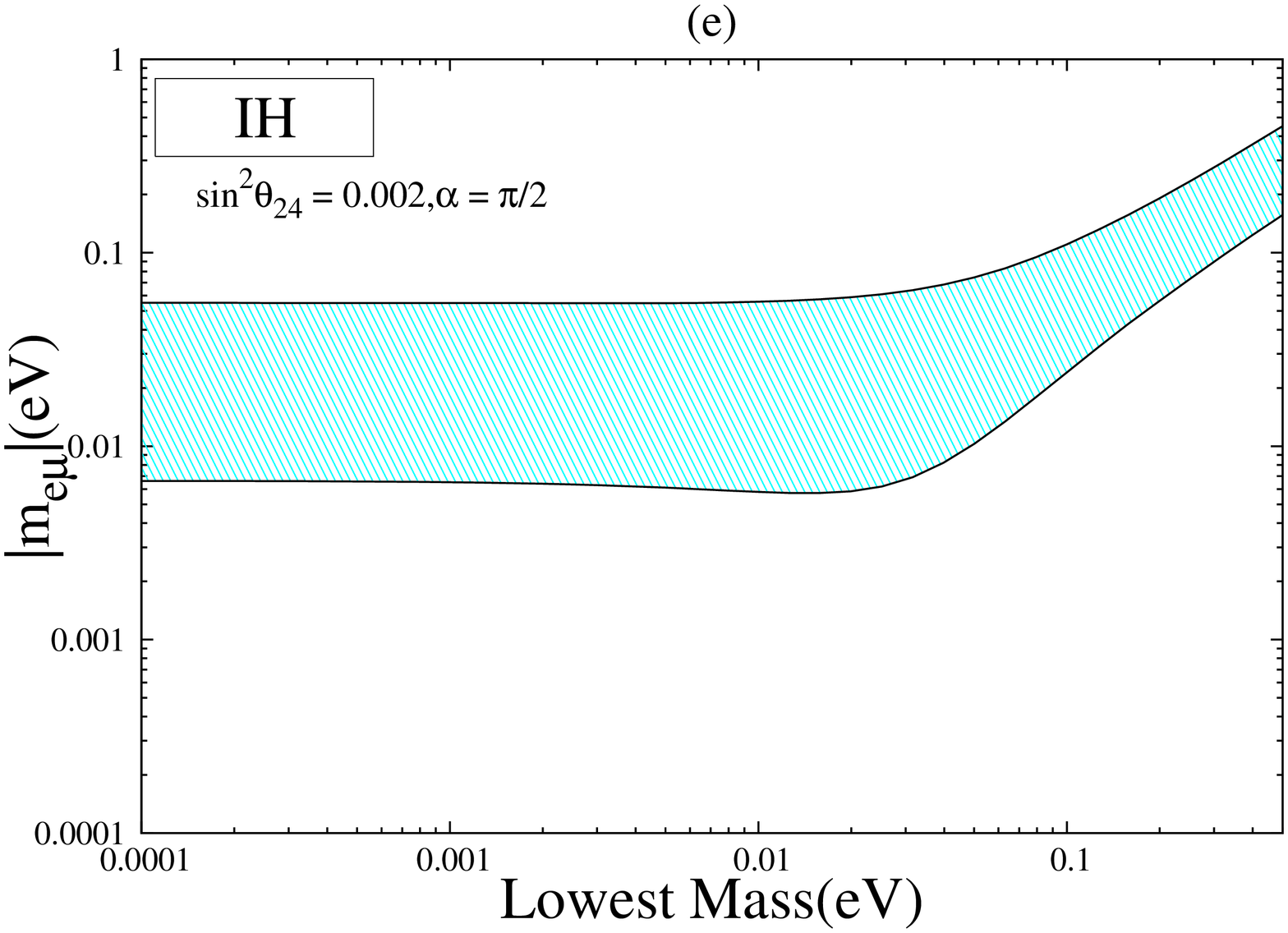}
\includegraphics[width=0.33\textwidth,angle=270]{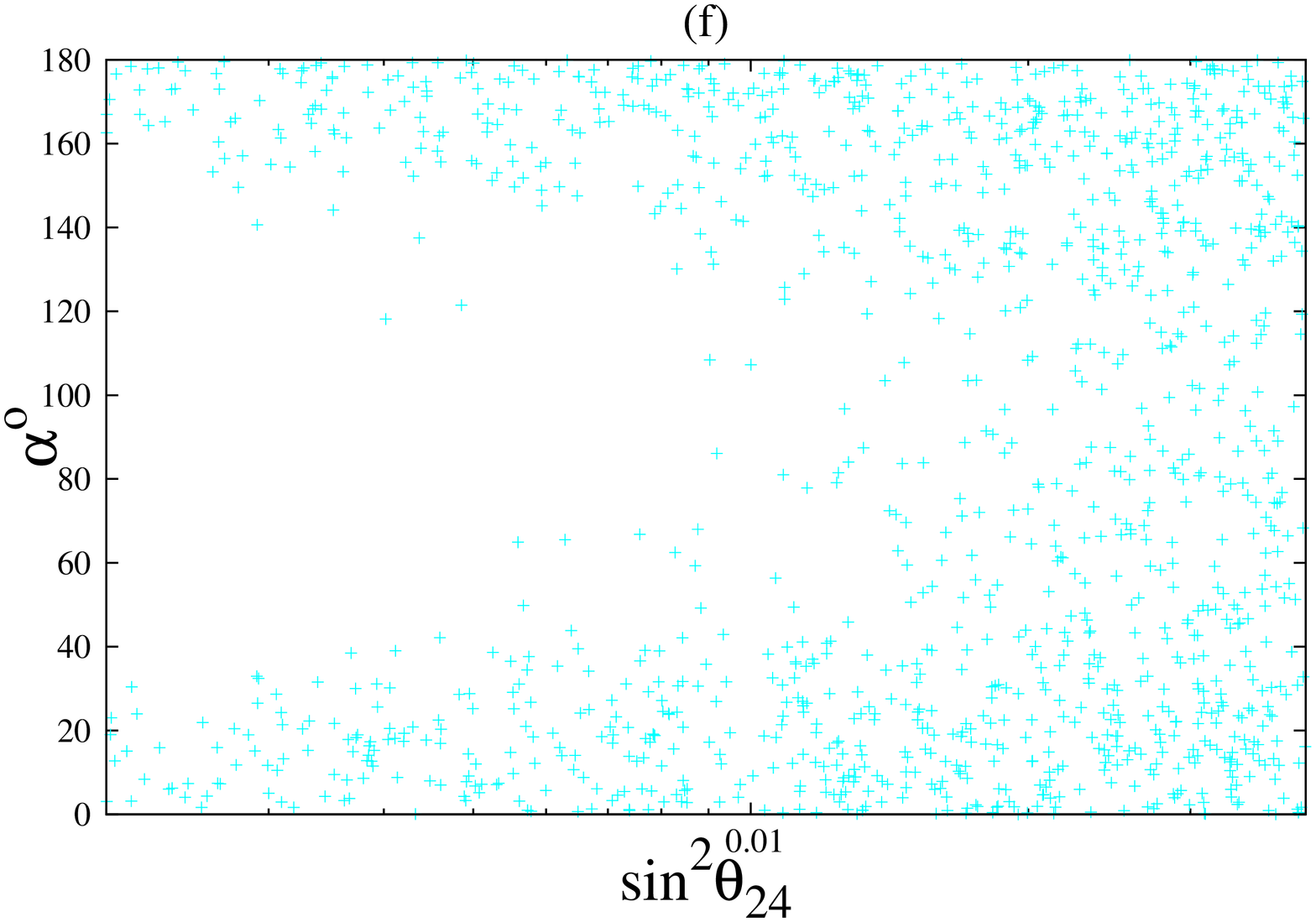}
\caption[Plots of $|m_{e\mu}|$ vs $m_3$ for inverted hierarchy.]{Plots of $|m_{e\mu}|$ vs $m_3$ for inverted hierarchy for
(a) three generation case (b) 3+1 case with all parameters varied
randomly in their full range. Panel (c), (d) and (e) are for specific values of $s_{24}^2$ and
$\alpha$. The panel (f) shows the correlation between $\alpha$ and $s_{24}^2$ when all other parameters are randomly varied.}
\label{fig5}
\end{center}
\end{figure}

For IH using the approximation Eq. \ref{xih} for the hierarchical
limit  we get the expression
\bea  \label{memih}
|m_{e \mu}| &\approx& |\sqrt{\Delta_{31}}\{c_{12}s_{12}c_{23}(e^{2i\alpha}-1)-e^{i\delta_{13}}(c_{12}^2+s_{12}^2e^{2i\alpha})s_{23}\chi_{13}\lambda \\ \nonumber
&-& e^{i(\delta_{14}-\delta_{24})}\lambda^2
\chi_{14}\chi_{24}(c_{12}^2 -e^{2i\gamma}\sqrt{\xi}+e^{2i\alpha}s_{12}^2)\}|.
\eea
To see the order of magnitude of the
various terms we consider the case when Majorana phases vanish and the
Dirac phases assume  the value $\pi$. Then we get for vanishing $m_{e\mu}$,
\bea
 s_{23}\lambda\chi_{13}-\lambda^2(1-\sqrt{\xi})\chi_{14}\chi_{24}=0.
\eea
In panel (a) of Fig. \ref{fig5} we display the plot of $|m_{e\mu}|$ with $m_3$ for the
3 generation scenario i.e., for $\sin^2\theta_{24} =0$ for IH.
In panel (b) we consider the 3+1 case with all the parameters varying
in their allowed range. Note that in the small
$m_3$ limit for $\alpha =0$ the leading order term vanishes.
For this case, for large active sterile mixing angle $\theta_{24}$, the
$\lambda^2$ term becomes large $\mathcal{O}$ (10$^{-1}$) and the cancellation with $\lambda$ term is not be possible.
When CP violating phase $\alpha$ is non zero, the leading order
term  can cancel the $\lambda^2$ term
even for large values of $s_{24}^2$.
These features are reflected in
panel (c) where we plot $|m_{e\mu}|$ for
$s_{24}^2 = 0.04$ and $ \alpha = 0$ (blue/dark region)
and by varying $\alpha$ in its full range (cyan/light region). As expected,
for $\alpha=0$, cancellation is not achieved for
smaller values of $m_3$.
Thus the condition $|m_{e\mu}|=0$ implies some correlation between
$m_3$ and $\alpha$ for IH.
Even if $\alpha$ is varied in its full range,
the absolute value of the
matrix element $|m_{e\mu}|$ can vanish only if the product $\chi_{14}\chi_{24}$ is small, i.e., $s_{14}^2$ and $s_{24}^2$ are simultaneously small .
This is because if they are large the $\lambda^2$ term becomes of the $\mathcal{O}$ (10$^{-1}$) and hence cancellation will not be possible.
This is seen from panel (d) where for
$s_{14}^2 = 0.06$ and $s_{24}^2 = 0.04$ the region where $m_3$ is small
gets disallowed.
Taking CP violating phase $\alpha=\pi/2$ makes the magnitude of
leading order term ($s_{12} c_{12}c_{23}\sqrt{\zeta}$) quite large and
smaller values of $\theta_{24}$ cannot give cancellation even for
large values of $m_3$ which can be seen from panel (e) of Fig. \ref{fig5}.
For the occurrence of cancellation $s_{24}^2$ has to be $\geq$ 0.01 for $\alpha=\pi/2$ as can be seen from panel (f) where we have
plotted the correlation between $\alpha$ and $s_{24}^2$ for $|m_{e\mu}| = 0$.

\subsubsection{The Mass Matrix Element $m_{e\tau}$}
The mass matrix element $m_{e\tau}$ in the presence of an extra sterile neutrino is given by
\bea
m_{e\tau}&=&c_{14}c_{24}e^{i(2\gamma + \delta_{14})}m_4s_{14}s_{34}+m_3c_{14}s_{13}e^{i(2\beta+\delta_{13})}(-c_{24}s_{13}s_{14}s_{34}
e^{i(\delta_{14}-\delta_{13})}\\ \nonumber
&+&c_{13}(c_{23}c_{34}-e^{i\delta_{24}}s_{23}s_{24}s_{34}))+m_2s_{12}c_{13}c_{14}e^{2i\alpha}(c_{12}(-c_{34}s_{23}-c_{23}s_{24}s_{34}e^{i\delta_{24}}) \\ \nonumber
&+&s_{12}(-c_{13}c_{24}s_{14}s_{34}e^{i\delta_{14}}-e^{i\delta_{13}}s_{13}(c_{23}c_{34}-e^{i\delta_{24}}s_{23}s_{24}s_{34}))) \\ \nonumber
&+&m_1c_{12}c_{13}c_{14}(-s_{12}(-c_{34}s_{23}-c_{23}s_{24}s_{34}e^{i\delta_{24}})+c_{12}(-c_{13}c_{24}s_{14}s_{34}e^{i\delta_{14}}
-e^{i\delta_{13}}s_{13}\\ \nonumber
&&(c_{23}c_{34}-e^{i\delta_{24}}s_{23}s_{24}s_{34}))).
\eea
As mentioned in the previous section, the elements $m_{e\tau}$ and $m_{e\mu}$ are related by
$\mu-\tau$   permutation symmetry and
it is found that in the limit of small $\theta_{24}$
the two active sterile mixing angles $\bar\theta_{24} \approx \theta_{34}$. (Eq. \ref{th24})
The same can be seen from Eq. \ref{th34} which gives
$\bar\theta_{34} \approx \theta_{24}$
for smaller values of the mixing angle $\theta_{34}$.
Thus, for these cases
the behaviour shown by $\theta_{24}$ in $m_{e\mu}$ ($m_{\mu\mu}$) is same
as shown by $\theta_{34}$ in $m_{e\tau}$ ($m_{\tau\tau}$).

\begin{figure}
\begin{center}
\includegraphics[width=0.33\textwidth,angle=270]{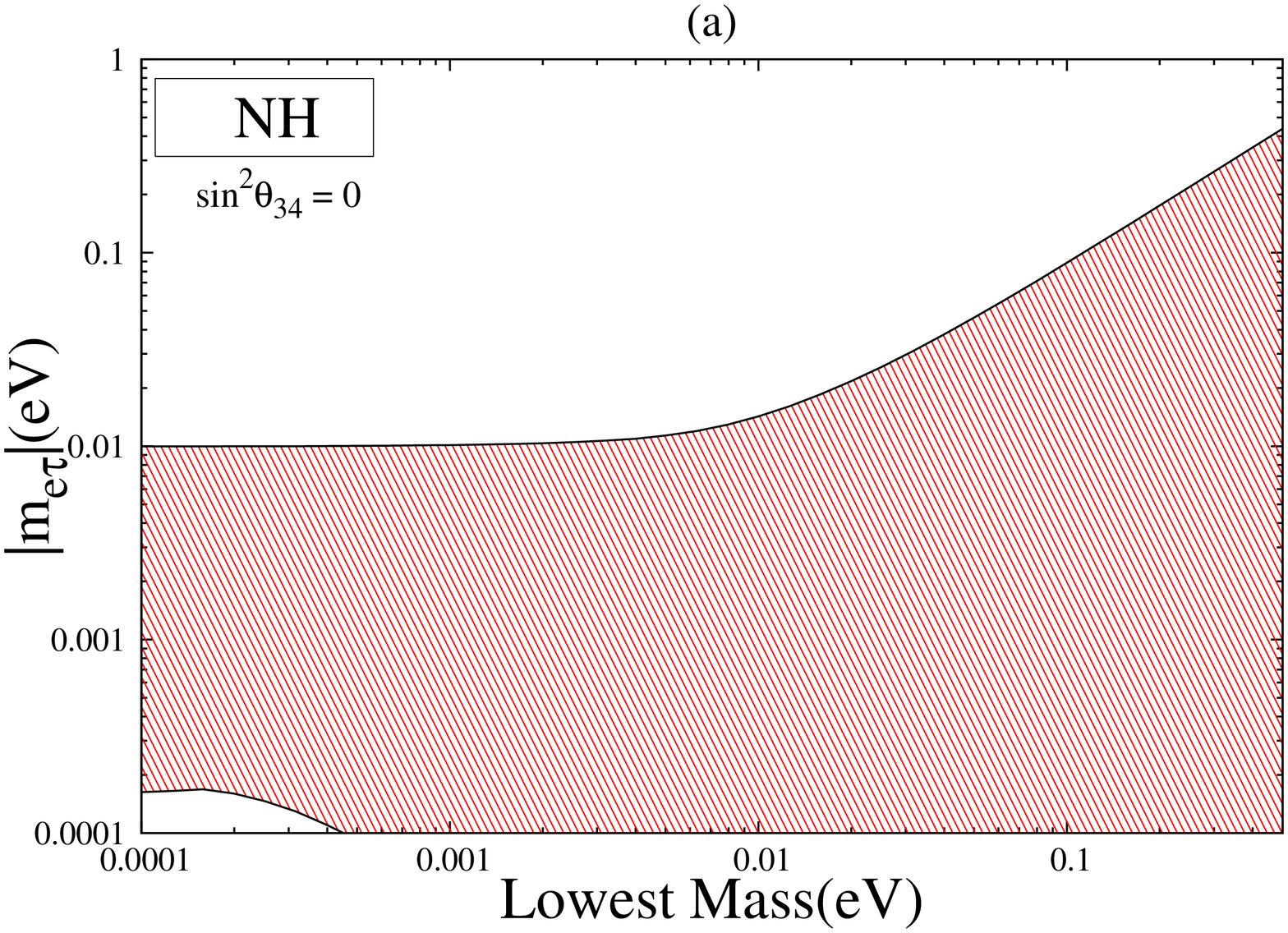}
\includegraphics[width=0.33\textwidth,angle=270]{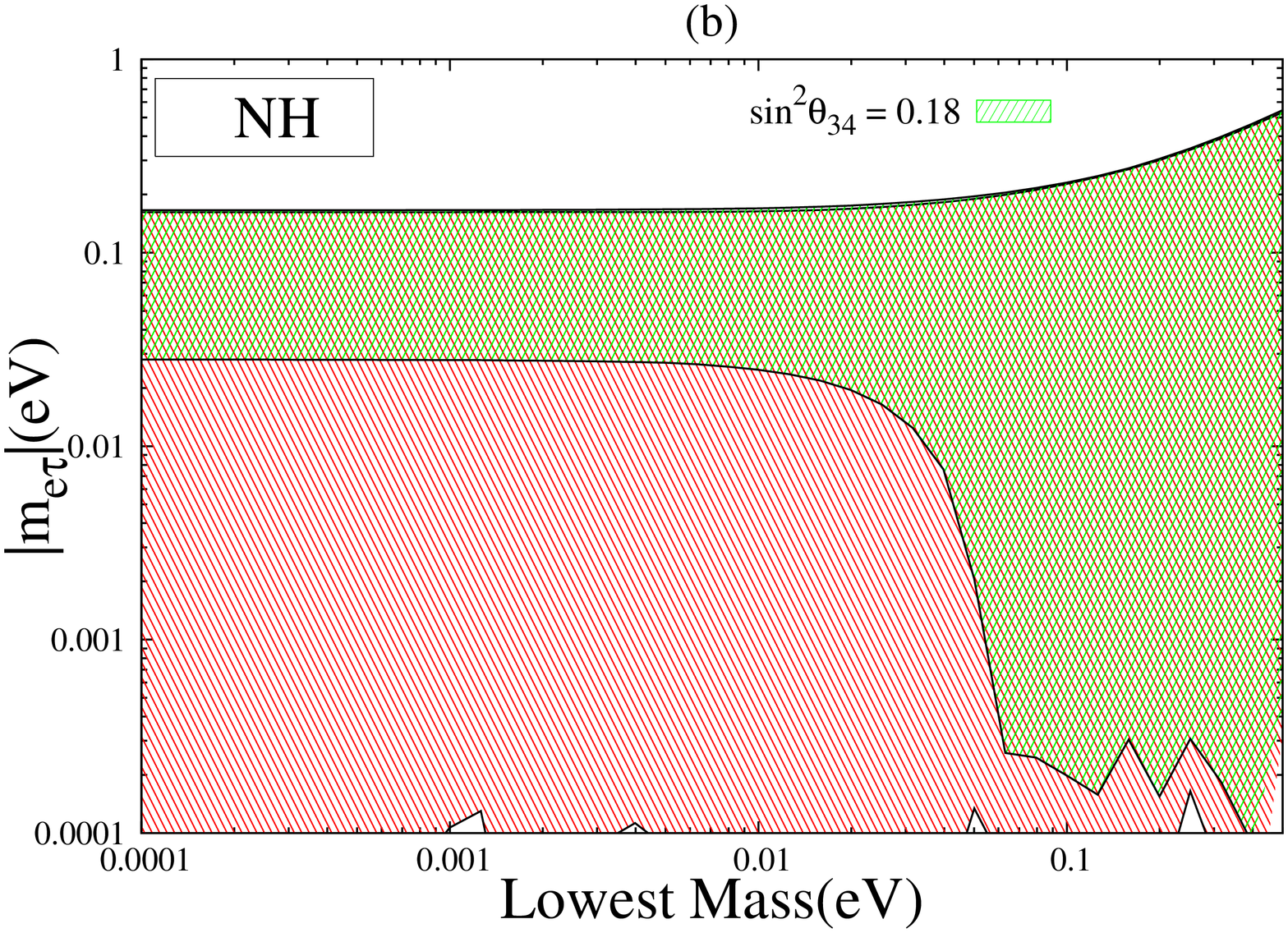}
\caption[Plots of $|m_{e\tau}|$ vs $m_1$ for normal hierarchy.]{Plots of $|m_{e\tau}|$ for normal hierarchy with lowest mass $m_1$.
 The panel (a) corresponds to three generation case. In (b) (red/light region) all the parameters are varied in their full allowed range and
 the green/dark region is for $s_{34}^2 = 0.18$ with all the other parameters covering their full range.}
\label{fig6}
\end{center}
\end{figure}
\begin{figure}[ht!]
\begin{center}
\includegraphics[width=0.33\textwidth,angle=270]{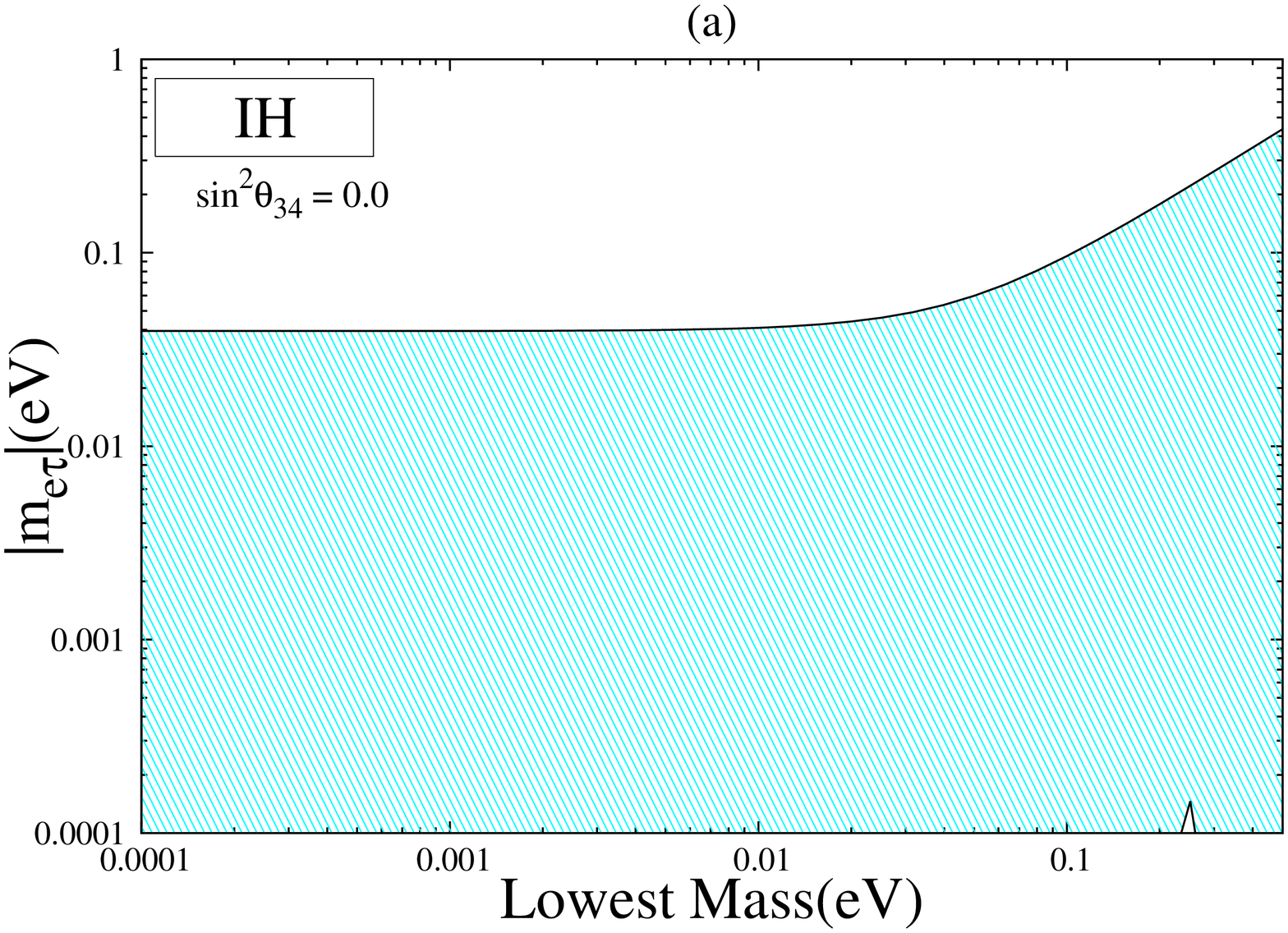}
\includegraphics[width=0.33\textwidth,angle=270]{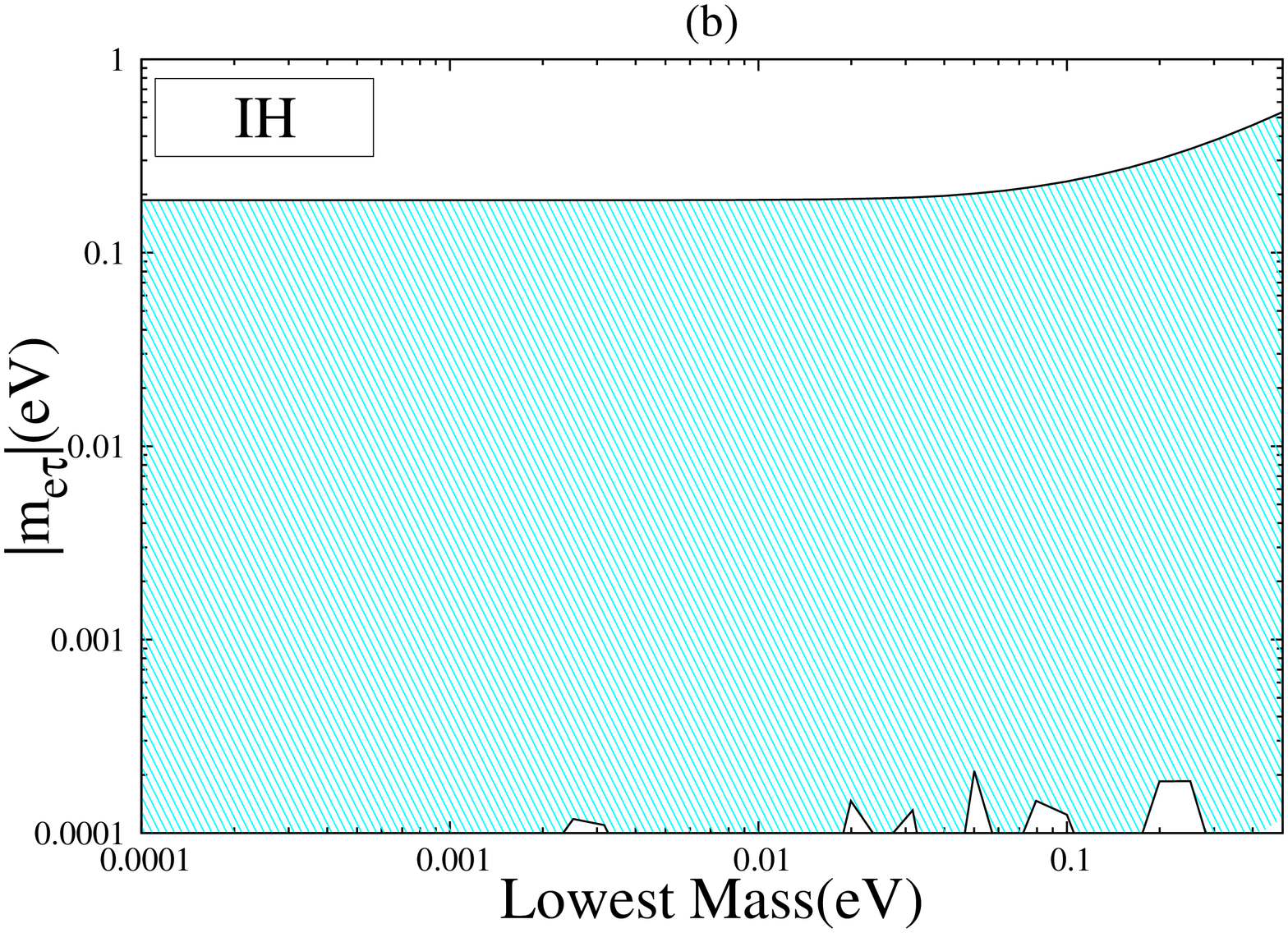}\\
\includegraphics[width=0.33\textwidth,angle=270]{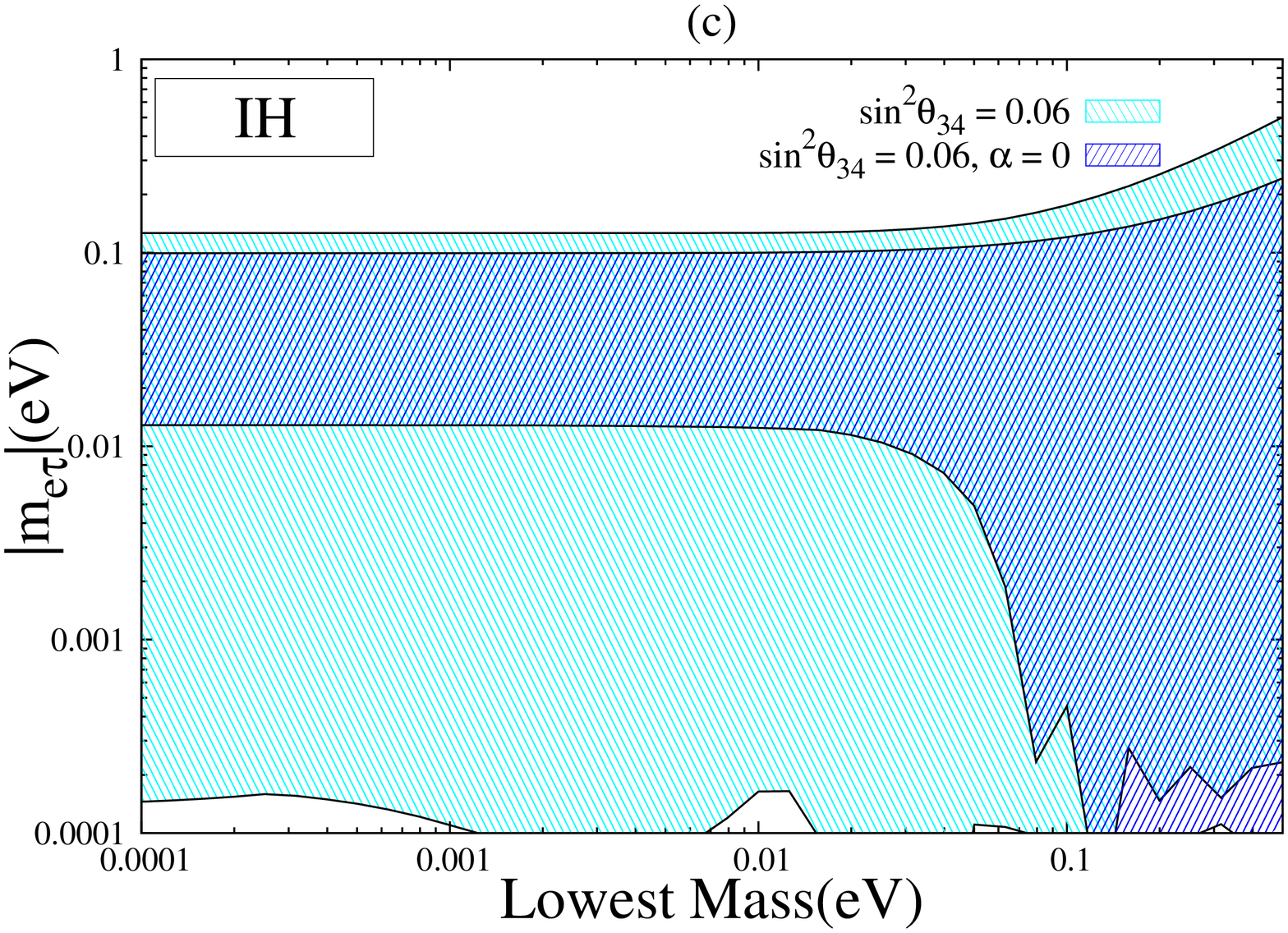}
\includegraphics[width=0.33\textwidth,angle=270]{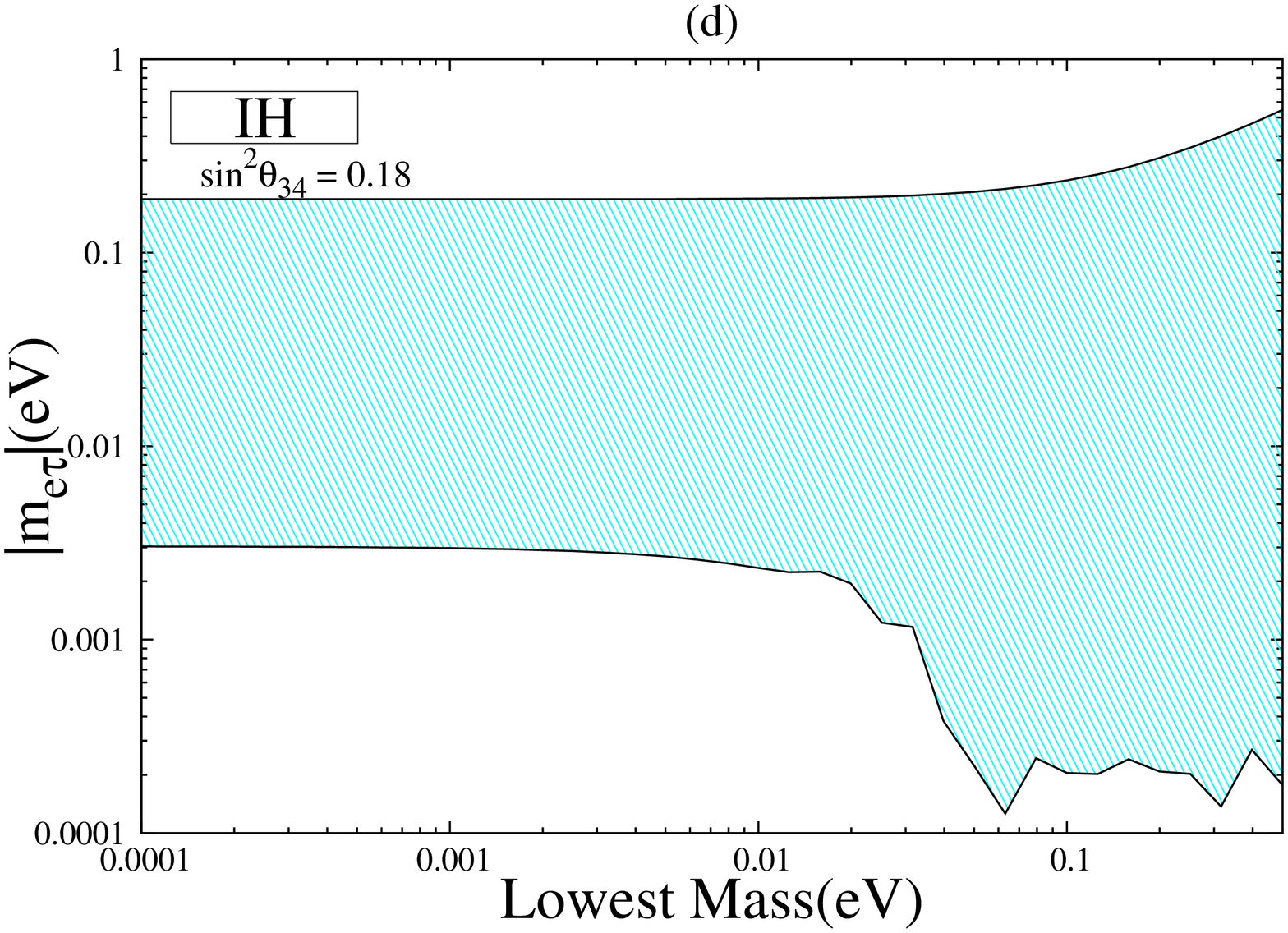}\\
\includegraphics[width=0.33\textwidth,angle=270]{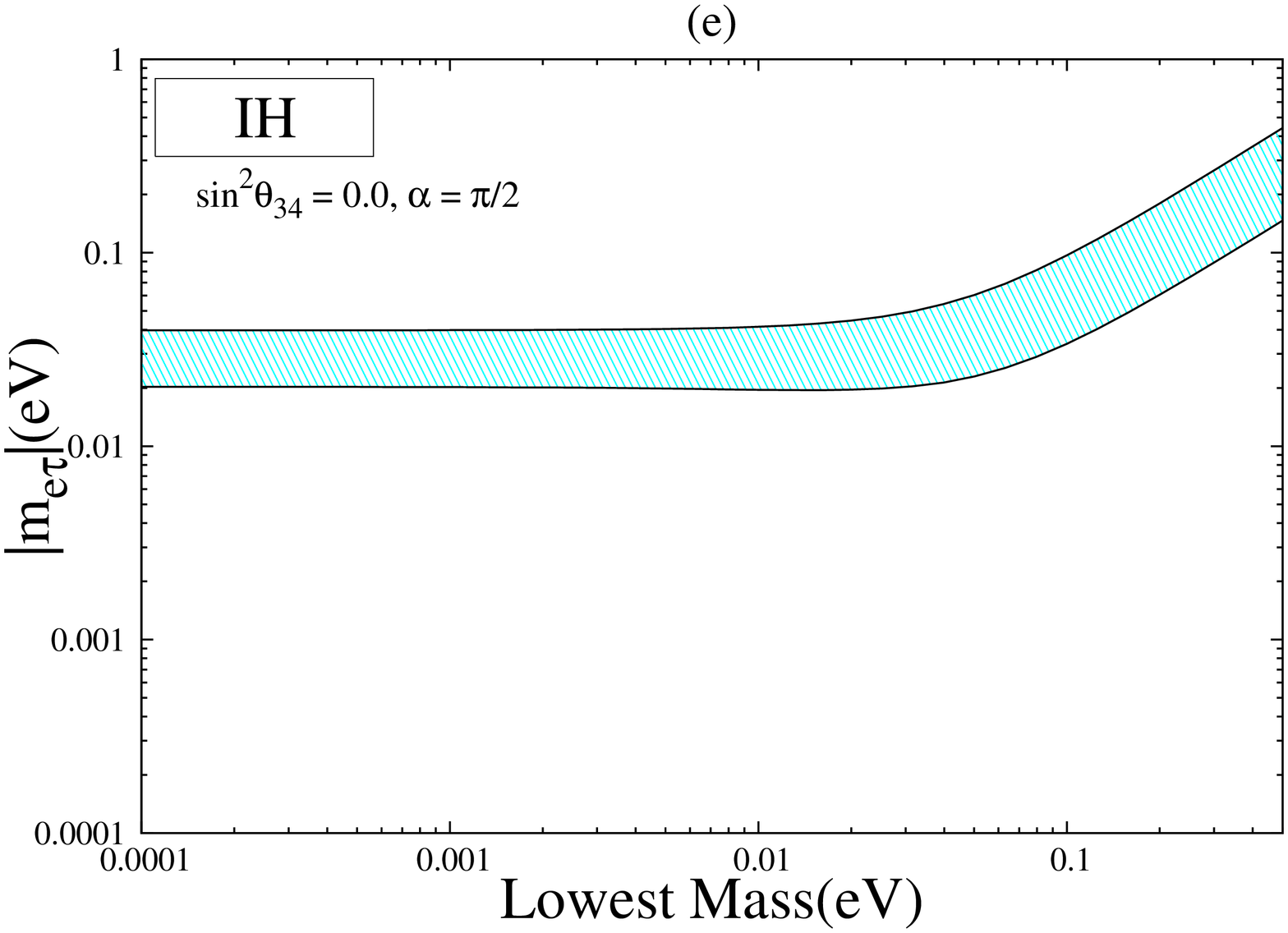}
\includegraphics[width=0.33\textwidth,angle=270]{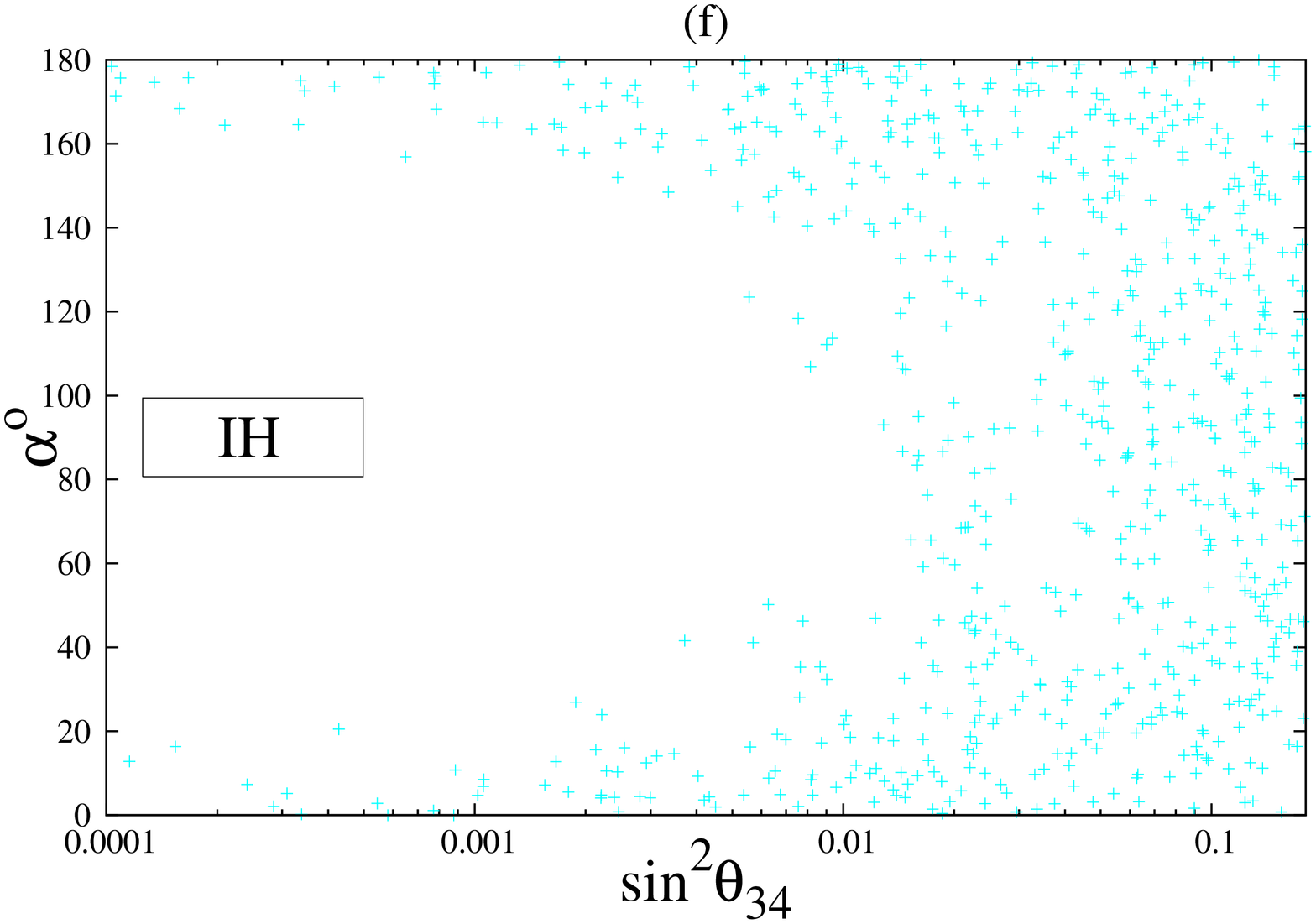}
\caption[Plots of $|m_{e\tau}|$ vs $m_3$ for inverted hierarchy.]{Plots of $|m_{e\tau}|$ for inverted hierarchy with lowest mass $m_3$.
The panel (a) corresponds to three generation case. In (b) all the parameters are varied in their full allowed range (3+1).
The panel (c), (d) is for specific value of $\theta_{34}$  and $\alpha$ with all the other parameters covering their full range.
The panel (f) shows correlation between $\alpha$ and $s_{24}^2$.}
\label{fig7}
\end{center}
\end{figure}

In the limit of vanishing active sterile mixing angle $\theta_{34}$ this element becomes
\bea
m_{e \tau}= c_{14}(m_{e \tau})_{3 \nu}. \nonumber
\eea
Using the approximation in Eq. \ref{xnh} for NH the above element  can be expressed as,
\bea
|m_{e\tau}|&\approx& |\sqrt{\Delta_{32}}\{- s_{12}s_{23}c_{12}c_{34}\sqrt{\zeta}e^{2i\alpha}+\lambda(c_{23}c_{34}e^{i(2\beta+\delta_{13})}\chi_{13}-c_{23}c_{34}s_{12}^2
\\ \nonumber &&\chi_{13}\sqrt{\zeta}e^{i(2\alpha+\delta_{13})}+e^{i(2\gamma+\delta_{14})}\sqrt{\xi}s_{34}\chi_{14}-
e^{i(2\alpha+\delta_{14})}\sqrt{\zeta}s_{12}^2s_{34}\chi_{14}-c_{12}c_{23}\\ \nonumber &&e^{i(2\alpha+\delta_{24})}s_{12}s_{34}\chi_{24}\sqrt{\zeta})
-e^{i(\delta_{13}+\delta_{24})}(e^{2i\beta}-e^{2i\alpha}s_{12}^2\sqrt{\zeta})s_{23}s_{34}\chi_{13}\chi_{24}\lambda^2\}|.
\eea
For the case of vanishing Majorana phases and Dirac phases having the value $\pi$, this element can vanish when
\bea \label{metau:fixedphase}
&-&c_{12}c_{34}\sqrt{\zeta}s_{12}s_{23}-(1-\sqrt{\zeta}s_{12}^2)s_{23}s_{34}\lambda^2\chi_{13}\chi_{24}+\lambda(-c_{23}c_{34}\chi_{13}+\\
\nonumber && \sqrt{\zeta}c_{23}s_{12}(c_{34}s_{12}\chi_{13}+c_{12}s_{34}\chi_{24})+s_{12}^2s_{34}\chi_{14}\sqrt{\zeta}-\sqrt{\xi}s_{34}\chi_{14})
=0.
\eea
For a vanishing active sterile mixing angle $\theta_{34}$
one recovers the 3 generation case. In this limit, from Eq. \ref{metau:fixedphase} one observes that the leading order term and the term with $\lambda$ are of the same
order  $\sim \mathcal{O}$ (10$^{-2}$) while the $\lambda^2$ term vanishes and hence
cancellation is possible excepting for very low values of the lightest mass. We can see this in panel (a) of Fig. \ref{fig6}.
In panel (b) (red/light region) all the parameters are varied randomly (3+1 case) and
cancellation is seen to be possible over the whole range of $m_1$.
In panel (b) (green/dark region) we also plot the element $|m_{e\tau}|$ for the upper limit of
$s_{34}^2 = 0.18$. In this case there is no cancellation for very low values of
the smallest mass. This is because
when $s_{34}^2$ is large,  the $\lambda$ term containing $\xi$ becomes large $\mathcal{O}$ (1) and there will be no cancellation.\\
For inverted hierarchy the element $m_{e\tau}$ using the approximation in Eq. \ref{xih} becomes
\bea \nonumber
|m_{e\tau}|&\approx& |\sqrt{\Delta_{31}}\{c_{12}c_{34}s_{12}s_{23}(-e^{2i\alpha}+1)+e^{i(\delta_{13}+\delta_{24})}
(c_{12}^2+e^{2i\alpha}s_{12}^2)s_{23}s_{34}\lambda^2\chi_{13}\chi_{24}\\ \nonumber
&-& \lambda(c_{23}c_{34}\chi_{13}e^{i\delta_{13}}(c_{12}^2+e^{2i\alpha}s_{12}^2)+e^{i\delta_{14}}s_{34}\chi_{14}(c_{12}^2+e^{i\alpha}s_{12}^2)\\
&-& e^{i(2\gamma+\delta_{14})}s_{34}\chi_{14}\sqrt{\xi}+c_{12}c_{23}s_{12}s_{34}\chi_{24}e^{i\delta_{24}}(e^{2i\alpha}-1))\}|
\eea
In the limit of vanishing Majorana phases and Dirac CP violating phases equal to $\pi$ this element becomes negligible when
\bea
\lambda(c_{23}c_{34}\chi_{13}+s_{34}\chi_{14}-s_{34}\chi_{14}\sqrt{\xi})
+s_{23}s_{34}\chi_{13}\chi_{24}\lambda^2 =0.
\eea
In panel (a) of Fig. \ref{fig7} the three generation case is reproduced by putting $s_{34}^2=0$ and in (b) all the parameters are varied in their allowed range (3+1 case). In both the
figures we can see that cancellation is permissible over the whole range of $m_3$ considered.
When the CP violating phase $\alpha=0$ we see that the leading order term ($\sin2\theta_{12}s_{23}c_{34}$) vanishes and
as a result for large values of $s_{34}^2$ the cancellation is not possible
because the term with coefficient $\lambda$ becomes large ($\mathcal{O}$(10$^{-1}$)).
For non zero values of the CP violating phase $\alpha$ this leading order term is non zero
and its contribution will be significant. So in this case high values of $\theta_{34}$
are also allowed because now the leading order and the term with coefficient $\lambda$ will
be of same magnitude. When we fix $s_{34}^2 = 0.06$ and $\alpha = 0$ the region where $m_3$
is small is disallowed (panel (c) blue/dark region) but when $\alpha$ varies within its full range the
disallowed regions become allowed (panel (c) cyan/light region).
When $s_{34}^2$ approaches its upper limit, the $\lambda$ term having $\xi$ becomes very
large and cancellation is not possible even for non zero values
of  $\alpha$ which can be seen from panel (d).
However, when $\alpha=\pi/2$, very
small values of $s_{34}^2$ cannot give cancellation as the leading order term becomes large (panel (e)).
$s_{34}^2$ has to be $\geq$ 0.01 for the term to vanish which can be seen from panel (f) where we plotted the correlation between $\alpha$
and $s_{34}^2$ for $|m_{e\tau}| = 0$.

\subsubsection{The Mass Matrix Element $m_{\mu\mu}$}
The (2,2) diagonal entry in neutrino mass matrix is given as
\bea
m_{\mu \mu} &=& e^{2 i(\delta_{14} - \delta_{24} + \gamma)} c_{14}^2 m_4 s_{24}^2 \\ \nonumber
&+& e^{2 i (\delta_{13} + \beta)} m_3(c_{13} c_{24} s_{23} - e^{i(\delta_{14} - \delta_{13} -\delta_{24})} s_{13} s_{14} s_{24})^2 \\ \nonumber
&+& m_1 \{-c_{23} c_{24} s_{12} + c_{12}(-e^{i \delta_{13}} c_{24} s_{13} s_{23} - e^{i(\delta_{14} - \delta_{24})} c_{13} s_{14} s_{24})\}^2 \\ \nonumber
&+& e^{2 i \alpha} m_2 \left\{c_{12} c_{23} c_{24} + s_{12}(-e^{i \delta_{13}} c_{24} s_{13} s_{23} - e^{i(\delta_{14} - \delta_{24})} c_{13} s_{14} s_{24})\right\}^2
\eea
This expression reduces to its three generation case if the
mixing angle $\theta_{24}$ vanishes.
Also we can see from the expression that there is no
dependence on the
mixing angle $\theta_{34}$.
Using the approximation in Eqs. \ref{xnh}
this element can be simplified to the form
\bea
 |m_{\mu \mu}| &\approx&|\sqrt{\Delta_{32}}\{ c_{12}^2 c_{23}^2 e^{2 i \alpha} \sqrt{\zeta} + e^{i(\delta_{13} + 2 \beta)} s_{23}^2 \\ \nonumber
 &-& 2 \lambda c_{12} c_{23} e^{i(\delta_{13} + 2 \alpha)} \sqrt{\zeta} s_{12} s_{23} \chi_{13} \\ \nonumber
 &+& \lambda^2\{e^{2i(\delta_{13} +\alpha)}\sqrt{\zeta}s_{12}^2s_{23}^2\chi^2_{13}+e^{i(\delta_{14}-\delta_{24})}(e^{i(2\gamma +
\delta_{14} -\delta_{24})}\sqrt{\xi}\chi_{24} \\ \nonumber
&-& 2 e^{2i\alpha}\sqrt{\zeta}c_{12}c_{23}s_{12}\chi_{14})\chi_{24}\}\}|.
\eea
For the case of Majorana CP phases having the value 0
and the Dirac phases having the value  $\pi$, this element vanishes when
\bea \label{mmmphase}
&& s_{23}^2 + c_{12}^2 c_{23}^2 \sqrt{\zeta} + c_{12} s_{12} \sin 2 \theta_{23}\sqrt{\zeta} \lambda \chi_{13} \\ \nonumber
&+&\lambda^2(s_{12}^2s_{23}^2\sqrt{\zeta}\chi_{13}^2-c_{23}\sin2\theta_{12}\sqrt{\zeta}\chi_{14}\chi_{24}+\sqrt{\xi}\chi_{24}^2)=0.
\eea
\begin{figure}[h]
\begin{center}
\includegraphics[width=0.33\textwidth,angle=270]{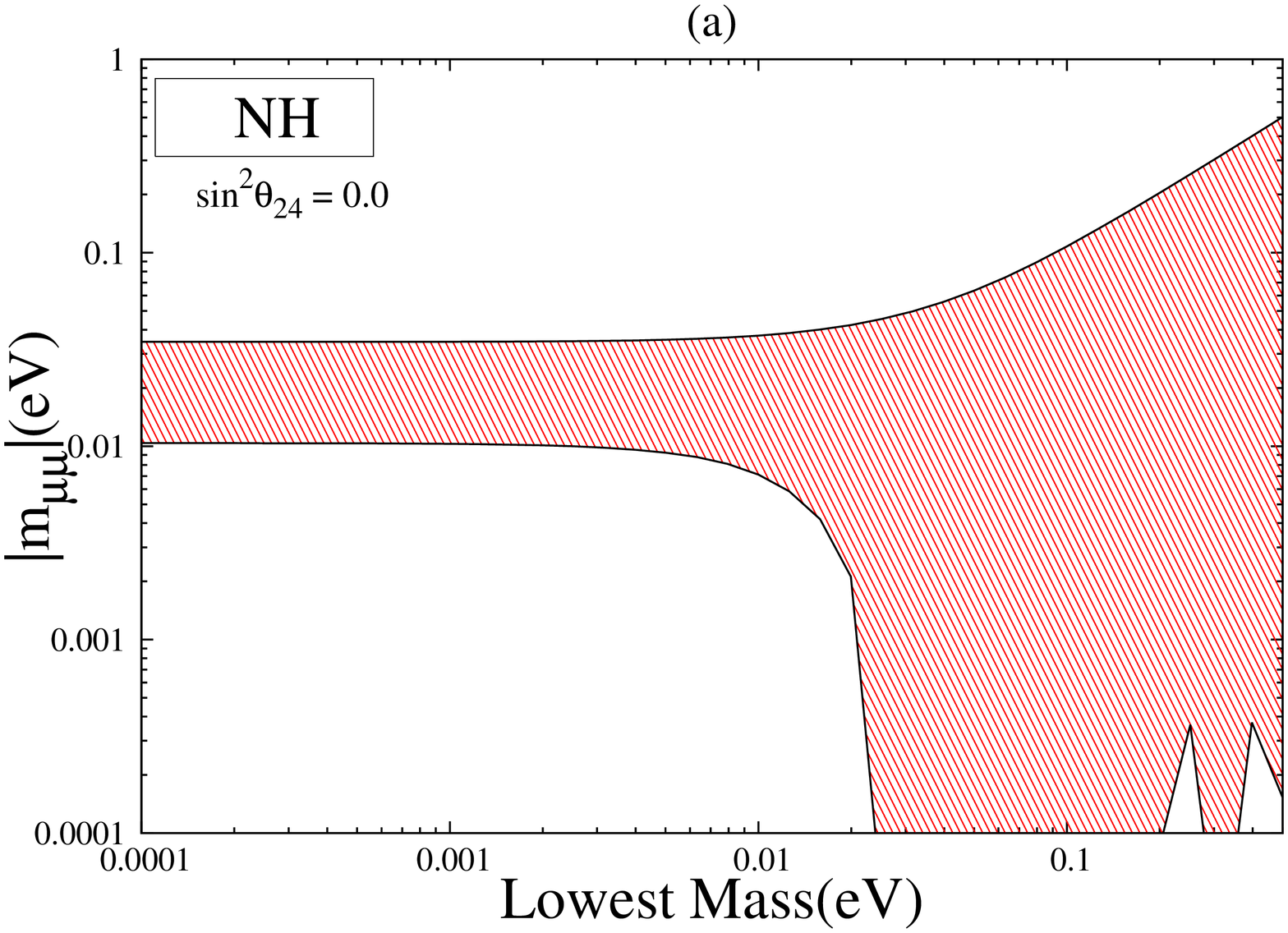}
\includegraphics[width=0.33\textwidth,angle=270]{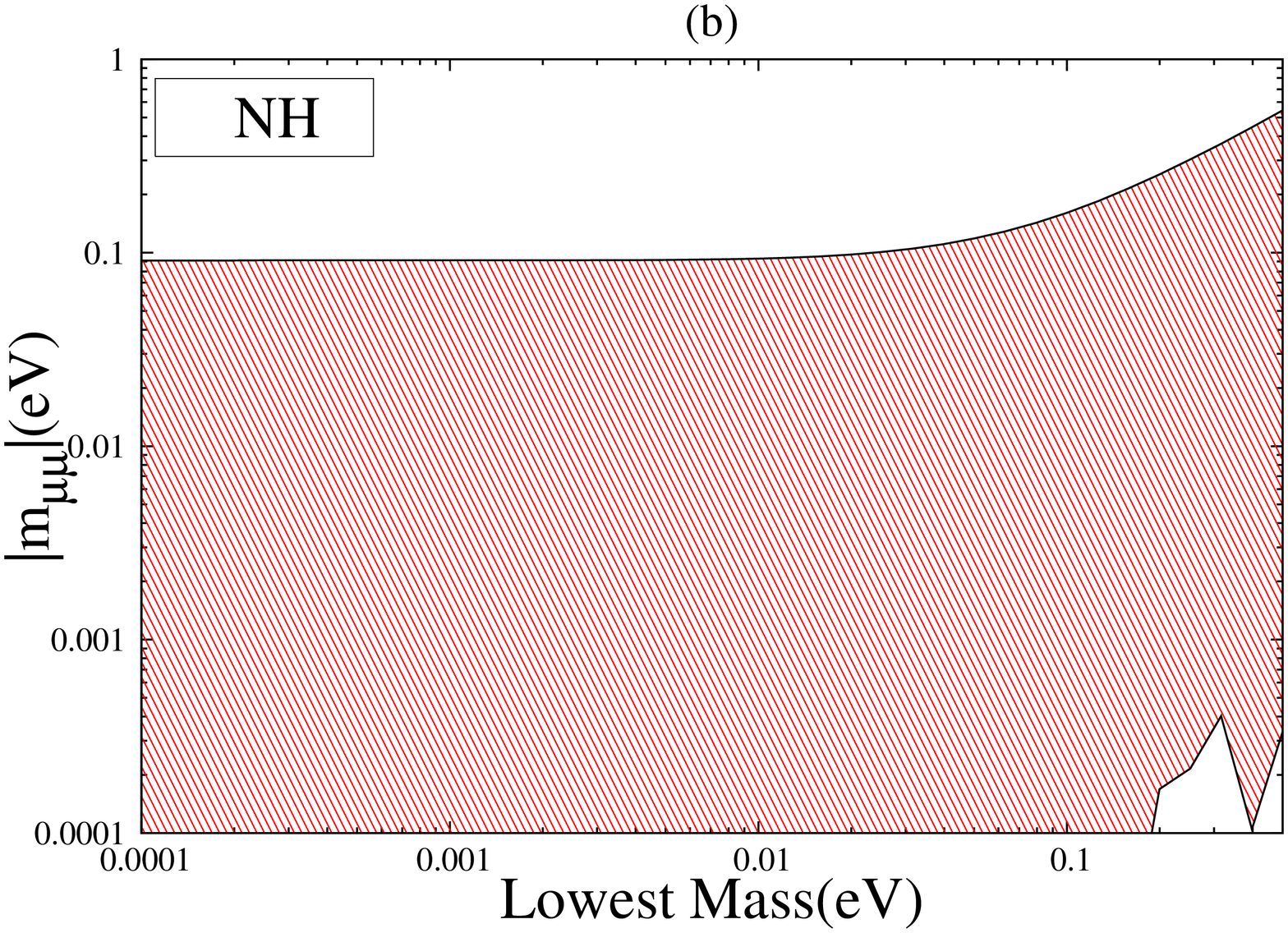}\\
\includegraphics[width=0.33\textwidth,angle=270]{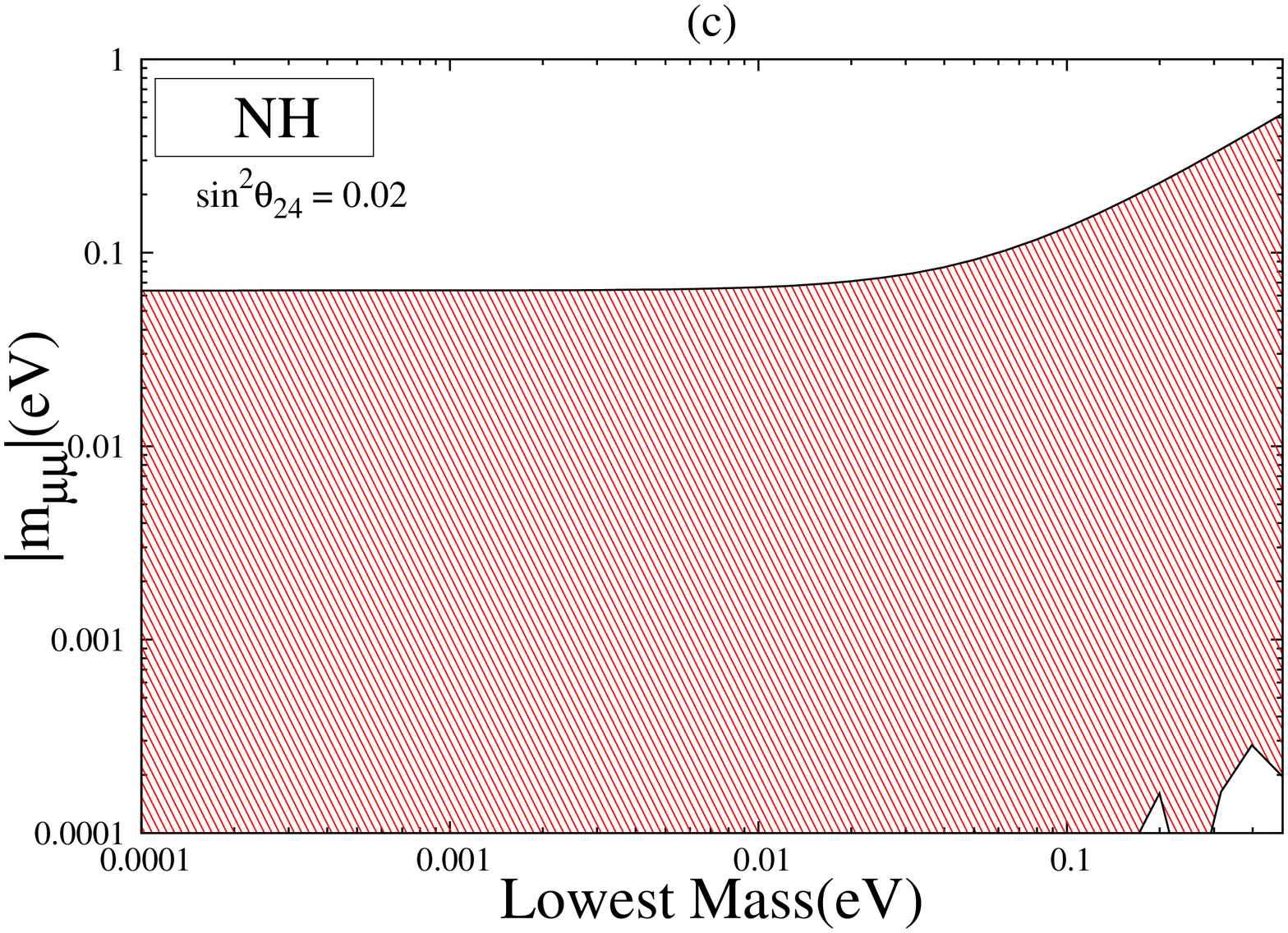}
\includegraphics[width=0.33\textwidth,angle=270]{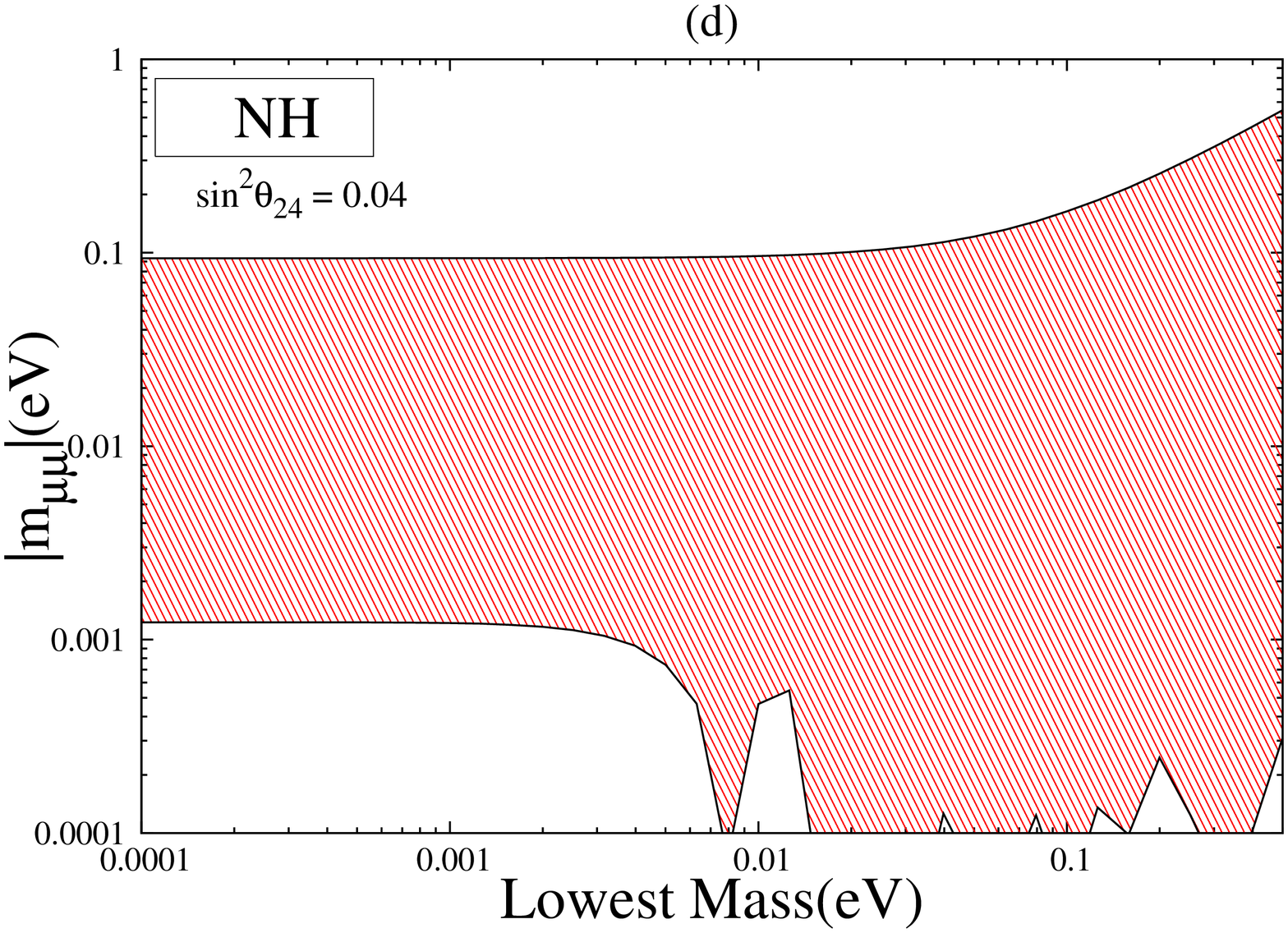}
\caption[Plots of $|m_{\mu\mu}|$ vs $m_1$ for normal hierarchy.]{Plots of vanishing $m_{\mu\mu}$ for normal hierarchy for different values of $\theta_{24}$
 when all other mixing angles are varied in their 3$\sigma$ ranges, Dirac CP phases are varied from 0 to $2\pi$ and Majorana phases from 0 to $\pi$.}
\label{fig8}
\end{center}
\end{figure}
We know that for the case of 3 generations,
the elements in the $\mu -\tau$ block are quite large and cannot vanish for
normal hierarchy. In panel (a) of Fig. \ref{fig8} we can see that $|m_{\mu \mu}|$
cannot vanish in small $m_1$ region for $s_{24}^2 = 0$ which is indeed the 3 generation case.
This is because the magnitude of the
first two terms in Eq. \ref{mmmphase} is quite large in this case,
$\sim \mathcal{O}$ (10$^{-1}$)
and for cancellation to occur the term with coefficient $\lambda^2$
has to be of the same order. This is not possible when $s_{24}^2$ is small.
However when $s_{24}^2 $ is varied in its full allowed range the contribution of the sterile
part is enhanced and this can cancel the active part as can be seen from panel (b).
Now to understand the dependence of $m_{\mu \mu}$ with $\theta_{24}$ we note that if we increase $s_{24}^2$ from its lower bound
then the two terms become of the same order.
So there will be regions in the limit of small $m_1$ for which this element
vanishes (panel (c)). We see in panel (d) of Fig. \ref{fig8} that when
$\theta_{24}$ acquires very large values, the magnitude of the $\lambda^2$ ($\sqrt{\xi}\chi_{24}^2$) term becomes large, thus leading to non cancellation
of the terms with the leading order first two terms. Hence, the region with very small $m_1$ is not allowed.
Using the approximation for inverted hierarchy the element $m_{\mu\mu}$ becomes
\bea
|m_{\mu \mu}| &\approx& |\sqrt{\Delta_{31}}\{c_{23}^2(s_{12}^2 + c_{12}^2 e^{2 i \alpha}) \\ \nonumber
&+& \frac{1}{2}
 \lambda \sin2\theta_{12} \sin2\theta_{23} e^{i \delta_{13}} ( 1 - e^{2 i \alpha}) \chi_{13} \\ \nonumber
 &+& \lambda^2[\sin2\theta_{12} c_{23} e^{i(\delta_{14} - \delta_{24})}(1 - e^{2 i \alpha}) \chi_{14} \chi_{24}
 + s_{23}^2 e^{2 i \delta_{13}}(c_{12}^2 + e^{ 2 i \alpha} s_{12}^2) \chi_{13}^2 \\ \nonumber
 &+& e^{2i(\gamma + \delta_{14} - \delta_{24})} \sqrt{\xi} \chi_{24}^2]\}|.
\eea
\begin{figure}[ht!]
\begin{center}
\includegraphics[width=0.33\textwidth,angle=270]{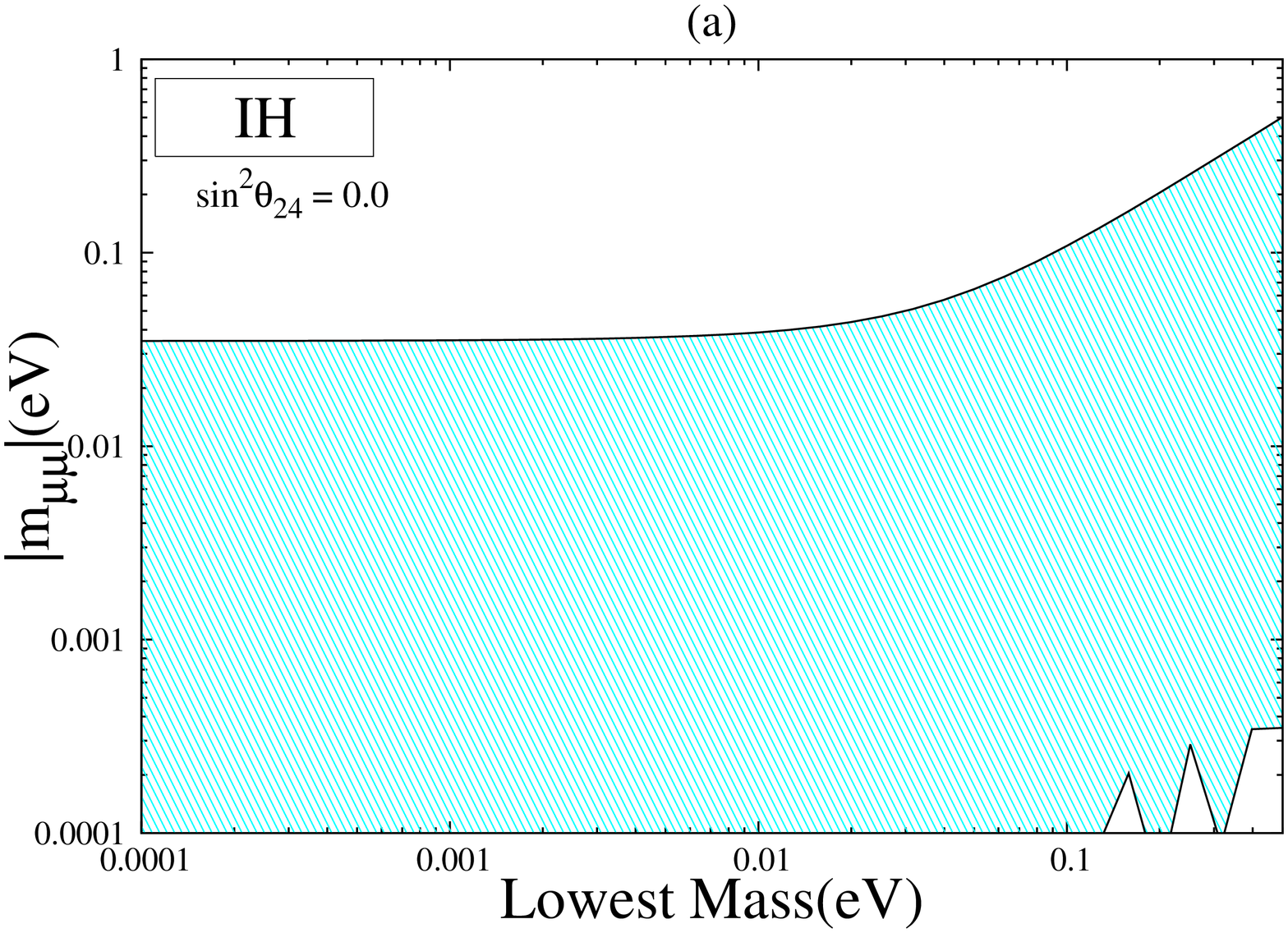}
\includegraphics[width=0.33\textwidth,angle=270]{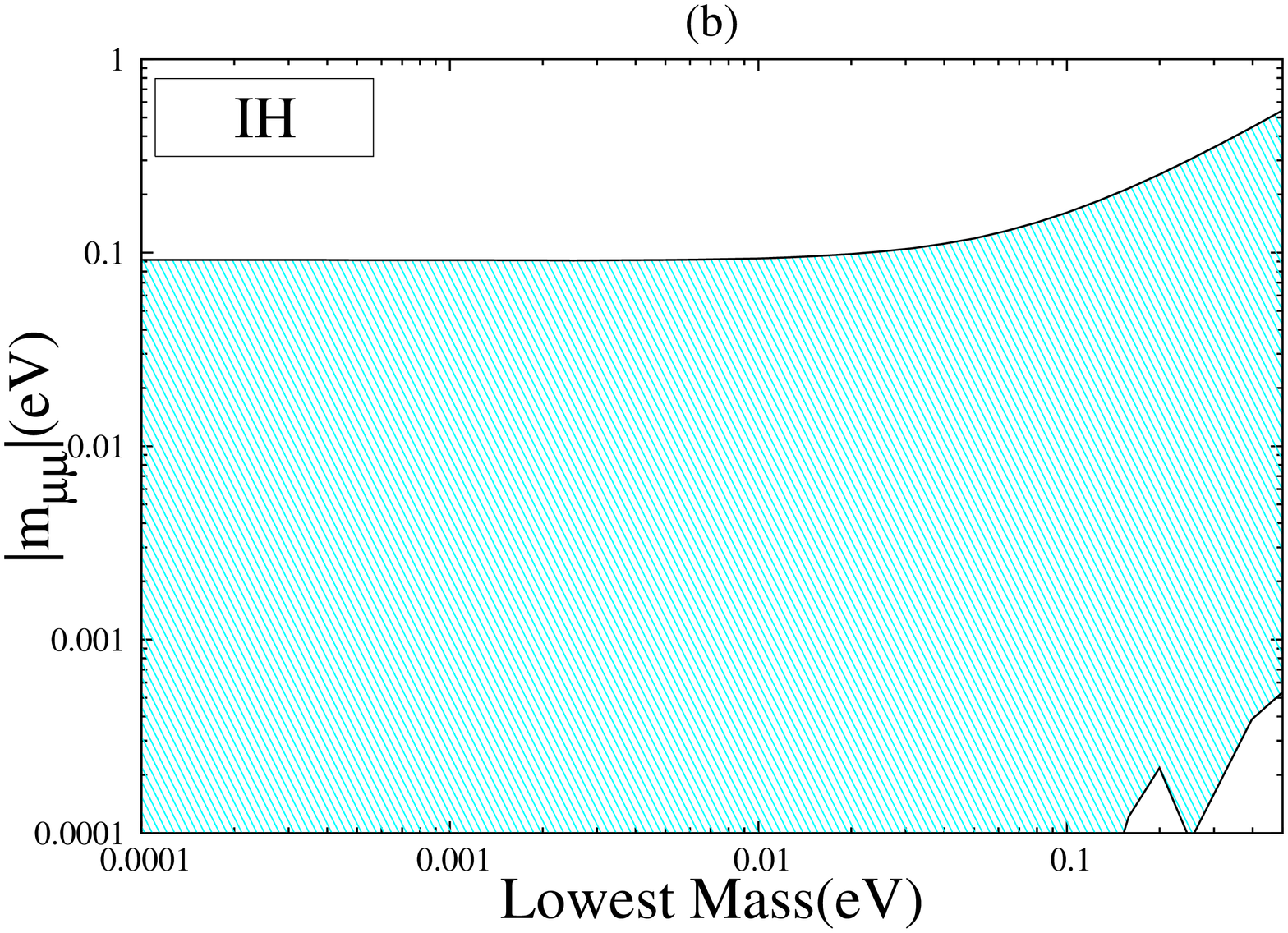}\\
\includegraphics[width=0.33\textwidth,angle=270]{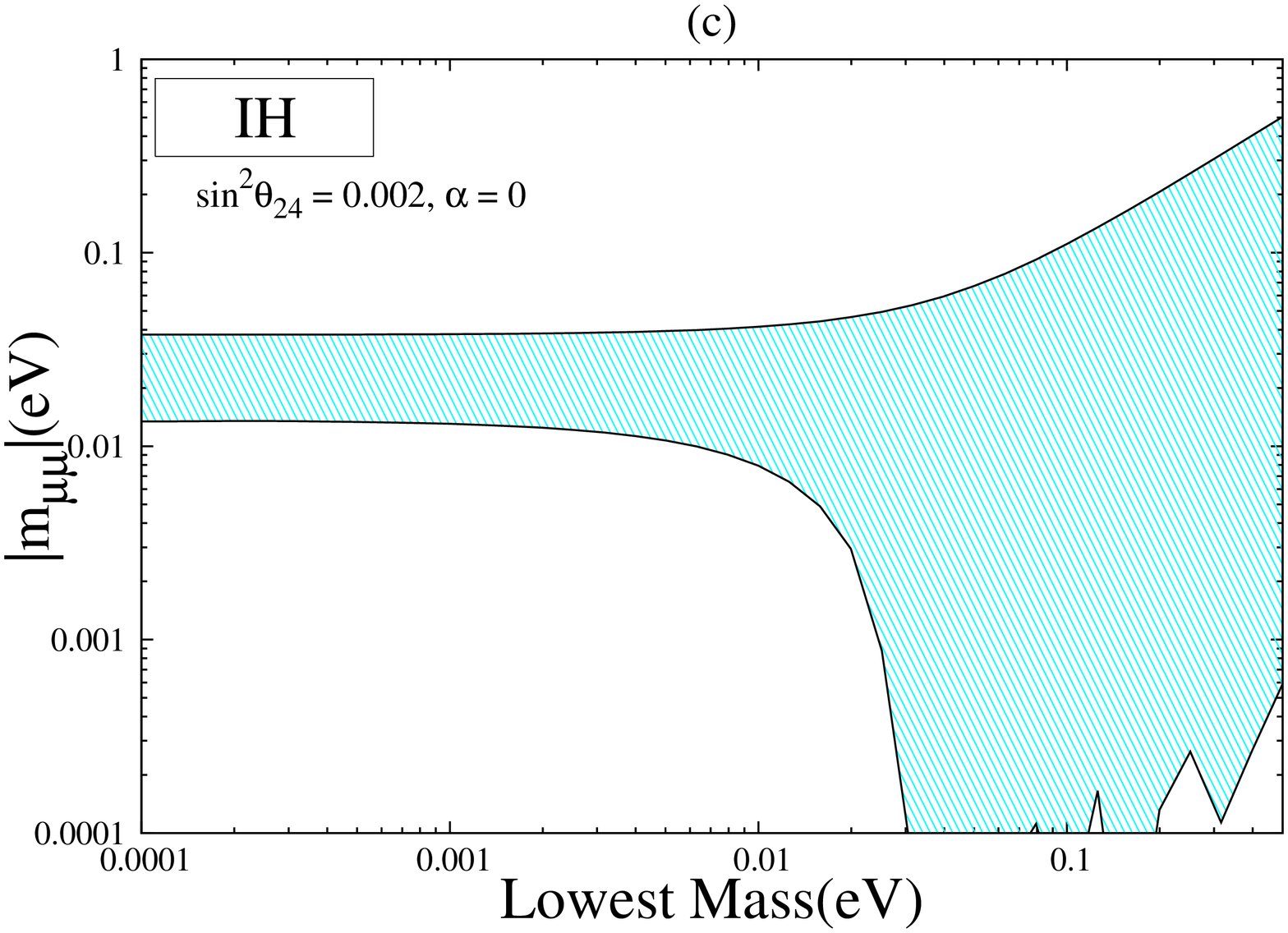}
\includegraphics[width=0.33\textwidth,angle=270]{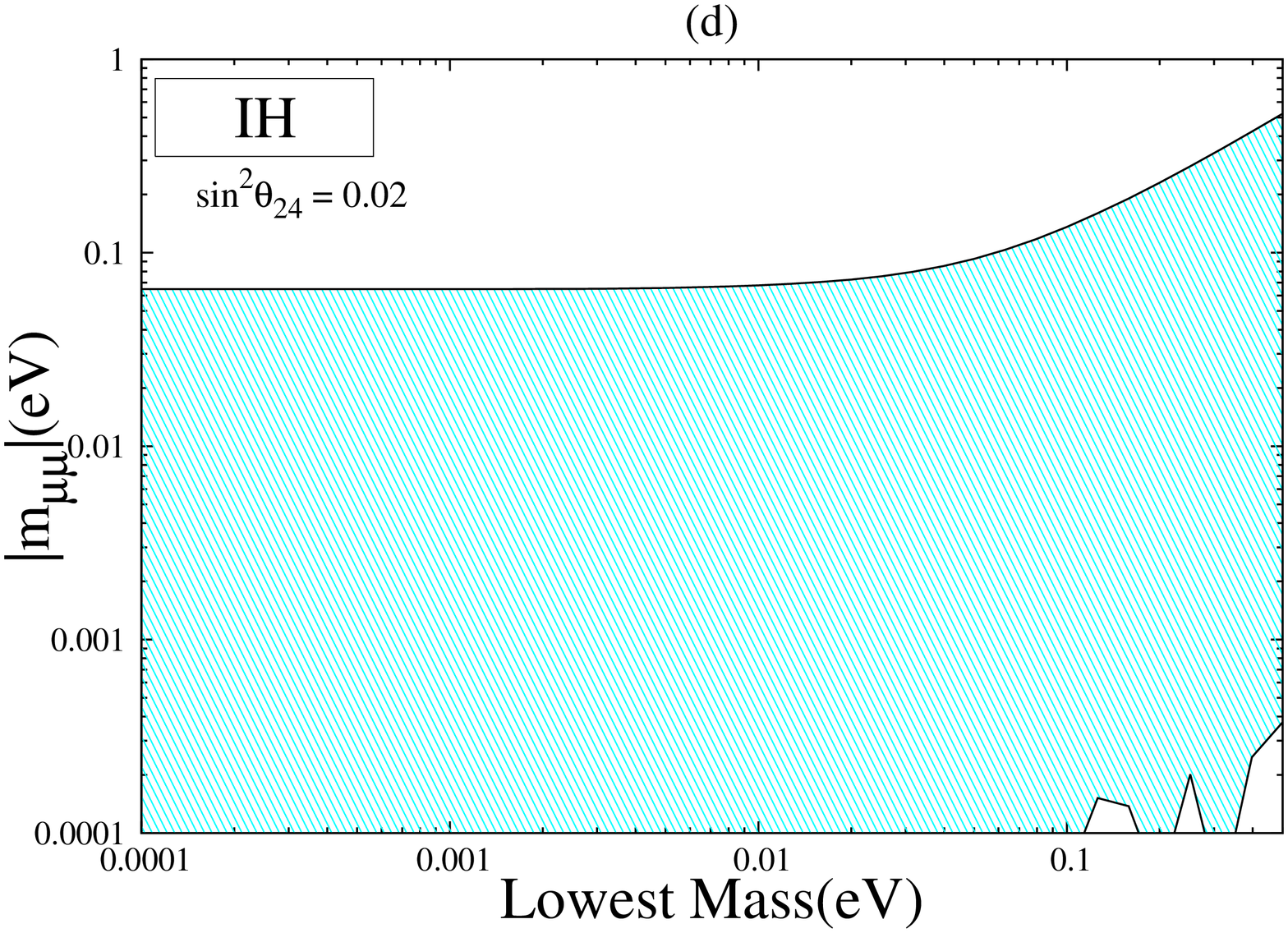}
\caption[Plots of $|m_{\mu\mu}|$ vs $m_3$ for inverted hierarchy.]{Plots of vanishing $m_{\mu\mu}$ for inverted hierarchy with lowest mass $m_3$.
Panel (a) for 3 generation case (b) all the parameters are varied in full allowed range (3+1).  Panel (c) and (d) are for
specific values of $\alpha$ and $s_{24}^2$ are taken with all other parameters covering their full range.}
\label{fig9}
\end{center}
\end{figure}
Assuming Majorana phases to be zero
 and Dirac phases having value $\pi$, this element can vanish when
\bea
&& c_{23}^2+\lambda^2(s_{23}^2\chi_{13}^2+\sqrt{\xi}\chi_{24}^2)=0.
\eea
In panel (a) of Fig. \ref{fig9} we plotted $|m_{\mu \mu}|$ for $s_{24}^2 = 0$ to reproduce 3 generation case whereas  in panel (b) all the parameters are varied in their allowed range in 3+1 scenario.
In both cases we can see that cancellation is possible for full range of $m_3$.
It can be noticed that unlike normal hierarchy, here cancellation is possible for small values of $s_{24}^2$
because in this case all the terms are of same order and there can always be cancellations. However,
  if we put $\alpha = 0$ then the term $\lambda$ ($\sin 2\theta_{12}s_{23}c_{23}\chi_{13}$) drops out from the equation and
the leading order term can not be canceled for small values of $s_{24}^2$. It can be seen from panel (c) that for $s_{24}^2 = 0.002$ and $\alpha = 0$ the regions where
$m_3$ is small is not allowed.
As the value of $\theta_{24}$ increases there is the possibility of cancellation of terms
for all the values of $\alpha$ as can be seen from panel (d) where we plot $|m_{\mu\mu}|$ with the lowest mass for $s_{24}^2=0.02$ when all the other mixing angles are
varied in 3$\sigma$ range and CP violating phases are varied in full range.
Now if we keep increasing $s_{24}^2$ then $\lambda^2$ term will become large and the chance of cancellation will be less.

\subsubsection{The Mass Matrix Element $m_{\mu\tau}$}
\begin{figure}[h]
\begin{center}
\includegraphics[width=0.33\textwidth,angle=270]{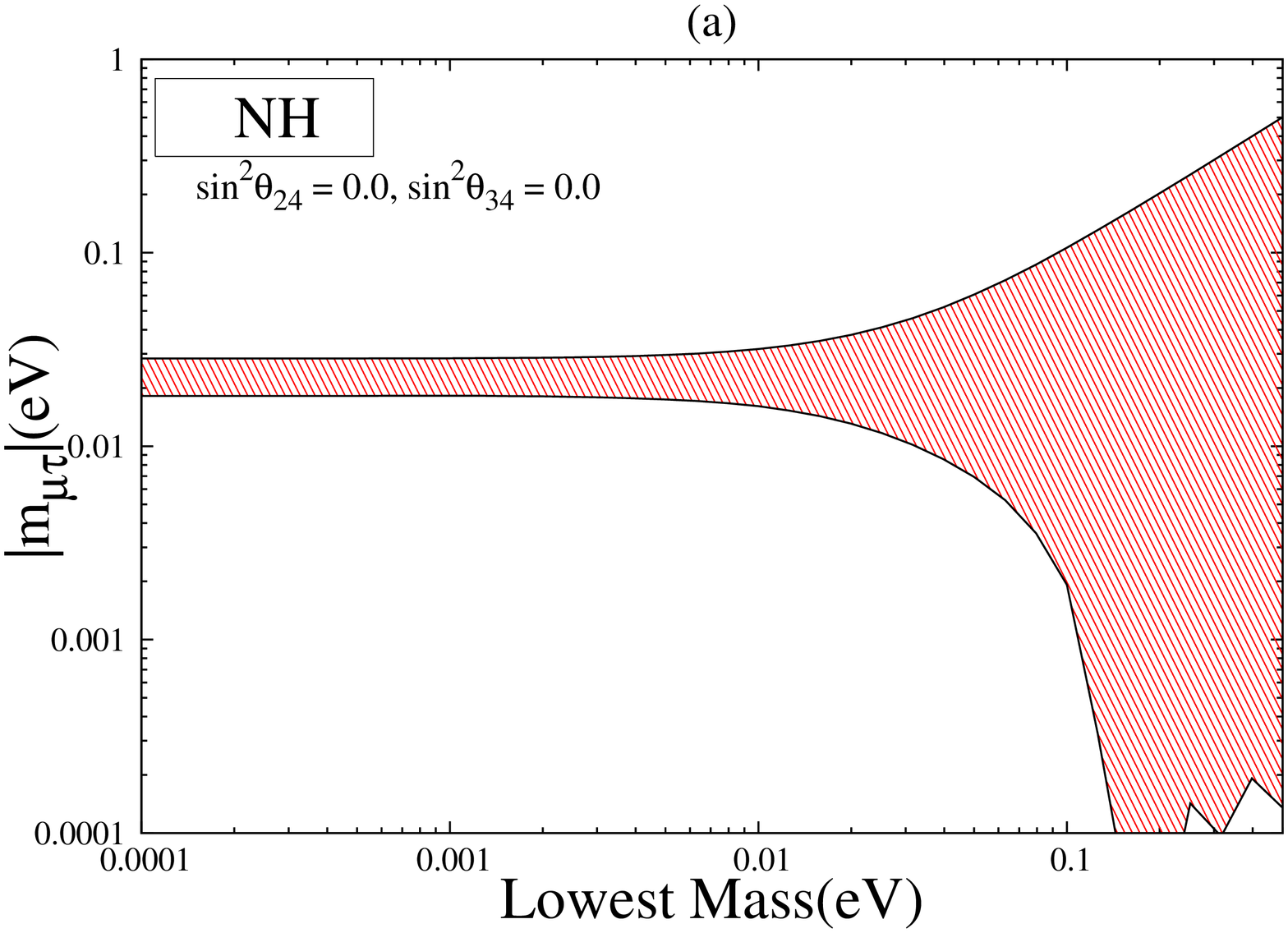}
\includegraphics[width=0.33\textwidth,angle=270]{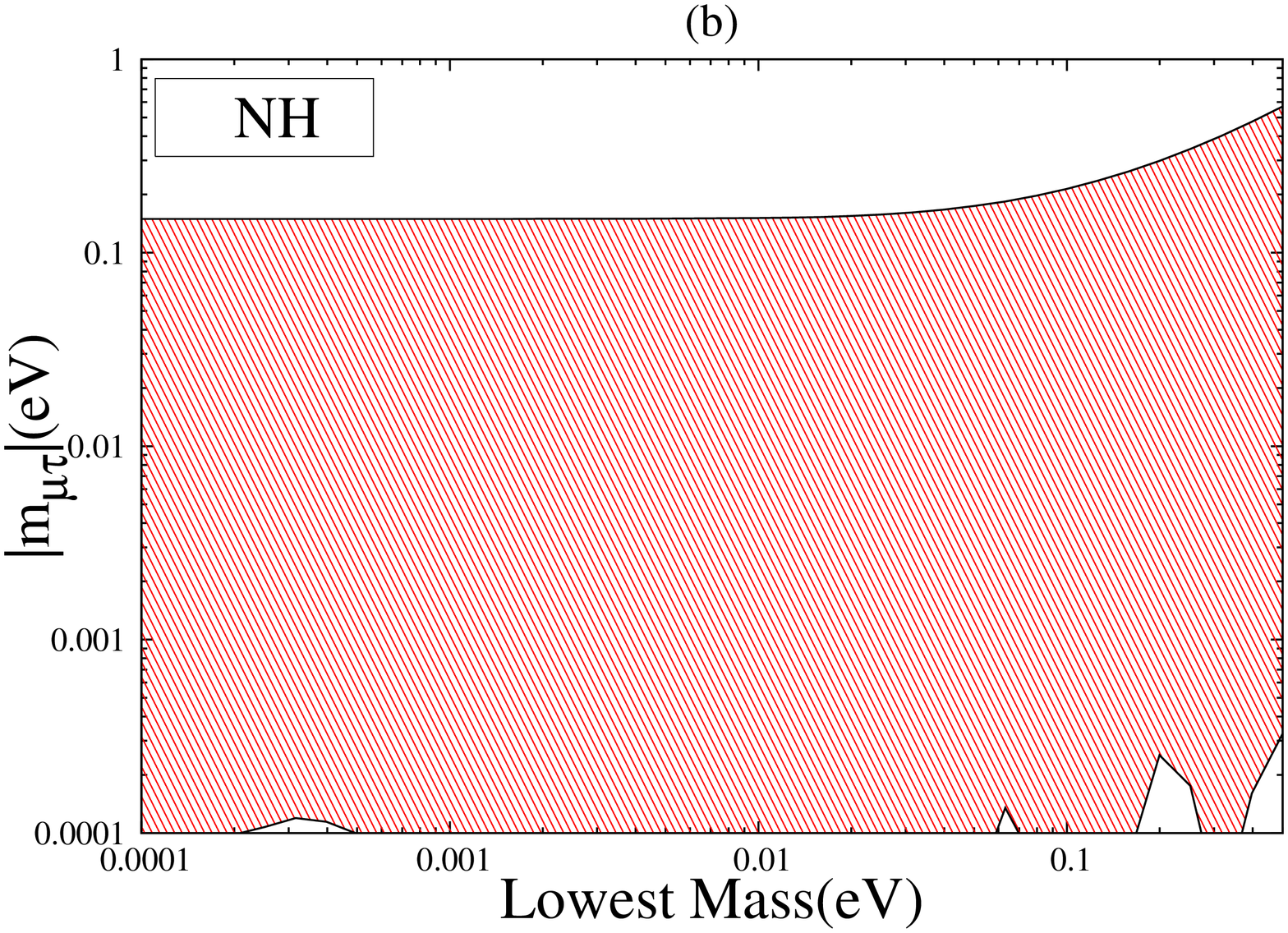}\\
\includegraphics[width=0.33\textwidth,angle=270]{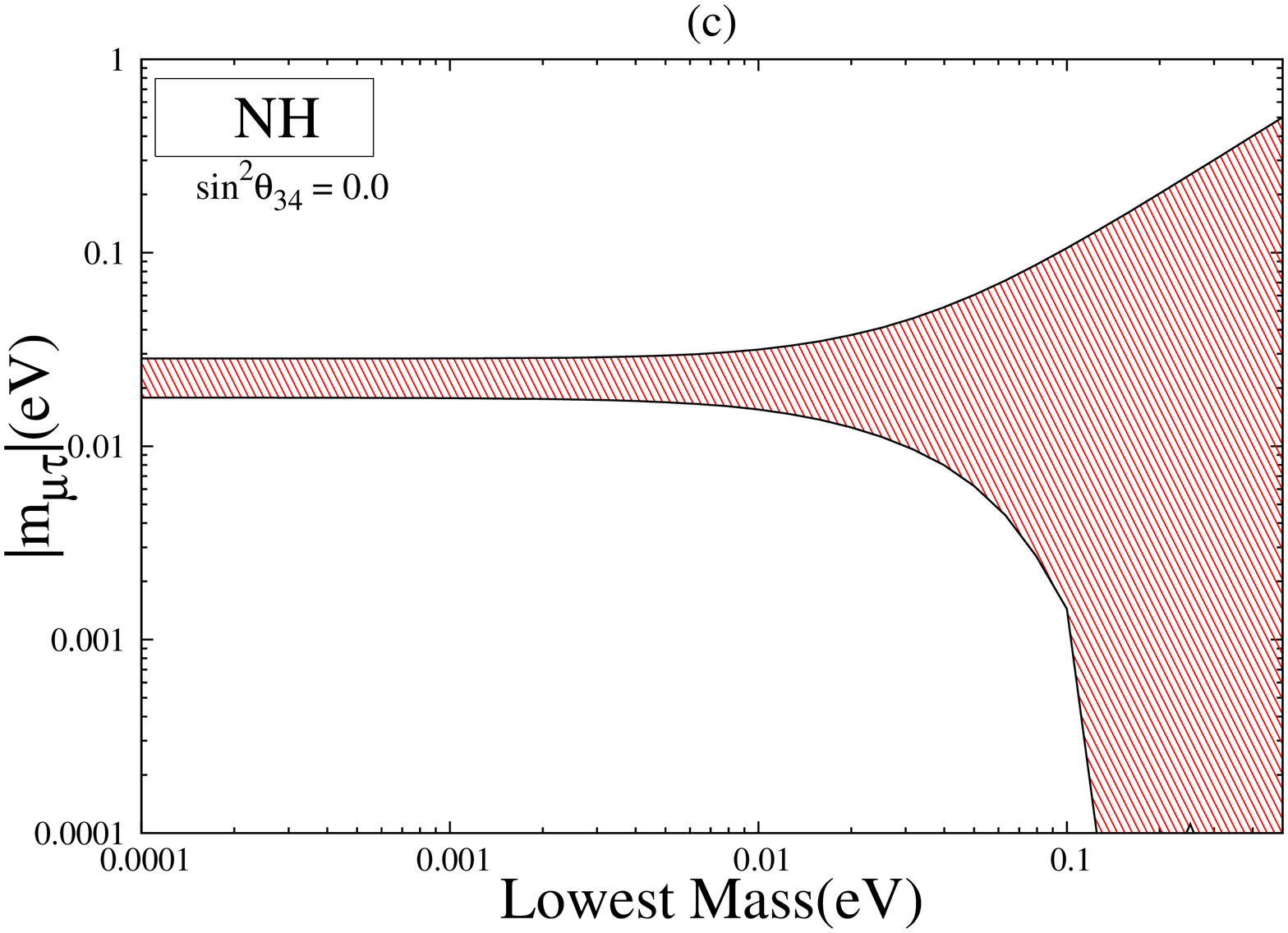}
\includegraphics[width=0.33\textwidth,angle=270]{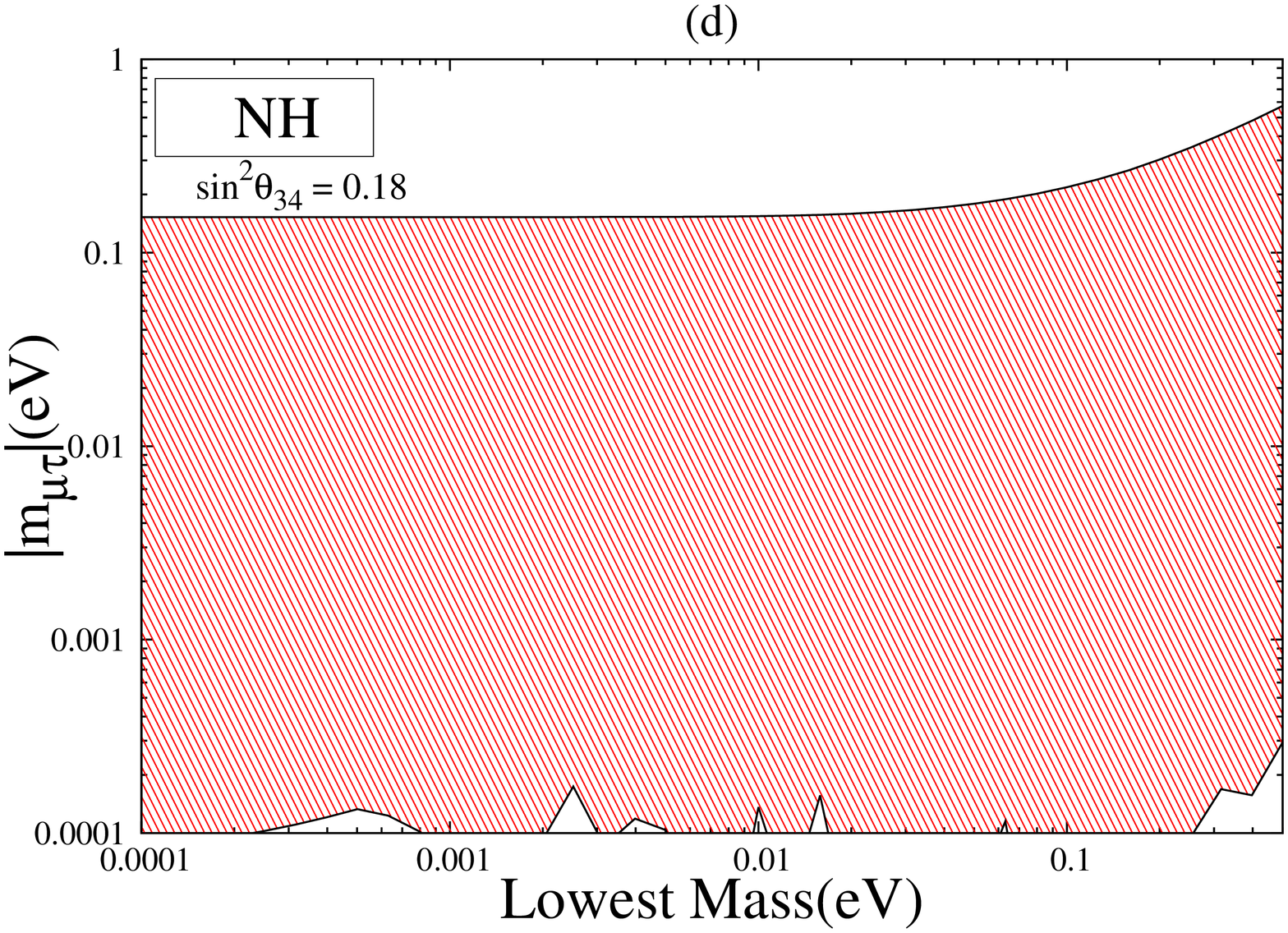}
\caption[Plots of $|m_{\mu\tau}|$ vs $m_1$ for normal hierarchy.]{Plots of vanishing $m_{\mu\tau}$ for normal hierarchy (a) for vanishing $\theta_{34}$ and $\theta_{24}$.
In panel (b) all parameters are varied in their full allowed range. Panel(c, d) are for specific values of $\theta_{34}$
when all other mixing angles are varied in their full range.}
\label{fig10}
\end{center}
\end{figure}

The (2,3) element of $M_{\nu}$ in the flavor basis becomes quite complicated in the presence of an extra sterile neutrino. The expression is
\bea
 m_{\mu \tau} &=& e^{i(2 \delta_{14} - \delta_{24} + 2 \gamma)} c_{14}^2 c_{24} m_4 s_{24} s_{34} \\ \nonumber
 &+& e^{2 i (\delta_{13} + \beta)} m_3 ( c_{13} c_{24} s_{23} - e^{i(\delta_{14} - \delta_{24} - \delta_{13})} s_{13} s_{14} s_{24}) \\ \nonumber
 && \{-e^{i(\delta_{14} - \delta_{13})} c_{24} s_{13} s_{14} s_{34}+ c_{13} (c_{23} c_{34} - e^{i \delta_{24}} s_{23} s_{24} s_{34})\} \\ \nonumber
 &+& m_1\{- c_{23} c_{24} s_{12} + c_{12}(-e^{i \delta_{13}} c_{24} s_{13} s_{23} - e^{i(\delta_{14} - \delta_{24})} c_{13} s_{14} s_{24})\} \\ \nonumber
 && [-s_{12} (-c_{34} s_{23} - e^{i \delta_{24}} c_{23} s_{24} s_{34}) \\ \nonumber
 &+& c_{12} \{ -e^{i \delta_{14}} c_{13} c_{24} s_{14} s_{34} - e^{i \delta_{13}} s_{13}(c_{23} c_{34} - e^{i\delta_{24}} s_{23} s_{24} s_{34})\}] \\ \nonumber
 &+& e^{2 i \alpha} m_2\{c_{12} c_{23} c_{24} + s_{12}(-e^{i \delta_{13}} c_{24} s_{13} s_{23} - e^{i(\delta_{14} - \delta_{24})} c_{13} s_{14} s_{24})\} \\ \nonumber
 && [c_{12}(-c_{34} s_{23} - e^{i \delta_{24}} c_{23} s_{24} s_{34}) \\ \nonumber
 &+& s_{12}\{-e^{i \delta_{14}} c_{13} c_{24} s_{14} s_{34} - e^{i \delta_{13}} s_{13}(c_{23} c_{34} - e^{i \delta_{24}} s_{23} s_{24} s_{34})\}].
\eea
It reduces to the 3 generation case when $\theta_{24} = \theta_{34} = 0$.
In the normal hierarchical region where $m_1$ can assume very small values and can be neglected, using approximations in
Eqs. \ref{xnh}, \ref{chi1} and \ref{chi2} we get
\bea
 |m_{\mu \tau}| &\approx & |\sqrt{\Delta_{32}}\{c_{23} c_{34}(e^{2 i (\beta + \delta_{13})} - e^{2 i \alpha} \sqrt{\zeta} c_{12}^2)s_{23} \\ \nonumber
 &-& \lambda[c_{12} c_{34} e^{i(2\alpha+ \delta_{13})} \sqrt{\zeta} s_{12} \cos2\theta_{23} \chi_{13} + e^{i \delta_{24}}(e^{2i \alpha}
 c_{12}^2 c_{23}^2 \sqrt{\zeta} + e^{2i(\beta + \delta_{13})} s_{23}^2) \chi_{24}s_{34} \\ \nonumber
 &-& e^{2 i \delta_{14}}(e^{i(2 \gamma - \delta_{24})} \sqrt{\xi} \chi_{24} - c_{12} c_{23} e^{2 i \alpha} \sqrt{\zeta} s_{12} \chi_{14}) s_{34}] \\ \nonumber
 &+& \lambda^2 [\sqrt{\zeta} e^{i(2 \alpha + \delta_{13})}(e^{i \delta_{14}} s_{12} \chi_{14} + 2 c_{12} c_{23} e^{i \delta_{24}} \chi_{24})s_{12} s_{23} s_{34} \chi_{13}\\ \nonumber
 &+& e^{i \delta_{14}}(e^{i(2 \alpha - \delta_{24})} c_{12} c_{34} \sqrt{\zeta} s_{12} \chi_{24} - e^{i(2 \beta + \delta_{13})} \chi_{13} s_{34}) \chi_{14} s_{23} \\ \nonumber
 &+& c_{23} c_{34} e^{2i(\alpha + \delta_{13})} s_{12}^2 \chi_{13}^2 \sqrt{\zeta} s_{23}]\}|.
\eea
To see the order of the terms we consider the case where Majorana CP phases vanish and Dirac phases have the value $\pi$. In this limit the element becomes negligible when
\bea
&& c_{23} c_{34}s_{23}(1-c_{12}^2\sqrt{\zeta})+\lambda\{(c_{12}c_{34}s_{12}\sqrt{\zeta}\chi_{13})\cos2\theta_{23} \\ \nonumber
 &+& \chi_{24}s_{34}(s_{23}^2+c_{12}^2c_{23}^2\sqrt{\zeta})+s_{34}(\sqrt{\xi}\chi_{24}+c_{12}c_{23}s_{12}\sqrt{\zeta}\chi_{14})\} \\ \nonumber
 &+& \lambda^2\{s_{12}\chi_{13}s_{23}s_{34}\sqrt{\zeta}(s_{12}\chi_{14}+2c_{12}c_{23}\chi_{24})+\chi_{14}s_{23}(c_{12}c_{34}s_{12}^2s_{23}\sqrt{\zeta}\chi_{13}^2)\}=0.
\eea
Being an element of $\mu \tau$ block, $m_{\mu \tau}$ shows the same behaviour that of $m_{\mu \mu}$ in normal hierarchy. In panel (a) of Fig. \ref{fig10} we plotted $|m_{\mu \tau}|$
for $s_{24}^2 = s_{34}^2 = 0$ which coincides
with the 3 generation case and we can see that cancellation is not possible in hierarchical region. However, when all
the parameters are varied in their allowed range in panel (b) it get contribution from the sterile part and cancellation is always possible.
It can also be seen from panel (c) of Fig. \ref{fig10} that for $s_{34}^2 = 0$ there is no
 cancellation in the region when $m_1$ is small and the figure is quite similar to that of 3 generation case.
However, as this active sterile mixing angle becomes larger
there is always a possibility of allowed region towards the lower values of $m_1$ as is evident from panel (d).
This is because for the vanishing value of $\theta_{34}$
the terms with $\lambda$ and $\lambda^2$ become very small and cannot
cancel the leading term $\mathcal{O}$ (10$^{-1}$). It can also be seen that in this case (i.e., $s_{34}^2 = 0$),
there is no $\chi_{24}$ term and this is why the figure is somewhat similar to the 3 generation case.
However, when $\theta_{34}$ increases
these two contributions become large and cancellation becomes possible.

\begin{figure}[h]
\begin{center}
\includegraphics[width=0.33\textwidth,angle=270]{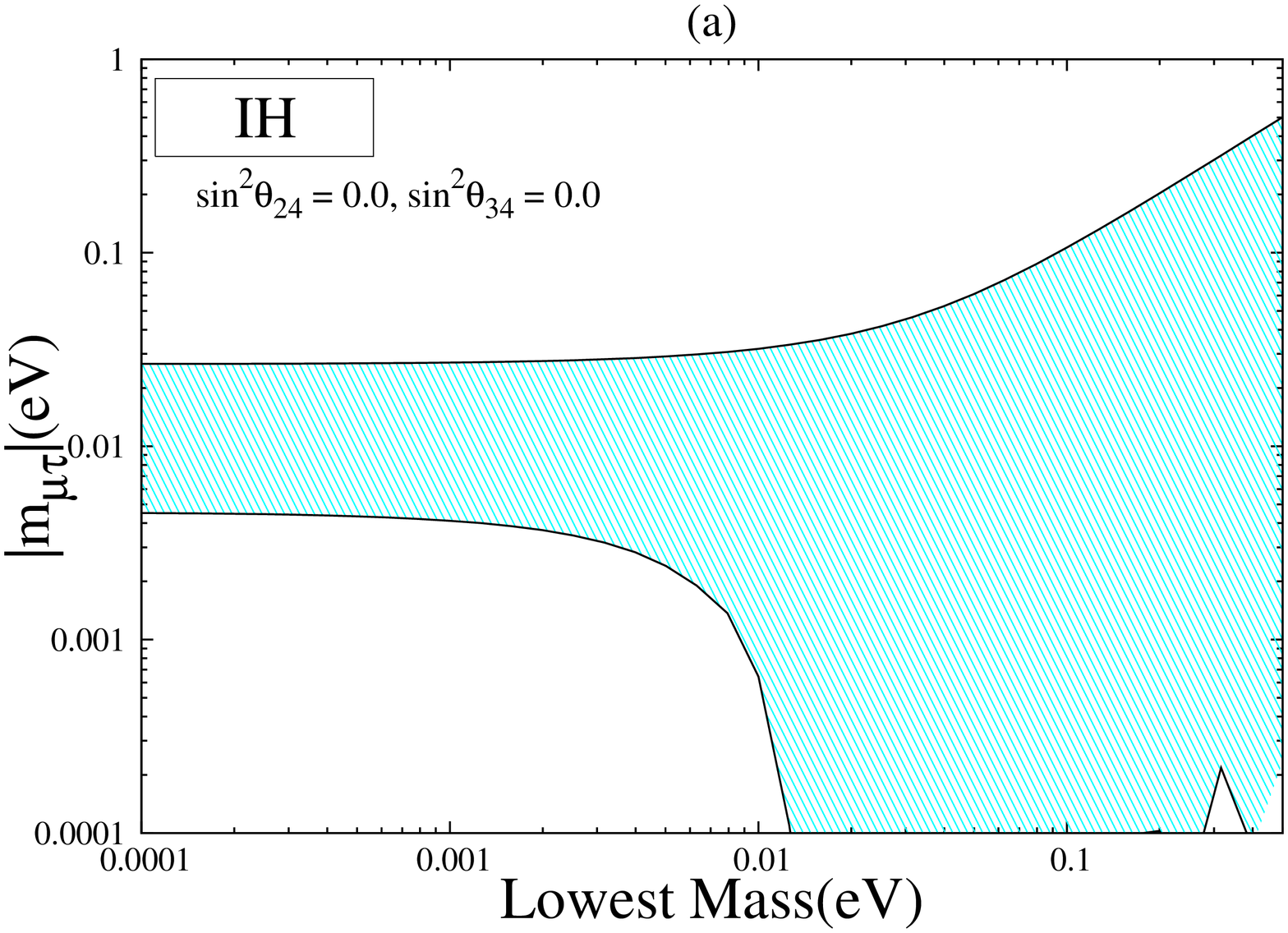}
\includegraphics[width=0.33\textwidth,angle=270]{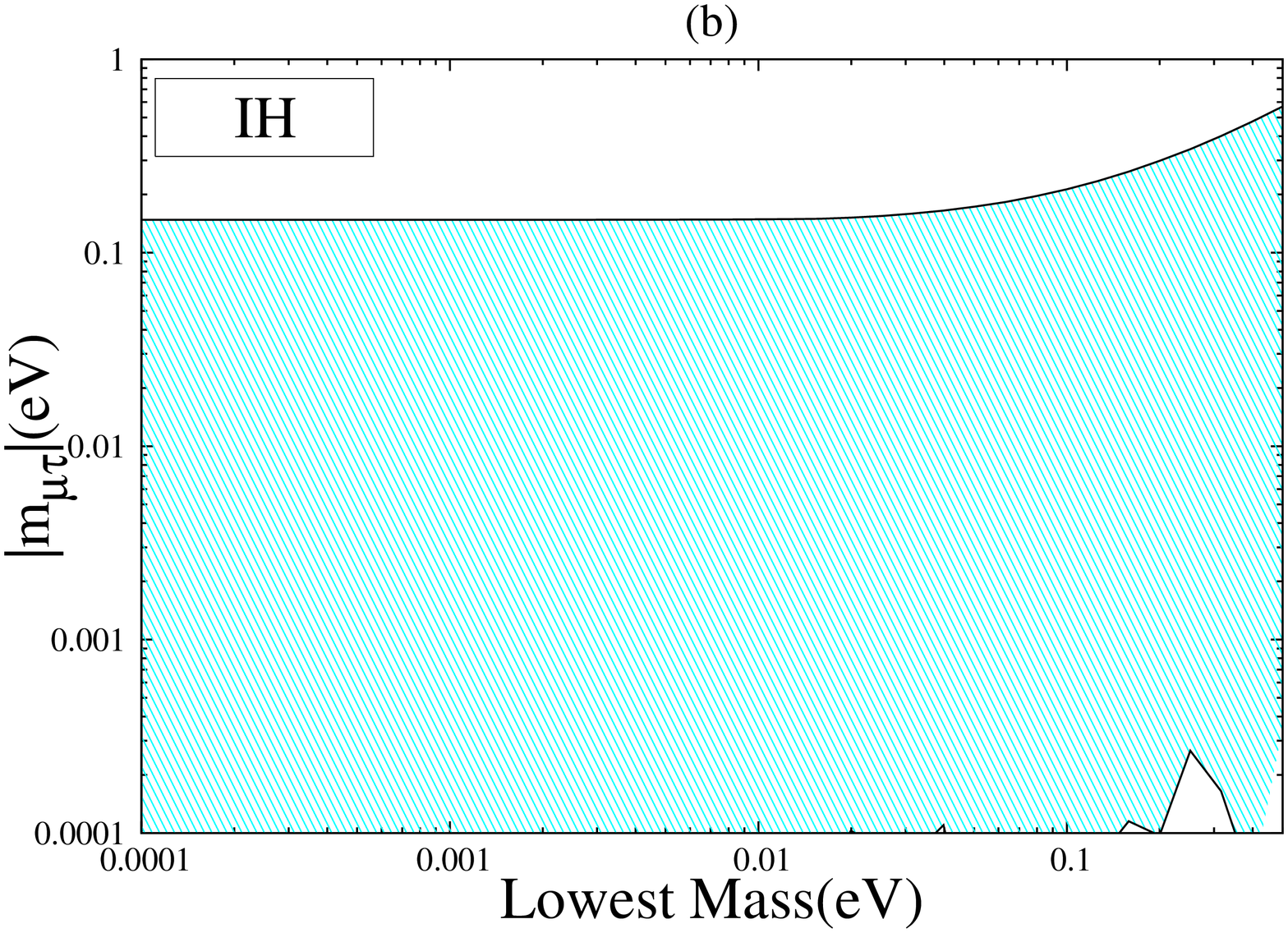}\\
\includegraphics[width=0.33\textwidth,angle=270]{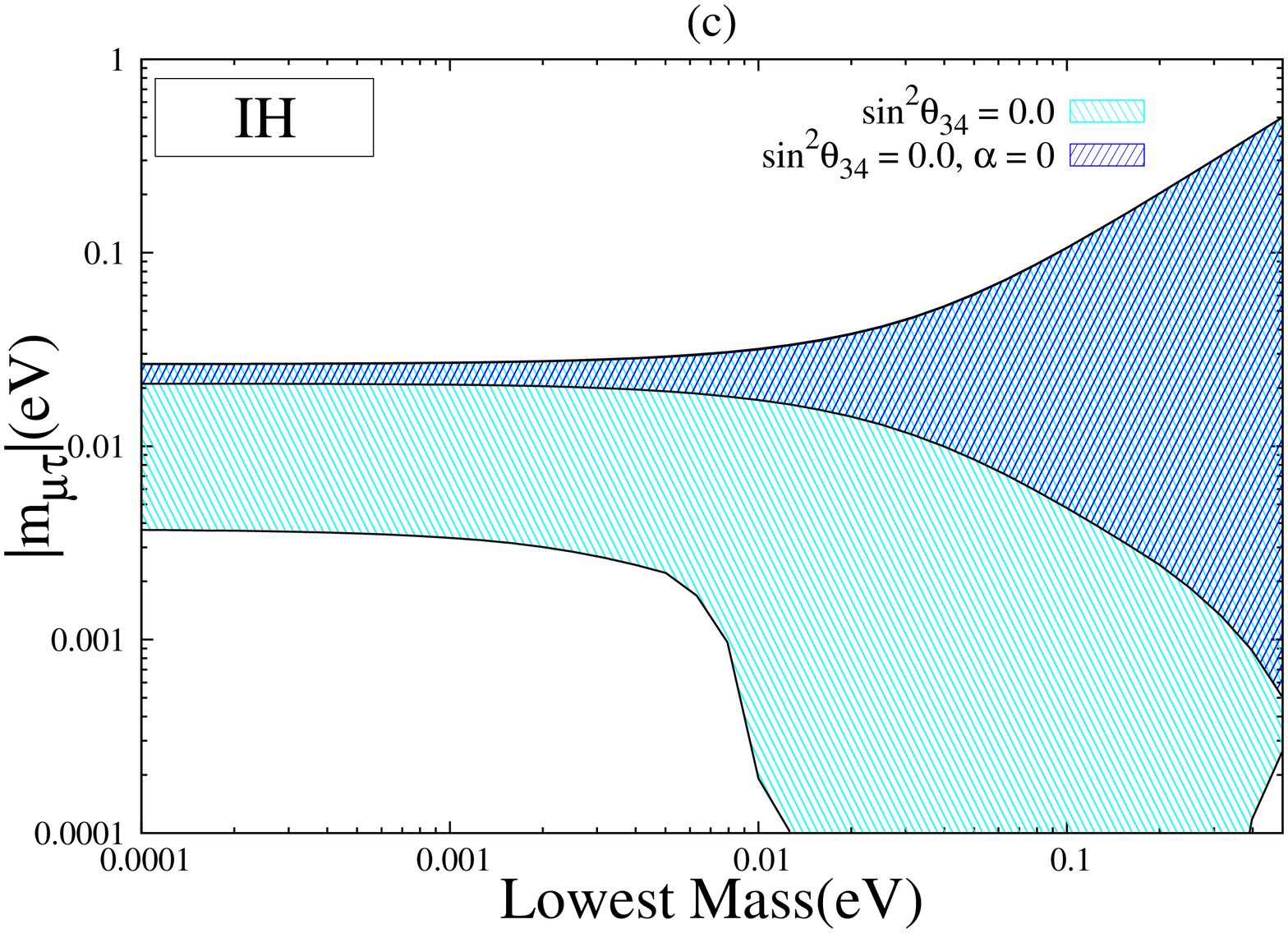}
\includegraphics[width=0.33\textwidth,angle=270]{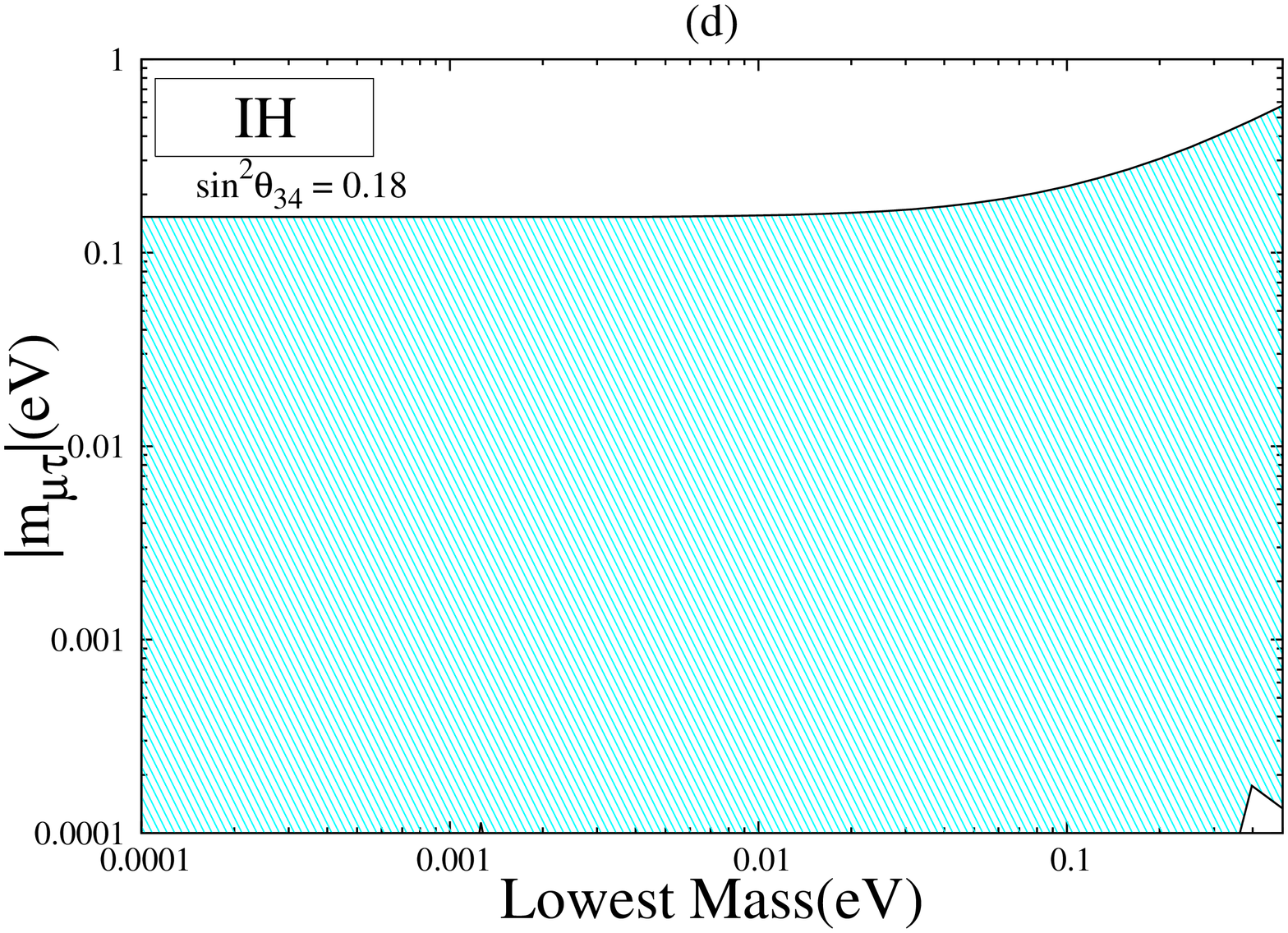}
\caption[Plots of $|m_{\mu\tau}|$ vs $m_3$ for inverted hierarchy.]{Plots of vanishing $m_{\mu\tau}$ for inverted hierarchy (a) for vanishing $\theta_{34}$ and
$\theta_{24}$ (3 generation). In panel (b) all parameters are varied in their full allowed range (3+1).
Panel (c) and (d) are for specific values of $\theta_{34}$ and $\alpha$ when all other mixing angles are varied in their full range.}
\label{fig11}
\end{center}
\end{figure}

For the case of inverted hierarchy where $m_3$ can have very small values, $m_{\mu\tau}$ becomes
\bea
 |m_{\mu \tau}| &\approx&|\sqrt{\Delta_{31}}\{ -c_{23} c_{34} s_{23}(c_{12}^2 e^{2 i \alpha} + s_{12}^2) \\ \nonumber
 &+& \lambda [c_{12} s_{12} (1 - e^{2 i \alpha})(c_{34} \cos2\theta_{23} e^{i \delta_{13}} \chi_{13} + c_{23} s_{34} e^{i \delta_{14}} \chi_{14}) \\ \nonumber
 &+& s_{34}\{e^{i(2 \gamma+ 2 \delta_{14} - \delta_{24})} \sqrt{\xi} - c_{23}^2 e^{i \delta_{24}}(s_{12}^2 + c_{12}^2 e^{2 i \alpha})\}\chi_{24}] \\ \nonumber
 &+& \lambda^2 [c_{23} c_{34} s_{23} e^{ 2 i \delta_{13}}(c_{12}^2 + e^{2 i \alpha} s_{12}^2) \chi_{13}^2 \\ \nonumber
 &+& c_{12} s_{12} s_{23} (e^{2 i \alpha} - 1)(c_{34} e^{i(\delta_{14} - \delta_{24})} \chi_{14} + 2 s_{34} c_{23} e^{i(\delta_{13} + \delta_{24})} \chi_{13})\chi_{24}]\}|.
\eea
To get an idea about the magnitude of the terms we take  vanishing Majorana phases and
Dirac CP phases to be of the order $\pi$. The expression in this case for vanishing $m_{\mu \tau}$ becomes
\bea
&&-c_{23}c_{34}s_{23}+\lambda(s_{34}\chi_{24}(c_{23}^2-\sqrt{\xi})) \\ \nonumber
&-&\lambda^2(-s_{34}\chi_{13}\chi_{14}-c_{23}c_{34}\chi_{13}^2)=0
\eea
In panel (a) of Fig. \ref{fig11},
where $|m_{\mu \tau}|$ for 3 generation is plotted,
we can see that unlike $m_{\mu \mu}$ there is
no cancellation in small $m_3$ region but when plotted for the full range it gets contribution from the sterile part
and there is cancellation for the full range of $m_3$ (panel (b)).
Clearly the cancellation of the terms do not become possible for small values of $\theta_{34}$ in strict hierarchical
region. This case is similar to the three generation case in IH (cyan/light region, panel (c)). This is because for $s_{34}^2 = 0$
the contribution of $s_{24}^2$ comes from the $\lambda^2$ term. If we put the CP violating phase $\alpha$ as zero then
cancellation is not possible for whole range of $m_3$ (blue/dark region panel (c)). However,
as the value of $s_{34}^2$ increases all the terms in the above equation becomes of the same order and
cancellation for very small values of $m_3$ is possible (panel (d)).

\subsubsection{The Mass Matrix Element $m_{\tau\tau}$}
This element is related to $m_{\mu\mu}$ by the $\mu-\tau$ symmetry.
As discussed earlier, in the limit when $\theta_{24}$ and $\theta_{34}$ are not very large,
the two mixing angles $\theta_{34}$ and $\theta_{24}$ will behave in the same way
in the textures related by $\mu-\tau$ symmetry.
The (3,3) element of the neutrino mass matrix in the presence of one sterile neutrino is given as
\bea
  m_{\tau \tau} &=& e^{2i(\delta_{14} + \gamma)} c_{14}^2 c_{24}^2 m_4 s_{34}^2 \\ \nonumber
  &+& e^{2i(\delta_{13} + \beta)} m_3 \{e^{i(\delta_{14} - \delta_{13})} c_{24} s_{13} s_{14} s_{34} +
  c_{13}(c_{23} c_{34} - e^{i \delta_{24}} s_{23} s_{24} s_{34})\}^2 \\ \nonumber
  &+& m_1[-s_{12}(-c_{34} s_{23} - e^{i \delta_{24}} c_{23} s_{24} s_{34}) \\ \nonumber
  &+& c_{12}\{-e^{i\delta_{14}} c_{13} c_{24} s_{14} s_{34} - e^{i \delta_{13}} s_{13}(c_{23} c_{34}
  - e^{i \delta_{24}} s_{23} s_{24} s_{34})\}]^2 \\ \nonumber
  &+& e^{2 i \alpha} m_2 [c_{12}(-c_{34} s_{23} - e^{i \delta_{24}} c_{23} s_{24} s_{34}) \\ \nonumber
  &+& s_{12}\{-e^{i \delta_{14}} c_{13} c_{24} s_{14} s_{34} - e^{i \delta_{13}} s_{13}(c_{23} c_{34}
  - e^{i \delta_{24}} s_{23} s_{24} s_{34})\}]^2.
 \eea
 It reduces to the 3 generation case for $\theta_{34} = 0$.
Using the approximation for normal hierarchy in Eqs. \ref{xnh}, \ref{chi1} and \ref{chi2} this becomes
\bea
|m_{\tau\tau}|&\approx& |\sqrt{\Delta_{32}}\{c_{23}c_{34}s_{23}(e^{2i\beta+\delta_{13}}-c_{12}^2\sqrt{\zeta}e^{2i\alpha})
+\lambda\{-e^{i(2\alpha+\delta_{13})}\sqrt{\zeta}\\ \nonumber
&&s_{12}c_{12}c_{34}\cos2\theta_{23}\chi_{13}-\sqrt{\zeta}c_{12}c_{23}s_{34}e^{2i\alpha}(s_{12}\chi_{14}e^{i\delta_{14}} +c_{12}c_{23}\chi_{24}
e^{2i\delta_{24}}) \\ \nonumber 
&+& s_{34}\chi_{24}(-s_{23}^2e^{2i(\beta + \delta_{13})+
i\delta_{24}}\\ \nonumber
&-&\sqrt{\xi}e^{2i(\gamma + \delta_{14})-i\delta_{24}})\}+\lambda^2\{\sqrt{\zeta} s_{12}^2s_{23}\chi_{13}e^{i(2\alpha +
\delta_{13})}(c_{23}c_{34}\chi_{13}e^{i\delta_{13}}+s_{34}\chi_{14}e^{i\delta_{14}})\\ \nonumber
&+&\sqrt{\zeta}c_{12}s_{12}s_{23}\chi_{24}e^{2i\alpha}(2c_{23}s_{34}\chi_{13}e^{i(\delta_{13}+\delta_{24})}+c_{34}\chi_{14}e^{i(\delta_{14}-\delta_{24})})\\ \nonumber
&-&s_{23}s_{34}\chi_{13}\chi_{14}e^{i(2\beta+\delta_{13}+\delta_{14})}\}\}|.
\eea
To get an idea of the order of the terms we consider the vanishing Majorana phases and the Dirac phases having the value equal to $\pi$. This element vanishes when
\bea
 && c_{23}c_{34}s_{23}(1-c_{12}^2\sqrt{\zeta}) + \\ \nonumber
 && \lambda\{\sqrt{\zeta}s_{12}c_{12}c_{34}\cos2\theta_{23}\chi_{13}
+\sqrt{\zeta}c_{12}c_{23}s_{34}(s_{12}\chi_{14} \\ \nonumber
&+& c_{12}c_{23}\chi_{24})-s_{34}\chi_{24}(s_{23}^2-\sqrt{\xi})\}+\lambda^2\{\sqrt{\zeta} s_{12}^2s_{23}\chi_{13}(c_{23}c_{34}\chi_{13}+s_{34}\chi_{14})\\ \nonumber
&+&\sqrt{\zeta}c_{12}s_{12}s_{23}\chi_{24}(2c_{23}s_{34}\chi_{13}+c_{34}\chi_{14})+s_{23}s_{34}\chi_{13}\chi_{14}\}=0.
\eea

\begin{figure}[ht!]
\begin{center}
\includegraphics[width=0.33\textwidth,angle=270]{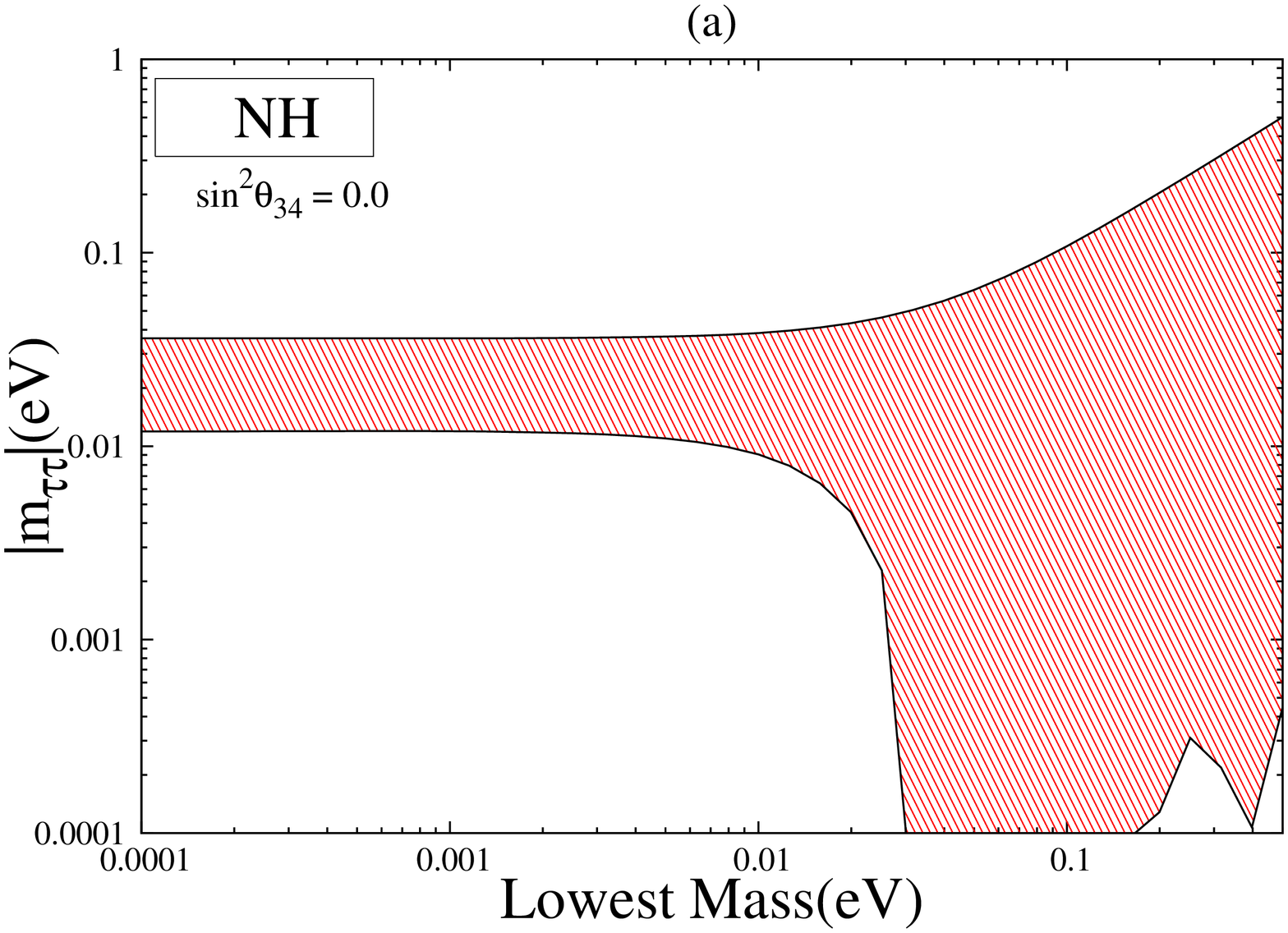}
\includegraphics[width=0.33\textwidth,angle=270]{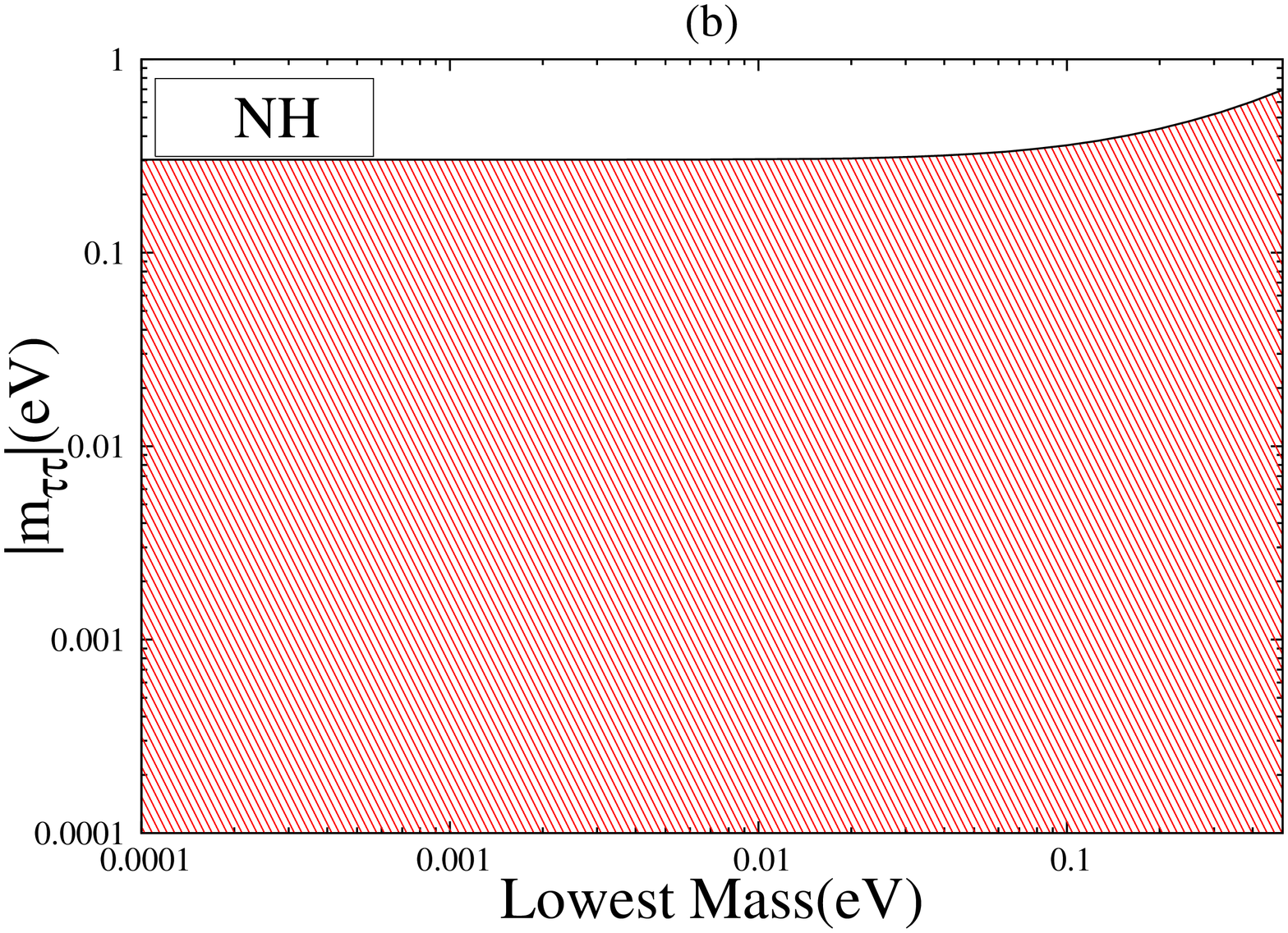}\\
\includegraphics[width=0.33\textwidth,angle=270]{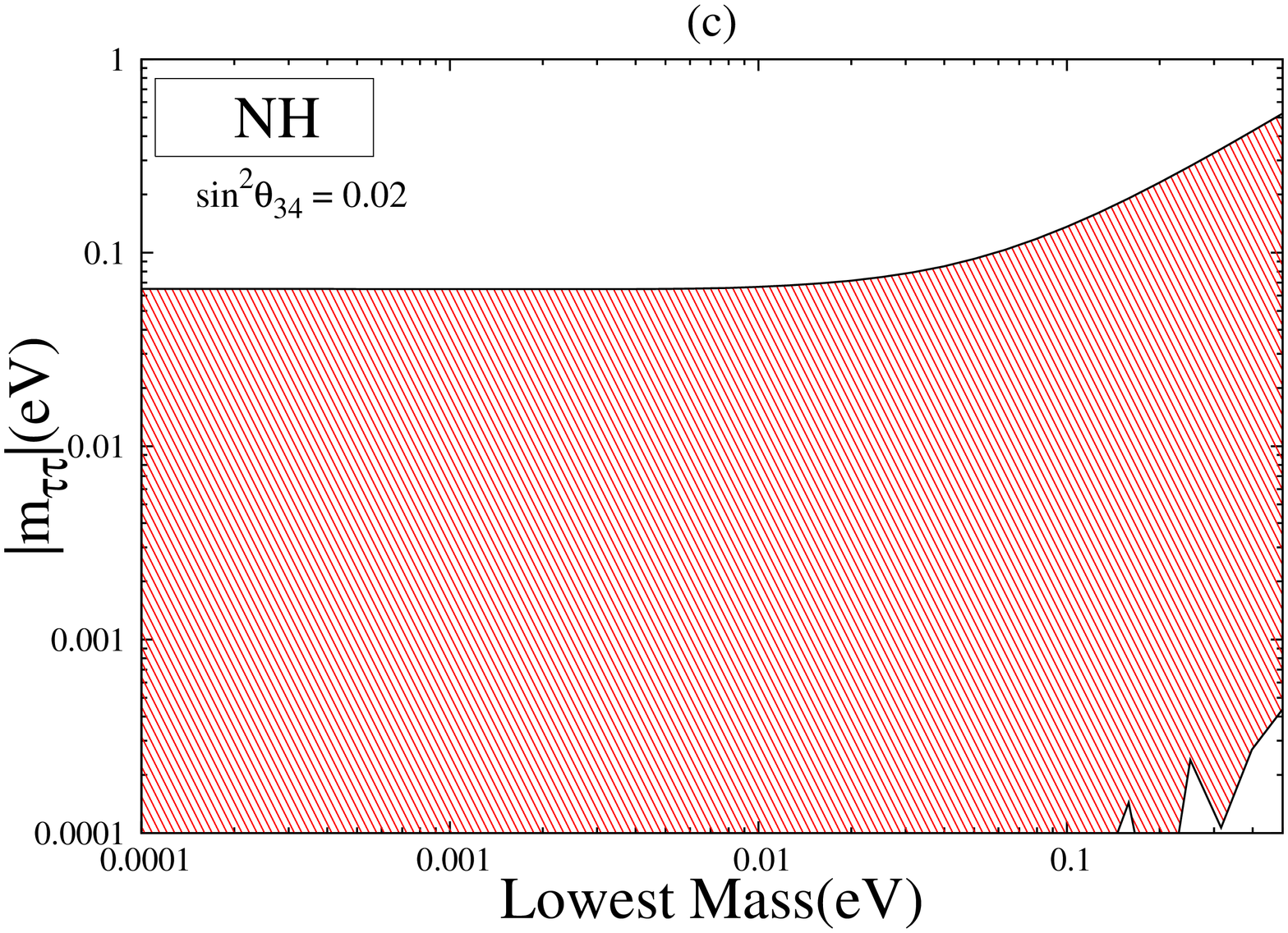}
\includegraphics[width=0.33\textwidth,angle=270]{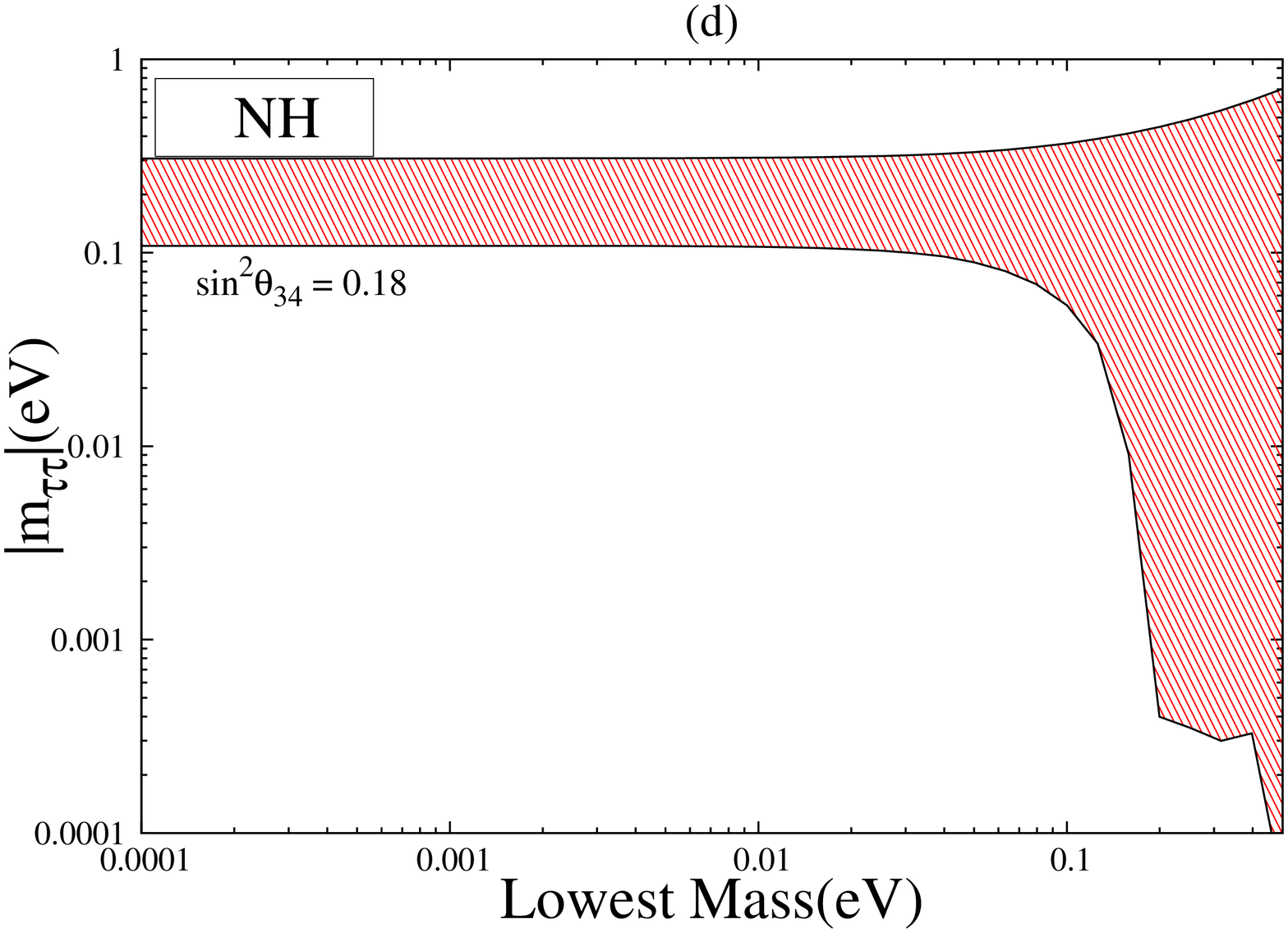}
\caption[Plots of $|m_{\tau\tau}|$ vs $m_1$ for normal hierarchy.]{Plots of vanishing $m_{\tau\tau}$ for normal hierarchy with lowest mass $m_1$.
Panel (a) correspond to three generation case. In panel (b) all the mixing angles are
varied in their full allowed range of parameters (3+1). Panel (c) and (d) are for some specific values of $\theta_{34}$.}
\label{fig12}
\end{center}
\end{figure}

For vanishing $\theta_{34}$, which is the case for 3 generation, $m_{\tau \tau} = 0$ is disallowed for small $m_1$ as can be seen from panel (a) of Fig. \ref{fig12}.
This is the generic behaviour of a element belonging to the $\mu -\tau$ block in normal hierarchy which
we mentioned previously. This is because for $\theta_{34}$ equal to zero the leading
order term is large ($\mathcal{O}$ (10$^{-1}$)). Here the term with $\lambda^2$ is quite small
(10$^{-3}$-10$^{-4}$) and hence will not have very significant role to play.
Thus, only terms with coefficient $\lambda$ can cancel the leading order term.
However, for vanishing $\theta_{34}$ this term is small
$\mathcal{O}$ (10$^{-3}$), and cannot cancel the leading order term.
In panel (b) when all the parameters are varied in their $3 \sigma$ range we can see that cancellation is possible over the whole range of $m_1$ (3+1 case).
Now, when $\theta_{34}$ starts increasing from its lowest value there exist a region for intermediate values where both the terms become approximately
of the same order and hence there can be cancellations (panel (c)).
Towards very large values of $\theta_{34}$ the term with coefficient $\lambda$ becomes larger than
the leading order term due to which this element cannot vanish.
For the cancellation very large values of $m_1$ is required as can be seen from panel (d) of Fig \ref{fig12}.
\begin{figure}[ht!]
\begin{center}
\includegraphics[width=0.33\textwidth,angle=270]{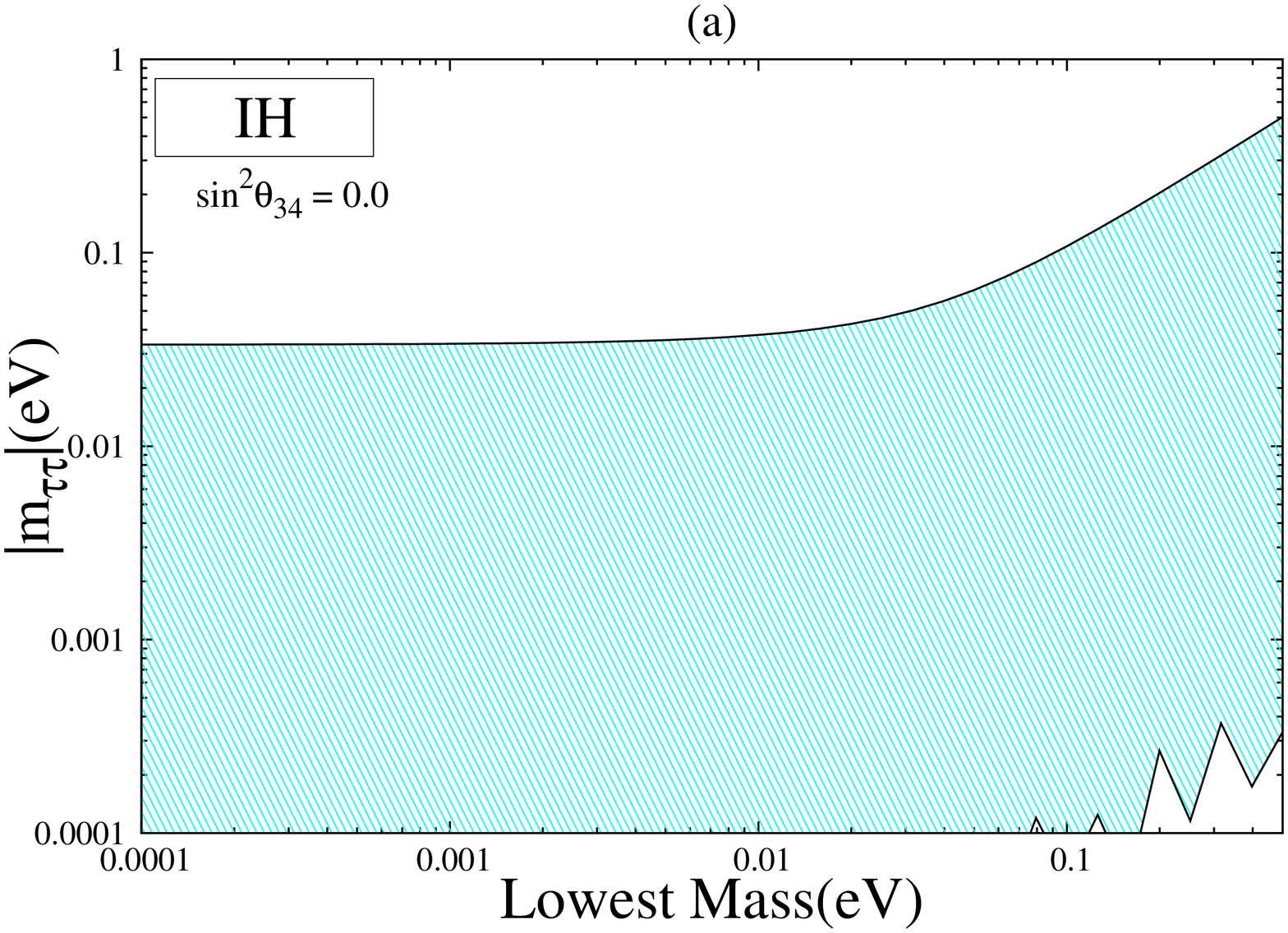}
\includegraphics[width=0.33\textwidth,angle=270]{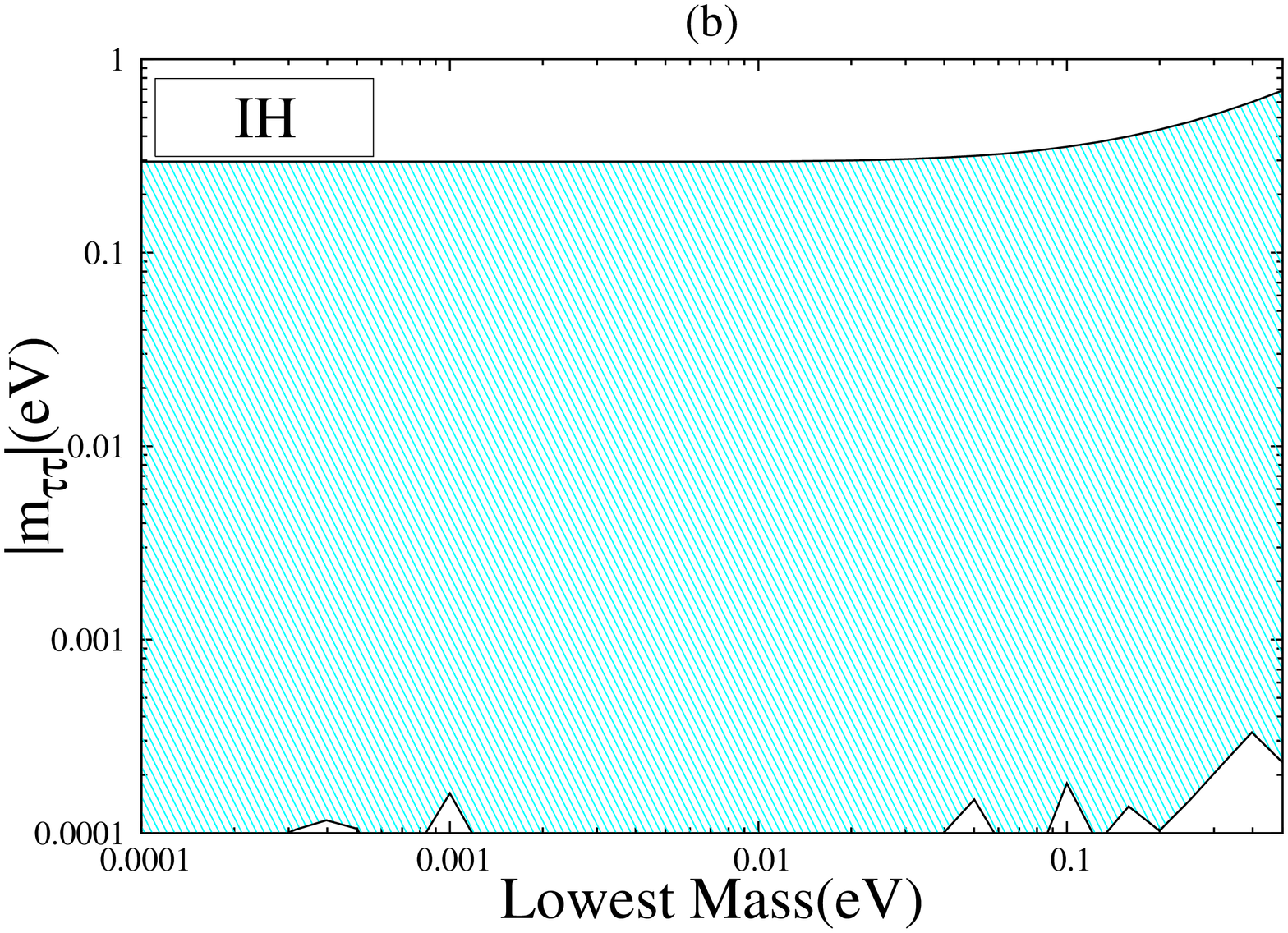}\\
\includegraphics[width=0.33\textwidth,angle=270]{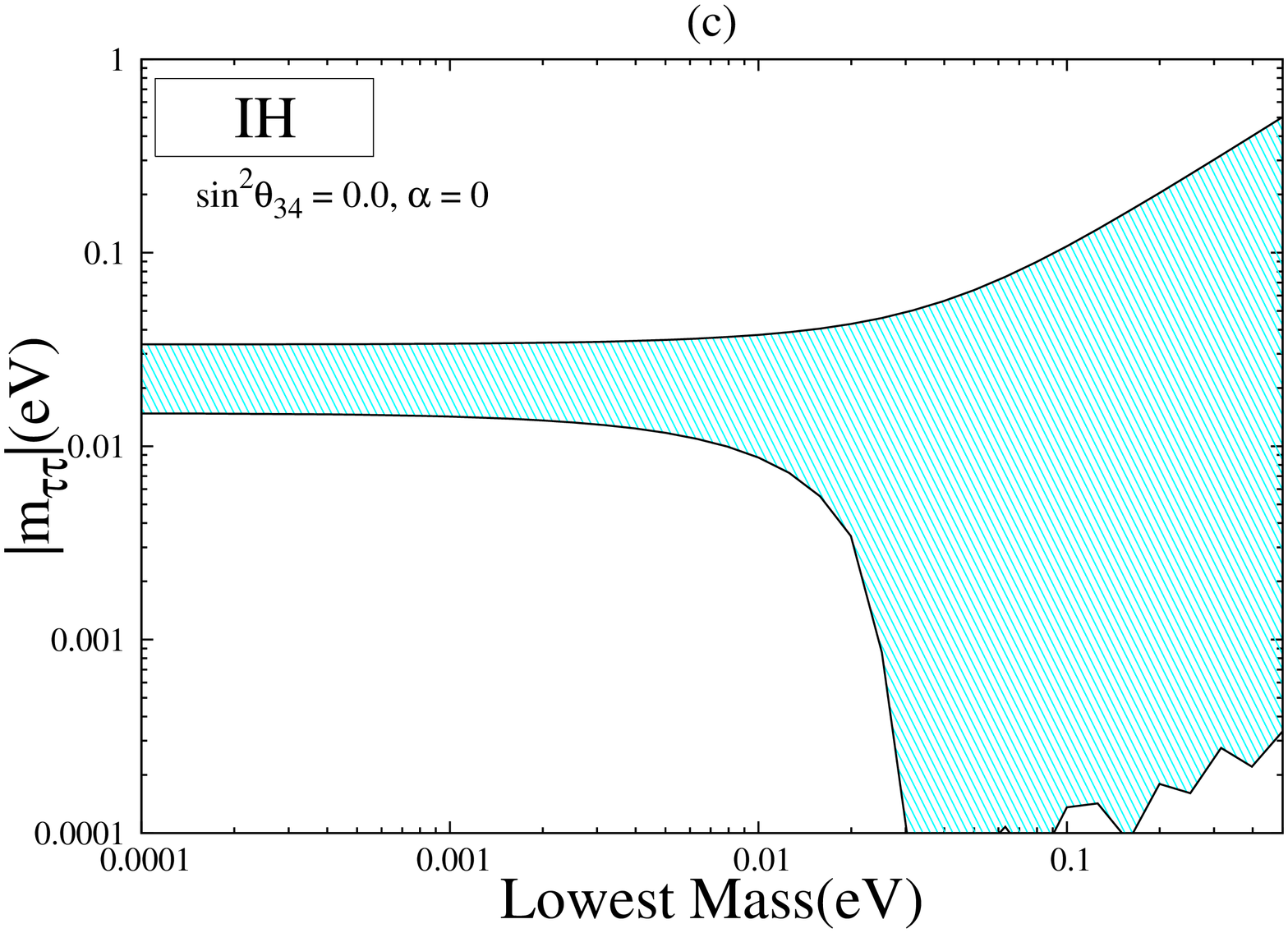}
\includegraphics[width=0.33\textwidth,angle=270]{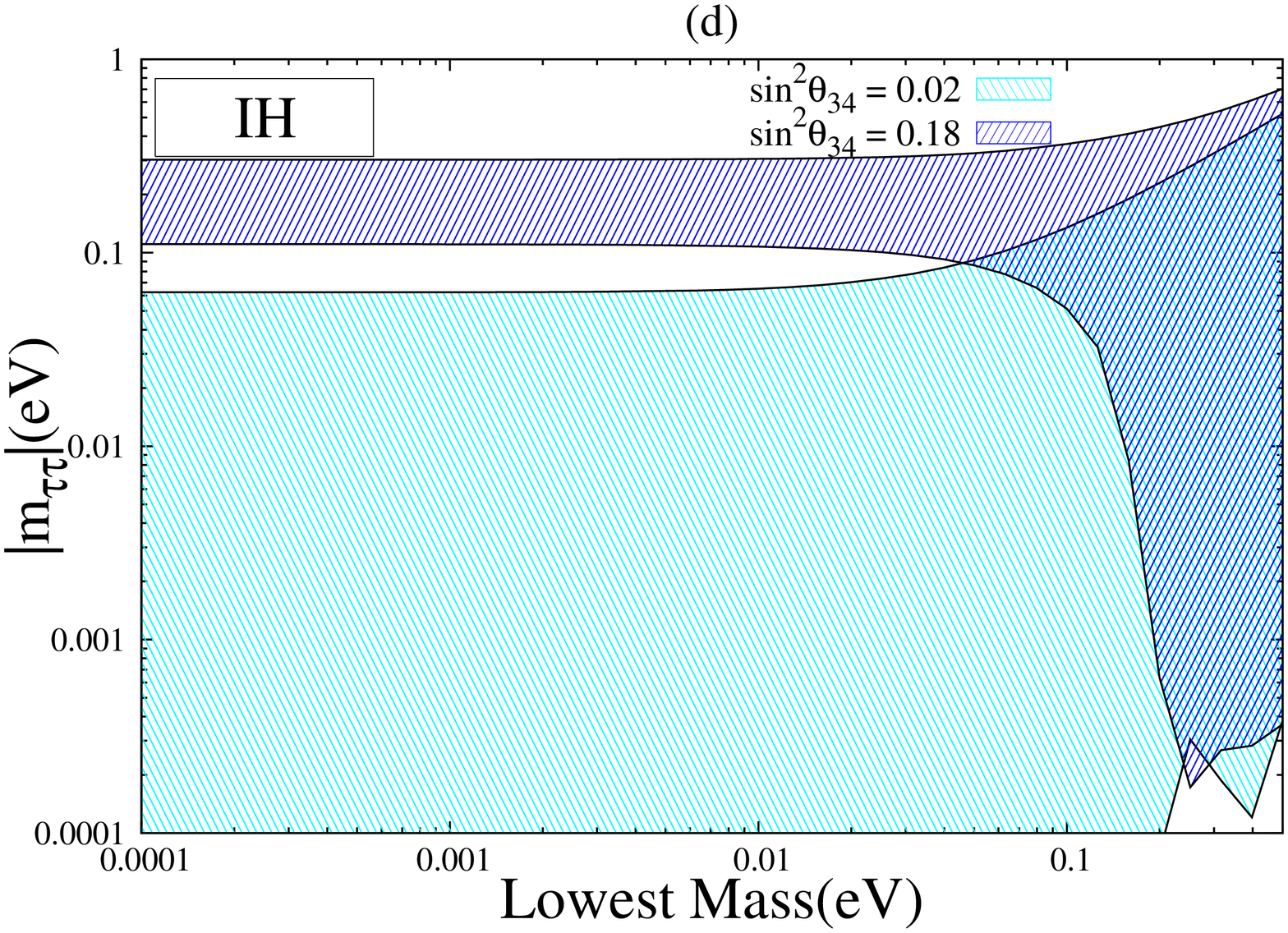}
\caption[Plots of $|m_{\tau\tau}|$ vs $m_3$ for inverted hierarchy.]{Plots of vanishing $m_{\tau\tau}$ for inverted hierarchy with lowest mass $m_3$. Panel (a) correspond
to three generation case. In panel (b) all the mixing angles are varied in their full allowed range of parameters
(3+1). Panel (c) and (d) are for some specific values of $\theta_{34}$ and $\alpha$.}
\label{fig13}
\end{center}
\end{figure}
For the case of inverted hierarchy where $m_3$ approaches small values we get the expression
\bea
 m_{\tau \tau} &\approx& c_{34}^2 s_{23}^2(c_{12}^2 e^{2 i \alpha} + s_{12}^2) + e^{2 i (\delta_{14} + \gamma)} \sqrt{\xi} s_{34}^2 \\ \nonumber
 &+& 2 \lambda[(e^{2 i \alpha} - 1)c_{12} c_{34} s_{12} s_{23}(c_{23} c_{34} s_{12} s_{23}(c_{23} c_{34} e^{ 2 i \delta_{13}} \chi_{13} + e^{ 2 i \delta_{14}} s_{34} \chi_{14}) \\ \nonumber
 &+& 2 c_{23} c_{34} e^{ 2 i \delta_{24}} s_{23} s_{34}(c_{12}^2 e^{ 2 i \alpha} + s_{12}^2) \chi_{24}] \\ \nonumber
 &+& \lambda^2[(c_{12}^2 + e^{ 2 i \alpha} s_{12}^2)\{c_{23} c_{34} \chi_{13} e^{ i \delta_{13}}(c_{23} c_{34} \chi_{13} e^{ i \delta_{13}} + 2 \chi_{14} s_{34}
 e^{ i \delta_{14}}) + e^{ 2 i \delta_{14}} \chi_{14}^2 s_{34}^2\} \\ \nonumber
 &+& (c_{12}^2 e^{ 2 i \alpha} + s_{12}^2)c_{23}^2 e^{ 2 i \delta_{24}} \chi_{24}^2 s_{34}^2 \\ \nonumber
 &+& 2 s_{12}(e^{2 i \alpha} - 1) e^{ i \delta_{24}}(c_{34} \chi_{13} \cos2\theta_{23} e^{ i \delta_{13}} + c_{12} c_{23} \chi_{14} s_{34}) s_{34} \chi_{24}].
\eea
For vanishing Majorana CP phases and Dirac phases having the value equal to $\pi$ this expression becomes
\bea
m_{\tau \tau} &\approx& -c_{23}c_{34}s_{23}+\lambda s_{34}\chi_{24}(c_{23}^2-\sqrt{\xi}) \\ \nonumber
&+& \lambda^2s_{23}\chi_{13}(c_{23}c_{34}\chi_{13}+s_{34}\chi_{14})
\eea
In panel (a) of Fig. \ref{fig13} we reproduced the 3 generation behaviour by plotting
$|m_{\tau \tau}|$ for $s_{34}^2 = 0$ and in panel (b) all the parameters are varied randomly (3+1).
In both the cases we can see that cancellations are possible for the whole range of $m_3$.
For $s_{34}^2 = 0$ all the terms are of same order and cancellations are always possible. But if we put $\alpha = 0$ then one term with coefficient $\lambda$
 and another term with coefficient $\lambda^2$ drops out from the equation and then small values of $s_{34}^2$ can not cancel the leading order term any more.
 This can be seen from panel (c) where cancellation is not possible for lower $m_3$ region.
However when $s_{34}^2$ increases to a value of about 0.02 this element can vanish (panel (d) the cyan region).
We see that when $\theta_{34}$ increase towards its upper bound the $\lambda$ term becomes large $\mathcal{O} (1)$. Hence, the other terms are
not able to cancel this term and we do not get small $m_3$ region allowed (Panel (d), blue region).

\subsubsection{The Mass Matrix Elements $m_{es}$, $m_{\mu s}$, $m_{\tau s}$ and $m_{ss}$ }
The elements $m_{es}$, $m_{\mu s}$, $m_{\tau s}$ and $m_{ss}$ are present in the fourth row and fourth column in the neutrino mass matrix.
They are the new elements that arises in 3+1 scenario due to the addition of one light sterile neutrino.
The expressions for $m_{es}$, $m_{\mu s}$ and $m_{\tau s}$ are given by
\begin{figure}[ht!]
\begin{center}
\includegraphics[width=0.33\textwidth,angle=270]{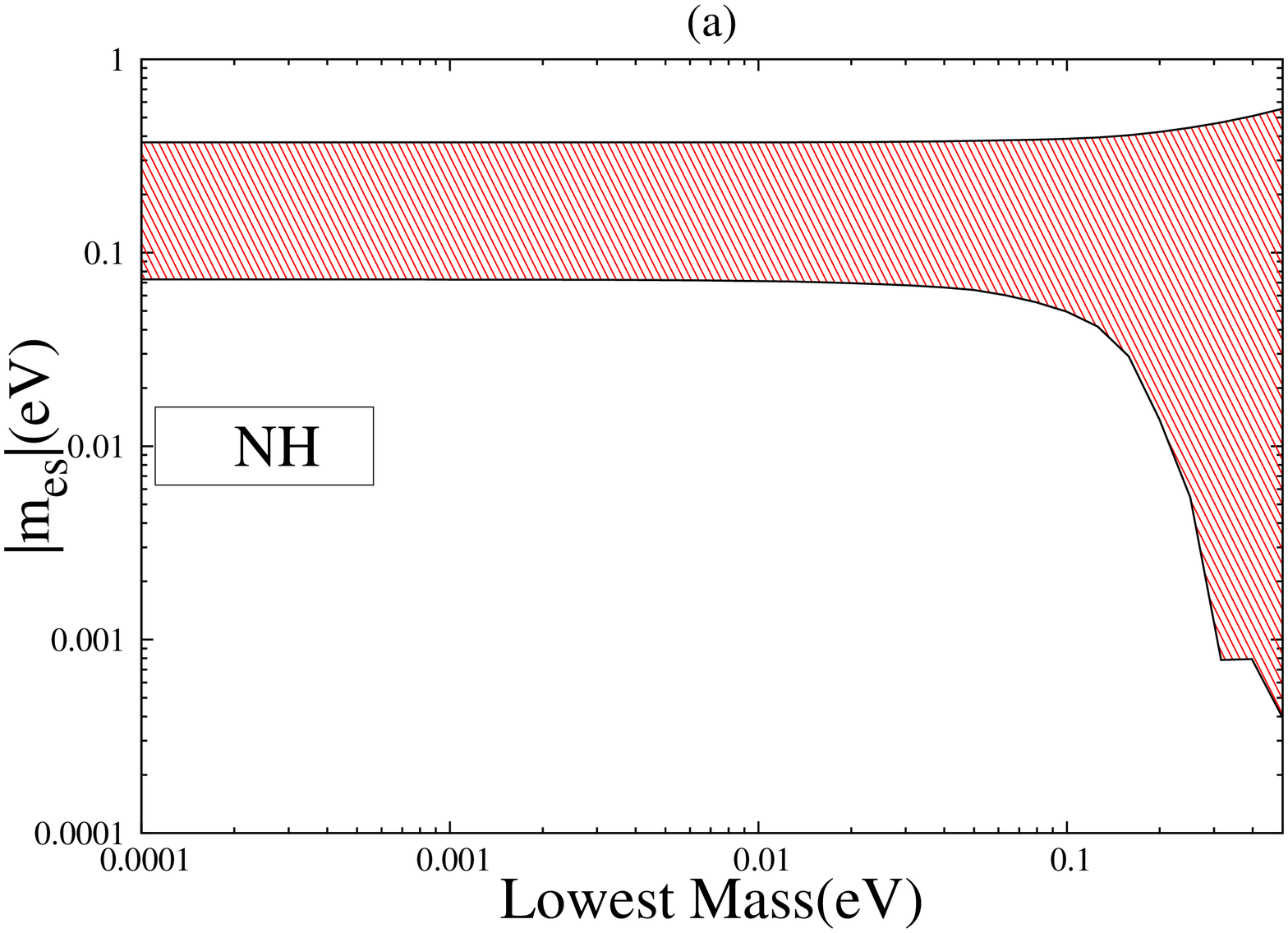}
\includegraphics[width=0.33\textwidth,angle=270]{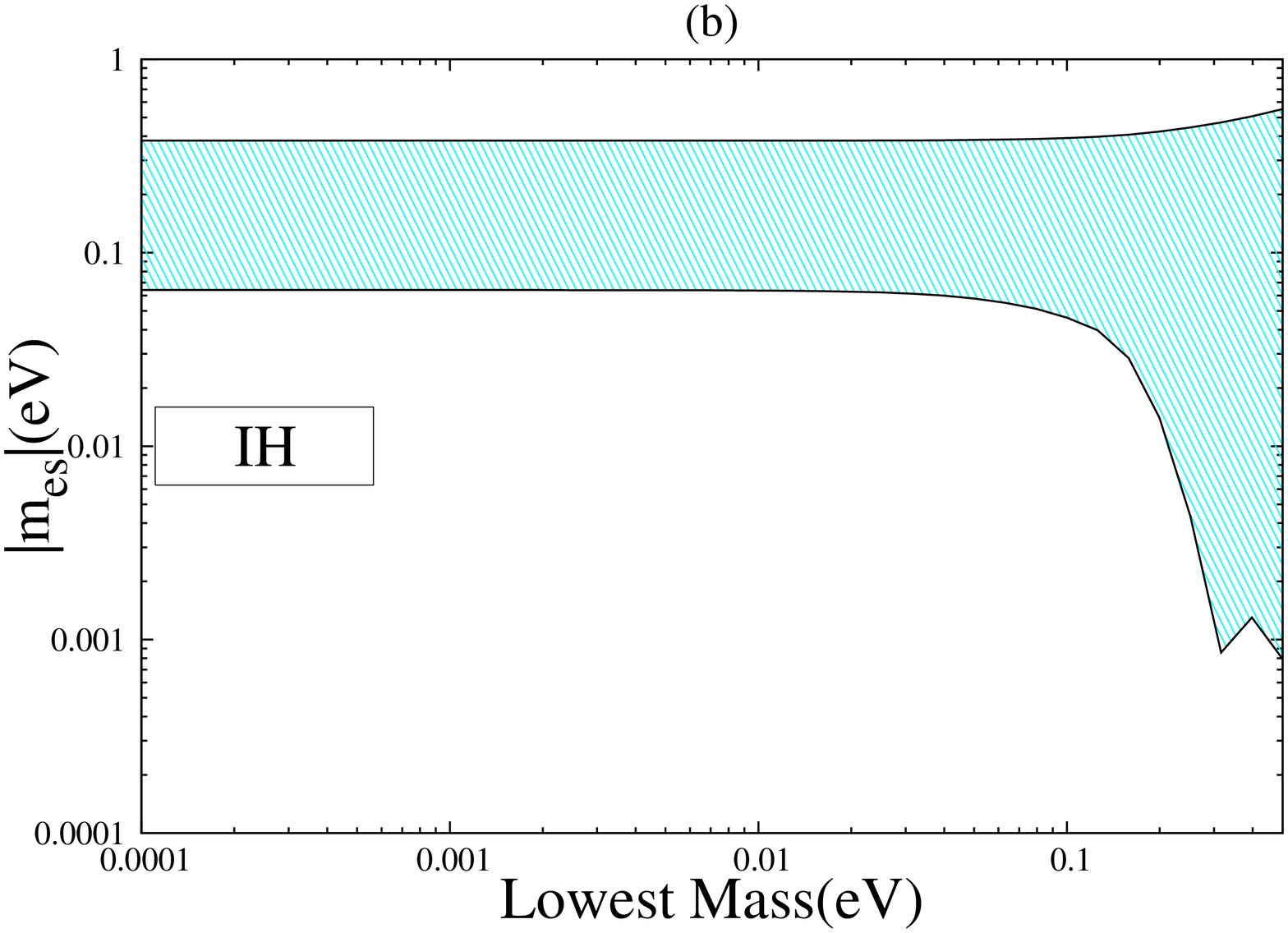} \\
\includegraphics[width=0.33\textwidth,angle=270]{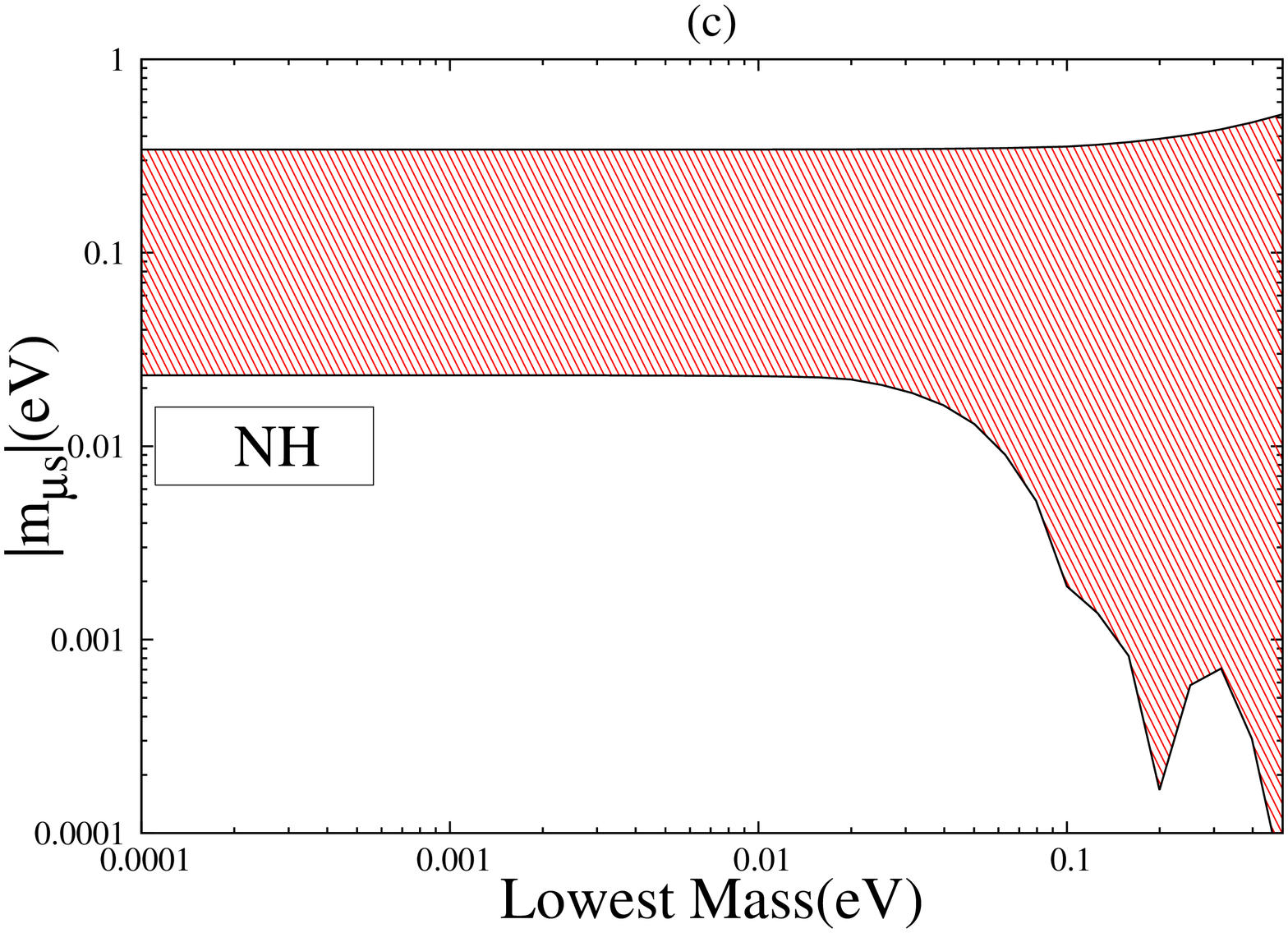}
\includegraphics[width=0.33\textwidth,angle=270]{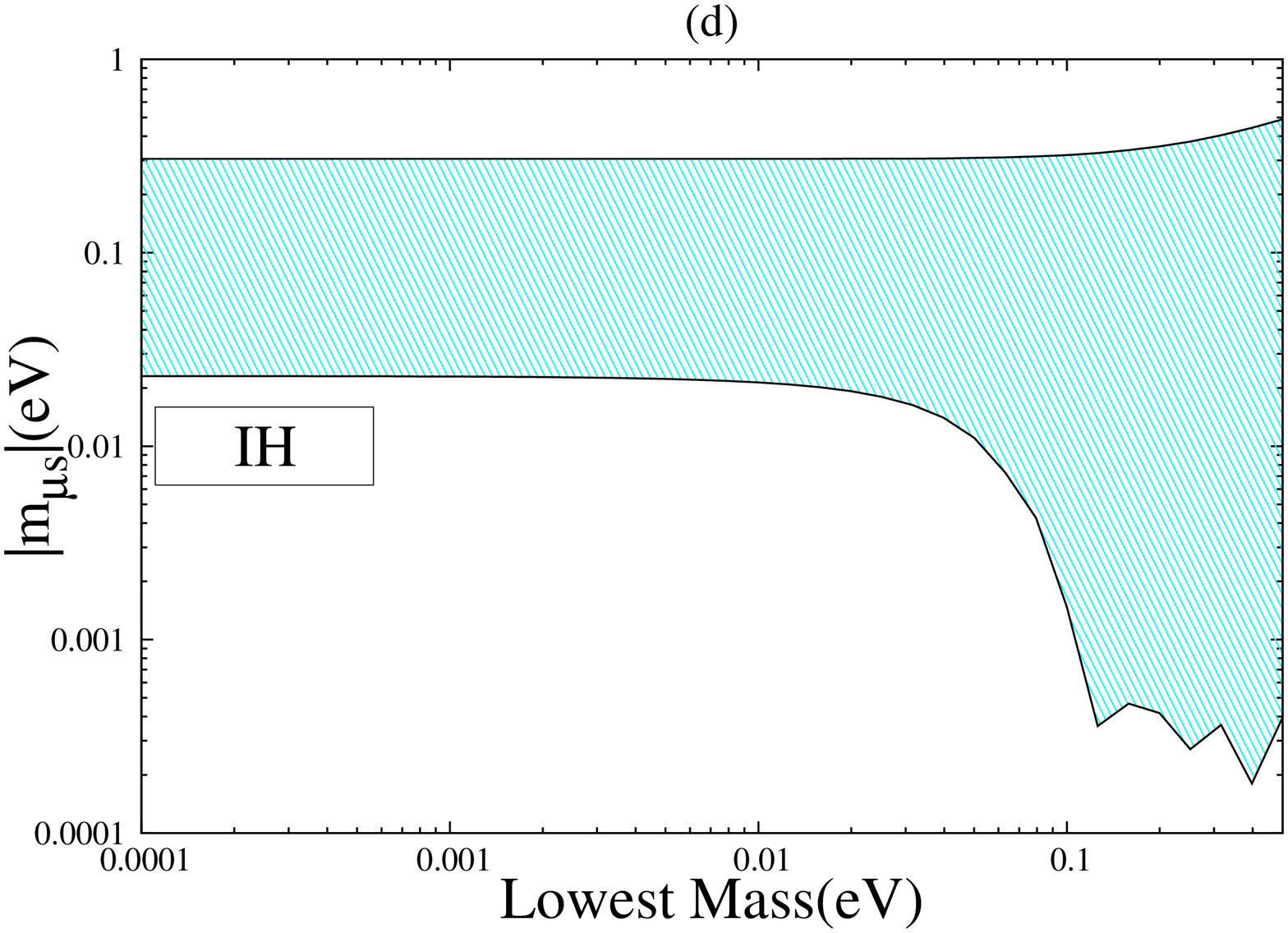} \\
\includegraphics[width=0.33\textwidth,angle=270]{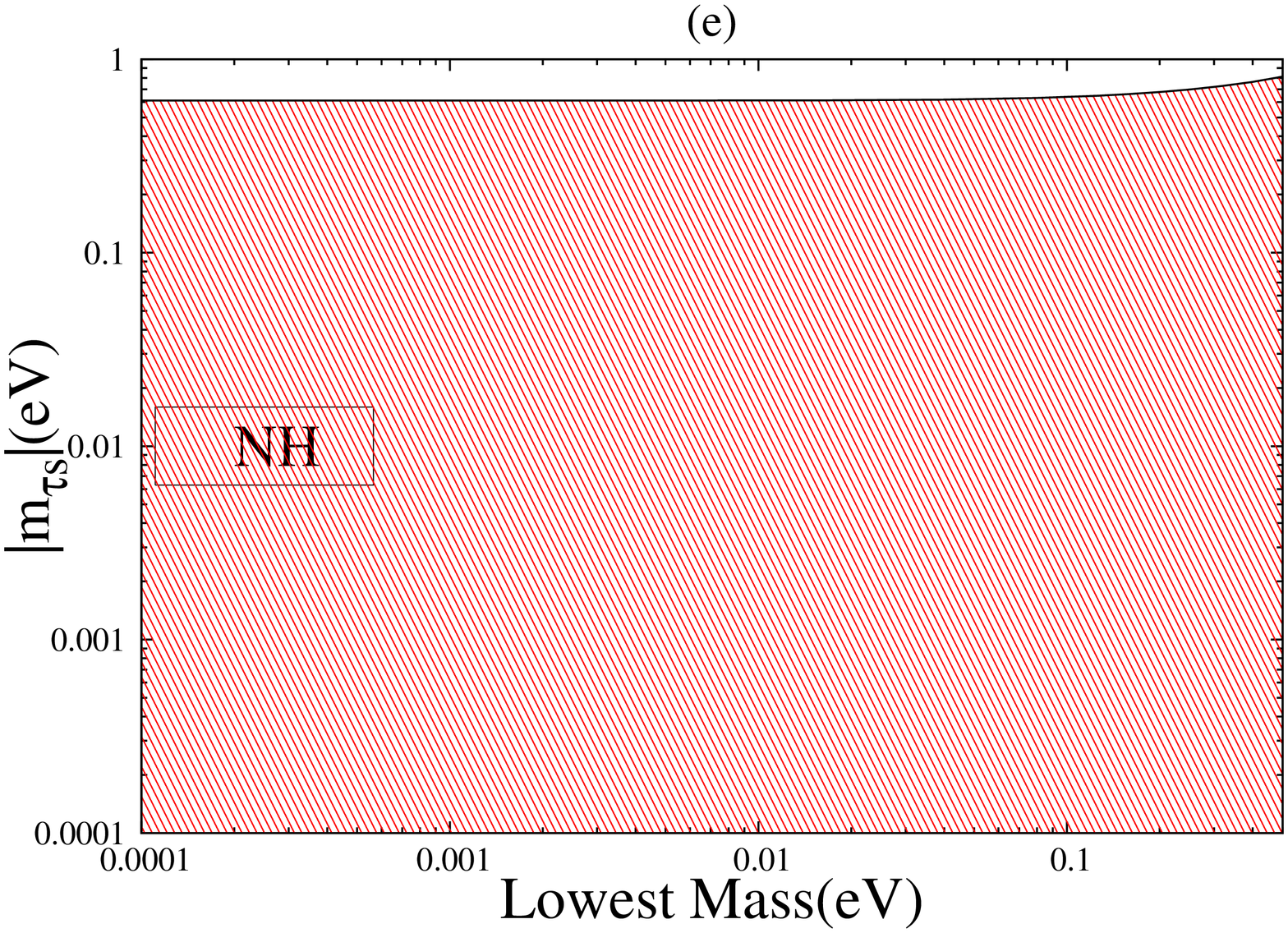}
\includegraphics[width=0.33\textwidth,angle=270]{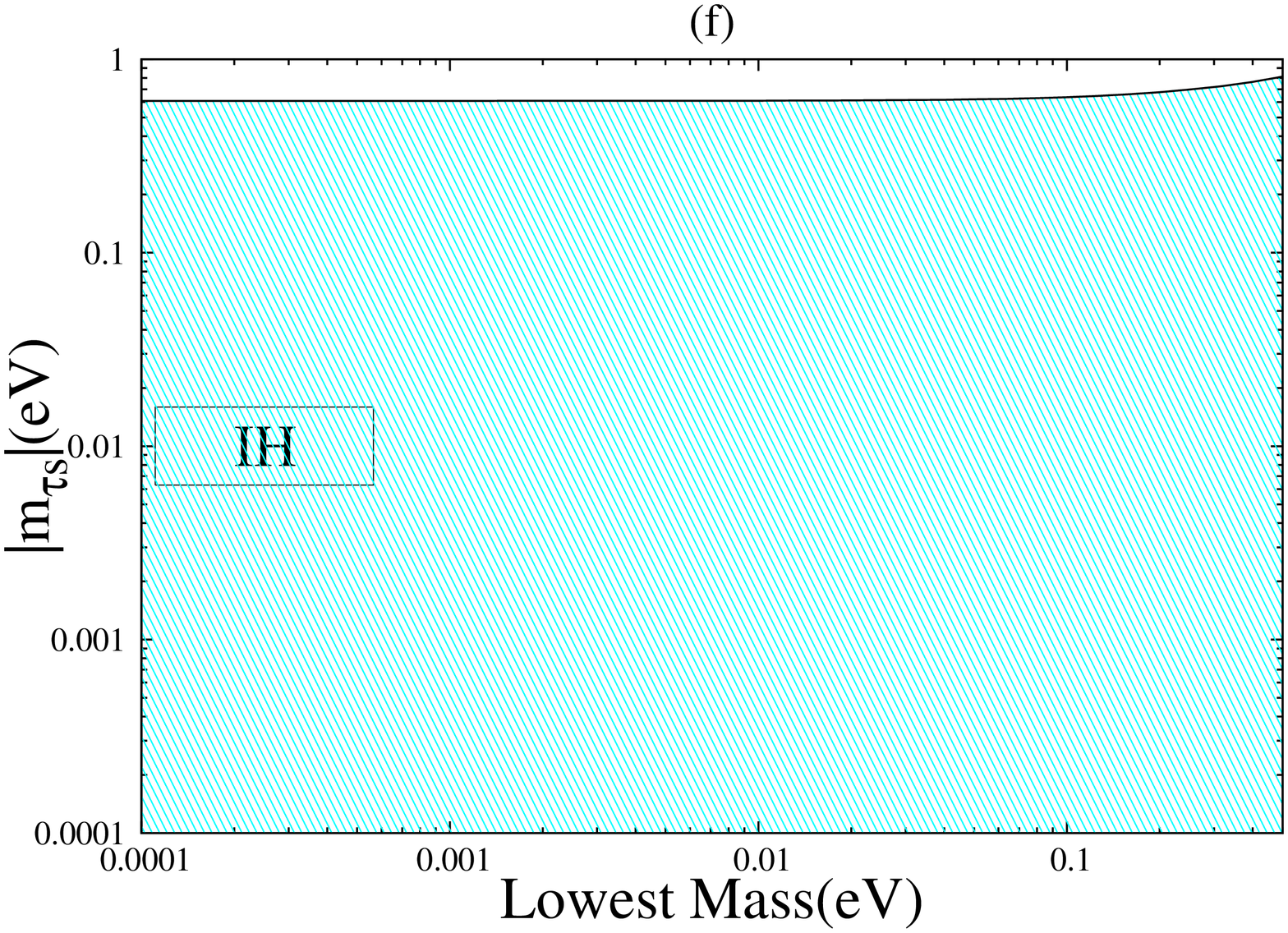} \\
\includegraphics[width=0.33\textwidth,angle=270]{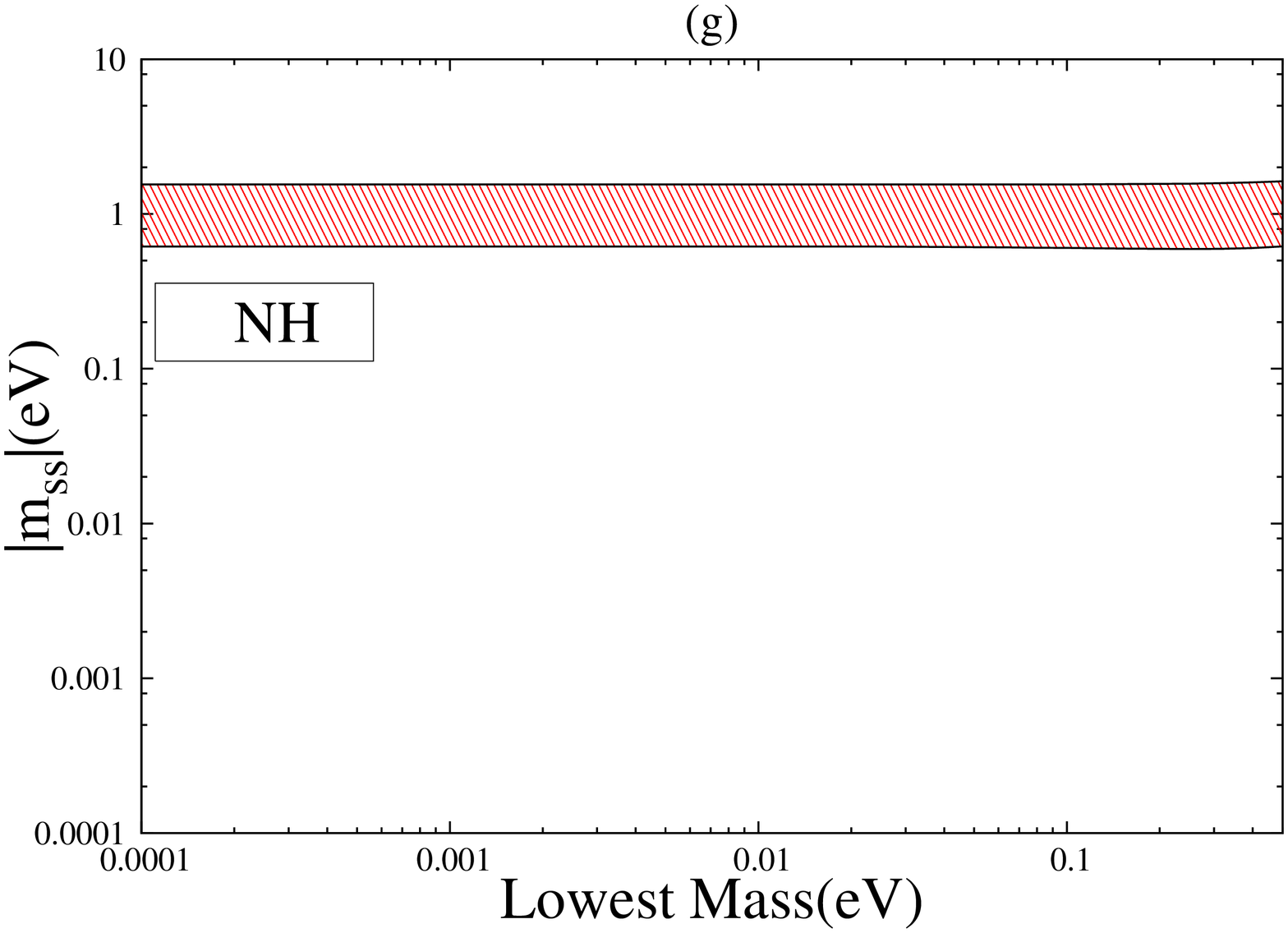}
\includegraphics[width=0.33\textwidth,angle=270]{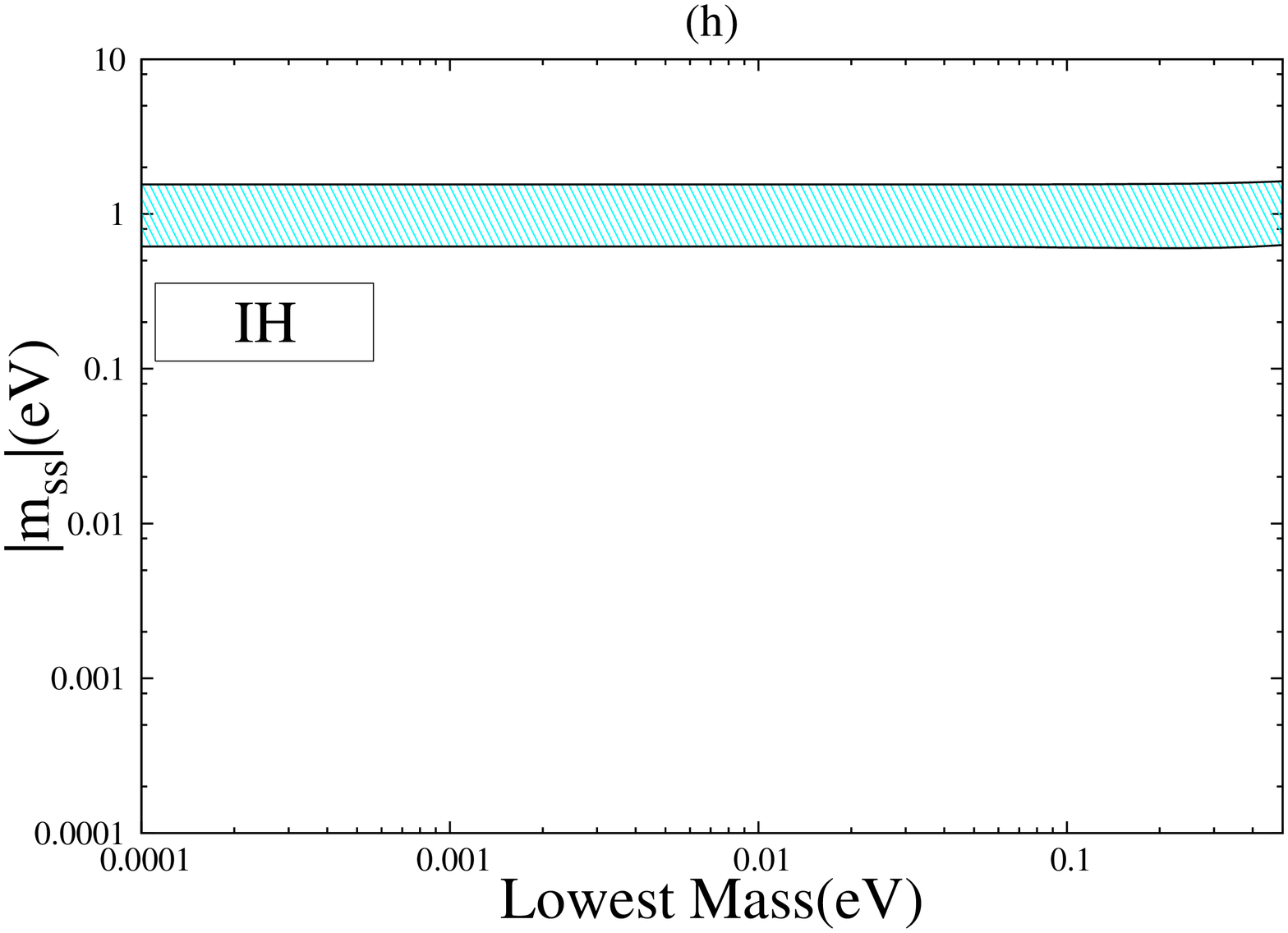}
\caption[ Plots of $|m_{ks}|$ versus the lowest mass.]{Correlation plots for vanishing $|m_{ks}|$ for both normal and inverted hierarchy.
In these plots all the mixing angles are varied in their 3 $\sigma$ allowed range, Dirac CP phases are varied from 0 to $2\pi$ and Majorana phases from 0 to $\pi$.}
\label{fig14}
\end{center}
\end{figure} 

\bea
  m_{es} &=& e^{i(2 \gamma + \delta_{14})} c_{14} c_{24} c_{34} m_4 s_{14} \\ \nonumber
  &+& e^{i(2 \beta + \delta_{13})} c_{14} m_3 s_{13}\{-e^{i(\delta_{14} - \delta_{13})} c_{24} c_{34} s_{13} s_{14} + c_{13}(-e^{ i \delta_{14}} c_{34} s_{23} s_{24} - c_{23} s_{34})\} \\ \nonumber
  &+& c_{12} c_{13} c_{14} m_1[-s_{12}(-e^{i \delta_{24}} c_{23} c_{34} s_{24} + s_{23} s_{34}) \\ \nonumber
  &+& c_{12}\{-e^{i \delta_{14}} c_{13} c_{24} c_{34} s_{14} - e^{i \delta_{13}} s_{13}(-e^{i \delta_{24}} c_{34} s_{23} s_{24} - c_{23} s_{34})\}] \\ \nonumber
  &+& e^{ 2 i \alpha} c_{13} c_{14} m_2 s_{12}[c_{12}(-e^{ i \delta_{24}} c_{23} c_{34} s_{24}+s_{23}s_{34})  \\ \nonumber
  &+& s_{12} \{- e^{i \delta_{14}} c_{13} c_{24} c_{34} s_{14} - e^{i \delta_{13}} s_{13}(-e^{i \delta_{24}} c_{34} s_{23} s_{24} - c_{23} s_{34})\}].
 \eea
 
\begin{eqnarray}
  m_{\mu s} &=& e^{i(2 \gamma + \delta_{14})} c_{14}^2 c_{24} c_{34} m_4 s_{24} \\ \nonumber
  &+&e^{i(2 \beta + \delta_{13})} m_3(c_{13} c_{24} s_{23} - e^{i(\delta_{14} - \delta_{13} - \delta_{24})} s_{13} s_{14} s_{24}) \\ \nonumber
 && \{-e^{i(\delta_{14} - \delta_{13})} c_{24} c_{34} s_{13} s_{14} + c_{13}(-e^{ i \delta_{24}} c_{34} s_{23} s_{24} - c_{23} s_{34})\} \\ \nonumber
 &+& m_1\{-c_{23} c_{24} s_{12} + c_{12}(-e^{i \delta_{13}} c_{24} s_{13} s_{23} - e^{i(\delta_{14} - \delta_{24})} c_{13} s_{14} s_{24})\} \\ \nonumber
 &&[-s_{12}(-e^{i \delta_{24}} c_{23} c_{34} s_{24} + s_{23} s_{34}) \\ \nonumber
 &+& c_{12}\{-e^{ i \delta_{14}} c_{13} c_{24} c_{34} s_{14} - e^{i \delta_{13}} s_{13}(-e^{ i \delta_{24}} c_{34} s_{23} s_{24} - c_{23} s_{34})\}] \\ \nonumber
 &+& e^{2 i \alpha} m_2 \{c_{12} c_{23} c_{24} + s_{12}(-e^{i \delta_{13}} c_{24} s_{13} s_{23} - e^{i(\delta_{14} - \delta_{24}} c_{13} s_{14} s_{24})\} \\ \nonumber
 && [c_{12}(-e^{i \delta_{24}} c_{23} c_{34} s_{24} + s_{23} s_{34}) \\ \nonumber
 &+& s_{12}\{-e^{i \delta_{14}} c_{13} c_{24} c_{34} s_{14} - e^{i \delta_{13}} s_{13}(-e^{ i \delta_{24}} c_{34} s_{23} s_{24} - c_{23} s_{34})\}].
 \end{eqnarray} 

\bea
  m_{\tau s} &=& c_{14}^2 c_{24}^2 c_{34} e^{ 2 i(\delta_{14} + \gamma)} m_4 s_{34} \\ \nonumber
  &+& e^{2 i (\beta + \delta_{13})} m_3\{-c_{24} c_{34} e^{i(\delta_{14} - \delta_{13})} s_{13} s_{14} + c_{13}(-c_{23} s_{34} - c_{34} e^{ i \delta_{24}} s_{23} s_{24})\} \\ \nonumber
  &&\{-c_{24} e^{i(\delta_{14} - \delta_{13})} s_{13} s_{14} s_{34} + c_{13}(c_{23} c_{34} - e^{i \delta_{24}} s_{23} s_{34}s_{24})\} \\ \nonumber
  &+& m_1[-s_{12}(-c_{23} c_{34} e^{i \delta_{24}} s_{24} + s_{23} s_{34}) \\ \nonumber
  &+& c_{12}\{-c_{13} c_{24} c_{34}  e^{ i \delta_{14}} s_{14} - e^{ i \delta_{13}} s_{13}(-c_{23} s_{34} - c_{34} e^{i \delta_{24}} s_{23} s_{34})\}] \\ \nonumber
  &&[-s_{12}(-c_{34} s_{23} - c_{23} e^{i \delta_{24}} s_{34}s_{24}) \\ \nonumber
  &+& c_{12}\{-c_{13} c_{24} e^{ i \delta_{14}} s_{14} s_{34} - e^{i \delta_{13}} s_{13}(c_{23} c_{34} - e^{ i \delta_{24}} s_{23} s_{34}s_{24})\}] \\ \nonumber
  &+& e^{ 2 i \alpha} m_2[c_{12}(-c_{23} c_{34} e^{i \delta_{24}} s_{34} + s_{23} s_{34}) \\ \nonumber
  &+& s_{12}\{-c_{13} c_{24} c_{34} e^{ i \delta_{14}} s_{14} - e^{ i \delta_{13}} s_{13}(-c_{23} s_{34} - c_{34} e^{ i \delta_{24}} s_{23} s_{24})\}] \\ \nonumber
  &&[c_{12}(-c_{34} s_{23} - c_{23} e^{ i \delta_{24}} s_{34}s_{24}) \\ \nonumber
  &+& s_{12}\{-c_{13} c_{24} e^{ i \delta_{14}} s_{14} s_{34} - e^{ i \delta_{13}} s_{13}(c_{23} c_{34} - e^{i \delta_{24}} s_{23} s_{34}s_{14})\}].
 \eea

Though the equations seem very complex, one can easily understand the properties of these elements by just looking at the $m_4$ terms.
The $m_4$ term in  $m_{es}$ is proportional to $s_{14}$. So in general it is quite large ($\mathcal{O}(1)$). For this
element to become negligible very small values of $s_{14}^2$ is required. But as this angle is bounded by the SBL experiments,
complete cancellations never occurs for both normal and inverted hierarchy (Panel (a), (b) of Fig. \ref{fig14}). Similar predictions are obtained for $m_{\mu s}$ element which
cannot vanish since $s_{24}^2$ has to be negligible which is not allowed by the data. This can be seen from Panel (c), (d) of Fig. \ref{fig14}.
For the element $m_{\tau s}$ the scenario is quite different.
In this case the $m_4$ term is proportional to $\theta_{34}$ and there is no lower bound on it from the SBL
experiments i.e. it can approach smaller values. As a result the term with
$m_4$ can be very small. Thus this matrix element can possibly vanish in both hierarchies for whole range of the lowest mass
(Panel (e), (f)).\\
The (4,4) element of the neutrino mass matrix is given as
\bea
 m_{ss} &=&  c_{14}^2 c_{24}^2 c_{34}^2 e^{2 i (\gamma + \delta_{14})} m_4 \\ \nonumber
 &+& e^{2 i(\beta+ \delta_{13})} m_3\{-c_{24} c_{34} e^{i(\delta_{14} - \delta_{13})} s_{13} s_{14} + c_{13}(-c_{23} s_{34} - c_{34} e^{ i \delta_{24}} s_{23} s_{24})\}^2 \\ \nonumber
 &+& m_1[-s_{12}(-c_{23} c_{34} e^{ i \delta_{24}} s_{24} + s_{23} s_{34}) \\ \nonumber
 &+& c_{12}\{-c_{13} c_{24} c_{34} e^{ i \delta_{14}} s_{14} - e^{ i \delta_{13}} s_{13}(-c_{23} s_{34} - c_{34} e^{ i \delta_{24}} s_{23} s_{24})\}]^2 \\ \nonumber
 &+& e^{ 2 i \alpha} m_2[c_{12}(-c_{23} c_{34} e^{ i \delta_{24}} s_{24} + s_{23} s_{34}) \\ \nonumber
 &+& s_{12}\{-c_{13} c_{24} c_{34} e^{ i \delta_{14}} s_{14} - e^{i \delta_{13}} s_{13}(-c_{23} s_{34} - c_{34} e^{i \delta_{24}} s_{23} s_{34})\}]^2
 \eea

The $m_4$ term for $m_{ss}$ is proportional to  $c_{14}^2 c_{24}^2 c_{34}^2$.
One can see that this term is of order one as a result this element can
never vanish as is evident from panel (g, h).

\section{Summary}
\label{sum}
 
In this chapter we have studied the structure of the low energy neutrino mass matrix in terms of texture zeros in the presence of
one light sterile neutrino.
In Section \ref{two_zero}, we have analyzed the two-zero textures in 3+1 scenario.
We find many distinctive features in this case as compared to the three
neutrino scenario. For the 3+1 case there can be 45 possible two
zero textures as opposed to 15 for the 3 generation case.
Among these 45 possible two-zero textures only 15 survive the
constraints from global oscillation data. Interestingly these
15 cases are the 15 two-zero textures that are possible
for three active neutrino mass matrices. While for the
three active neutrino case only 7 of these were allowed
addition of one sterile neutrino make all 15 cases allowed
as the sterile contribution can be instrumental for additional
cancellations leading to zeros. All the allowed textures admit NH.
IH is permissible in all the textures excepting those in the $A$ class.
The classes $B$, $C$, $F$ also allow
QD solutions.
The results are summarized in Table \ref{2results}. 

If we vary the mass and mixing parameters normally peaked at the best-fit value and 1$\sigma$ error as the width then we find solutions for smaller 
values of $m_1$(NH) and $m_3$(IH). 
In this case 
for the textures with vanishing $m_{ee}$ i.e., A class and E class
we obtain correlations between the mixing angles $\sin^2\theta_{14}$,
$\sin^2\theta_{24}$, $\sin^2\theta_{34}$
and the lowest mass scale $m_1$. For these textures the
effective mass responsible for neutrinoless double beta decay is zero.
For the other allowed textures
we present
the effective mass measured in
neutrinoless double beta decay as a function of the smallest
mass scale.
If however, the known oscillation parameters 
are varied randomly in their allowed
3$\sigma$ range  then although the main conclusions 
deduced above regarding the allowed mass spectra in various
textures remain the same  the allowed parameter space 
reduces in size. In particular,  we obtain a
bound on the lowest mass as $m_{\rm lowest} > 0.01$ eV 
and completely hierarchical neutrinos are no longer allowed.  

In Section \ref{one_zero} we have analyzed systematically the one-zero textures
of the $4\times4$ mass matrix in presence of a sterile neutrino by varying all the oscillation parameters randomly in their allowed $3\sigma$ range.
Assuming  neutrinos to be Majorana particles,
this is a symmetric matrix with 10
independent entries.
We find that $|m_{ee}| =0$ is possible for NH only for higher values
of the  smallest mass $m_1$ while for IH it is possible even for
lower values. This is in sharp contrast with the 3 generation case
where complete cancellation can never take place for IH.
$|m_{e\mu}|$ can vanish over the whole range of the smallest mass
for both 3 and 3+1 neutrino scenarios. However for larger values of
the mixing angle $s_{24}^2$, cancellation is not achieved for smaller
$m_1$ for NH.
For IH
the cancellation condition depend on the Majorana phase
$\alpha$ and the mixing angle $\theta_{24}$.
Cancellation is achieved for the element $m_{e\tau}$ for
the full range of the lowest mass in the 3+1 scenario.
The element $m_{e\mu}$ is related to the
element $m_{e\tau}$ by $\mu-\tau$ symmetry.
For small
values of $\theta_{24}$ we get $\overline\theta_{24} = \theta_{34}$
in these textures. Consequently the role played by $\theta_{24}$
for $m_{e\mu}$ is played by $\theta_{34}$ in $m_{e \tau}$ in this limit.
Thus in this case cancellation is not achieved for larger values
of $s_{34}^2$ in the hierarchical regime for NH.
For IH we obtain correlations between
$\alpha$ and $\sin^2\theta_{34}$ for fulfilling the condition
for cancellations.
The elements $m_{\mu \mu}$ and $m_{\tau \tau}$ are related by
$\mu-\tau$ symmetry.
For these cases, cancellation is not possible in the
hierarchical zone for IH, in the 3 generation case.
However the extra contribution coming from the sterile part
helps in achieving cancellation in this region.
For IH one can obtain correlations between the Majorana phase
$\alpha$ and the mixing angle $\theta_{24}$($\theta_{34}$) for
$|m_{\mu \mu}|=0$($|m_{\tau \tau}|=0$).
For $m_{\mu \tau}$ element cancellation was possible for three
generation case only for higher values of the lightest mass.
However if one includes the sterile neutrino then this element
can vanish over the whole range of the lightest neutrino mass
considered.
With the current constraints on sterile parameters it is not possible
to obtain $m_{ss}=0$ while $m_{es}$ and $m_{\mu s}$ can only vanish
in the QD regime of the active neutrinos.
However, the element $m_{\tau s}$ can  be vanishingly small
in the whole mass range.

%% file: Conclusion.tex

In this thesis, we have focused on two very important areas of current research in neutrino physics. In the first part of our work, we have studied 
the potential of the present/future generation neutrino oscillation experiments to determine the remaining unknowns of neutrino oscillation parameters namely: 
the neutrino mass hierarchy, the octant of $\theta_{23}$ and the leptonic CP phase $\dcp$.
The prospective experiments to study these parameters that have been considered in this thesis are: the accelerator based current generation long-baseline (LBL) experiments
T2K and \nova, the future generation proposed LBL  experiments LBNO and LBNE, 
the future atmospheric experiment ICAL@INO and the ongoing ultra high energy neutrino experiment IceCube. 
First we have analysed the CP sensitivity of T2K, \nova\ and ICAL.
It is found that due to the presence of parameter degeneracies, the current generation long baseline experiments T2K and \nova\
have CP sensitivity only in a limited region of the parameter space.
With the precise measurement of $\theta_{13}$, two types of degeneracies namely the hierarchy-$\dcp$ degeneracy and octant-$\dcp$ degeneracy need to be considered.
This gives rise to wrong hierarchy and wrong octant solutions.
The nature of hierarchy-$\dcp$ degeneracy is same for both neutrinos and antineutrinos but octant-$\dcp$  
degeneracy behaves differently in these two cases. Due to this reason, the wrong octant solutions can be largely resolved
by a balanced neutrino-antineutrino run of these experiments but the wrong hierarchy solutions can not be resolved by using data from T2K/\nova\ alone.
This affects the CP sensitivity since the wrong hierarchy solution gets preference over the true solution.
To solve this problem, we suggested that in this case adding data from atmospheric neutrino experiments can help.
Taking ICAL@INO as 
the representative detector, we showed that the atmospheric neutrino data can be instrumental in removing the wrong hierarchy solutions
of T2K/\nova\ giving an improved CP sensitivity for these experiments. 
We find that with the projected exposures of T2K and \nova\ i.e., an exposure corresponding to $8 \times 10^{21}$ POT for T2K and 3 years running of \nova\ in
both neutrino and antineutrino mode gives $2\sigma$ CP violation (CPV) discovery sensitivity for only 24\% true $\dcp$ values for
$\theta_{\mu\mu}^{tr} = 39^\circ$, $\sin^2 2\theta_{13} = 0.1$ and true NH.
But with the addition of 10 yr (500 kt yr) of ICAL data, $2\sigma$ CPV discovery can be achieved for $58\%$ of true $\dcp$ values.
Note that CP sensitivity of ICAL is washed out because of the angular smearing but due to large earth matter effects and 
it has good hierarchy sensitivity which is independent of $\dcp$. 
These results are very important because, if nature has chosen such unfavorable combinations of parameters, 
then it is the addition of atmospheric neutrino data to T2K+\nova\ which may give us the first signal of CP violation.
Next we have done a chronological study of $\dcp$ taking different exposures of these setups.
We found that the CP sensitivity of T2K(3+2) (which implies 3 years of running in neutrino mode and 2 years of running in antineutrino mode) 
is better than T2K(5+0) but with the addition of \nova(3+3) data,
T2K(5+0) and T2K(3+2) gives similar CP sensitivity.
We also go beyond the projected exposure of these experiments to assess the effect of enhanced statistics.
We find that, with the combination of T2K(5+5)+\nova(5+5)+ICAL(500 kt yr), it is possible to have 
$2\sigma$ ($3\sigma$) CPV discovery sensitivity for $62\%$ ($36\%$) of the total $\dcp$ values for
$\theta_{\mu\mu}^{tr} = 39^\circ$, $\sin^2 2\theta_{13} = 0.1$ and true NH. In our study 
we also show that if hierarchy is unknown then T2K(5+5) gives better CP sensitivity than \nova(5+5).
Our results show that the current best-fit of $\theta_{13}$ lies in a region where the CP sensitivity is maximum for these experiments.
We also found that for T2K(5+0)+\nova(5+5), there are wrong octant solutions in the region $40^\circ < \theta_{\mu \mu} < 49^\circ$ when hierarchy is known.
Due to addition of 10 years of ICAL data, this range is restricted to $41^\circ < \theta_{\mu \mu} < 48^\circ$.  
We note that the results discussed in this part of the thesis can be of importance and interest to other atmospheric and/or reactor experiments
sensitive to the mass hierarchy and can initiate
similar studies. These results have valuable ramifications for current experiments. 

Though the addition of ICAL data improves the CP sensitivity of T2K/\nova\ largely, from our results we see that
these current generation experiments are capable of measuring $\dcp$ only up to $2\sigma$ to $3\sigma$
with their projected exposures. The conclusions are also similar for the case of determination of hierarchy and octant.
In \cite{t2knova} it was shown that even an optimistic exposure of T2K+\nova\ can only resolve hierarchy at 90\% C.L.
In \cite{suprabhoctant} it was shown that a combination of T2K and \nova\ can resolve octant for $\sin^2\theta_{23} < 0.43$ or $> 0.58$
only at $2\sigma$. The sensitivities of T2K/\nova\ are limited because of their comparative shorter baseline ($< 1000$ km) and less statistics.
These experiments are also not sensitive to the second oscillation maximum because of their narrow band beam profile.
Thus to determine the unknown oscillation parameters at a greater confidence level, one needs to explore neutrino oscillations at longer baselines 
with large number of event samples. The proposed LBL experiments LBNO and LBNE are the examples of such initiatives. They will also have wide band flux to explore
the sensitivity coming from the second oscillation maximum. As the exact configurations of these experiments are not decided yet, a comprehensive study to estimate their 
sensitivities are extremely important. 
In our work we have studied how the
results from T2K, \nova\ and ICAL
can economise the configurations of LBNO and LBNE. We find that the
required adequate exposure of these experiments in establishing mass hierarchy/true octant at $5 \sigma$ and discovering CP violation at $3\sigma$
can be reduced significantly by using data of these three experiments.
For LBNO, among the several proposed baselines,
we have considered the following three baselines in our study: CERN-Pyh\"{a}salmi (2290 km), CERN-Slanic (1500 km) and CERN-Fr\'{e}jus (130 km).
We have shown that ICAL data
helps remarkably in reducing the required exposures of these experiments in determining hierarchy and octant for a given C.L. For the baseline 130 km, the matter effect
is not sufficient to break the degeneracy and ICAL data in this case is even more useful.
We have found that for the baselines 2290 km and 1500 km, ICAL data has no role to play in the determination of $\dcp$. This conclusion is also
same for LBNE. This is because, the matter effect in these baselines are sufficient enough to break the hierarchy-$\dcp$ degeneracy
that occurs for relatively short baseline experiments T2K/\nova\ and thus for these experiments there are no wrong hierarchy solutions in the 
unfavourable half planes. As ICAL has no CP sensitivity of its own, addition of ICAL data in this case does not play any role.  
For LBNE, we have also examined the role of a near detector, effect of the second oscillation maximum and the contribution from antineutrinos. 
We find that with the inclusion of a near detector there is a significant reduction in the systematic errors as compared to the only far detector case
and the second oscillation maximum plays a non-trivial role in the mass hierarchy sensitivity. 
The antineutrino data in LBNE helps in resolving the octant degeneracy and the combined synergy between neutrino and antineutrino runs improve the octant and CP sensitivity.
The ‘adequate’ exposures calculated in our studies can
be attained by various combinations of beam power, runtime and detector mass.
These minimal values can be used to set up the first phase of LBNO/LBNE, if an incremental/staged
approach is being followed. Our studies show that the synergies between
the existing and upcoming LBL and atmospheric experiments can play an important role
and should be taken into consideration in planning economised future facilities.

Apart from studying the physics potential of the long-baseline and atmospheric neutrinos in determining unknown oscillation parameters, in this thesis we have
also studied
how the oscillation of ultra high energy neutrinos depend on the leptonic CP phase $\dcp$.
For this we have used the recently detected PeV neutrino events at IceCube.
As mentioned Section \ref{sec4}, these high energy neutrinos can originate in extragalactic sources via one of these three processes: $\pi S$, $\mu DS$ and $nS$.  
As these extragalactic neutrinos travel a long distance in vacuum,
these oscillations can be described by the averaged vacuum oscillation and thus they provide an interesting opportunity
to measure $\dcp$ independent of hierarchy-$\dcp$ degeneracy. We find that, in this scenario the result depends strongly on the initial 
sources of these neutrinos. For $\pi S$ and $\mu DS$ source the best-fit value of $\dcp$ occurs at $0^\circ$ and for $nS$ source the best-fit is $180^\circ$. 
In this context we have also tried to constrain the astrophysical sources taking the effect of backgrounds. 
Our results are important in the sense that if $\dcp$
is measured by other LBL/atmospheric experiments, then IceCube data
can be used to determine the production mechanism of
these neutrinos. Similar analyses can also be carried out
for other parameters/scenarios related to neutrino physics and astrophysics.

In the second part of our work we have studied the neutrino mass matrix in terms of texture zeros in presence of one extra light sterile neutrino (3+1 scenario). 
Texture zeros constitute an important aspect in the study of the neutrino mass matrix. In general the low energy neutrino mass matrix depends on 
oscillation parameters and masses. As all of them have not been measured in the experiments, it is impossible to fully determine the neutrino mass matrix.
Texture zeros imply that some elements of the neutrino mass matrix are smaller than the other elements. In our work, we have studied two-zero and one-zero textures in 3+1
scenario and compared our results with that of the three generation case. For three generation it is well known that there are 15 possible two-zero textures and among
them only 7 are phenomenologically allowed. But in 3+1 case, there are 45 possibilities. We find that among 45 only 15 are allowed and remarkably these 15 correspond to the 
possible 15 two-zero texture cases that arise for the low energy neutrino mass matrix in three generation. 
So the disallowed textures in the 3 generation case become allowed by including an extra sterile neutrino. 
For one-zero textures
we find that the correlations among the various oscillation parameters are different as compared to the 3 generation case.
The most remarkable contrast with the 3 generation results is the prediction for the element $m_{ee}$ which is the effective mass
governing neutrinoless double beta decay ($0 \nu \beta \beta$).
At present there are bounds on this element from various ongoing experiments. Future experiments can lower this bound and probe the IH region.
For three generation $m_{ee}$ could vanish only in NH but in 3+1 case it can vanish for both NH and IH. 
Thus if the existence of sterile neutrinos is confirmed by future experiments then
it may be difficult to probe the hierarchy from $0\nu\beta\beta$ alone.
These results  can be useful in probing underlying flavour symmetries and
also for  obtaining textures of Yukawa matrices
in presence of light sterile neutrinos. 
These are also useful for building models for light
sterile neutrinos and shed light on the underlying new physics
if future experiments and analysis reconfirm the explanation of
the present anomalies in terms of  sterile neutrinos.

%% file: appendix1.tex

In this appendix we will provide the expressions for oscillation probabilities for three generations of neutrinos. 
In view of the current experimental scenario, the most important channels are : the electron disappearance channel $P_{ee}$, the muon appearance channel $P_{e \mu}$
and muon disappearance channel $P_{\mu \mu}$. The accelerator based long-baseline experiments will mainly use the $P_{\mu e}$\footnote{$P_{\mu e}$ is the
time reversal state of $P_{e \mu}$ and can be obtained by replacing $\dcp$ by $-\dcp$.} channel for determining
hierarchy, octant and $\dcp$ and the $P_{\mu \mu}$ channel for the precision measurement of $\theta_{23}$ and $|\Delta_{31}|$. For atmospheric neutrino experiments (like ICAL),
the  main sensitivity comes form $P_{\mu \mu}$ channel whereas the rector experiments are sensitive to the $P_{ee}$ channel.
In the following subsections, first we will give the expressions for the above mentioned channels for OMSD approximation. 
Next we will give the corresponding expressions for the
$\alpha-s_{13}$ approximation. 
Then we write down
the same set of expressions for vacuum by putting the matter term zero in the $\alpha-s_{13}$ approximation. 

\section{Expressions for OMSD Approximation}

In the OMSD approximation, the smaller mass squared difference $\Delta_{21}$ is neglected as compared to $\Delta_{31}$.
Under this approximation, the effect of $\theta_{12}$ and $\dcp$ also becomes inconsequential and one obtains:

\begin{eqnarray}
 P_{ee} &=& 1 - \sin^22\theta_{13}^M \sin^2\left(\frac{\Delta_{31}^M L}{4E}\right), \\
 P_{\mu e} &=& \sin^2\theta_{23} \sin^22\theta_{13}^M \sin^2\left(\frac{\Delta_{31}^M L}{4E}\right), \\
 P_{\mu \mu} &=& 1 - \cos^2\theta_{13}^M \sin^22\theta_{23} \sin^2\left(\frac{\Delta_{31} + A + \Delta_{31}^M}{4E}\right)L \\ \nonumber
             && ~ - \sin^2\theta_{13}^M \sin^2\theta_{23} \sin^2\left(\frac{\Delta_{31} + A - \Delta_{31}^M}{4E}\right)L \\ \nonumber
             && ~ - \sin^4\theta_{23} \sin^22\theta_{13}^M \sin^2\left(\frac{\Delta_{31}^M L}{4E}\right),
\end{eqnarray}

where 
\begin{eqnarray}
 \tan2\theta_{13}^M = \frac{\Delta_{31} \sin2\theta_{13}}{\Delta_{31} \cos2\theta_{13} - A},
\end{eqnarray}

and
\begin{eqnarray}
\Delta_{31}^M = \sqrt{(\Delta_{31} \cos2\theta_{13} - A)^2 + (\Delta_{31} \sin2\theta_{13})^2},
\end{eqnarray}

with 
\begin{eqnarray}
 A = 2\sqrt{2} G_F N_e E.
\end{eqnarray}

\section{Expressions for $\alpha-s_{13}$ Approximation}

In the $\alpha-s_{13}$ approximation, the diagonalisation of the Hamiltonian is performed 
in the mass hierarchy parameter $\alpha = \frac{\Delta_{21}}{\Delta_{31}}$ and $s_{13}=\sin\theta_{13}$
up to their second order and one obtains:

\begin{eqnarray}
 P_{ee} &=& 1 - \alpha^2 \sin^22\theta_{12} \frac{\sin^2\hat{A} \Delta}{\hat{A}^2} - 4 s_{13}^2 \frac{\sin^2(\hat{A}-1)\Delta}{(\hat{A} - 1)^2}, \\
 P_{\mu e} &=& \alpha^2 \sin^22\theta_{12} c_{23}^2 \frac{\sin^2\hat{A}\Delta}{\hat{A}^2} + 4 s_{13}^2 s_{23}^2 \frac{\sin^2(\hat{A} - 1) \Delta}{(\hat{A} - 1)^2} \\ \nonumber
           && ~ + 2 \alpha s_{13} \sin2\theta_{12} \sin2\theta_{23} \cos(\Delta + \delta_{CP}) \frac{\sin\hat{A}\Delta}{\hat{A}} \frac{\sin(\hat{A} - 1)\Delta}{\hat{A} - 1}, \\
 P_{\mu \mu} &=& 1 - \sin^22\theta_{23} \sin^2{\Delta} + \alpha c_{12}^2 \sin^22\theta_{23} \Delta \sin2\Delta \\ \nonumber
              && ~ - \alpha^2 \sin^22\theta_{12}c_{23}^2 \frac{\sin^2\hat{A} \Delta}{\hat{A}^2} - \alpha^2 c_{12}^2 \sin^22\theta_{23} \Delta^2 \cos2\Delta \\ \nonumber 
              && ~ + \frac{1}{2\hat{A}} \alpha^2 \sin^22\theta_{12} \sin^22\theta_{23} \\ \nonumber
              && ~ \times \big(\sin\Delta \frac{\sin\hat{A}\Delta}{\hat{A}} \cos(\hat{A} - 1)\Delta - \frac{\Delta}{2} \sin2\Delta \big) \\ \nonumber
              && ~ - 4 s_{13}^2 s_{23}^2 \frac{\sin^2(\hat{A} - 1)\Delta}{(\hat{A} - 1)^2} - \frac{2}{\hat{A} - 1} s_{13}^2 \sin^22\theta_{23} \\ \nonumber
              && ~ \times \big(\sin\Delta \cos\hat{A}\Delta \frac{\sin(\hat{A} - 1)\Delta}{\hat{A} - 1} - \frac{\hat{A}}{2} \Delta \sin2\Delta \big) \\ \nonumber
              && ~ - 2 \alpha s_{13} \sin2\theta_{12} \sin2\theta_{23} \cos\delta_{CP} \cos\Delta \frac{\sin\hat{A}\Delta}{\hat{A}} \frac{\sin(\hat{A} - 1)\Delta}{\hat{A} - 1} \\ \nonumber
              && ~ + \frac{2}{\hat{A} - 1} \alpha s_{13} \sin2\theta_{12} \sin2\theta_{23} \cos2\theta_{23} \cos\delta_{CP} \sin\Delta \\ \nonumber
              && ~ \times \big(\hat{A} \sin\Delta - \frac{\sin\hat{A}\Delta}{\hat{A}} \cos(\hat{A} - 1) \Delta \big),
\end{eqnarray}
with $\alpha = \Delta_{21}/\Delta_{31}$, $\hat{A} = A/\Delta_{31}$, $A= 2\sqrt{2} G_F N_e E$ and $\Delta = \Delta_{31}/4E$.
Where  $\Delta_{ij} = m_i^2 - m_j^2$, $G_F$ is the Fermi constant, $N_e$ is the electron number density of the medium and $E$ is the energy of the neutrinos.

\section{Expressions in Vacuum}

The vacuum oscillation probabilities up to second order in $\alpha$ and $s_{13}$ can be readily derived with the approximation $\hat{A} \rightarrow 0$ in the 
above set of equation.
\begin{eqnarray}
 P_{ee} &=& 1 - \alpha^2 \sin^22\theta_{12} \Delta^2 - 4 s_{13}^2 \sin^2 \Delta, \\
 P_{\mu e} &=& \alpha^2 \sin^22\theta_{12} c_{23}^2 \Delta^2 + 4 s_{13}^2 s_{23}^2 \sin^2\Delta \\ \nonumber
 && ~ + 2 \alpha s_{13} \sin2\theta{12} \sin2\theta_{23} \cos(\Delta + \delta_{CP}) \Delta \sin\Delta, \\  
 P_{\mu \mu} &=& 1 - \sin^22\theta_{23} \sin^2\Delta + \alpha c_{12}^2 \sin^2\theta_{23} \Delta \sin2\Delta \\ \nonumber
             && ~ - \alpha^2 \Delta^2 \big[\sin^22\theta_{12} c_{23}^2 + c_{12}^2 \sin^22\theta_{23} \big(\cos2\Delta - s_{12}^2 \big) \big] \\ \nonumber
             && ~ + 4 s_{13}^2 s_{23}^2 \cos2\theta_{23} \sin^2\Delta \\ \nonumber
             && ~ - 2 \alpha s_{13} \sin2\theta_{12} s_{23}^2 \sin2\theta_{23} \cos\delta_{CP} \Delta \sin2\Delta,
\end{eqnarray}

The expressions corresponding to both the OMSD and the $\alpha-s_{13}$ approximations are derived in matter of constant density. 

%% file: appendix2.tex

In this appendix we will discuss the method of $\chi^2$ analysis in detail. Let us first consider a superbeam experiment where
muon neutrinos are produced at the source.
In the detector one gets electron and muon events\footnote{These muon neutrinos will also oscillate to tau neutrinos but
the current superbeam experiments are not sensitive to them.} governed by the neutrino oscillation probability $P_{\mu e}$ and survival probability $P_{\mu \mu}$.   
The number of oscillated electron events at the detector contain the contribution of the appearance channel
$P_{\mu e}(x)$ which is a function of oscillation parameters $x$.
If we will denote the $\nu_\mu$ flux at the detector as $\Phi_\mu$, then the number of electron neutrinos at the detector can be written as
\begin{eqnarray}
 N_e = \Phi_\mu P_{\mu e}(x) \sigma_e \epsilon,
\end{eqnarray}
where $\sigma_e$ is the cross section of the electron neutrinos and $\epsilon$ is the detector efficiency.
Let us discuss how are electron events are generated on the detector.
In the detector, neutrinos undergo charge current interactions to produce electrons.
\begin{eqnarray} \nonumber
\nu_e + n \rightarrow p + e^-.
\end{eqnarray}
The signal produced by that electron is considered as the signature of the detected neutrino and this is called an `event'.
In this case the true energy of the neutrinos $E_t$ is reconstructed from the measured energy $E_m$ and the expression for total number of events (which is a function of 
$E_m$) can be obtained by 
\begin{eqnarray}
 N_e(E_m) = \int \Phi_\mu(E_t) P_{\mu e}(x, E_t) R(E_t, E_m) \sigma_e(E_t) \epsilon dE_t,
\end{eqnarray}
where $R(E_t, E_m)$ is the Gaussian resolution (smearing) function, which gives the spread of $E_m$ for a given $E_t$ and can be written as
\begin{eqnarray}
 R(E_t, E_m)  = C_1 {\rm exp}\bigg[-\frac{(E_t - E_m)^2}{2 \sigma_E^2}\bigg],
\end{eqnarray}
with normalisation coefficient $C_1$ and 
\begin{eqnarray}
 \sigma_E = \alpha E + \beta \sqrt{E} + \gamma.
\end{eqnarray}
The values of $\alpha$, $\beta$ and $\gamma$ are different for different detectors. To incorporate the energy dependence of the oscillation probability,
 we divide the total measured energy range into several bins and the events in the $i^{th}$ bin is given by
 \begin{eqnarray}
 (N_e)_i = \int \int \Phi_\mu(E_t) P_{\mu e}(x, E_t) R(E_t, E_m) \sigma_e(E_t) \epsilon dE_t dE_m.
\end{eqnarray}
The above expression is for the superbeam experiments. Now let us briefly discuss about the atmospheric experiments.
As atmospheric neutrino flux consists of both $\nu_e$ and $\nu_\mu$, the electron events at the detector get contribution from both $P_{\mu e}$ and $P_{ee}$ channels. 
In this case
the flux and the oscillation probabilities are also functions of
the direction of the incoming neutrinos i.e., the zenith angle $\theta$. Thus the events rates for atmospheric neutrinos in $i^{th}$ energy bin and $j^{th}$ zenith angle bin
can be expressed as
\begin{eqnarray}\nonumber
 (N_e)_{ij} &=& \int \int \int \int \left[\Phi_\mu(E_t, \Omega_t) P_{\mu e}(x, E_t,\Omega_t) + \Phi_e(E_t, \Omega_t) P_{e e}(x, E_t,\Omega_t)\right]  R(E_t, E_m) \nonumber \\
        &&~~~~~~~~~~~~ \times R(\Omega_t, \Omega_m) \sigma_e(E_t) \epsilon dE_t dE_m d\Omega_t d\Omega_m, 
\end{eqnarray}
where $\Phi_e$ is the $\nu_e$ flux at the detector. The angular smearing function $R(\Omega_t, \Omega_m)$ given by,
\begin{eqnarray}
R(\Omega_t, \Omega_m) = C_2 {\rm exp}\bigg[\frac{(\theta_t - \theta_m)^2 + \sin^2\theta_t(\phi_t - \phi_m)^2}{2 \sigma_\Omega^2}\bigg],
\end{eqnarray}
where $C_2$ is the normalisation constant and $\phi$ is the azimuthal angle. 

After calculating the event rates, the statistical $\chi^2$ for each energy bin (for the case of accelerator experiment) can be obtained by either using the Gaussian formula i.e.,
\begin{eqnarray}
 \chi^2_{{\rm stat}} = \sum_i \frac{((N_e)_i^{{\rm th}} - (N_e)_i^{{\rm exp}})^2}{(N_e)_i^{{\rm exp}}},
\end{eqnarray}
or by using the Poisson formula \footnote{If the events per bin are very small ($<5$) then one should use the Poisson formula.} i.e.,
\begin{eqnarray}
 \chi^2_{{\rm stat}} = \sum_i 2 \bigg[(N_e)_i^{{\rm th}} - (N_e)_i^{{\rm exp}} - (N_e)_i^{{\rm exp}} \log\bigg(\frac{(N_e)_i^{{\rm th}}}{(N_e)_i^{{\rm exp}}}\bigg) \bigg].
\end{eqnarray}

After calculating the statistical $\chi^2$, we need to incorporate the systematic errors in the experiments which can arise from the uncertainties in the
neutrino flux, cross-sections, direction of the neutrino etc. In our analysis
we have included the effect of systematics by the method of pull \cite{pulls_gg,pull_lisi} in the following way. Suppose we want to incorporate a 5$\%$ overall
normalisation error in our analysis. For this we modify the events $N_i^{{\rm th}}$ as (in the linearised approximation)
\begin{eqnarray}
 N_i^{\rm th} \rightarrow  N_i^{\rm th}(1+ 0.05 \xi),
 \label{modi}
\end{eqnarray}
where $\xi$ is called the pull variable. After modifying the events, the combined statistical and systematic $\chi^2$ is given by
\begin{eqnarray}
\chi^2_{\rm stat+sys} = \chi^2_{{\rm stat}} + \xi^2,
 \label{sys}
\end{eqnarray}
where the $\xi^2$ term is due to penalty for deviating $N_i^{{\rm th}}$ from its mean value. Up to now we have implemented only one source of systematics. For $k$ number of
sources of systematics, Eqn. \ref{modi} and \ref{sys} becomes
\begin{eqnarray}
 N_i^{{\rm th}} &\rightarrow&  N_i^{\rm th}\bigg(1+ \sum_k c_i^k \xi_k \bigg), \\
  \chi^2_{{\rm stat+sys}} &=& \chi^2_{{\rm stat}} + \sum_k \xi_k^2.
\end{eqnarray}
Finally $\chi^2_{{\rm pull}}$ is obtained by varying $\xi_k$ from $-3$ to $+3$, corresponding to their $3\sigma$ ranges and minimising over $\xi_k$
\begin{eqnarray}
 \chi^2_{{\rm pull}} = {\rm min}\{\xi_k\} \big[\chi^2_{\rm stat+sys}\big].
\end{eqnarray}
Now let us discuss the marginalisation procedure. The $\chi^2_{{\rm pull}}$ defined above is for a particular set of oscillation parameters in $N_i^{{\rm th}}$ and $N_i^{{\rm exp}}$.
To incorporate the uncertainties in the
oscillation parameters in our analysis, 
we calculate $\chi^2$ by varying all the oscillation parameters in their $3\sigma$ range in the test
events and then minimise the resultant $\chi^2$ with respect to this test parameter set: 
\begin{equation}
 \chi^2 = {\rm min}\{x_{\rm th}\} \big[\chi^2_{\rm pull} + \chi^2_{ \rm prior}\big], 
\end{equation}
where $\chi^2_{\rm prior}$ is the penalty for deviating the oscillation parameters from their true values. In principle, we should add a prior term for each oscillation parameters.
But in our analysis we have added prior only for $\theta_{13}$ in the form
\begin{eqnarray}
  \chi^2_{{\rm prior}} = 
 \left(\frac{{\sin^2 2\theta_{13}^{{\rm exp}}} - \sin^2 2\theta_{13}^{\rm th}}
 {\sigma(\sin^2 2\theta_{13})}\right)^2.
\end{eqnarray}

In our analysis of the long-baseline and atmospheric neutrino experiments, 
we have not varied the solar parameters as they are already measured very precisely. For the atmospheric parameters we did not add any prior 
because, in all our analysis we have considered measurements from disappearance channel and disappearance channel is itself capable of measuring the atmospheric parameters
$\theta_{23}$ and $\Delta_{31}$\footnote{However for the our analysis of the ultra high energy neutrinos we have taken priors for all the three mixing angles.}. 

To determine the combined sensitivity of different experiments, $\chi^2_{{\rm pull}}$ for each experiment is calculated corresponding to a given test parameter value.
Then the resultant $\chi^2_{{\rm pull}}$'s are added and minimised over the test parameter set after adding the priors.

%% file: appendix3.tex

In this appendix we explain the extractions of the sterile mixing parameters from the global fit results of the short-baseline (SBL)
neutrino oscillation experiments.
In the 3+1 scenario (three active neutrinos, one sterile neutrino), the mixing matrix $U$ can be parametrised as 
\begin{equation}
U={R_{34}}\tilde R_{24}\tilde R_{14}R_{23}\tilde R_{13}R_{12},
\end{equation}
where $R_{ij}$ denotes rotation matrices in the \textit{ij} generation space
and is expressed as,
\begin{center}
$R_{34}$=$\left(
\begin{array}{cccc}
1~ &~0 & 0 & 0 \\  0~ &~ 1 & 0 & 0 \\ 0~ & ~0 & c_{34}& s_{34} \\0 ~& ~0 & -s_{34} & c_{34}
\end{array}
\right)$ , $\tilde R_{14}$=$\left(
\begin{array}{cccc}
c_{14}~ & ~0 &~ ~0 &~s_{14}e^{-i \delta_{14}} \\ 0 ~ & ~ 1&~~ 0 & 0 \\ 0 ~& ~0 &~~ 1 & 0 \\-s_{14}e^{i \delta_{14}}  & ~ 0& ~~0 &c_{14}
\end{array}
\right)$ \\
\end{center}
with $s_{ij}=\sin\theta_{ij}$ and $c_{ij}=\cos\theta_{ij}$. In this case there are three new mixing angles $\theta_{14}$, $\theta_{24}$ and $\theta_{34}$
and three Dirac type phases $\delta_{13}$, $\delta_{14}$ and $\delta_{24}$
which describe the oscillations of the sterile neutrinos. The explicit forms of the $U_{\alpha4}$ ($\alpha=e$, $\mu$ and $\tau$) elements under this parametrisation are given by
\begin{eqnarray}
 U_{e4} &=& e^{−i \delta_{14}} s_{14}, \\
U_{\mu 4} &=& e^{−i \delta_{24}} c_{14} s_{24}, \\
U_{\tau 4} &=& c_{14} c_{24} s_{34}. 
\end{eqnarray}

In 3+1 scheme the effective transition and survival probabilities relevant for the SBL experiments are given by
\begin{eqnarray}
 P_{\alpha \beta} = \sin^22\theta_{\alpha \beta} \sin^2\left(\frac{\Delta_{41}L} {4E}\right), \\
 P_{\alpha \alpha} = \sin^22\theta_{\alpha \alpha} \sin^2\left(\frac{\Delta_{41}L} {4E}\right),
\end{eqnarray}
where $\Delta_{41} = m_4^2 - m_1^2$, with
\begin{eqnarray}
 \sin^22\theta_{\alpha \beta} &=& 4 |U_{\alpha 4}|^2 |U_{\beta 4}|^2, \\
 \sin^22\theta_{\alpha \alpha} &=& 4 |U_{\alpha 4}|^2 \left(1 - |U_{\alpha 4}|^2\right). 
\end{eqnarray}
The reactor experiments Bugey \cite{Declais:1994su}, ROVNO \cite{Kuvshinnikov:1990ry} and Gosgen \cite{Zacek:1985kr}
give bound in the $\sin^22\theta_{ee} - \Delta_{41}$ plane via electron disappearance channel measurement. 
The bound for $\sin^22\theta_{\mu\mu}$ and $\Delta_{41}$
comes from the $\nu_\mu$ disappearance channel measurements of CDHSW \cite{Dydak:1983zq} and the atmospheric neutrinos \cite{Giunti:2011gz}.
The experiments LSND \cite{Aguilar:2001ty}, MiniBooNE \cite{Aguilar-Arevalo:2013pmq}, NOMAD \cite{Astier:2003gs} and 
KARMEN \cite{Armbruster:2002mp} give constraints on $\sin^22\theta_{\mu e}$ and $\Delta_{41}$ from the $\nu_e$ appearance channel.

The global analysis of the data from all these experiments give constraints on $|U_{e4}|^2$ and $|U_{\mu 4}|^2$ \cite{Giunti:2011gz} from which one can calculate 
$\theta_{14}$ and $\theta_{24}$. The mixing angle $\theta_{34}$ is constrained from the MINOS neutral current (NC) data \cite{Adamson:2011ku}.
It is important to note that the SBL data do not give any constraint on the phases.
